\documentclass[aps,a4paper,showkeys,nofootinbib,longbibliography,notitlepage,twocolumn,superscriptaddress]{revtex4-2}
\usepackage[utf8]{inputenc}
\usepackage[T1]{fontenc}
\usepackage{lmodern}
\usepackage{microtype}
\usepackage{filecontents}

\usepackage{natbib}

\usepackage[usenames,svgnames]{xcolor}
\usepackage{natbib}
\usepackage[hyperindex,breaklinks]{hyperref}
\hypersetup{
     colorlinks=true,       		
     linkcolor=Navy,          	
     citecolor=Navy,            
     filecolor=Navy,      		
     urlcolor=Navy,           	
    runcolor=cyan,
 }
\newcommand{\bra}[1]{\langle #1 |}
\newcommand{\ket}[1]{| #1 \rangle}

\newcommand{\half}[1]{{ \rm h}}

\usepackage{graphicx}
\usepackage{amsfonts}
\usepackage{amssymb}
\usepackage{amsmath}
\usepackage{enumitem}
\usepackage{booktabs}
\usepackage{placeins}

\def\beq{\begin{equation}}
\def\eeq{\end{equation}}
\def\nbeq{\begin{equation*}}
\def\neeq{\end{equation*}}
\def\<{\langle}
\def\>{\rangle}

\usepackage{mathtools}
\def\multiset#1#2{\ensuremath{\left(\kern-.3em\left(\genfrac{}{}{0pt}{}{#1}{#2}\right)\kern-.3em\right)}}

\graphicspath{{figures/}{./}}
\begin{document}

\title{Optimal low-rank compression of quantum dynamics}

\author{Hugo Mackay}
\affiliation{Department of Physics, Harvard University, Cambridge, MA 02138, USA}
\affiliation{Analytical Quantum Complexity RIKEN Hakubi Research Team, RIKEN Center for Quantum Computing (RQC), Wako, Saitama 351-0198, Japan}

\author{Donghoon Kim}
\affiliation{Analytical Quantum Complexity RIKEN Hakubi Research Team, RIKEN Center for Quantum Computing (RQC), Wako, Saitama 351-0198, Japan}

\author{Tan Van Vu}
\affiliation{Yukawa Institute for Theoretical Physics, Kyoto University, Kitashirakawa Oiwakecho, Sakyo-ku, Kyoto 606-8502, Japan}

\author{Tomotaka Kuwahara}
\affiliation{Analytical Quantum Complexity RIKEN Hakubi Research Team, RIKEN Center for Quantum Computing (RQC), Wako, Saitama 351-0198, Japan}
\affiliation{RIKEN Pioneering Research Institute (PRI), Wako, Saitama 351-0198, Japan}

\begin{abstract}
Local quantum interactions generate dynamics in an exponentially large Hilbert space, yet locality and entanglement restrict the information that can spread and accumulate. Bounds on propagation and entanglement growth, together with tensor networks, exploit these restrictions to discard information unnecessary for describing the evolution. This leads to a sharper question: how much information must any low-rank representation retain? Here we determine these limits, up to logarithmic factors, for short-range interactions. For time-independent evolution, the optimal rank obeys $\log D=\widetilde O(t+\sqrt{\log(1/\epsilon)})$, with matching lower bounds fixing the accuracy exponent $1/2$; for arbitrary driving, matching fixed-time bounds instead give $2/3$. Correspondingly, the corresponding dynamical entanglement spectra exhibit distinct small-$\alpha$ R\'enyi laws, $\alpha^{-1}$ and $\alpha^{-2}$. In one dimension, the static limit is constructively attained by an explicit MPO algorithm, with an analogous extension to Liouvillian dynamics. Together, these results determine the irreducible information required to represent local quantum evolution and uncover distinct entanglement structures in static and driven dynamics.
\end{abstract}

\maketitle

\section{Introduction}
\label{sec:main-introduction}

A central question in quantum many-body physics is how complex the quantum structures generated by simple local interactions can be. Quantum dynamics provides a direct setting for this question. It is also closely related to the continuous-time counterpart of the quantum circuit model: both build complex states and operations by combining local quantum operations~\cite{Lloyd1996,HHKL2021}. For this reason, information propagation, entanglement generation, quantum simulation, and circuit complexity have long been studied within the shared task of understanding the structure of quantum dynamics~\cite{0034-4885-75-2-022001,Lewis-Swan2019,PAECKEL2019167998,doi:10.1142/S0217979222300079,Chen_2023}. A basic question running through these areas is what structure local time evolution actually creates in a vast Hilbert space, and how much information is truly needed to describe it.

Previous work has developed several principles for extracting only the information needed to describe quantum dynamics. Lieb--Robinson bounds restrict the influence of a local perturbation to a finite effective region~\cite{ref:LR-bound72,PhysRevLett.97.050401,NachtergaeleSimsYoung2019}. They allow distant degrees of freedom to be removed from the description of the evolution with a controlled error~\cite{PhysRevLett.97.157202,PRXQuantum.2.040331}. Bounds on entanglement-generation rates, such as small-incremental-entangling~\cite{PhysRevA.76.052319,PhysRevLett.111.170501,10.1063/1.4901039}, constrain the quantum correlations created across a boundary and quantify how quickly nonlocal information can accumulate. Through tensor-network methods, these constraints on locality and entanglement have also led to low-rank representations that efficiently describe time evolution using a small number of effective degrees of freedom~\cite{PhysRevLett.93.040502,PhysRevLett.93.076401,SchuchEtAl2008Dynamics,SCHOLLWOCK201196,PAECKEL2019167998}. Although these approaches grew from different questions, they share a common aim: \textit{to extract, from an exponentially large quantum state space, only the information needed to reproduce the dynamics.} Taking this goal to its limit leads to a natural question: what is the minimum complexity needed to represent quantum dynamics at a given time and accuracy? The goal is therefore not only to show that compression is possible, but to identify how far it can go and where further compression becomes impossible in principle.

\begin{figure*}[!t]
  \centering
  \includegraphics[width=1\textwidth]{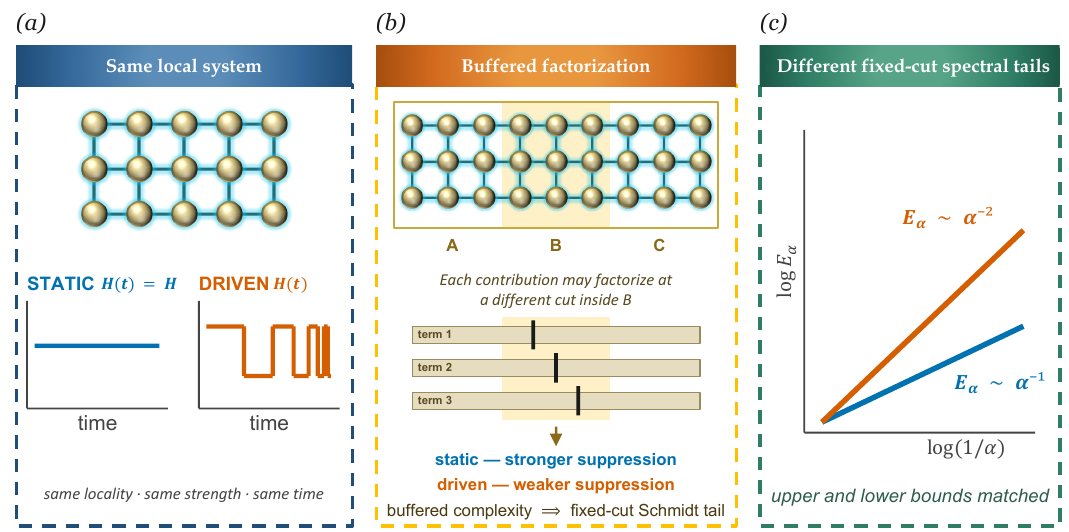}
\caption{Arbitrary temporal control can change the extremal compression law.
(a) We compare time-independent and arbitrarily driven evolution under the same spatial locality, local interaction strength, and total evolution time.
(b) Keeping a finite buffer \(B\) between endpoint regions \(A\) and \(C\) leads to the buffered Schmidt rank (BSR), which is the minimum number of product terms when different terms may factorize across different cuts within the buffer $B$.
Our bounds show stronger suppression of this buffered complexity for static dynamics than in the worst case allowed by arbitrary driving; converting the BSR bounds to a prescribed cut controls the corresponding Schmidt-spectrum tail.
(c) At fixed time and boundary size, before finite-size saturation, the resulting extremal small-\(\alpha\) R\'enyi scaling is \(E_\alpha=\widetilde{\Theta}(\alpha^{-1})\) for time-independent dynamics and \(E_\alpha=\widetilde{\Theta}(\alpha^{-2})\) for the arbitrarily driven class.
Here, \(E_\alpha\) is the R\'enyi entanglement entropy across a fixed cut of a pure state evolved from a product input across that cut.
 The characteristic powers are matched by upper and lower bounds up to logarithmic factors (Supplementary Theorems~6 and 9; Supplementary Lemma~39). The matching constructions use exponentially-decaying interactions.}
  \label{fig:Overview}
\end{figure*}

The sharp compression boundary for local dynamics is directly linked to the extremal structure of the entanglement that time evolution can generate~\cite{Calabrese_2005}. 
Let \(D_\epsilon\) be the minimum low-rank complexity needed to represent the evolution to accuracy \(\epsilon\). Determining its dependence on \(\epsilon\) sets resource limits for tensor-network representations and, when the evolution acts on an initial state, directly constrains the tail of the generated Schmidt coefficients~\cite{PhysRevB.73.094423,PhysRevLett.100.030504}. The full accuracy dependence therefore captures the fine structure of the dynamical entanglement spectrum, including the small Schmidt coefficients probed by low-order R\'enyi entropies~\cite{PhysRevA.109.042404,ycdh-z8zf,KimEtAl2025SpectralSIE}.  Low-rank compression and entanglement-spectrum tails thus provide two descriptions of the same boundary: how little information must be retained, and how much entanglement structure local dynamics can generate.

A basic question along this sharp dynamical entanglement boundary is whether it depends on the temporal structure of the generator, such as a Hamiltonian or a Liouvillian (Fig.~\ref{fig:Overview} a). In detail, can temporal structure alone change the optimal law while all other local conditions are kept the same? Lieb--Robinson bounds and entanglement-rate bounds are formulated in terms of instantaneous local interaction data and therefore do not introduce a qualitative distinction between fixed and time-dependent generators~\cite{NachtergaeleSimsYoung2019,PhysRevLett.111.170501}. This lack of distinction is also suggested at a formal level by constructions that embed time-dependent evolution into time-independent dynamics on a larger system~\cite{MooneyEtAl2025}. 
Such embeddings enlarge the local auxiliary space in a way that depends on the simulation resources, and therefore do not remain within the fixed-local-dimension class considered here.
Using a quantum computer, time-dependent quantum simulation matches the static query scaling in time and accuracy up to logarithmic factors~\cite{LowWiebe2018,KieferovaSchererBerry2019,ChenGaoWangZhou2026}.
It has therefore remained unclear whether the same optimal law survives without this assumption. The question is not merely whether an existing method can be extended to time-dependent dynamics, but whether temporal structure itself changes the optimal complexity law under otherwise identical local constraints.

Here we characterize, through upper and matching lower bounds, the
minimum information that a low-rank representation must retain for
local quantum dynamics with finite-range or exponentially decaying
interactions. Writing $L_\epsilon=\log(1/\epsilon)$, for
time-independent dynamics, we obtain
\begin{align}
\log D_\epsilon
=
\widetilde O\!\left(t+\sqrt{L_\epsilon}\right),
\label{eq:main-intro-static}
\end{align}
up to logarithmic corrections. 
For comparison, the previous best constructive MPO bound for static one-dimensional dynamics scales as
\(\log D_{\epsilon}=\widetilde O(t+\sqrt{tL_\epsilon})\) at inverse-polynomial accuracy~\cite{PhysRevX.11.011047,PRXQuantum.2.040331}, whereas Eq.~\eqref{eq:main-intro-static}separates the time and accuracy dependences as \(\widetilde O(t+\sqrt L_\epsilon)\).

At fixed time, however, arbitrary time-dependent driving changes the optimal accuracy dependence from
the static law
$\log D_\epsilon=\widetilde\Theta(L_\epsilon^{1/2})$ to
\begin{align}
\log D_\epsilon
=
\widetilde\Theta\!\left(L_\epsilon^{2/3}\right).
\label{eq:main-intro-driven}
\end{align}
Matching lower bounds show that these characteristic powers cannot
be improved uniformly over their respective classes. Equivalently,
the associated dynamical entanglement spectra exhibit distinct
small-$\alpha$ R\'enyi laws, scaling as $1/\alpha$ for static dynamics
and $1/\alpha^2$ for driven dynamics, up to logarithmic corrections (Fig.~\ref{fig:Overview} c).

The static--driven separation is naturally exposed by a three-region
structure that is not resolved by a single fixed cut. We retain a
finite buffer \(B\) between endpoint regions \(A\) and \(C\) and
introduce the buffered Schmidt rank (BSR), which allows different
contributions to factorize across different cuts inside the buffer
(Fig.~\ref{fig:Overview} b). A fixed generator repeatedly explores the same local
operator structure, whereas temporal control can access different
operator directions at different times, weakening the resulting
suppression of buffered complexity. Converting the BSR bounds back to
a prescribed bipartition yields the distinct low-rank and
entanglement-spectrum laws above.

%

Finally, for time-independent dynamics, these compression bounds are also
constructive. In one-dimensional finite-range systems, we give an
explicit MPO algorithm that attains the same asymptotic compression
scale, with a construction time set by the size of the resulting
representation up to polynomial factors. The same construction
principle extends to time-independent Liouvillian dynamics, yielding
an explicit Liouville-space MPO approximation in diamond norm.
For static local dynamics, this closes the gap that can generally exist between the size of an optimal representation and the classical cost of finding it.


\section{Setup}
\label{sec:main-setup}

We consider a quantum system on a finite lattice $\Lambda$ in arbitrary spatial dimension, with on-site Hilbert-space dimensions bounded by a constant $d_0$. We describe closed and open systems using local Markovian dynamics. For clarity, we write the generator in two-body form,
\begin{align}
\mathcal L^{(\tau)}
&=\sum_i\mathcal L_i^{(\tau)}
+\sum_{i<j}\mathcal L_{ij}^{(\tau)}, \notag\\
\mathcal E_{0\to t}
&=\mathcal T\exp\!\left[\int_0^t\mathcal L^{(\tau)}\,d\tau\right].
\label{eq:main-setup-dynamics}
\end{align}
The full generator is of Lindblad form~\cite{Lindblad1976,GoriniKossakowskiSudarshan1976}, so $\rho_t=\mathcal E_{0\to t}(\rho_0)$ is a completely positive, trace-preserving (CPTP) evolution. The superscript $(\tau)$ denotes explicit time dependence, $t$ is the total evolution time, and $\mathcal T$ denotes time ordering. Interactions involving any fixed number $k$ of sites are treated in Supplementary Sections~I.B and I.D.

We assume uniformly bounded on-site terms and exponentially decaying interactions,
\begin{align}
\sup_{i,\tau}
\sum_{\substack{j\ne i\\ \operatorname{dist}(i,j)\ge r}}
\|\mathcal L_{ij}^{(\tau)}\|_\diamond
\le J e^{-\mu r},
\qquad r\ge0,
\label{eq:main-setup-decay}
\end{align}
where $J,\mu>0$ are independent of system size and $\|\cdot\|_\diamond$ is the diamond norm. The distance is measured on the lattice. This summed bound includes finite-range interactions. For dissipative compression, we also assume that the background obtained by retaining only terms within individual spatial layers generates CPTP evolution, as specified in Supplementary Section~I.D.

Hamiltonian dynamics is the special case
$\mathcal L_H^{(\tau)}(X)=-i[H^{(\tau)},X]$, where
$H^{(\tau)}=\sum_i h_i^{(\tau)}+\sum_{i<j}h_{ij}^{(\tau)}$
has Hermitian terms obeying analogous locality bounds in the operator norm $\|\cdot\|$. The corresponding unitary evolution is
\begin{align}
U_{0\to t}
&=\mathcal T\exp\!\left[-i\int_0^t H^{(\tau)}\,d\tau\right],
\label{eq:main-setup-unitary}
\end{align}
with $\mathcal E_{0\to t}(X)=U_{0\to t}XU_{0\to t}^\dagger$. Time-independent dynamics are obtained by taking the generator constant.

Fix a spatial bipartition \(A|A^{\mathrm c}\). For clarity, throughout the main text, we consider a \(d\)-dimensional rectangular lattice partitioned by planar layers, with the target cut chosen perpendicular to the layering. We denote by \(|\partial A|\) the cross-sectional area of this cut. In one dimension, \(|\partial A|=O(1)\). Supplementary Sections~I.A and I.C treat more general layered geometries, where \(|\partial A|\) is replaced by the maximal elementary cross-section.

For operators and linear maps, respectively, $\operatorname{SR}_A(O)$ and $\operatorname{SR}^{\mathrm L}_A(\Phi)$ are the smallest numbers of product terms across $A|A^{\mathrm c}$ in
\begin{align}
O&=\sum_{j=1}^{D}O_A^{(j)}\otimes O_{A^{\mathrm c}}^{(j)}, \quad 
\Phi=\sum_{j=1}^{D}\Phi_A^{(j)}\otimes\Phi_{A^{\mathrm c}}^{(j)}.
\label{eq:main-setup-Schmidt-rank}
\end{align}
The map factors are arbitrary linear maps. We quantify compression at accuracy $0<\epsilon<1$ by
\begin{align}
D_{\epsilon,A}^{\mathrm U}(t)
&:=\min_{\|U_{0\to t}-\widetilde U\|\le\epsilon}
\operatorname{SR}_A(\widetilde U), \notag\\
D_{\epsilon,A}^{\mathrm L}(t)
&:=\min_{\|\mathcal E_{0\to t}-\widetilde{\mathcal E}\|_\diamond\le\epsilon}
\operatorname{SR}^{\mathrm L}_A(\widetilde{\mathcal E}).
\label{eq:main-setup-approximate-rank}
\end{align}
Thus, we compress the unitary operator itself for closed systems and the quantum channel for open systems. The approximants need not be unitary or CPTP. We write $D_\epsilon$ when the context is clear.

For a normalized product input across the cut, the Schmidt coefficients $\lambda_1\ge\lambda_2\ge\cdots$ of the unitarily evolved state satisfy
\begin{align}
\sum_{s>D_{\epsilon,A}^{\mathrm U}(t)}\lambda_s^2\le\epsilon^2.
\label{eq:main-setup-rank-tail}
\end{align}
Thus, the approximate operator rank directly controls the dynamically generated Schmidt tail (Supplementary Proposition~36).
We quantify the entanglement of $\ket{\psi}$ by the R\'enyi entanglement:
$
E_\alpha(\psi)
:=\frac{1}{1-\alpha}\log\operatorname{tr}\rho_A^\alpha,
$
where $\rho_A=\operatorname{tr}_{A^c}\ket{\psi}\bra{\psi}$ for $0<\alpha<1$.

\subsection{Buffered Schmidt rank}
\label{sec:main-bsr}

Consider two end regions $A$ and $C$ separated by a buffer $B$, so that $\Lambda=A\sqcup B\sqcup C$. Following the ordered spatial decomposition in Supplementary Sections~I.A and V.A, the buffer consists of $\ell-1$ consecutive layers. These determine $\ell$ admissible cuts separating $A$ from $C$, including the two boundaries of the buffer.

The usual operator Schmidt rank uses one fixed cut. Buffered Schmidt rank (BSR), denoted by $\operatorname{BSR}_{A:C\mid B}(O)$, allows individual terms to factorize across different admissible cuts. For an operator $O$ on $ABC$, consider representations
\begin{align}
O
&=\sum_{j=1}^{D}
X^{(j)}_{AB_j}\otimes Y^{(j)}_{\overline B_j C},
\qquad B=B_j\sqcup\overline B_j .
\label{eq:main-bsr-definition}
\end{align}
For each term, $B_j$ consists of the first $\ell_j-1$ buffer layers and $\overline B_j$ contains the remaining layers, for some $\ell_j\in\{1,\ldots,\ell\}$. The smallest total number of terms $D$ is the buffered Schmidt rank $\operatorname{BSR}_{A:C\mid B}(O)$. Restricting the admissible family to a single cut recovers the usual operator Schmidt rank.

For a linear map $\mathcal M$, the Liouville buffered Schmidt rank $\operatorname{BSR}_{\mathrm{Liouv}}(\mathcal M)$ is defined by replacing the operator factors with arbitrary linear maps on the corresponding operator spaces. These decomposed factors need not be completely positive or trace-preserving. We use these definitions to approximate $U_{0\to t}$ and $\mathcal E_{0\to t}$, respectively. Converting all terms to one prescribed cut gives an ordinary Schmidt-rank bound with a dimension factor determined by the buffer, as detailed in Supplementary Propositions~9 and 10.

BSR also admits a restricted information-theoretic interpretation. If the buffer is initially unentangled, quantum information is encoded in \(A\), and the receiver has access only to \(C\), a unitary of BSR \(D\) cannot perfectly transmit more than \(\log_2 D\) qubits from \(A\) to \(C\). More generally, an approximate low-BSR representation bounds the corresponding entanglement fidelity, as shown in Supplementary Proposition~11 and Supplementary Corollary~12. Thus, BSR quantifies not only a three-region factorization complexity but, in this operational sense, the complexity required for coherent quantum information to traverse the buffer.

\section{Main results}
\label{sec:main-results}

We first consider time-independent dynamics. We bound the
three-region complexity quantified by BSR and then convert it into
ordinary Schmidt-rank compression across a prescribed bipartition.
For readability, write $L_\epsilon:=\log(1/\epsilon)$. Local
dimensions and interaction scales are held fixed, while
$\widetilde O$ and $\widetilde\Omega$ suppress logarithmic factors.
Precise bounds and geometric assumptions are given in
Supplementary Theorem~2 and Supplementary Section~I.

The operational viewpoint above makes it useful to first state the
result in a large-deviation-like form. 
For an approximation
$\widetilde U_t$, define its logarithmic buffered Schmidt rank by
$
\mathcal Q
:=
\log \operatorname{BSR}_{A:C\mid B}(\widetilde U_t).
$
For the family of approximations constructed below, in the regime
$\mathcal Q = \Omega(|\partial A|t)$, a time-independent Hamiltonian
satisfies
\begin{align}
\|e^{-itH}-\widetilde U_t\|
\le
\exp\!\left[-\widetilde\Omega(\ell\mathcal Q)\right].
\label{eq:main-bsr-tail-static}
\end{align}
Thus, increasing the buffer produces a large-deviation-like
suppression of the error at fixed nonlocal complexity. In the
endpoint-transfer setting described above, the same complexity also
limits the coherent quantum information that can be transmitted from
$A$ to $C$. Equivalently, achieving accuracy $\epsilon$ requires only
$\mathcal Q 
=\widetilde O(|\partial A|t+L_\epsilon/\ell)$.
An analogous large-deviation-type bound was obtained in the context of the quantum particle transport~\cite{27qs-vlrn}.

A low-BSR approximation can be converted to the prescribed
bipartition by moving each contribution to the target cut. The buffer
contains $O(|\partial A|\ell)$ sites, so this conversion costs at most
$O(|\partial A|\ell)$ in the logarithmic Schmidt rank. Optimizing over
$\ell$ therefore gives
\begin{align}
\log D_{\epsilon,A}^{\mathrm U}(t)
=
\widetilde O\!\left(
|\partial A|t+
\sqrt{|\partial A|L_\epsilon}
\right).
\label{eq:main-fixed-cut-static}
\end{align}
In one dimension this reduces to
$\log D_{\epsilon,A}^{\mathrm U}(t)
=\widetilde O(t+\sqrt{L_\epsilon})$ (Supplementary Lemma~37).
At fixed time and boundary scale, the accuracy exponent $1/2$ is
optimal up to logarithmic corrections: matching constructions with
exponentially decaying interactions give the corresponding lower
bound before finite-size saturation. The construction is given in
Methods and Supplementary Section~IX.A.

The static compression bound is also constructive. For a
one-dimensional finite-range Hamiltonian on $n$ sites, we give an
explicit algorithm producing an MPO approximation
$\widetilde U_t^{\rm MPO}$ with
$\|e^{-itH}-\widetilde U_t^{\rm MPO}\|\le\epsilon$ and
\begin{align}
D_{\rm MPO}
&\le
\exp\!\left[
\widetilde O\!\left(
t+\sqrt{\log(n/\epsilon)}
\right)
\right], \notag\\
T_{\rm construct}
&\le
n\,
\exp\!\left[
\widetilde O\!\left(
t+\sqrt{\log(n/\epsilon)}
\right)
\right].
\label{eq:main-mpo-static}
\end{align}
Thus, the optimal static compression scale is attained by an explicit
matrix-product representation, with construction time polynomial in
the size of the materialized representation (Supplementary Corollary~50). The algorithm
constructively implements the same differential structure underlying
the compression bound, rather than introducing a separate
variational tensor-network ansatz.

For a product input across the target cut, the rank--tail relation
in Eq.~\eqref{eq:main-setup-rank-tail} converts the fixed-cut bound
into a constraint on the full dynamical Schmidt spectrum. In
particular, at fixed time and boundary scale,
$E_\alpha=\widetilde O(\alpha^{-1})$ as $\alpha\to0$, and the
corresponding scaling is optimal up to logarithmic corrections
(Supplementary Theorem~6 and Supplementary Lemma~39).

The same static compression mechanism extends to time-independent
local Liouvillian dynamics. Using diamond-norm error, one obtains
the corresponding BSR and fixed-cut compression bounds for the
quantum channel under the assumptions stated in Supplementary
Section~VI.B (Supplementary Theorem~3). In one-dimensional finite-range systems, we also give
an explicit MPO representation of the superoperator in Liouville
space, with the same output-sensitive structure (Supplementary Theorem~13). These statements
concern channel complexity, whereas the Schmidt-spectrum and
R\'enyi statements above concern unitarily evolved pure states.

Classical approaches to open-system dynamics combine quasilocality~\cite{BarthelKliesch2012} with matrix-product descriptions of mixed-state evolution~\cite{PhysRevLett.93.207204,PhysRevLett.93.207205}, while recent stochastic methods combine quantum trajectories with MPS~\cite{SanderEtAl2025}. Complementary quantum algorithms simulate Markovian dynamics~\cite{ChildsLi2017,DingLiLin2024}, including query-optimal Lindblad evolution~\cite{WangYe2026}.

\subsection{Driven dynamics with static-type compression}

Arbitrary time dependence alone is not sufficient to produce the driven extremal law. 
Quadratic fermionic systems admit a description in terms of correlation matrices~\cite{Peschel_2009}. For local quadratic fermionic dynamics, the static-type $E_\alpha= O(t+\alpha^{-1})$ behavior persists even under arbitrary driving (see Supplementary Proposition~63). 
For general interacting Hamiltonians, there are also broad classes of driven dynamics for which the temporal structure can be represented with only a modest overhead, and the same dependence on time and accuracy as in the static case is recovered. The precise conditions and bounds are given in Supplementary Section~VII.B.

A first example is discrete driving. Suppose that
$H^{(\tau)}=H_j$ for $t_{j-1}\le \tau<t_j$ ($j=1,\ldots,N$), 
with no commutativity assumed between the Hamiltonians $H_j$.
The temporal complexity of such a schedule enters the compression
bound only logarithmically in the number of intervals: compared with
the static result, the rank exponent acquires only a $\log N$ overhead,
up to logarithmic corrections already present in the bound.
Importantly, this dependence is insensitive to the shortest interval
length. Thus, arbitrarily short pulses are allowed, provided that the
number of temporal pieces remains controlled (Supplementary Corollary~33).

The same mechanism extends to regular continuous driving.
Suppose that, on a controlled number of time intervals, every local
interaction can be approximated in the local interaction norm by a
polynomial of degree $\nu$. After truncating the local commutator
expansion, the resulting dressed interaction can be represented using
a common operator basis whose size is polynomial in $\nu$ and in the
logarithmic target accuracy. Consequently, the polynomial degree enters
the logarithmic rank only through $\log \nu$, together with the same
logarithmic accuracy overhead already present in the construction
(Supplementary Theorem~5).

This condition includes a broad class of analytic drives. Standard
Chebyshev approximation theory shows that a function admitting a
uniform analytic continuation to a fixed Bernstein ellipse has a
degree-$\nu$ polynomial approximation with error exponentially small
in $\nu$~\cite{Trefethen2019}. Hence an accuracy $\delta$
requires only $\nu=O(\log(1/\delta))$. The same scaling is obtained
using Chebyshev interpolation, allowing the polynomial representation
to be constructed from a finite set of time samples. Thus piecewise
analytic local drives with uniform analyticity parameters fall naturally
within the static-type compression class; precise approximation
conditions are given in Supplementary Corollary~34.

The same conclusion is constructive. In one-dimensional finite-range systems, these temporal representations also yield MPO constructions with the static output-sensitive scaling, up to the temporal-description overhead (see Supplementary Section~XI.A and Supplementary Corollary~52 for precise assumptions and bounds). Thus, these broad, practically relevant classes obey the static compression law, with accuracy dependence that is optimal up to logarithmic corrections because each class contains static dynamics.

A related role of temporal structure also appears in quantum simulation, where periodic and multiperiodic drives with a fixed number of Fourier components and bounded frequencies admit query complexity matching the static scaling up to logarithmic factors~\cite{PhysRevResearch.5.033067,MizutaFujii2023}.

This raises a natural question: does the same static law continue to hold for completely arbitrary time dependence?

\subsection{Arbitrary driving}

At first sight, there is a natural reason to expect the static law to
survive even for completely arbitrary time dependence. Divide the
evolution into $N$ intervals of duration $\Delta t=t/N$. On each
interval, a short-time Magnus expansion~\cite{BLANES2009151} suggests an effective-static
representation
$
U \simeq e^{-i\Delta t H_{{\rm eff}}},
$
with an error exponentially small in $1/\Delta t$ when the expansion
is truncated near its optimal order~\cite{KUWAHARA201696,Abanin2017}.
Schematically, the accumulated error can therefore be reduced below
$\epsilon$ using only
$N=\widetilde O(tL_\epsilon)$ intervals. Since the compression cost of
a discrete schedule grows only logarithmically with $N$, the preceding
results might then appear to imply the static compression law for an
arbitrary drive as well.

The obstruction is the loss of fixed $k$-locality. 
The static compression theorem uses not
only spatial decay, but also the fixed $k$-local structure of the
microscopic generator. Higher-order Magnus terms are nested
commutators whose support and body order grow with the expansion
order. Thus, although the effective Hamiltonian remains quasi-local
in norm, it no longer belongs uniformly to the local class covered by
the static theorem. The apparent reduction of an arbitrary drive to a
controlled sequence of static evolutions therefore breaks down
precisely at this point.

This limitation is not merely technical. For completely arbitrary
time dependence, the optimal compression law is in fact different.
Under the same spatial-locality assumptions, we obtain an
approximation $\widetilde U_t$ satisfying
$\|U_{0\to t}-\widetilde U_t\|\le\epsilon$ and
\begin{align}
\mathcal Q=
\widetilde O\!\left[
|\partial A|t+
\sqrt{
\left(
|\partial A|t+\frac{L_\epsilon}{\ell}
\right)L_\epsilon
}
\right],
\label{eq:main-bsr-driven}
\end{align}
where we have set $\mathcal Q:=\log \operatorname{BSR}_{A:C\mid B}(\widetilde U_t)$
(Supplementary Theorem~4).
At a fixed time, this corresponds to the large-deviation-like relation
\[
\|U_{0\to t}-\widetilde U_t\|
\le
\exp[-\widetilde\Omega(\sqrt{\ell}\,\mathcal Q)],
\]
in contrast to the static
$\exp[-\widetilde\Omega(\ell\mathcal Q)]$ behavior.

Converting the buffered approximation to a prescribed cut and
optimizing over $\ell$ gives
\begin{align}
&\log D_{\epsilon,A}^{\rm U}(t) \notag \\
&=
\widetilde O\!\left(
|\partial A|t+
\sqrt{|\partial A|tL_\epsilon}
+
|\partial A|^{1/3}L_\epsilon^{2/3}
\right).
\label{eq:main-fixed-cut-driven}
\end{align}
In one dimension and at fixed time, the accuracy dependence is therefore
$\widetilde O(L_\epsilon^{2/3})$, rather than the static
$\widetilde O(L_\epsilon^{1/2})$ law (Supplementary Lemma~37).

This change of exponent is intrinsic. Before finite-size saturation,
there exist arbitrarily driven Hamiltonians with exponentially
decaying interactions for which
\begin{align}
\log D_{\epsilon,A}^{\rm U}(t)
=
\Omega\!\left(
|\partial A|t+
|\partial A|^{1/3}L_\epsilon^{2/3}
\right).
\label{eq:main-driven-lower}
\end{align}
Hence, at fixed time and boundary scale,
\[
\log D_\epsilon
=
\widetilde\Theta(L_\epsilon^{2/3}),
\]
so the accuracy exponent $2/3$ is optimal up to logarithmic
corrections. The construction uses bounded local interaction strengths
but allows arbitrarily rapid temporal control; its details are given
in Methods and Supplementary Section~X.A.

The accuracy exponent is therefore optimal at fixed time and boundary size. For joint variation of time and accuracy, Eqs.~\eqref{eq:main-fixed-cut-driven} and \eqref{eq:main-driven-lower} leave a possible additional mixed contribution \((|\partial A|tL_\epsilon)^{1/2}\). 


\begin{figure*}[!t]
\centering
\includegraphics[width=1\textwidth]{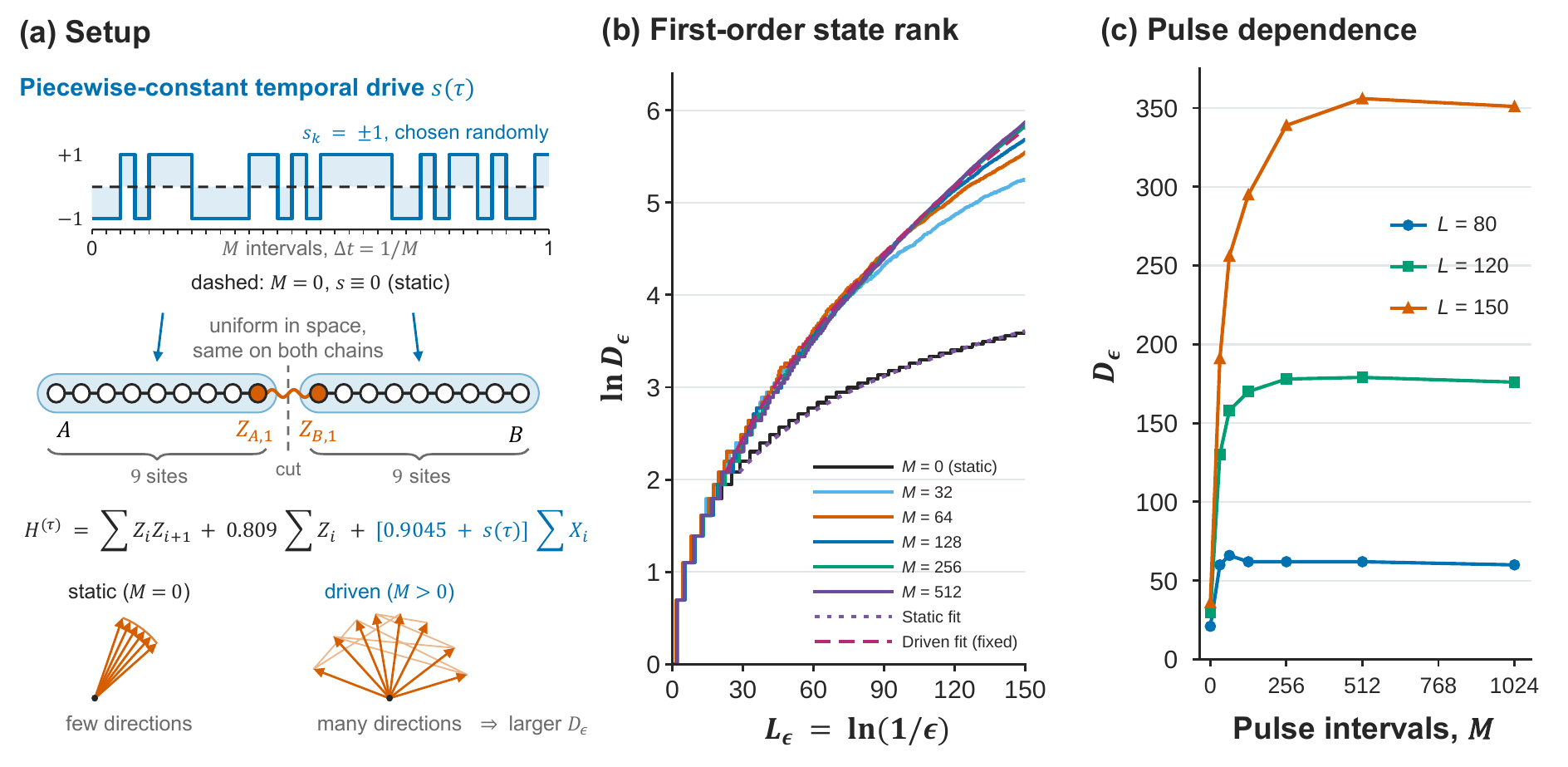}
\caption{Temporal entanglement-capacity amplification of a boundary interaction.
(a) We numerically probe the first-order contribution in Eq.~\eqref{eq:driven-dressed-boundary} for two identical open spin-$1/2$ chains, \(A\) and \(B\), of length \(b=9\) (i.e., $18$ spins in total).
The uncoupled evolution is generated by
\(H_0^{(\tau)}=H_A^{(\tau)}+H_B^{(\tau)}\), with
$
H_A^{(\tau)}=H_B^{(\tau)}
=\sum_{i=1}^{b-1}Z_iZ_{i+1}
 +0.809\sum_{i=1}^{b}Z_i
 +(0.9045+s(\tau))\sum_{i=1}^{b}X_i ,
$
in units \(J=\hbar=1\), and the boundary interaction is
\(V=Z_{A,1}\otimes Z_{B,1}\).
The total evolution time is \(t=1\).
For \(M>0\), \(s(\tau)=\pm1\) is spatially uniform and piecewise
constant over \(M\) equal intervals, with the same random sequence
applied to both chains; \(M=0\) denotes the static case \(s(\tau)=0\).
For the product state
\(|\psi_0\rangle=|+_x\rangle_A^{\otimes b}
\otimes|+_x\rangle_B^{\otimes b}\), we evaluate
$
|\chi\rangle
=
\int_0^1
U_{A|B,0\to\tau}^{\dagger}
V
U_{A|B,0\to\tau}
|\psi_0\rangle\,d\tau ,
$
where
\(U_{A|B,0\to\tau}=U_{A,0\to\tau}\otimes U_{B,0\to\tau}\).
This probes the first-order mechanism, rather than the compression rank of the full propagator.
For the descending Schmidt coefficients \(\lambda_s\) of this
unnormalized state, we define
\(\epsilon_D=(\sum_{s>D}\lambda_s^2)^{1/2}\) and
\(D_\epsilon=\min\{D:\epsilon_D\le\epsilon\}\), with
\(L_\epsilon=\ln(1/\epsilon)\).
(b) Required rank \(\ln D_\epsilon\) versus \(L_\epsilon\)
for different \(M\).
Driving substantially increases the effective rank relative to the
static case. The guide curves use independently calibrated forms
\(D_\epsilon\sim L_\epsilon\) for the static case and
\(D_\epsilon\sim\exp[c\sqrt{L_\epsilon}]\) for the driven case,
motivated by the respective static and arbitrary-drive rank scalings;
they are not refitted to the displayed \(b=9\) data and are intended
only as finite-range guides to the theoretically predicted functional
forms.
(c) \(D_\epsilon\) versus the number of pulse intervals \(M\)
at fixed \(L_\epsilon=80,120,150\).
The required rank rises rapidly above its static value and then shows
only weak dependence on further increases of \(M\) between \(M=512\) and $1024$. Each point corresponds to a single random drive realization without
disorder averaging. Here \(M\) controls the physical pulse sequence,
rather than the numerical integration resolution (Supplementary Section~XIII).}
 \label{fig:numerical}
\end{figure*}

\subsection{Mechanism: temporal amplification of entangling capacity}
\label{sec:mechanism}

The entanglement-generating capacity of an interaction provides a useful viewpoint~\cite{DurEtAl2001,BennettEtAl2003}.
The static--driven separation originates from a simple physical
mechanism: temporal control can amplify the effective entangling
capacity of a fixed boundary interaction.

To see this, consider a boundary interaction $V$ across a bipartition.
In the interaction picture, its leading contribution for a static
Hamiltonian takes the form
\begin{align}
\int_0^t e^{iH_0x} V e^{-iH_0x}\,dx,
\qquad
H_0=H_A+H_B .
\label{eq:static-dressed-boundary}
\end{align}
All times are generated by the same fixed Hamiltonian, so the dressed
copies of $V$ are strongly constrained.

Under arbitrary temporal control, the corresponding contribution becomes
\begin{align}
\int_0^t
U_{A|B,0\to x}^{\dagger}
V
U_{A|B,0\to x}\,dx,
\label{eq:driven-dressed-boundary}
\end{align}
where
$U_{A|B,0\to x}
=
U_{A,0\to x}\otimes U_{B,0\to x}$.
Although each instantaneous term has exactly the same Schmidt rank as
$V$, the local unitaries can rotate it through different operator
directions as time progresses. Their coherent integral can therefore
have an arbitrarily large exact Schmidt rank. Temporal control does not
increase the instantaneous strength or Schmidt rank of the interaction;
it increases the number of nonlocal operator directions that the same
interaction can access over time.

At finite accuracy, however, locality sharply limits how much of this
enlarged operator space can remain relevant. At accuracy $\epsilon$,
one such dressed boundary interaction requires only
\[
\exp\!\left[
\widetilde O\!\left(\sqrt{L_\epsilon}\right)
\right]
\]
effective Schmidt rank. By contrast, in the static case the
corresponding cost grows only as
$\exp[\widetilde O(\log L_\epsilon)]$
(see Supplementary Lemma~18 and Supplementary Proposition~29).
Temporal control can therefore parametrically increase the effective
entangling capacity of a single boundary interaction, while locality
still prevents the unrestricted growth possible for a general
operator-valued function.

In Fig.~\ref{fig:numerical}, we numerically demonstrate this temporal amplification in a simple one-dimensional spin chain with a randomly modulated transverse field.

Crucially, this temporal organization is invisible to instantaneous
entanglement-rate bounds such as SIE~\cite{PhysRevA.76.052319,PhysRevLett.111.170501}, which depend only on the local
interaction data at each time and not on how different operator
directions are coordinated over time. The driven compression bounds
quantify the residual constraints imposed by locality on this enlarged
operator space.

The matching lower-bound construction exploits precisely this
amplification mechanism multiple times. By applying it across many
independent block pairs, temporal control converts the enhanced
capacity of individual dressed boundary interactions into the
$L_\epsilon^{2/3}$ Schmidt-rank scaling. The quantitative construction
is given in Methods.

\section{Outlook}
\label{sec:Outlook}
Our results identify temporal control as a resource that can change the extremal structure of dynamical entanglement spectra. A first open question is to quantify the temporal resources needed to depart from the static-type \(\alpha^{-1}\) regime. Broad classes of temporally structured drives retain the static law, whereas the matching driven construction exploits arbitrarily rapid switching; the finite-size example in Fig.~\ref{fig:numerical} already suggests a strong sensitivity to temporal modulation. 
It remains to determine how the optimal entanglement-spectrum bounds depend on switching rate, bandwidth, or temporal variation. Whether increasing these temporal resources produces a smooth crossover or, in an appropriate scaling limit, a sharp transition between the \(\alpha^{-1}\) and \(\alpha^{-2}\) regimes is also an intriguing future problem. A second challenge is to extend the present boundary-compression theory to efficiently constructible tensor-network representations in higher dimensions, where low-rank structures across different cuts must be made mutually compatible. 
Constructive PEPO approximations for Gibbs operators~\cite{PhysRevB.91.045138} provide a useful precedent.

\section{Methods}
\subsection{Differential formalism and cut-count decomposition}
\label{meth:diff-cut-count}

For clarity, in this Methods section, we explain the core argument for a one-dimensional finite-dimensional system with nearest-neighbor two-body interactions.  The extensions to finite-range and exponentially decaying interactions, and to higher dimensions, use the same structure and are given in Supplementary Sections~I, IV, VI and VII.  Consider a region containing $\ell$ consecutive cuts and write
\begin{align}
H=H_0+\sum_{p=1}^{\ell}V_p .
\label{eq:method-H-decomp-EN}
\end{align}
Here $V_p$ is the nearest-neighbor interaction crossing the $p$th cut, whereas $H_0$ collects all local terms that cross none of these $\ell$ cuts.

Our starting point is the Schmidt-rank counting idea used in the approach to one-dimensional area laws~\cite{PhysRevB.85.195145,arad2013area}.  When a power $H^m$ is expanded, locality strongly restricts how many interaction factors need to cross any one spatial cut, leading to much smaller Schmidt-rank bounds than a naive product estimate.  The limitation is that this statement is naturally formulated for powers, or more generally polynomials, of $H$, whereas here we need the same counting principle for time-evolution operators, Duhamel expansions, and later Liouvillian propagators.  
We separate the combinatorics of how often each interaction is used from the particular function of $H$ being considered.

Introduce one auxiliary variable for each boundary interaction,
\begin{align}
H(\boldsymbol\lambda)
=H_0+\sum_{p=1}^{\ell}\lambda_pV_p,
\qquad
\mathcal D=\sum_{p=1}^{\ell}\partial_{\lambda_p}.
\label{eq:method-H-lambda-EN}
\end{align}
For an analytic operator-valued function $\Phi[H]$,
\begin{align}
\Phi[H]
&=
\left.e^{\mathcal D}\Phi[H(\boldsymbol\lambda)]\right|_{\boldsymbol\lambda=0}
\notag\\
&=
\sum_{s_1,\ldots,s_\ell\ge0}
\frac{\left.
\partial_{\lambda_1}^{s_1}\cdots
\partial_{\lambda_\ell}^{s_\ell}
\Phi[H(\boldsymbol\lambda)]
\right|_{\boldsymbol\lambda=0}}{s_1!\cdots s_\ell!} .
\label{eq:method-multivariate-Taylor-EN}
\end{align}
This is simply a multivariable Taylor expansion, but its interpretation is useful.  
A differential term is labeled by the multi-index $\boldsymbol s=(s_1,\ldots,s_\ell)$, and $s_p$ counts how many times the interaction $V_p$, which crosses the $p$th cut, is inserted in that term.  Thus, the derivative orders directly record how often the term crosses each spatial cut.
A similar expansion was utilized in studying the quantum Gibbs states~\cite{PhysRevLett.124.220601,PRXQuantum.4.020340}.

If the total differential order is truncated at $m$, every retained
multi-index $\boldsymbol{s}=(s_1,\ldots,s_\ell)$ satisfies
\begin{align}
\sum_{p=1}^{\ell}s_p\le m,
\end{align}
and therefore contains at least one cut for which
$s_p\lesssim m/\ell$.  We call a cut chosen in this way the
\emph{native cut} of the term.

The nontrivial point is that this observation does not by itself give
a useful decomposition of the Taylor series.  For a fixed cut $p$, one
would like to collect the terms for which $s_p$ is among the smallest
crossing counts.  Such a restricted sum cannot simply be written as
$(\partial_{\lambda_p})^{s_p}
(\mathcal D-\partial_{\lambda_p})^{m-s_p}$, because the latter also
contains monomials for which another cut has a smaller crossing count.
Thus the selection of a native cut imposes correlated constraints on
the Taylor indices.

Supplementary Theorem~1 resolves this obstruction by expressing these
constrained sums as a finite linear combination of unconstrained
differential exponentials with auxiliary unit-modulus phases.  In the
present nearest-neighbor setting, its useful form is
\begin{align}
&\Phi[H] \notag \\
&\approx
\sum_{p=1}^{\ell}
\sum_{s=0}^{\lfloor 2m/\ell\rfloor}
\sum_{\boldsymbol z\in\mathcal Z_p}
C_{p,\boldsymbol z}
\left.
\frac{(z_p\partial_{\lambda_p})^s}{s!}
e^{\mathcal D_{\boldsymbol z,\neq p}}
\Phi[H(\boldsymbol\lambda)]
\right|_{\boldsymbol\lambda=0},
\label{eq:method-cut-count-toolkit-EN}
\end{align}
where
$
\mathcal D_{\boldsymbol z,\neq p}
:=
\sum_{q\neq p}z_q\partial_{\lambda_q}$ ($|z_q|=1$).
The number of phase branches obeys
\begin{align}
|\mathcal Z_p|
\le
(2m+1)^{\lceil\log_2\ell\rceil},
\qquad
|C_{p,\boldsymbol z}|\le1,
\end{align}
so the resulting overhead is subleading and does not affect the leading-order scaling (Supplementary Proposition~4).

The meaning of Eq.~\eqref{eq:method-cut-count-toolkit-EN} is especially transparent after defining
\begin{align}
A_{p,\boldsymbol z}
:=H_0+\sum_{q\ne p}z_qV_q,
\qquad
B_{p,\boldsymbol z}:=z_pV_p.
\label{eq:method-phase-twisted-background-EN}
\end{align}
The differential exponential resums all nonselected interactions, and each branch in Eq.~\eqref{eq:method-cut-count-toolkit-EN} becomes
\begin{align}
\left.
\frac{1}{s!}\frac{d^s}{dx^s}
\Phi[A_{p,\boldsymbol z}+xB_{p,\boldsymbol z}]
\right|_{x=0}.
\label{eq:method-native-branch-EN}
\end{align}
For a nearest-neighbour chain, every interaction with $q\ne p$ lies entirely on one side of the selected cut, so
\begin{align}
A_{p,\boldsymbol z}
=A^L_{p,\boldsymbol z}\otimes\mathbf 1
+\mathbf 1\otimes A^R_{p,\boldsymbol z}.
\label{eq:method-background-split-EN}
\end{align}
Hence, the only interaction left explicitly across the selected cut is $B_{p,\boldsymbol z}$, and it appears only to order $O(m/\ell)$.  This is the central simplification provided by the cut-count decomposition.

A further advantage over a direct polynomial Schmidt-rank estimate is that approximation errors are tracked within the same expansion.  The approximation error in~\eqref{eq:method-cut-count-toolkit-EN} is the norm tail produced by truncating the total differential order and can be quantitatively controlled for time evolution.  For exponentially decaying interactions, an interaction spanning $r$
cuts is counted as $r$ crossing insertions.  Long-range terms, therefore, enter the cut-count tail with a weight proportional to their range.
Combined with the exponential decay of their norms, this allows them to be discarded within the same quantitative error estimate.

Buffered Schmidt rank (BSR) is designed to retain this native-cut structure: different summands are allowed to use different cuts inside the buffer, and their Schmidt ranks are then added.  Equation~\eqref{eq:method-cut-count-toolkit-EN} therefore reduces the BSR problem to estimating the Schmidt rank of a single native-cut branch.

For a time-independent Hamiltonian, this remaining rank estimate is simple.  Set $\Phi[H]=e^{-itH}$.  Since the total strength of the $\ell$ boundary interactions is $O(\ell)$, the total differential order required for accuracy $\epsilon$ is, up to logarithmic factors,
$m=\widetilde O\!\left(t\ell+L_\epsilon\right)$ with $L_\epsilon:=\log(1/\epsilon)$. 
The number of explicit boundary insertions at one native cut is therefore
$s\le 2m/\ell=\widetilde O\!\left(t+L_\epsilon/\ell\right)$. 

For a fixed branch, abbreviate $A=A_{p,\boldsymbol z}$ and $B=B_{p,\boldsymbol z}$.  Equation~\eqref{eq:method-background-split-EN} gives $A=A_L\otimes\mathbf1+\mathbf1\otimes A_R$.  Since $B$ is a nearest-neighbor two-body interaction, its operator Schmidt rank is a constant $\mathcal{R}=O(1)$, and we may write
$
B=\sum_{\alpha=1}^{\mathcal{R}}B^L_\alpha\otimes B^R_\alpha.
$
In the interaction picture, we consider
\begin{align}
B(\tau)=e^{i\tau A}Be^{-i\tau A}
=\sum_{q=0}^{\infty}
\frac{(i\tau)^q}{q!}\operatorname{ad}_A^q(B),
\label{eq:method-static-commutator-expansion-EN}
\end{align}
where $\operatorname{ad}_A(\cdot):=[A,\cdot]$. 
Because $A_L$ and $A_R$ act on different sides of the cut, their
commutator actions commute.  Hence
\begin{align}
\operatorname{ad}_{A}^{q}(B)
=
\sum_{a=0}^{q}
\binom qa
\operatorname{ad}_{A_L}^{a}
\operatorname{ad}_{A_R}^{q-a}(B),
\end{align}
which yields $\operatorname{SR}_p\!\left(\operatorname{ad}_{A}^{q}(B)\right)
\le (q+1)\mathcal{R}$. 

Consequently, on a constant-duration time interval, we can truncate the expansion~\eqref{eq:method-static-commutator-expansion-EN} and its time integral can be approximated, to the required accuracy,
with only polylogarithmic Schmidt rank:
\begin{align}
\operatorname{SR}_p\!\left(
\int d\tau\, B_\epsilon(\tau)
\right)
=
\widetilde O(1),
\label{eq:method-first-insertion-rank-EN}
\end{align}
where $B_\epsilon(\tau) \approx B(\tau)$ is defined by truncating the expansion in~\eqref{eq:method-static-commutator-expansion-EN} up to $q= O(L_\epsilon)$. 
Thus, each explicit boundary insertion contributes only a
polylogarithmic factor to the Schmidt rank.  An $s$th-order
differential branch therefore obeys
\begin{align}
\operatorname{SR}_p(\text{one native-cut branch})
\le
[\widetilde O(1)]^s.
\label{eq:method-static-branch-rank-EN}
\end{align}
For arbitrary $t$, we divide the evolution into $O(1+t)$
constant-duration intervals.  The additional ways of distributing the
$s$ insertions among these intervals contribute only logarithmic
factors to the exponent and are absorbed into the
$\widetilde O(s)$ notation (Supplementary Proposition~20).

Finally, BSR is bounded by the sum of these native-cut ranks.  Combining Eqs.~\eqref{eq:method-cut-count-toolkit-EN} and \eqref{eq:method-static-branch-rank-EN} gives schematically
\begin{align}
\operatorname{BSR}(\widetilde U_t)
=
\exp\!\left[\widetilde O\!\left(\frac{m}{\ell}\right)\right]
\label{eq:method-static-BSR-before-simplification-EN} ,
\end{align}
where the phase-family size $|\mathcal{Z}_p|$, the sum over cuts $p$, and the sum over $s$ contribute only logarithmic corrections.  
Using $m=\widetilde O\!\left(t\ell+L_\epsilon\right)$, we obtain the desired estimation
$\log\operatorname{BSR}(\widetilde U_t)
=
\widetilde O\!\left(
 t+\frac{L_\epsilon}{\ell}
\right).$
Thus, the differential formalism converts the original global compression problem into a local one: after the cut-count decomposition, it is enough to control the Schmidt-rank growth generated by only $O(t+L_\epsilon/\ell)$ explicit boundary insertions at one selected cut (Supplementary Theorem~2).

\subsection{Arbitrarily time-dependent dynamics}
\label{meth:driven}

The cut-count reduction carries over verbatim to a time-dependent
Hamiltonian $H(\boldsymbol\lambda,t)=H_0(t)+\sum_{p=1}^{\ell}\lambda_p V_p(t)$: assigning the same auxiliary variable $\lambda_p$ to
$V_p(t)$ at all times makes $s_p$ count its total number of insertions
throughout the evolution.  Hence, each native-cut branch again contains
only $s=\widetilde O\!\left(
t+L_\epsilon/\ell
\right)$
explicit insertions across the selected cut.  The only new issue is
their Schmidt-rank cost.

For a fixed native cut $p$, we adopt a similar notation to Eq.~\eqref{eq:method-phase-twisted-background-EN}:
$
A(t)
=
A_L(t)\otimes\mathbf 1+
\mathbf 1\otimes A_R(t),
$
and the crossing interaction $B(t)$ by omitting the indices $p, {\bf z}$.  Let $W_{0\to x}$ be
the propagator generated by $A(t)$.  Already at first order the
relevant object is
\begin{align}
\int_0^{t_c}\widehat B(x)\,dx,
\qquad
\widehat B(x)
=
W_{0\to x}^{-1}B(x)W_{0\to x},
\label{eq:method-driven-first-order}
\end{align}
where $t_c=O(1)$ is a constant time interval.
At every fixed time $\widehat B(x)$ has $O(1)$ Schmidt rank, because
the background evolution $W_{0\to x}$ acts separately on the two sides of the cut.
This does not imply that the integral has also small rank: under arbitrary driving,
the Schmidt directions of $\widehat B(x)$ may change continuously
with time, and the integral may contain many independent directions.
This is the first point at which arbitrary temporal control differs
from a static Hamiltonian.

To obtain a common operator representation over the whole interval,
divide $[0,t_c]$ into windows of width $\delta t$.  For
$x=a+y$ within one window,
\begin{align}
\widehat B(a+y)
=
W_{0\to a}^{-1}
\left[
W_{a\to a+y}^{-1}B(a+y)W_{a\to a+y}
\right]
W_{0\to a}.
\notag 
\end{align}
The outer evolution is a product of left and right operators and hence
does not increase Schmidt rank.  Inside the brackets, only a time
$\delta t$ has appeared.  
Truncating the short-time Dyson expansion at order $q$ gives an error of order
$(C\delta t)^{q+1}$.

Moreover, by locality, the inner evolution $W_{a\to a+y}^{-1}B(a+y)W_{a\to a+y}$ can be restricted, up to exponentially
small error, to $O(L_\epsilon)$ sites around the cut.  There are at most $O(L_\epsilon^k)$ independent operator directions
supported on at most $k$ sites within this region and acting entirely
on one side of the cut. 
Expanding the short-time evolution to order $q$ therefore produces at most
$\operatorname{poly}(L_\epsilon)^q$ nested-commutator sequences.
Since each nested-commutator sequence produces a single product operator across the cut, 
the truncated expansion in one window can be written as a sum of at most \(\operatorname{poly}(L_\epsilon)^q\) product operators.

Choosing $\delta t=e^{-\Theta(q)}$ and $q\asymp\sqrt{L_\epsilon}$ makes the short-time truncation error exponentially small in
$L_\epsilon$, while the total number of operator strings over all
windows is
\begin{align}
\delta t^{-1}\operatorname{poly}(L_\epsilon)^q
=
\exp[\widetilde O(\sqrt{L_\epsilon})].
\end{align}
Consequently, the integrated first-order interaction~\eqref{eq:method-driven-first-order} admits an
approximation up to an error $\epsilon$ within the Schmidt rank of $\exp[\widetilde O(\sqrt{L_\epsilon})]$
(Supplementary Proposition~29 and Supplementary Corollary~30).

For an $s$th-order differential branch, however, we should not apply
this first-order estimate independently at accuracy $\epsilon$ to all
$s$ insertions, which would give the unnecessarily large cost
$\exp[\widetilde O(s\sqrt{L_\epsilon})]$.
Instead, we resolve each dressed interaction into accuracy shells,
\begin{align}
\widehat B(x)=\sum_{j\ge0}\Delta \widehat B^{[j]}(x),
\qquad
\|\Delta \widehat B^{[j]}(x)\|
\lesssim e^{-cj},
\end{align}
where, over the whole integration interval, the $j$th shell can be
expanded in a common set of
$\exp[\widetilde O(\sqrt{j+1})]$ product operators, with all time
dependence contained in scalar coefficients.

In an $s$-fold time-ordered integral, a shell tuple
$(j_1,\ldots,j_s)$ is therefore suppressed as
$e^{-c(j_1+\cdots+j_s)}$.  We may discard all tuples with
$j_1+\cdots+j_s>\bar{j}$, with $\bar{j}=\widetilde O(L_\epsilon+s)$, while the
retained terms satisfy
$
\sum_{a=1}^s\sqrt{j_a+1}
\le
\sqrt{s(\bar{j}+s)}.
$
Consequently, up to logarithmic and combinatorial factors,
$
\log\operatorname{SR}_p(\widetilde U_t)
=
\widetilde O\!\left(s+\sqrt{sL_\epsilon}\right).
$
Using
$s=\widetilde O(t+L_\epsilon/\ell)$ then gives
\begin{align}
\log\operatorname{BSR}(\widetilde U_t)
=
\widetilde O\!\left[
t+
\sqrt{\left(t+\frac{L_\epsilon}{\ell}\right)L_\epsilon}
\right],
\end{align}
which is the driven bound quoted in the main text
(Supplementary Proposition~31 and Supplementary Theorem~4).

\subsection{Time dependence preserving the static compression law}
\label{meth:structured}

The improvement in the preceding arbitrary-drive bound originates from the
number of independent operator directions explored in time.  If the
temporal dependence itself admits a finite representation, the
static compression law is recovered up to logarithmic factors.

Consider first a piecewise-constant drive with $N$ time intervals.
Splitting each time integral at the switching times reduces every
piece to the static problem.  For a branch with $s$ crossing
insertions, there are at most $N^s$ assignments of the insertions to
the time intervals.  Thus the only additional contribution to the
logarithmic Schmidt rank is
$
s\log N .
$
In particular, if $N=\operatorname{poly}(1+t+L_\epsilon)$, this is
absorbed into the logarithmic factors of the static bound
(Supplementary Corollary~33).

A similar conclusion holds for polynomially approximable continuous
driving.  Approximate the local Hamiltonian on each $O(1)$ time
interval by a polynomial of degree $\nu$.  Truncating the corresponding
commutator expansion at order $r$ gives a polynomial approximation
$\widehat B_r(x)$ of degree at most $r(\nu+1)$.  Hence, using
$P=r(\nu+1)+1$ interpolation points,
\begin{align}
\widehat B_r(x)
=
\sum_{j=1}^{P} f_j(x)\widehat B_r(x_j).
\end{align}
Since $\widehat B_r(x_j)$ approximates $\widehat B(x_j)$, we may replace
the sampled values by the exact operators $\widehat B(x_j)$, up to the
same controlled accuracy.  Each $\widehat B(x_j)$ has $O(1)$ Schmidt
rank, so the whole interval admits a common representation using only
$O(P)$ product operators.

For a uniformly analytic drive, polynomial degree
$\nu=O(L_\epsilon)$ suffices on each $O(1)$ time interval, and the
required commutator cutoff $r$ is likewise polynomial in $L_\epsilon$.
Thus
$
P=r(\nu+1)+1=\operatorname{poly}(L_\epsilon)
$
sampled operators suffice per interval.  Since there are only
$O(1+t)$ such intervals, the whole evolution requires an order of 
$(1+t)P$ temporal operator directions.
An $s$-fold integral, therefore, incurs only a factor
$[(1+t)P]^s$, or
$
s\log[(1+t)P]=\widetilde O(s)
$
in the logarithmic rank.
Using
$
s=\widetilde O\!\left(t+L_\epsilon/\ell\right),
$
we recover the static-type bound (Supplementary Theorem~5 and Supplementary Corollary~34).

\subsection{Protocols for the matching lower bounds}

\begin{figure}[!t]
  \centering
  \includegraphics[width=0.47\textwidth]{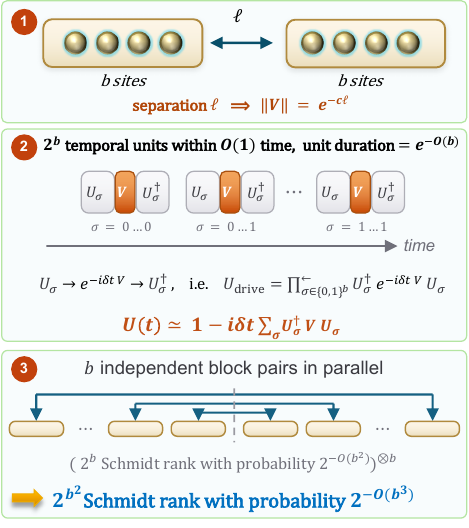}
\caption{Driven lower-bound protocol underlying the $2/3$ accuracy exponent.
Step 1: Weak interaction between distant blocks.
Consider two $b$-site blocks on opposite sides of the target cut, separated by a distance $\ell$ and coupled by a weak interaction $V$. 
For exponentially decaying interactions, the coupling strength is suppressed as $\|V\|\sim e^{-c\ell}$.
The blocks otherwise evolve independently, so local control on the two sides can change the operator direction of $V$ without increasing its instantaneous Schmidt rank. \\
Step 2:  Temporal amplification within one block pair.
For each bit string $\sigma\in\{0,1\}^b$, local encoding operations
$U_\sigma=U_{\sigma,A}\otimes U_{\sigma,B}$ rotate the same weak interaction into a different direction, producing one temporal unit
$U_\sigma^\dagger e^{-i\delta t V}U_\sigma$.
The $2^b$ units are made exponentially short in $b$, so all of them fit within $O(1)$ total evolution time.
For sufficiently short boundary pulses,
$
U(t)\simeq
1-i\delta t\sum_\sigma U_\sigma^\dagger VU_\sigma ,
$
and the $2^b$ encoded components therefore add coherently rather than incoherently.
Encoding one complete \(b\)-bit string requires \(b\) successive local encoding steps; their cumulative attenuation gives a squared weight \(e^{-O(b^2)}\).
The resulting state contains an approximately flat Schmidt sector of rank $2^b$ with total squared weight $e^{-O(b^2)}$.
Thus, temporal control reuses a single weak interaction to create exponentially many independent entangling directions. \\
Step 3:  Parallelization and the $2/3$ law.
Run the same protocol synchronously on $b$ independent block pairs.
The pairs can be arranged at separations $O(b^2)$, whose exponential interaction suppression remains within the one-pair $e^{-O(b^2)}$ scale.
The Schmidt ranks multiply to $D=2^{b^2}$, while the squared weights multiply to
$p=e^{-O(b^3)}$.
Since the corresponding vector-error scale is $\epsilon\sim\sqrt p$, one has
$L_\epsilon=\log(1/\epsilon)=\Theta(b^3)$ and $\log D=\Theta(b^2)$, yielding
$\log D_\epsilon=\Omega(L_\epsilon^{2/3})$.
No postselection is involved: the small-weight Schmidt sector is part of the full evolved state (Supplementary Section~X.A).}
  \label{fig:Fig.3_protocol}
\end{figure}

The lower bounds are obtained at the state level.  Starting from a
product state across the target cut, an operator of Schmidt rank $D$
can produce a state of Schmidt rank at most $D$.  Hence a Schmidt-tail
lower bound for one evolved product state, together with
Eq.~\eqref{eq:main-setup-rank-tail}, directly gives the same lower
bound for operator-norm approximation of the propagator.

For static dynamics, place $m$ independent weakly entangled pairs
across the cut, with exponentially decreasing couplings.  Their tensor
product contains $2^m$ Schmidt coefficients, while the smallest
relevant squared weights scale as
$\exp[-\Theta(m^2)]$.  Thus
\begin{align}
L_\epsilon=\Theta(m^2),
\qquad
\log D_\epsilon=\Omega(m)
=\Omega(\sqrt{L_\epsilon}),
\end{align}
which gives the static accuracy exponent
(Supplementary Lemma~39 and Supplementary Proposition~40).

The driven construction is summarized in Fig.~\ref{fig:Fig.3_protocol}.
It exploits the temporal-amplification mechanism of Eq.~\eqref{eq:driven-dressed-boundary}:
one block pair produces an approximately flat Schmidt sector of
rank $2^b$ with squared weight $e^{-O(b^2)}$, and parallelizing
$b$ independent pairs gives rank $2^{b^2}$ with squared weight
$e^{-O(b^3)}$. Hence
\begin{align}
L_\epsilon=\Theta(b^3),
\qquad
\log D_\epsilon=\Omega(b^2)
=\Omega(L_\epsilon^{2/3}).
\end{align}
The complete microscopic construction and finite-size estimates are
given in Supplementary Theorems~7 and 8 (see also Supplementary Corollary~43).

Higher-dimensional examples follow by placing independent
one-dimensional copies across the boundary. Sharing the logarithmic
accuracy scale among $O(|\partial A|)$ copies gives the
boundary dependence in Eq.~\eqref{eq:main-driven-lower}.

The construction above uses only an $O(1)$ time window and produces
the accuracy-dependent part of Eq.~\eqref{eq:main-driven-lower}. For general
$t$, one uses the remaining $O(t)$ evolution to generate entanglement
deterministically on independent local degrees of freedom, producing
a flat Schmidt spectrum of rank $e^{\Omega(t)}$ per one-dimensional
copy. Combining the two contributions multiplies the required ranks
and hence adds $\Omega(|\partial A|t)$ to $\log D_\epsilon$.

\subsection{Constructive MPO for static unitary dynamics}

\begin{figure*}[!t]
  \centering
  \includegraphics[width=1\textwidth]{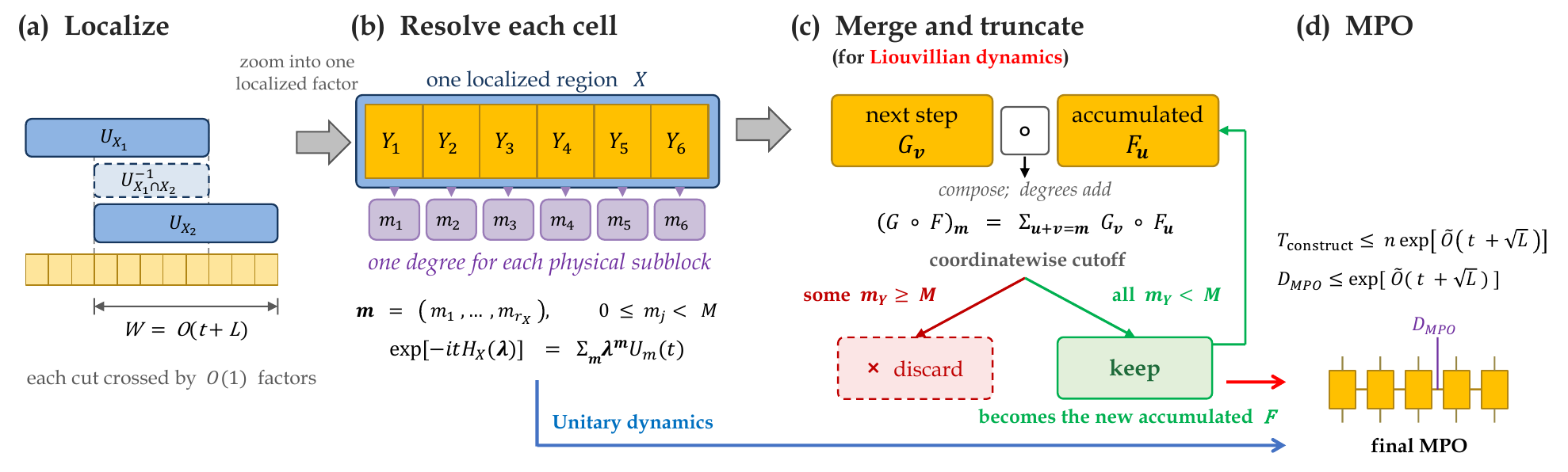}
\caption{Constructive MPO realization of the static compression bound.
(a) Localization. For unitary dynamics, the HHKL decomposition reduces the full evolution to a product of localized unitaries supported on intervals of width \(W=O(t+L)\), where \(L=\log[Cn(1+t)/\epsilon]\). 
For each cut, the decomposition has bounded multiplicity: every microscopic cut is crossed by only \(O(1)\) localized unitaries. It is therefore sufficient to construct an MPO for one localized evolution \(U_X(t)=e^{-itH_X}\), where $|X|=O(t+L)$.
(b) Resolution by local degrees. A localized region \(X\) is partitioned into coarse physical subblocks \(Y_j\) of width \(b\asymp\sqrt L\), grouping each local interaction into one subblock so that \(H_X=\sum_jG_j\).
 Introducing one auxiliary variable \(\lambda_j\) for each subblock, define \(H_X(\boldsymbol\lambda)=\sum_j\lambda_jG_j\),  so that
$e^{-itH_X(\boldsymbol\lambda)} =\sum_{\mathbf m\ge0}\boldsymbol\lambda^{\mathbf m}U_{\mathbf m}(t), $
where \(m_j\) counts the number of interaction insertions belonging to \(Y_j\). The construction retains the full local degree pattern \(\mathbf m=(m_1,\ldots,m_{r_X})\), rather than only its total degree, subject to the coordinatewise cutoff \(0\le m_j< M=O(tb+L)\), where $r_X=O\!\left(|X|/b\right)= O(t/\sqrt{L} + \sqrt{L})$. 
(c) Liouvillian extension. For Liouvillian dynamics, long-time inverse maps are avoided by dividing the evolution into \(O(1+t)\) short-time steps, each approximated by an HHKL-type product of local maps, where the localized cell width is chosen as $O(L)$. 
 The degree of each physical subblock is shared across all of its occurrences and across time steps. If two successive degree-resolved MPOs are written as \(F(\boldsymbol\lambda)=\sum_{\mathbf u}F_{\mathbf u}\boldsymbol\lambda^{\mathbf u}\) and \(G(\boldsymbol\lambda)=\sum_{\mathbf v}G_{\mathbf v}\boldsymbol\lambda^{\mathbf v}\), their composition is resolved by
$ (G\circ F)_{\mathbf m} =\sum_{\mathbf u+\mathbf v=\mathbf m}G_{\mathbf v}\circ F_{\mathbf u}. $
Since all degrees are nonnegative, any sector for which some \(m_Y>M\) can be discarded immediately after each composition, while the retained sectors form the accumulated MPO for the next step.
(d) Final MPO. Combining the localized factors gives an explicit MPO with
$ D_{\rm MPO}\le \exp[\widetilde O(t+\sqrt L)]$ and $T_{\rm construct}\le n\,\exp[\widetilde O(t+\sqrt L)],$
with the analogous output-sensitive bound for the Liouville-space MPO.}
  \label{fig:Fig.4_MPO}
\end{figure*}

The preceding fixed-cut compression argument does not directly yield
an efficient MPO algorithm: the low-rank representation may depend on
the chosen cut, whereas an MPO must realize compatible low-rank
structures across all cuts simultaneously.  We achieve this by
constructing the retained differential coefficients globally, while
keeping the degree of each spatial subblock as a separate label.

The constructive scheme is summarized in Fig.~\ref{fig:Fig.4_MPO}. We first use the Haah--Hastings--Kothari--Low (HHKL) decomposition and resolve each localized factor by the local degree vector \(\mathbf m\) defined there. The coordinatewise cutoff is \(0\le m_j< M=O(tb+L)\) with 
\begin{align}
L
:=
\log\frac{C n(1+t)}{\epsilon},
\label{eq:method-mpo-L}
\end{align}
where $C$ is chosen sufficiently large that $L\ge1$.
We now explain why every retained coefficient has a small exact MPO rank and how the retained coefficients can be constructed efficiently.

%

The key is to retain this full vector of local degrees rather than
grouping the expansion only by the total Taylor order
$|\boldsymbol m|_1$.  We impose the coordinatewise cutoff
\begin{align}
0\le m_j< M=O(tb+L)
\label{eq:method-coordinatewise-cutoff}
\end{align}
for every $j$ (Supplementary Proposition~48).
The total degree $\sum_jm_j$ may therefore be much larger than $M$.
What matters for compression across a given cut is only the degrees
of the spatial subblocks that can contribute interactions near that
cut.

Consider a window containing $b$ consecutive microscopic cuts.
Because the interactions are nearest-neighbor and each $G_j$ is
associated with a coarse subblock of width $b$, only $O(1)$ of the
$G_j$ can contain interactions crossing any cut in this window.
For a retained coefficient $U_{\boldsymbol m}$, each such $G_j$
appears fewer than $M$ times.  Hence the total number of crossings,
summed over all $b$ cuts in the window, is $O(M)$.
By averaging over the $b$ cuts, at least one cut is therefore crossed
only
\begin{align}
O\!\left(\frac{M}{b}\right)
=
O(t+\sqrt L)
\label{eq:method-mpo-crossing-budget}
\end{align}
times.

Consequently, each retained coefficient operator 
$U_{\boldsymbol m}(t)$ has an exact MPO
representation with
\begin{align}
D_{\rm coeff}
\le
\exp\!\left[
\widetilde O\!\left(
b+\frac{M}{b}
\right)
\right]
=
\exp[\widetilde O(t+\sqrt L)].
\label{eq:method-mpo-coeff-rank}
\end{align}
Here ``exact'' means that, once the differential coefficient has been
retained, no further approximation is introduced in its MPO
compression.  Uniform exact Schmidt-rank bounds across all cuts can
be converted into an MPO by successive operator-Schmidt, or
equivalently exact rank, factorizations along the chain
(Supplementary Proposition~49).

The remaining issue is how to construct these $U_{\boldsymbol m}(t)$ without
enumerating all noncommutative words.  From the definition $U_{\boldsymbol m}(t):=[\boldsymbol\lambda^{\boldsymbol m}]e^{-itH_X(\boldsymbol\lambda)}$, we give the exact recursion
\begin{align}
U_{\boldsymbol 0}(t)&=\mathbf 1,
\\
U_{\boldsymbol m}(t)
&=
\frac{-it}{|\boldsymbol m|_1}
\sum_{j:m_j\ge1}
U_{\boldsymbol m-\boldsymbol e_j}(t)\,G_j,
\qquad
\boldsymbol m\neq\boldsymbol0 ,
\label{eq:method-mpo-dp}
\end{align}
where $\boldsymbol e_j$ is the unit vector in the $j$th coordinate.
The terms in the sum simply classify all operator words according to
their last interaction group.  Thus the retained multi-indices can be
processed in increasing total degree, without ever listing the
individual words.

Suppose that all predecessor coefficients $U_{\boldsymbol m}(t)$ have already been reduced
to the bound in Eq.~\eqref{eq:method-mpo-coeff-rank}.  
Since each $G_j$ is a sum of only $O(b)$ nearest-neighbor local terms,
it has an exact MPO representation of bond dimension $O(b)$.
Hence
$U_{\boldsymbol m-\boldsymbol e_j}G_j$
has a bond dimension at most $O(b)D_{\rm coeff}.$
Moreover, there are at most $r_X$ predecessors in
Eq.~\eqref{eq:method-mpo-dp}, so even the partially accumulated sum
has a bond dimension at most
$
r_X O(b) D_{\rm coeff}.
$
After the full update is completed, the successive exact rank factorizations reduce it back to the exact Schmidt rank bound $D_{\rm coeff}$.
Hence, the arithmetic cost of constructing one retained $U_{\boldsymbol m}(t)$ is
\begin{align}
T_{\rm coeff}
\le 
\operatorname{poly}
\!\left(
D_{\rm coeff}
\right)
=
\exp[\widetilde O(t+\sqrt L)].
\label{eq:method-single-coeff-cost}
\end{align}

It remains to count how many coefficients have to be constructed.
The rectangular cutoff
\eqref{eq:method-coordinatewise-cutoff} retains
\begin{align}
N_{\rm ind}
=
M^{r_X} = (tb+L)^{O(t/\sqrt{L} + \sqrt{L})}
\end{align}
multi-indices.
Thus, the sum of all retained coefficients can be represented by a
direct-sum MPO with
$
D_{\rm loc}
\le
N_{\rm ind}D_{\rm coeff},
$
and hence
\begin{align}
\log D_{\rm loc}
=
\widetilde O(t+\sqrt L).
\label{eq:method-mpo-local-rank}
\end{align}
Likewise, constructing all $N_{\rm ind}$ coefficients by
Eq.~\eqref{eq:method-mpo-dp} gives
\begin{align}
T_{\rm loc}
\le
N_{\rm ind}T_{\rm coeff}
=
\exp[\widetilde O(t+\sqrt L)].
\end{align}

Finally, because the HHKL decomposition has bounded cut multiplicity, multiplying the localized MPOs does not change the leading exponent of the bond dimension.  
Summing their spatial construction costs gives
$
D_{\rm MPO}
\le
\exp[\widetilde O(t+\sqrt L)]$,
and
$T_{\rm construct}
\le
n\,\exp[\widetilde O(t+\sqrt L)].
$
Thus, the same differential structure used to prove low-rank
compressibility also yields an explicit MPO algorithm whose running
time is, up to polynomial factors, set by the size of the
representation it constructs (Supplementary Theorem~11 and Supplementary Corollary~50).

\subsection{Extension to Liouvillian dynamics}

For local Liouvillian dynamics, the unitary construction must be
modified because inverse maps appearing in a long-time localization
need not be bounded in diamond norm.  We therefore divide the
evolution into $R=O(1+t)$ short-time steps and use, for each step, a
depth-two HHKL-type local-cell approximation $\Gamma_\Delta(\boldsymbol\lambda)$.  Its local correction factors are
controlled by the same short-time composition-commutator mechanism
used in imaginary-time constructions for local Gibbs operators~\cite{PhysRevX.11.011047}.  We
choose the cell width $B=\Theta(L)$, which makes the accumulated
localization error over all time steps smaller than the target
accuracy (Supplementary Proposition~53).

The first new issue is the global degree cutoff.  We divide each cell
into physical subblocks $Y$ of width $b\asymp\sqrt L$ and assign one
variable $\lambda_Y$ to each subblock, shared by all of its
occurrences and by all time steps.  Setting $\lambda_Y=0$ for any
collection $S$ removes the same complete local Lindblad terms from
every occurrence.  The resulting short-time product approximates the
corresponding deleted Lindblad evolution uniformly in $S$; since the
exact deleted evolution is CPTP,
\begin{align}
\|\Gamma_\Delta(0_S,1_{S^c})\|_\diamond
\le 1+\delta_H(B),
\end{align}
where \(\delta_H(B)\) denotes the one-step localization error.
Together with Cauchy estimates and inclusion--exclusion over the
discarded events $m_Y>M$, this gives a coordinatewise cutoff
\begin{align}
0\le m_Y< M,
\qquad
M=O(tb+L),
\label{eq:method-liouvillian-M}
\end{align}
for the complete time evolution.  More precisely, the localization
error is exponentially small in $B$ up to polynomial prefactors,
whereas the coordinatewise Taylor tail is exponentially small in
$M-O(tb)$.  Thus $B=\Theta(L)$ and $M=O(tb+L)$ make both errors smaller
than the target accuracy (Supplementary Propositions~55 and 56).

The second issue is the construction of one short-time
degree-resolved MPO.  The differential coefficients of each
elementary exponential are constructed exactly as in the unitary
case.  The only additional bookkeeping is that the same physical
subblock can occur several times---at most three times in the present
one-dimensional depth-two construction.  Their temporary degrees
$m_{Y,f}$ are merged locally into
\begin{align}
m_Y=\sum_f m_{Y,f}.
\end{align}
The resulting short-time MPO is resolved by the collection
$\boldsymbol m=(m_Y)_Y$, stored as classical block indices.
Each HHKL cell contains $O(B/b)$ physical subblocks, and its
coefficient MPO is correspondingly block-resolved by their joint
degree pattern.  This is the same multidegree structure as in the
unitary construction; the only new point is that repeated occurrences
of the same physical subblock share a single degree $m_Y$
(Supplementary Proposition~57).

Finally, successive short-time MPOs are merged degree by degree.  If
\begin{align}
F(\boldsymbol\lambda)
=\sum_{\boldsymbol u}F_{\boldsymbol u}
 \boldsymbol\lambda^{\boldsymbol u},
\qquad
G(\boldsymbol\lambda)
=\sum_{\boldsymbol v}G_{\boldsymbol v}
 \boldsymbol\lambda^{\boldsymbol v},
\end{align}
then
\begin{align}
(G\circ F)_{\boldsymbol m}
=
\sum_{\boldsymbol u+\boldsymbol v=\boldsymbol m}
G_{\boldsymbol v}\circ F_{\boldsymbol u}.
\label{eq:method-liouvillian-convolution}
\end{align}
Because all degrees are nonnegative, sectors with some $m_Y>M$ may
be discarded after every composition.  Moreover, the degree labels
are block indices, so the subblocks can be merged sequentially:
merging one $m_Y$ leaves all unresolved degree labels unchanged, and
exact MPO rank reduction is performed independently within each
remaining block sector (Supplementary Proposition~60).

Each degree merge therefore reduces to a local block-sparse MPO
contraction followed by sector-preserving exact rank reduction on the
affected $O(B)$-site region.  Its cost is polynomial in the local
region size and in the current MPO bond dimension
$D=\exp[\widetilde O(t+\sqrt L)]$.  Consequently, all merges can be
performed within the same output-sensitive complexity as in the
unitary case (Supplementary Theorem~13), yielding
$
D_{\rm Liouville\text{-}MPO}
=
\exp[\widetilde O(t+\sqrt L)],
$ 
and 
$
T_{\rm construct}
= n\,\exp[\widetilde O(t+\sqrt L)].
$

\section*{Acknowledgements}

This work was supported by the RIKEN Hakubi Fellows Program.
D.~K. is supported by the Japan Society for the Promotion of Science through its Grants-in-Aid for Scientific Research program (JSPS KAKENHI Grant Number JP26K17060).
T.~V.~V. is supported by JSPS KAKENHI Grant Numbers JP23K13032, JP26K00019, JP26K00022, and JP26H02015.
T.~K. is supported by the Japan Science and Technology Agency through its Exploratory Research for Advanced Technology program (JST ERATO Grant Number JPMJER2302) and by JSPS KAKENHI Grant Numbers JP23K25796, JP25K24674, JP26K00019, JP26K00020, and JP26H02015.
This research was partially conducted during the internship of H.~M. at RIKEN, under the supervision of T.~K.

\def\bibsection{\section*{References}}

\bibliography{Short_range_Entanglement}

\end{document}




\title{Supplementary Information for \\ ``Optimal low-rank compression of quantum dynamics''}

\author{Hugo Mackay}
\affiliation{Department of Physics, Harvard University, Cambridge, MA 02138, USA}
\affiliation{Analytical Quantum Complexity RIKEN Hakubi Research Team, RIKEN Center for Quantum Computing (RQC), Wako, Saitama 351-0198, Japan}

\author{Donghoon Kim}
\affiliation{Analytical Quantum Complexity RIKEN Hakubi Research Team, RIKEN Center for Quantum Computing (RQC), Wako, Saitama 351-0198, Japan}

\author{Tan Van Vu}
\affiliation{Yukawa Institute for Theoretical Physics, Kyoto University, Kitashirakawa Oiwakecho, Sakyo-ku, Kyoto 606-8502, Japan}

\author{Tomotaka Kuwahara}
\affiliation{Analytical Quantum Complexity RIKEN Hakubi Research Team, RIKEN Center for Quantum Computing (RQC), Wako, Saitama 351-0198, Japan}
\affiliation{RIKEN Pioneering Research Institute (PRI), Wako, Saitama 351-0198, Japan}

%
%
%
%
%
%
%

\maketitle

\tableofcontents

\section{Setup}
\label{sec:setup}

This section fixes the notation and assumptions used repeatedly later in the paper.
Geometry-dependent auxiliary quantities are kept local to the statements in
which they are needed.  In particular, the dynamical estimates below are
formulated in terms of the three layer-strength parameters
$K_r$, $K_\eta$, and $J_{\mathrm{off}}$; spatial decay is used only to
supply sufficient bounds on these quantities.

\subsection{System and layered geometry}
\label{subsec:setup-system}

Let $\Lambda$ be a finite subset of a locally finite graph, endowed with the
graph distance $\dist(\cdot,\cdot)$.  No fixed spatial dimension is assumed
in the abstract results.  Each site $x\in\Lambda$ carries a finite-dimensional
Hilbert space $\mathcal H_x$, and
\begin{align}
    \mathcal H_X:=\bigotimes_{x\in X}\mathcal H_x,
    \qquad
    D_X:=\dim\mathcal H_X
    \le d_0^{|X|}
    \qquad (X\subseteq\Lambda)
\label{eq:local-dimension-bound}
\end{align}
for a system-size-independent local dimension bound $d_0$.
For nonempty $X,Y\subseteq\Lambda$, write
\begin{align}
    \dist(X,Y)&:=\min_{x\in X,\,y\in Y}\dist(x,y),
    &
    \diam(X)&:=\max_{x,y\in X}\dist(x,y).
\end{align}
When a $d$-dimensional specialization is invoked, we assume polynomial
volume growth,
\begin{align}
    \sup_{x\in\Lambda}|B_R(x)|\le C(1+R)^d,
    \qquad
    B_R(x):=\{y\in\Lambda:\dist(x,y)\le R\},
\label{eq:volume-growth}
\end{align}
with $C$ independent of $|\Lambda|$.  No volume-growth assumption is needed
for the abstract layer-strength statements.

Fix a disjoint ordered partition
\begin{align}
    \Lambda=\bigsqcup_{j=1}^{\bar\ell}\Lambda_j
\label{eq:ordered-elementary-partition}
\end{align}
that is compatible with the geometry.  After a fixed blocking, if needed,
we choose the unit of the layer index so that
\begin{align}
    \dist(\Lambda_j,\Lambda_{j'})\ge |j-j'|
    \qquad (j\ne j').
\label{eq:ordered-partition-distance}
\end{align}
This covers sites or contiguous blocks in one dimension and ordered slabs or
shells in higher dimensions.  More general spacings can be treated by
replacing the integer separation below by the corresponding graph distance;
we do not introduce a separate notation for that variant.

Choose integers $0\le\ell_0\le\bar\ell$ and
$1\le\ell\le\bar\ell-\ell_0+1$, and define
\begin{align}
    L_0&:=\bigsqcup_{1\le j\le\ell_0}\Lambda_j,
    \notag\\
    L_r&:=\Lambda_{\ell_0+r}
    \qquad (r=1,\ldots,\ell-1),
    \notag\\
    L_\ell&:=\bigsqcup_{j\ge\ell_0+\ell}\Lambda_j.
\label{L_Layer_definition}
\end{align}
Empty unions are allowed.  Thus the definition includes $\ell=1$ and empty
endpoint regions, and
\begin{align}
    \Lambda=\bigsqcup_{r=0}^{\ell}L_r.
\label{eq:coarse-layer-partition}
\end{align}
The endpoint layers $L_0$ and $L_\ell$ may be macroscopically large.  We use
\begin{align}
    |\partial\Lambda|:=\max_{1\le j\le\bar\ell}|\Lambda_j|
\label{eq:boundary-size}
\end{align}
for the maximal elementary cross-section; equivalently,
$\partial\Lambda$ may denote any elementary block attaining the maximum.
Consequently,
\begin{align}
    |L_r|\le |\partial\Lambda|
    \qquad (1\le r\le\ell-1),
\label{eq:interior-layer-size}
\end{align}
but no such bound is imposed on $L_0$ or $L_\ell$.
For $0\le p\le q\le\ell$, let
\begin{align}
    L_{[p,q]}:=\bigsqcup_{r=p}^{q}L_r,
\end{align}
with $L_{[p,q]}=\varnothing$ for $p>q$, and set
$[\ell]:=\{1,\ldots,\ell\}$.

For $p\in[\ell]$, the cut $(p-1,p)$ is the bipartition
$L_{[0,p-1]}:L_{[p,\ell]}$.  The operator Schmidt rank across this cut is~\cite{PhysRevA.67.052301}
\begin{align}
    \SR_p(O)
    :=\min\left\{
       R:\ O=\sum_{a=1}^{R}O_{<p}^{(a)}\otimes O_{\ge p}^{(a)}
    \right\}.
\label{eq:setup-operator-Schmidt-rank}
\end{align}
For any $p,q\in[\ell]$, expansion in an operator basis on the layers between
the two cuts gives
\begin{align}
    \SR_q(O)
    &\le
    D_{L_{[\min\{p,q\},\max\{p,q\}-1]}}^2\SR_p(O)
    \notag\\
    &\le
    d_0^{2|p-q||\partial\Lambda|}\SR_p(O).
\label{trivial_Schmidt_ineq}
\end{align}
Only interior layers occur between admissible cuts, so the aggregate endpoint
dimensions do not enter this estimate.

\subsection{Local Hamiltonians and layer strengths}
\label{subsec:setup-Hamiltonian}

We consider a possibly time-dependent $k$-local Hamiltonian
\begin{align}
    H(t)=\sum_{\substack{\varnothing\ne Z\subseteq\Lambda\\|Z|\le k}}h_Z(t),
    \qquad
    h_Z(t)=h_Z(t)^\dagger,
\label{prop1:klocal}
\end{align}
where $k=O(1)$, and the maps $t\mapsto h_Z(t)$ are strongly measurable and
locally integrable.  Unless stated otherwise, $\|\cdot\|$ denotes the
operator norm.  We assume the uniform single-site bound
\begin{align}
    \sup_t\sum_{Z\ni x}\|h_Z(t)\|\le g
    \qquad (x\in\Lambda).
\label{eq:g-extensivity}
\end{align}

The decay profile $\bar J:\mathbb Z_{\ge0}\to\mathbb R_{\ge0}$ bounds
the one-point diameter tail:
\begin{align}
    \sup_t\sup_{x\in\Lambda}
    \sum_{\substack{Z\ni x\\\diam(Z)\ge R}}
    \|h_Z(t)\|
    \le \bar J(R),
    \qquad R\in\mathbb Z_{\ge0}.
\label{prop2:short/range}
\end{align}
We take $\bar J$ nonincreasing, set $\bar J(0)=g$ without loss of
generality, and assume the weighted sums
$\sum_{R\ge1}e^{\eta R}\bar J(R)$ are finite for the values of $\eta>0$
used later.  We also set
\begin{align}
    \bar g_1:=\sum_{R=0}^{\infty}\bar J(R).
\label{eq:bar-g-one}
\end{align}
Equation~\eqref{prop2:short/range} is deliberately stronger than a pairwise
bound on terms containing two prescribed sites.  A pairwise bound may be
used only after the graph-shell growth has been summed into the profile
$\bar J$.

Every nonempty support $Z$ has unique leftmost and rightmost coarse layers.
We therefore define
\begin{align}
    H_{p,q}(t)
    :=\sum_{\substack{Z:\ p=\min\{r:Z\cap L_r\ne\varnothing\}\\
                          q=\max\{r:Z\cap L_r\ne\varnothing\}}}
      h_Z(t),
    \qquad 0\le p\le q\le\ell.
\label{definition_H/i_i'}
\end{align}
Then every local term appears exactly once and
\begin{align}
    H(t)=\sum_{0\le p\le q\le\ell}H_{p,q}(t).
\label{eq:exact-H-layer-decomposition}
\end{align}
Set
\begin{align}
    V_0(t)&:=\sum_{p=0}^{\ell}H_{p,p}(t),
    &
    \mathcal E_\ell&:=\{(p,q):0\le p<q\le\ell\}.
\label{eq:setup-H-block-notation}
\end{align}
For $e=(p,q)\in\mathcal E_\ell$, write
\begin{align}
    H_e(t):=H_{p,q}(t),
    \qquad
    r_e:=q-p.
\end{align}
The three strength parameters used in the dynamical sections are
\begin{align}
    K_r(t)
    &:=\sum_{\substack{e\in\mathcal E_\ell\\r_e=r}}\|H_e(t)\|,
    \qquad r\in[\ell],
    \notag\\
    K_\eta(t)
    &:=\sum_{r=1}^{\ell}e^{\eta r}K_r(t),
    \notag\\
    J_{\mathrm{off}}(t)
    &:=\sum_{r=1}^{\ell}K_r(t)
      =\sum_{e\in\mathcal E_\ell}\|H_e(t)\|.
\label{eq:setup-H-length-strengths}
\end{align}
For an interval $[s,t]$, we abbreviate
\begin{align}
    K_\eta[s,t]:=\int_s^t K_\eta(\tau)\,d\tau.
\label{eq:integrated-K-eta}
\end{align}
For a time-independent Hamiltonian, all time arguments are omitted.

If no interaction meets non-neighboring coarse layers, then
\begin{align}
    H(t)=\sum_{p=0}^{\ell}H_{p,p}(t)
    +\sum_{p=0}^{\ell-1}H_{p,p+1}(t).
\label{eq:finite_Ham_int}
\end{align}

\subsection{Geometric bounds on the layer strengths}
\label{subsec:setup-layer-bounds}

The following lemma is the only point at which the microscopic decay
profile is converted into a boundary-size estimate.

\begin{lemma}[Layer-block bounds]
\label{lem:setup-layer-block-bounds}
For every time $t$,
\begin{align}
    \|H_{p,p}(t)\|
    &\le g|L_p|,
    \qquad 0\le p\le\ell,
\label{eq:H-block-diagonal-bound}
\\
    \|H_{p,q}(t)\|
    &\le |\partial\Lambda|\,\bar J(q-p),
    \qquad
    0\le p<q\le\ell,
    \quad (p,q)\ne(0,\ell),
\label{eq:H-block-offdiagonal-bound}
\\
    \|H_{0,\ell}(t)\|
    &\le |\partial\Lambda|
       \sum_{R=\ell}^{\infty}\bar J(R).
\label{eq:H-endpoint-block-bound}
\end{align}
Consequently,
\begin{align}
    K_r(t)
    &\le(\ell-r+1)|\partial\Lambda|\,\bar J(r)
    \qquad (1\le r<\ell),
    \notag\\
    K_\ell(t)
    &\le |\partial\Lambda|
       \sum_{R=\ell}^{\infty}\bar J(R),
\label{Ineq_total/sum_p_p':r}
\end{align}
and
\begin{align}
    J_{\mathrm{off}}(t)
    &\le \bar g_1\,\ell|\partial\Lambda|,
\label{Ineq_total/sum_p_p'}
\\
    K_\eta(t)
    &\le \ell|\partial\Lambda|
       \sum_{R=1}^{\infty}e^{\eta R}\bar J(R).
\label{sum_H_p_p'_upp}
\end{align}
\end{lemma}

\textit{Proof.}
The diagonal estimate follows from
\begin{align}
    \|H_{p,p}(t)\|
    \le\sum_{x\in L_p}\sum_{Z\ni x}\|h_Z(t)\|
    \le g|L_p|.
\end{align}
Let $p<q$ and $(p,q)\ne(0,\ell)$.  At least one of the two endpoint layers
is an elementary block of size at most $|\partial\Lambda|$.  Every support
assigned to $(p,q)$ meets that elementary block and the opposite coarse
layer, so Eqs.~\eqref{eq:ordered-partition-distance} and
\eqref{prop2:short/range} give
\begin{align}
    \|H_{p,q}(t)\|
    \le |\partial\Lambda|\,\bar J(q-p).
\end{align}
For $H_{0,\ell}$, anchor successively in the elementary blocks
$\Lambda_j\subseteq L_0$.  A support reaching $L_\ell$ has diameter at least
$\ell+\ell_0-j$, and hence
\begin{align}
    \|H_{0,\ell}(t)\|
    \le
    \sum_{j\le\ell_0}|\Lambda_j|\,
    \bar J(\ell+\ell_0-j)
    \le
    |\partial\Lambda|\sum_{R=\ell}^{\infty}\bar J(R).
\end{align}
This also covers an empty endpoint.  Summing at fixed $r$ proves
Eq.~\eqref{Ineq_total/sum_p_p':r}.  Summing over $r$, and using
$e^{\eta\ell}\le e^{\eta R}$ for $R\ge\ell$, proves
Eqs.~\eqref{Ineq_total/sum_p_p'} and \eqref{sum_H_p_p'_upp}.
\hfill$\square$

In particular, if $\bar J(R)\le J_0e^{-\mu R}$ with $\mu>0$, then for some
$\kappa=O(J_0)$ independent of $t$, $\ell$, and $|\Lambda|$
(with the implicit constant allowed to depend on $\mu$),
\begin{align}
    K_r(t)\le\kappa\ell|\partial\Lambda|e^{-\mu r}
    \qquad (r\in[\ell]).
\label{eq:setup-exponential-Kr}
\end{align}
This is the exponential length profile used in the geometric corollaries;
the abstract theorems retain $K_r$, $K_\eta$, and $J_{\mathrm{off}}$
explicitly.

\subsection{Hamiltonian and Liouvillian dynamics}
\label{subsec:setup-dynamics}

For $t\ge s$, the Hamiltonian propagator is
\begin{align}
    U_{s\to t}
    :=\mathcal T\exp\left[-i\int_s^tH(\tau)\,d\tau\right],
\label{eq:H-propagator}
\end{align}
where $\mathcal T$ denotes chronological ordering.  It solves
\begin{align}
    \partial_tU_{s\to t}=-iH(t)U_{s\to t},
    \qquad
    U_{s\to s}=\mathbbm 1,
\end{align}
where the differential equation holds for almost every $t$.  The propagator
is unitary, so $\|U_{s\to t}\|=1$.  The corresponding unitary channel
$\mathcal U_{s\to t}(X):=U_{s\to t}XU_{s\to t}^\dagger$ is completely
positive and trace preserving (CPTP) and obeys
$\|\mathcal U_{s\to t}\|_\diamond=1$.

A local Liouvillian is written
\begin{align}
    \mathcal L(t)
    =\sum_{\substack{\varnothing\ne Z\subseteq\Lambda\\|Z|\le k}}
      \mathcal L_Z(t),
\label{eq:local-Liouvillian}
\end{align}
where $\mathcal L_Z(t)$ is supported on $Z$, and the maps
$t\mapsto\mathcal L_Z(t)$ are strongly measurable and locally integrable.
The full generator is assumed to be of GKLS form at each
time~\cite{Gorini1976,Lindblad1976} and therefore generates a CPTP propagator;
the individual local summands need not themselves generate trace-preserving
evolutions.  We use the diamond norm and assume
\begin{align}
    \sup_t\sum_{Z\ni x}\|\mathcal L_Z(t)\|_\diamond
    \le g_{\mathrm L}
    \qquad (x\in\Lambda).
\label{eq:L-g-extensivity}
\end{align}
To avoid introducing a second decay-profile symbol, all later Liouvillian
spatial assumptions are stated directly in terms of the layer strengths
below.

Using the same unique extreme-layer assignment, define
$\mathcal L_{p,q}(t)$ and
\begin{align}
    \mathcal L_0(t)&:=\sum_{p=0}^{\ell}\mathcal L_{p,p}(t),
    &
    \mathcal L_e(t)&:=\mathcal L_{p,q}(t)
    \quad (e=(p,q)\in\mathcal E_\ell).
\end{align}
The Liouvillian strength parameters are
\begin{align}
    K_r^{(\mathrm L)}(t)
    &:=\sum_{\substack{e\in\mathcal E_\ell\\r_e=r}}
      \|\mathcal L_e(t)\|_\diamond,
    \notag\\
    K_\eta^{(\mathrm L)}(t)
    &:=\sum_{r=1}^{\ell}e^{\eta r}K_r^{(\mathrm L)}(t),
    \notag\\
    J_{\mathrm{off}}^{(\mathrm L)}(t)
    &:=\sum_{r=1}^{\ell}K_r^{(\mathrm L)}(t).
\label{eq:setup-L-length-strengths}
\end{align}
For a time-independent Liouvillian, the time arguments are omitted.
Whenever contractivity of the diagonal background is used later, we
separately assume that $\mathcal L_0(t)$ itself generates a CPTP evolution.

The Liouvillian propagator
\begin{align}
    \mathcal T_{\mathcal L}(t,s)
    :=\mathcal T\exp\left[\int_s^t\mathcal L(\tau)\,d\tau\right]
\label{eq:L-propagator}
\end{align}
is CPTP and satisfies
$\|\mathcal T_{\mathcal L}(t,s)\|_\diamond=1$.
For the Hamiltonian Liouvillian
$\mathcal L_H(t)=-i[H(t),\,\cdot\,]$, the local choice
$(\mathcal L_H)_Z=-i[h_Z,\,\cdot\,]$ gives
\begin{align}
    g_{\mathrm L}&\le2g,
    &
    K_r^{(\mathrm L)}(t)&\le2K_r(t),
    &
    K_\eta^{(\mathrm L)}(t)&\le2K_\eta(t),
    &
    J_{\mathrm{off}}^{(\mathrm L)}(t)&\le2J_{\mathrm{off}}(t).
\label{eq:H-L-norm-relation}
\end{align}

\subsection{Norm conventions and useful inequalities}
\label{subsec:setup-norms}

For $1\le p<\infty$, let
$\|X\|_p:=(\operatorname{tr}|X|^p)^{1/p}$, with
$\|X\|_\infty=\|X\|$.  For a linear map $\Phi$, let
\begin{align}
    \|\Phi\|_\diamond
    :=\sup_{m\ge1}\|\Phi\otimes\operatorname{id}_m\|_{1\to1}.
\end{align}
The following standard bounds will be used without further comment; see
Refs.~\cite{Bhatia1997,Watrous2018} for Schatten norms and completely
bounded trace norms:
\begin{align}
    \|X\|_q
    &\le\|X\|_p
    \le(\operatorname{rank}X)^{1/p-1/q}\|X\|_q,
    &&1\le p\le q\le\infty,
\label{eq:Schatten-comparison}
\\
    \|AXB\|_p
    &\le\|A\|\,\|X\|_p\,\|B\|,
    &
    \|X\otimes Y\|_p&=\|X\|_p\|Y\|_p,
\label{eq:Schatten-product-bounds}
\\
    \|\operatorname{tr}_B X\|_p
    &\le D_B^{1-1/p}\|X\|_p,
    &&1\le p\le\infty,
\label{eq:partial-trace-Schatten}
\\
    \|[A,X]\|_p
    &\le2\|A\|\,\|X\|_p,
    &
    \|-i[A,\,\cdot\,]\|_\diamond&\le2\|A\|,
\label{eq:commutator-bounds}
\\
    \|\mathcal D_L\|_\diamond
    &\le2\|L\|^2,
    &
    \mathcal D_L(X)&:=LXL^\dagger-\tfrac12\{L^\dagger L,X\},
\label{eq:dissipator-bound}
\\
    \|\Phi(X)\|_1
    &\le\|\Phi\|_\diamond\|X\|_1,
    &
    \|\Phi\circ\Psi\|_\diamond
    &\le\|\Phi\|_\diamond\|\Psi\|_\diamond,
\label{eq:diamond-basic-bounds}
\\
    \|\Phi\otimes\operatorname{id}\|_\diamond
    &=\|\Phi\|_\diamond.
\end{align}
In particular, the partial trace is trace-norm contractive and every CPTP
map has diamond norm one.  If $\mathcal T_A(t,s)$ is generated by a locally
integrable family $\mathcal A(t)$, then every compatible submultiplicative
norm for which the identity has norm one satisfies
\begin{align}
    \|\mathcal T_A(t,s)\|_*
    \le
    \exp\left[\int_s^t\|\mathcal A(\tau)\|_*\,d\tau\right].
\label{eq:propagator-growth-bound}
\end{align}
Duhamel's formula implies, for two Hermitian Hamiltonians $H$ and $G$,
\begin{align}
    \|U^H_{s\to t}-U^G_{s\to t}\|
    \le\int_s^t\|H(\tau)-G(\tau)\|\,d\tau,
\label{eq:unitary-Duhamel-bound}
\end{align}
and, for two Liouvillian evolutions with CPTP propagators,
\begin{align}
    \|\mathcal T_{\mathcal L}(t,s)-\mathcal T_{\mathcal M}(t,s)\|_\diamond
    \le\int_s^t\|\mathcal L(\tau)-\mathcal M(\tau)\|_\diamond\,d\tau.
\label{eq:Liouvillian-Duhamel-bound}
\end{align}

\subsection{Notation table}
\label{subsec:setup-notation-table}

Table~\ref{tab:main-notation} lists only symbols whose meaning is fixed
across multiple sections.  Section-local parameters are deliberately omitted.

\begin{table}[t]
\caption{Main notation whose meaning is fixed across the paper.}
\label{tab:main-notation}
\centering
\small
\setlength{\tabcolsep}{5pt}
\renewcommand{\arraystretch}{1.18}
\begin{tabular}{@{}p{0.25\linewidth}p{0.67\linewidth}@{}}
\toprule
Symbol & Meaning \\
\midrule
$\Lambda$, $d$, $\mathcal H_X$, $D_X$, $d_0$
& Finite graph region, optional polynomial-growth dimension, Hilbert space on $X$, its dimension, and the uniform local-dimension bound. \\
$\ell$, $L_0,\ldots,L_\ell$, $|\partial\Lambda|$
& Number of layer cuts, coarse layers, and maximal elementary cross-section.  The endpoint layers may be macroscopic. \\
$k$, $g$, $\bar J(R)$, $\bar g_1$
& Hamiltonian locality, one-site strength, upper bound on the one-point diameter tail, and $\bar g_1=\sum_{R\ge0}\bar J(R)$. \\
$H_{p,q}$, $V_0$, $\mathcal E_\ell$, $r_e$
& Exact layer block, diagonal background, off-diagonal block index set, and number of cuts crossed by block $e$. \\
$K_r$, $K_\eta$, $J_{\mathrm{off}}$
& Total norm of length-$r$ Hamiltonian blocks, its exponential moment, and total off-diagonal strength. \\
$K_\eta[s,t]$
& Integrated exponential moment $\int_s^tK_\eta(\tau)d\tau$ for a driven Hamiltonian. \\
$\kappa$, $\mu$
& Constants in the standard specialization $K_r(t)\le\kappa\ell|\partial\Lambda|e^{-\mu r}$. \\
$U_{s\to t}$, $\mathcal U_{s\to t}$
& Hamiltonian propagator and the associated unitary channel. \\
$\mathcal L(t)$, $g_{\mathrm L}$, $\mathcal L_{p,q}$, $\mathcal L_0$
& Local Liouvillian, its one-site diamond-norm strength, its layer blocks, and diagonal background. \\
$K_r^{(\mathrm L)}$, $K_\eta^{(\mathrm L)}$, $J_{\mathrm{off}}^{(\mathrm L)}$
& Diamond-norm counterparts of the Hamiltonian layer strengths. \\
$\mathcal T_{\mathcal L}(t,s)$
& CPTP propagator generated by $\mathcal L(t)$. \\
$\SR_p(O)$
& Operator Schmidt rank of $O$ across the cut $(p-1,p)$. \\
$\|\cdot\|$, $\|\cdot\|_p$, $\|\cdot\|_\diamond$
& Operator, Schatten, and diamond norms. \\
\bottomrule
\end{tabular}
\end{table}
\FloatBarrier

\section{Overview of the main results}
\label{sec:overview-main-results}

A Lieb--Robinson bound limits how rapidly a local perturbation can influence a
distant observable~\cite{LiebRobinson1972,NachtergaeleSimsYoung2019}.  The question studied here is finer: how much
\emph{nonlocal structure} can a local evolution create inside its light cone?
We measure this structure through low-rank operator approximations, the decay
of the Schmidt spectrum, R\'enyi entanglement, and the cost of constructing
matrix-product representations.  This section gives the conceptual map; the
precise hypotheses and constants appear in the later sections.
Locality-based bounds on correlation generation and classical simulation
provide the starting context~\cite{PhysRevLett.97.050401,PhysRevLett.97.157202}.

\paragraph{The organizing idea: keep the cut selected by each term.}
Fix a buffer made of $\ell-1$ interior layers between two endpoint regions.
The differential decomposition in Secs.~III and IV rewrites the evolution as
a sum of terms.  Each term is assigned to a cut at which only a small number
of interaction insertions remain explicit.  A conventional Schmidt-rank
argument would immediately move every term to one prescribed cut, paying for
the entire region between the selected and prescribed cuts.  Buffered
Schmidt rank (BSR), introduced in Sec.~V, postpones that payment: different
summands may use different cuts inside the same buffer.  
In this sense, BSR is algebraically a bookkeeping device for retaining native
cuts, but it also admits a restricted endpoint quantum-transfer interpretation
developed in Sec.~V.D.  It is not a new physical entanglement measure, and it is
not generally submultiplicative under operator products.

Let $0<\epsilon<1$ and $L_\epsilon:=\log(1/\epsilon)$.  We use the
one-dimensional or uniformly layered geometric specializations of
Secs.~VI--VIII, including their uniform bounds on the retained crossing
neighborhoods.  For bounded $|\partial\Lambda|$, and suppressing constants
fixed by the local interaction data, the main compression results have the
following informal form.  For a time-independent Hamiltonian, there is an operator
$\widetilde U_t$ such that
\begin{align}
 \|e^{-itH}-\widetilde U_t\|&\le\epsilon,
 &
 \log\operatorname{BSR}(\widetilde U_t)
 &\le \widetilde O\left(t+\frac{L_\epsilon}{\ell}\right).
 \label{eq:overview-static-bsr-informal}
\end{align}
For an arbitrarily driven Hamiltonian, in the uniformly stable or
accuracy-dominated regimes made precise in Sec.~VII,
\begin{align}
 \|U_{0\to t}-\widetilde U_t\|&\le\epsilon,
 &
 \log\operatorname{BSR}(\widetilde U_t)
 &\le \widetilde O\left(
 t+\sqrt{\left(t+\frac{L_\epsilon}{\ell}\right)L_\epsilon}
 \right).
 \label{eq:overview-driven-bsr-informal}
\end{align}
The tilde hides polylogarithmic factors, including logarithmic contributions
from the polynomial local-channel counts.  The
exact static statement is Theorem~\ref{thm:static-unitary-low-BSR}; the exact
driven statement is Theorem~\ref{thm:time-dependent-low-BSR}.  The driven
theorem uses the transported localization estimate derived from the
setup in Lemma~\ref{lem:td-transported-localization}.  The
extra square-root term in Eq.~\eqref{eq:overview-driven-bsr-informal} is the
cost of allowing the drive to explore new local operator directions at
different times.  A static Hamiltonian repeatedly uses the same operator
directions and therefore admits the stronger precision dependence in
Eq.~\eqref{eq:overview-static-bsr-informal}.

\paragraph{Structured time dependence recovers the static scale.}
Section~\ref{subsec:td-static-type-compression} identifies driven systems
that retain the static precision dependence.  These include piecewise-constant
schedules with $N=\operatorname{poly}(1+t+L_\epsilon)$ intervals, and local
interactions admitting piecewise-polynomial approximations at the required
local tolerance with polynomially controlled interval count and degree, as
specified in Theorem~\ref{thm:reg-polynomial-compression}.  Hamiltonians at
different times need not commute.  Under the stated spatial assumptions,
the BSR bound and, after optimizing the buffer width, the fixed-cut operator
Schmidt rank bound are
\begin{align}
 \log\operatorname{BSR}(\widetilde U_t)
 &\le\widetilde O\left(
 1+|\partial\Lambda|t+\frac{L_\epsilon}{\ell}\right),
 \notag\\
 \log R_\epsilon^{\rm structured}
 &\le\widetilde O\left(
 |\partial\Lambda|(1+t)+\sqrt{|\partial\Lambda|L_\epsilon}\right),
 \label{eq:overview-structured-ranks}
\end{align}
for operator-norm error at most $\epsilon$.  Examples include fixed-degree
polynomial ramps and uniformly analytic local driving with the uniform bounds
specified in Sec.~\ref{subsec:td-static-type-compression}, such as a fixed
finite collection of bounded-frequency tones.  The proof approximates the
local dressed interaction families in common product-operator bases whose
sizes are polynomial in the logarithm of the required inverse error.  Local
commutator bounds suffice for this construction: neither the
transported-localization estimate used in the general driven proof nor
an accuracy-dominated regime is needed here.  Efficient construction additionally requires a computable
temporal representation, supplied in Secs.~XI.A and XII.A.

\paragraph{From BSR to an ordinary Schmidt spectrum.}
BSR becomes a standard fixed-cut statement through the elementary inequality
\begin{align}
 \operatorname{SR}_q(O)\le \chi_q\operatorname{BSR}(O),
 \qquad
 \chi_q\le d_0^{2\ell|\partial\Lambda|},
 \label{eq:overview-bsr-to-sr}
\end{align}
proved in Proposition~\ref{prop:basic-bsr-properties}.  Thus the buffer width
$\ell$ trades two costs against one another: a larger buffer reduces the
number of explicit crossings in Eqs.~\eqref{eq:overview-static-bsr-informal}
and \eqref{eq:overview-driven-bsr-informal}, but increases the conversion
factor in Eq.~\eqref{eq:overview-bsr-to-sr}.  Restoring the boundary-size
dependence, optimizing this tradeoff gives the following leading scaling
terms for a fixed-cut operator Schmidt rank $R_\epsilon$, up to logarithmic
factors:
\begin{align}
 \log R_\epsilon^{\rm static}
 &\le \widetilde O\left(
 \bar g_1|\partial\Lambda|t
 +\sqrt{|\partial\Lambda|L_\epsilon}
 \right),
 \label{eq:overview-static-sr-informal}
 \\
 \log R_\epsilon^{\rm driven}
 &\le \widetilde O\left(
 \bar g_1|\partial\Lambda|t
 +\sqrt{\bar g_1|\partial\Lambda|tL_\epsilon}
 +|\partial\Lambda|^{1/3}L_\epsilon^{2/3}
 \right).
 \label{eq:overview-driven-sr-informal}
\end{align}
These expressions describe the leading small-error scaling at fixed time
and boundary size.  Section~VIII gives uniform estimates that also retain
additive boundary-size terms.  The fixed-cut arbitrary-drive estimates use the general localization
and stability bounds derived in Sec.~VII.  Choosing the buffer width
jointly with the stability budget, as in Sec.~VIII, removes the need
for an additional stability assumption in these fixed-cut estimates.  If the initial state is a normalized
pure product state across the target cut and its evolved Schmidt coefficients
are $\lambda_1\ge\lambda_2\ge\cdots$, an operator approximation with
operator-norm error at most $\epsilon$ and operator Schmidt rank at most
$R_\epsilon$ across that cut implies
\begin{align}
 \sum_{j>R_\epsilon}\lambda_j^2\le\epsilon^2.
 \label{eq:overview-rank-to-tail}
\end{align}
Consequently, the static and general driven bounds have different
small-$\alpha$ R\'enyi dependences.  For fixed time, boundary size,
and local data, Sec.~VIII gives, as $\alpha\downarrow0$,
\begin{align}
 E_\alpha^{\rm static}
 &\le O\left(\alpha^{-1}\log^2(1/\alpha)\right),
 &
 E_\alpha^{\rm driven}
 &\le O\left(\alpha^{-2}\log^3(1/\alpha)\right),
 \label{eq:overview-renyi-powers}
\end{align}
with the full time- and boundary-dependent expressions stated there.
For the structured drives of Sec.~\ref{subsec:td-static-type-compression},
if the temporal approximation assumptions hold uniformly at every accuracy,
the same rank-to-spectrum argument gives the static dependence
$E_\alpha=\widetilde O(\alpha^{-1})$ at fixed time and boundary size.

\paragraph{Why the exponents are essentially optimal.}
The lower-bound constructions address the complete Schmidt spectrum rather
than only the largest coefficient.  The asymptotic statements below concern
families whose system size may increase as $\epsilon\downarrow0$ or
$\alpha\downarrow0$, with time, boundary size, and local interaction bounds
fixed.  The time-independent, exponentially decaying construction gives
\begin{align}
 \log D_\epsilon=\Omega\!\left(\sqrt{\log(1/\epsilon)}\right),
 \qquad
 E_\alpha=\Omega(\alpha^{-1}),
 \label{eq:overview-static-lower-informal}
\end{align}
whereas the construction with arbitrarily fast piecewise-constant driving gives
\begin{align}
 \log D_\epsilon=\Omega\!\left(\log^{2/3}(1/\epsilon)\right),
 \qquad
 E_\alpha=\Omega(\alpha^{-2}).
 \label{eq:overview-driven-lower-informal}
\end{align}
Here $D_\epsilon$ denotes the minimum Schmidt rank across the chosen cut
needed to approximate the constructed state in vector norm with error at most
$\epsilon$.  The corresponding MPS/MPO statements and the distinctions among
state-vector error, squared state error, and operator-norm error are made
explicit in Secs.~IX and X, together with the finite-size conditions and
thermodynamic-limit statements.  Equations
\eqref{eq:overview-static-lower-informal} and
\eqref{eq:overview-driven-lower-informal} match the characteristic powers in
the upper bounds, up to logarithmic factors and the explicit time and geometry
terms.  The separate flagged-sector constructions show that a small-amplitude
part of the wave function may also contain a state of very large circuit
depth.  This computational statement is complementary to, and should not be
identified with, the entanglement-spectrum lower bound.

\paragraph{Representation size and construction time.}
For one-dimensional finite-range, time-independent unitary dynamics on $n$
sites, the compression can be made algorithmic.  Section~XI constructs an MPO
approximating the propagator to operator-norm error at most $\epsilon$ and
satisfying
\begin{align}
 D_{\rm MPO}
 &\le \exp\left[\widetilde O\left(
 t+\sqrt{\log(n/\epsilon)}\right)\right],
 &
 T_{\rm construct}
 &\le n\exp\left[\widetilde O\left(
 t+\sqrt{\log(n/\epsilon)}\right)\right]
 \label{eq:overview-static-algorithmic}
\end{align}
in an exact-arithmetic model.  For the explicit representation constructed
in Sec.~XI, the construction time is polynomial in the materialized MPO bond
dimension, up to polynomial factors in the displayed parameters.
Section~\ref{subsubsec:alg-U-driven-extension} extends both bounds in
Eq.~\eqref{eq:overview-static-algorithmic} to structured time dependence.
The temporal input consists of exact piecewise-polynomial local interactions
or certified, efficiently computable polynomial approximants with the stated
supports and local-strength bounds.  Let $N$ be the number of intervals in a
common partition and $\nu$ the maximum polynomial degree.  Under the
time-dependent HHKL localization input, if $N$ and $\nu$ are polynomial in
$1+t+\log(n/\epsilon)$ and the input-preparation cost fits the displayed
construction budget, both the bond dimension and construction time have the
same scaling as in Eq.~\eqref{eq:overview-static-algorithmic}; see
Corollary~\ref{corol:alg-U-driven-global}.  A common coefficient recursion
across the entire drive controls the ranks of intermediate MPOs and avoids a
separate leading precision cost for each temporal segment.  This includes
piecewise-constant schedules, polynomial ramps, and uniformly analytic local
driving when the required polynomial data and error certificates are
computable within the stated budget.  The interval count, degree, and
preprocessing cost remain explicit for larger temporal descriptions.

\paragraph{Open-system dynamics.}
The algebraic mechanism is not restricted to unitaries.  Section~VI proves a
time-independent Liouville-space version in diamond norm when the diagonal
background generates a CPTP evolution.  For one-dimensional finite-range generators
decomposed into complete local Lindblad terms, Sec.~XII gives an explicit
Liouville-MPO construction with the same output-sensitive scaling, with $t$
replaced by $g_{\rm L}t$.  The general driven BSR analysis has the analogous
Liouville-space form when the diagonal background generates a
CPTP evolution and the local generator obeys the diamond-norm
microscopic diameter-decay condition stated in Sec.~VII; the transported
localization estimate is then derived there.
Section~\ref{subsubsec:alg-L-driven-extension} proves the corresponding
constructive extension for structured Lindblad driving.  The original
real-time local terms satisfy the assumptions above on complete local Lindblad terms,
and the temporal input consists of exact piecewise-polynomial data or
certified, efficiently computable local polynomial approximants in diamond
norm.  With
$L_{\mathrm L}:=\max\{1,\log[n(1+g_{\mathrm L}t)/\epsilon]\}$,
Theorem~\ref{thm:alg-L-driven-global} gives a linear-map MPO with diamond-norm
error at most $\epsilon$ and
\begin{align}
 \log D_{\rm MPO}
 &\le\widetilde O\left(g_{\mathrm L}t+\sqrt{L_{\mathrm L}}\right),
 &
 T_{\rm construct}
 &\le n\exp\left[\widetilde O\left(
 g_{\mathrm L}t+\sqrt{L_{\mathrm L}}\right)\right],
 \label{eq:overview-structured-liouvillian-algorithmic}
\end{align}
provided $N$ and $\nu$ are polynomial in
$1+g_{\mathrm L}t+L_{\mathrm L}$ and preprocessing fits this cost.
Examples include piecewise-constant Lindblad schedules, linear ramps between
local Lindbladians, and uniformly analytic local generator paths with the
required uniform bounds and computable approximants.  Both constructive
time-dependent extensions use the same exact-arithmetic model as their static
counterparts.  These statements concern channel or Choi-operator complexity.
A R\'enyi entanglement statement
for a mixed output state requires choosing a mixed-state entanglement measure
and does not follow merely by replacing operator norms with diamond norms.

\begin{figure}[!t]
 \centering
 \includegraphics[width=0.97\textwidth]{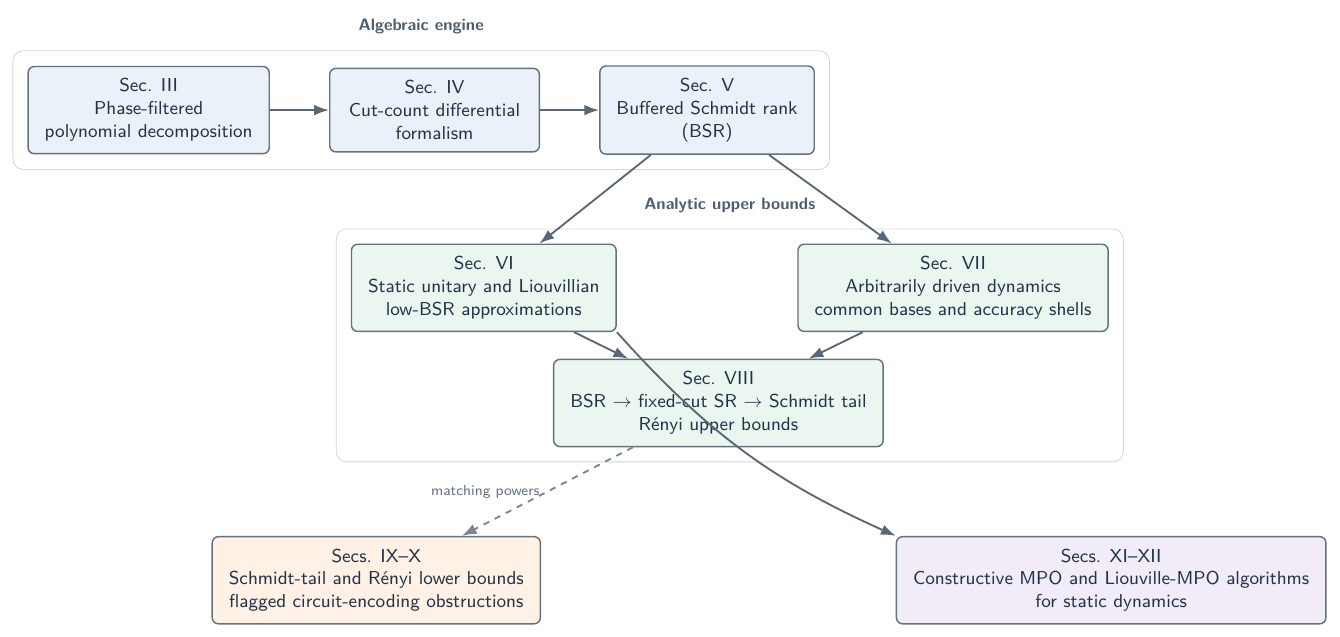}
 \caption{Road map of the paper.  Sections~III--V provide the algebraic
 engine.  Sections~VI and VII turn it into static and driven low-BSR
 approximations; Proposition~\ref{prop:basic-bsr-properties} and Sec.~VIII
 convert them into fixed-cut Schmidt-spectrum and R\'enyi bounds.  Sections~IX
 and X give matching entanglement-spectrum lower bounds and independent
 flagged-circuit obstructions.  Section~VII.B identifies structured drives
 with static-type compression bounds.  Sections~XI and XII give explicit
 unitary and Liouville-MPO algorithms, with the extensions in Secs.~XI.A and
 XII.A recovering the static scaling for efficiently represented structured
 driving.}
 \label{fig:main-results-roadmap}
\end{figure}

Figure~\ref{fig:main-results-roadmap} summarizes how the technical sections
fit together.

\FloatBarrier

\section{Preliminaries: polynomial decomposition}

Fix integers $r\ge 1$ and $Q\ge 0$. For every integer $m\ge1$, write
$\mathbb C_1^m:=\{\vec z\in\mathbb C^m:|z_j|=1\ \forall j\in[m]\}$.
Throughout the paper, a finite phase set or phase family appearing in a coefficient sum is understood as an indexed family of phase vectors; repeated vectors are retained as distinct entries, and its size counts entries. Expressions such as $\mZ\subseteq\mathbb C_1^m$ and $\sum_{\vec z\in\mZ}$ are interpreted in this indexed sense.
All partial exponential sums are interpreted as absolutely convergent subseries.
This is automatic for the scalar exponential at every fixed $\vec x$.

Consider $r$ variables $\vec{x}=(x_1,\ldots,x_r)$ on a given manifold
$\mathbb{R}^{r}_\mX\subseteq\mathbb R^r$.
We start from the expansion
\begin{align}
e^{\mathcal{X}}
= \sum_{\vec s} \mu_{\vec{s}}\  x_1^{s_1} \cdots x_r^{s_r} ,\quad
\mathcal{X}:= \sum_{j\in [r]} x_j , \quad  \mu_{\vec{s}}\ = \frac{1}{s_1!\cdots s_r!} , 
\label{eq:multinomial-expansion//0}
\end{align}
where $\vec{s}=(s_1,s_2,\ldots,s_r)\in \mathbb{Z}_{\ge 0}^{r}$.

We consider the following constraints on the exponents $\vec{s}$:
\begin{align}
\label{Def:check_mG-total}
\mS_Q:=\{\vec s  \in \mathbb{Z}_{\ge 0}^{r} \mid \sum_{j=1}^{r} s_j \le Q\}.
\end{align}
We also define 
\begin{align}
\delta_Q :=\sup_{\vec{x}\in \mathbb{R}^{r}_\mX} \sum_{\vec s \notin \mS_Q} \frac{\abs{x_1^{s_1} \cdots x_r^{s_r}}}{s_1!\cdots s_r!} ,
\end{align}
and
\begin{align}
\label{e^mathcal_X_mS_Q}
\br{ e^{\mathcal{X}} }_{\mS_Q} 
= \sum_{\vec s \in \mS_Q} \mu_{\vec{s}}\  x_1^{s_1} \cdots x_r^{s_r} . 
\end{align}

The uniform error bound below is finite if $\delta_Q<\infty$; this holds,
for example, when $\mathbb R^r_\mX$ is bounded.  Pointwise absolute convergence
alone does not imply that the supremum defining $\delta_Q$ is finite.
We prove the following theorem.
\begin{theorem} \label{thm:exp_decomposition_main}
For an appropriate choice of phase vectors $\vec{z}= (z_{1},z_{2},\ldots, z_{r})$, one can approximate $e^{\mathcal{X}}$ for $\vec{x}=(x_1,\ldots, x_r) \in \mathbb{R}^{r}_\mX$ by
\begin{align}
&e^{\mathcal{X}} = \sum_{j=1}^r \sum_{s=0}^{\floor{2Q/r}}   \sum_{\vec{z} \in \mZ_j}C_j(\vec{z}) \frac{(z_j x_j)^{s}}{s!} 
e^{\mX_{\vec{z}, \neq j}} + \delta(\vec{x})
\label{thm:exp_decomposition_main_approx}
\end{align}
where $\mZ_j$ is a finite indexed family of phase vectors in $\mathbb{C}_1^r$ satisfying
\begin{align}
\abs{\mZ_j} \le (2Q+1)^{\ceil{\log_2(r)}} ,\quad \abs{C_j(\vec{z})} \le 1 , 
\end{align}
and 
\begin{align}
\mathcal{X}_{\vec{z}, \neq j} :=\sum_{j' \in[r]: j' \neq j} z_{j'} x_{j'} . 
\end{align}
Moreover, the error $\delta(\vec{x})$ is upper-bounded by 
\begin{align}
\label{thm:exp_decomposition_main_error}
\abs{\delta(\vec{x})} \le 2 r (2Q+1)^{\ceil{\log_2(r)}} \delta_Q .
\end{align}
\end{theorem}

We also consider a generalization used in the later analysis.
Fix an integer $\tilde r\ge 1$ and subsets $u_p\subseteq[r]$ for $p\in[\tilde r]$.
Define the sums $\mathcal{X}_{u_p}$ and $S_{u_p}(\vec s)$ by
\begin{align}
\label{u_p/set_mathcal_X}
\mathcal{X}_{u_p} := \sum_{j\in u_p} x_j ,\quad S_{u_p}(\vec{s})  = \sum_{j\in u_p} s_j, \quad p=1,2,\ldots, \tilde{r}
\end{align}
where $S_{u_p}$ depends on $\vec{s}$. 
The subsets $\{u_p\}_{p=1}^{\tilde r}$ need not be pairwise disjoint. 

We also define
\begin{align}
\label{Def:check_mG-generalized}
\tilde{\mS}_Q:=\{\vec s  \in \mathbb{Z}_{\ge 0}^{r} \mid \sum_{p=1}^{\tilde r} S_{u_p}(\vec{s}) \le Q\}.
\end{align}
and 
\begin{align}
\tilde{\delta}_Q :=\sup_{\vec{x}\in \mathbb{R}^{r}_\mX}  \sum_{\vec s \notin \tilde{\mS}_Q} \frac{\abs{x_1^{s_1} \cdots x_r^{s_r}}}{s_1!\cdots s_r!} 
\end{align}
As above, the uniform error bound below is finite if $\tilde\delta_Q<\infty$.
We generalize Theorem~\ref{thm:exp_decomposition_main} as follows.
\begin{corol} \label{corol:exp_decomposition_main}
Under the same setup as in Theorem~\ref{thm:exp_decomposition_main}, we obtain
\begin{align}
\label{corol:exp_decomposition_main_equation}
&e^{\mathcal{X}} = \sum_{p=1}^{\tilde r} \sum_{s=0}^{\floor{2Q/\tilde{r}}}   \sum_{\vec z \in \tilde{\mZ}_p} \tilde C_{p}(\vec{z})
\frac{(\mathcal{X}_{\vec{z}, u_p})^s}{s!} 
e^{\mX_{\vec{z}, u_p^\co}} + \tilde{\delta}(\vec{x})
\end{align}
with
\begin{align}
\mathcal{X}_{\vec{z}, u_p} :=\sum_{j\in u_p} z_{j} x_j ,\quad
\mathcal{X}_{\vec{z}, u_p^\co} :=\sum_{j\in [r] \setminus u_p} z_{j} x_j ,
\end{align}
where $\tilde{\mZ}_p$ is a finite indexed family of phase vectors in $\mathbb{C}_1^r$ satisfying
\begin{align}
\abs{\tilde{\mZ}_p} \le (2Q+1)^{\ceil{\log_2(\tilde r)}} ,\quad \abs{\tilde C_p(\vec{z})} \le 1 , 
\end{align}
Moreover, the error $\tilde{\delta}(\vec{x})$ is upper-bounded by 
\begin{align}
\abs{\tilde{\delta}(\vec{x})} \le 2 \tilde{r} (2Q+1)^{\ceil{\log_2(\tilde r)}} \tilde{\delta}_Q .
\end{align}
\end{corol}

\subsection{Linear constraints and partial sums}

A primary task is to sum only those monomials whose index $\vec s$ satisfies linear constraints.
For an arbitrary $\mathcal S\subseteq\mathbb Z_{\ge0}^r$, we use the notation
$\br{e^{\mathcal X}}_{\mathcal S}:=\sum_{\vec s\in\mathcal S}\mu_{\vec s}x_1^{s_1}\cdots x_r^{s_r}$ whenever this subseries converges absolutely.
Given integers $\{c_j\}_{j=1}^r$, define the constrained index set
\begin{align}
\mathcal{G} := \Bigl\{ \vec s\in \mathcal{S} \ \Big|\ \sum_{j=1}^r c_j s_j \in \mathcal{I}  \subset \mathbb{Z} \Bigr\},
\label{eq:def-G-single}
\end{align}
where $\mathcal{S}$ is a subset of $\mathbb{Z}_{\ge 0}^r$ for which the indicated partial exponential converges absolutely.
Assume that there exist integers $S_{\min}\le S_{\max}$ such that
\begin{align}
\label{Def:parameter_S_max_min}
 \sum_{j=1}^r c_j s_j  \in [S_{\min}, S_{\max}]  \for \vec s\in \mathcal{S}.
\end{align}
Replacing $\mathcal I$ by $\mathcal I\cap[S_{\min},S_{\max}]$ does not change $\mathcal G$, so we assume without loss of generality that $\mathcal{I}\subseteq [S_{\min}, S_{\max}]$.
The corresponding partial sum is
\begin{align}
\br{ e^{\mathcal{X}} }_{\mathcal{G} } =  \sum_{\vec s \in \mathcal{G} } \mu_{\vec{s}}\  x_1^{s_1}\cdots x_r^{s_r}.
\label{eq:partial-sum}
\end{align}

More generally, let $\bar q\ge1$ be an integer.  For $\bar q$ linear constraints
with integer coefficients $\{c^{(q)}_j\}_{j\in[r]}$ and integer sets
$\mathcal I^{(q)}$, $q\in[\bar q]$, set
\begin{align}
&\mathcal{G} := \Bigl\{ \vec s\in \mathcal{S} \ \Big|\
\sum_{j=1}^r c^{(q)}_j s_j \in \mathcal{I}^{(q)} \subset \mathbb{Z} \ \ \forall q\in[\bar q] \Bigr\},\notag \\
&\sum_{j=1}^r c^{(q)}_j s_j \in [S_{\min}, S_{\max}]   \for \vec s\in \mathcal{S}  \quad \forall q \in[\bar{q}].
\label{eq:def-G-multi}
\end{align}
As above, we replace each $\mathcal I^{(q)}$ by its intersection with $[S_{\min},S_{\max}]$ without changing $\mathcal G$.

\subsection{Phase reweighting lemma: removing the linear constraints}

The key idea is to replace the constrained sum~\eqref{eq:partial-sum} by a
\emph{linear combination of partial exponentials} over $\mathcal S$, with the
additional linear constraints removed, using the phase-twisted sums
\begin{align}
\mathcal{X}_{\vec z}:=\sum_{j\in[r]} z_j x_j,\quad |z_j|=1.
\end{align}
This converts a hard combinatorial filter into a manageable number of evaluations of
$\br{e^{\mathcal{X}_{\vec z}}}_{\mS}$.

\begin{lemma}[Phase-reweighting decomposition]\label{Hugo's lemma}
For $\bar q\ge1$, assume the data and bounded-range condition in Eq.~\eqref{eq:def-G-multi}, and set $M= S_{\max}- S_{\min} + 1$. Then there exist an indexed family of phase vectors
\begin{align}
\vec z_w=(z_{w,1},\ldots,z_{w,r})\in\mathbb{C}^r \quad \textrm{with $|z_{w,j}|=1$},
\end{align}
and coefficients $C_{\mG,\vec z_w}$, indexed by $w=(t_1,\ldots,t_{\bar q})\in\{0,1,\ldots,M-1\}^{\bar q}$ (so in total $M^{\bar q}$ terms), such that
\begin{align}
\br{ e^{\mathcal{X}} }_{\mathcal{G} }
= \sum_{w\in\{0,1,\ldots,M-1\}^{\bar q}} C_{\mG, \vec z_w}  \br{e^{\mX_{\vec{z}_w}}}_{\mS},
\quad
\mX_{\vec{z}_w}:=\sum_{j\in[r]} z_{w,j} x_j,
\label{eq:efficient_decomp_constraint}
\end{align}
with $|C_{\mG,\vec z_w}|\le 1$ for all $w$.
The case $\bar q=0$ is understood as the identity decomposition with one entry, $\vec z=\vec 1$ and coefficient $1$.
\end{lemma}

{\bf Remark.}
Equation~\eqref{eq:efficient_decomp_constraint} is a \emph{discrete Fourier filter}: the constraints are enforced by averaging over roots of unity. The price is a multiplicative overhead of $M^{\bar q}$ evaluations of the phase-twisted partial exponential over $\mathcal S$, but each evaluation is structurally simple.

Writing $w\equiv(t_1,\ldots,t_{\bar q})$ with $t_q\in\{0,1,\ldots,M-1\}$ and $z_{t}:=e^{2\pi i t/M}$, one can choose
\begin{align}
z_{w,j}=\prod_{q=1}^{\bar q} z_{t_q}^{ c_j^{(q)}} \quad\text{and}\quad
C_{\mG,\vec z_w}=\frac{1}{M^{\bar q}}\prod_{q=1}^{\bar q}\br{\sum_{\kappa_q\in \mI^{(q)}} z_{t_q}^{-\kappa_q}}.
\end{align}
The bound $|C_{\mG,\vec z_w}|\le 1$ follows from the triangle inequality and $|\mI^{(q)}|\le M$.

\subsection{Proof of Lemma~\ref{Hugo's lemma}}
We first treat the single-constraint case ($\bar q=1$). Let us define the $M$-th roots of unity $z_t:=e^{2\pi i t/M}$ for $t=0,1,\ldots,M-1$. For any integer $S \in [S_{\min}, S_{\max}]$, the following discrete Fourier identity holds:
\begin{align}
\sum_{\kappa \in \mI} \sum_{t=0}^{M-1} e^{-2\pi i \kappa t /M}  e^{2\pi i S t/M}
= \begin{cases} M & \text{if } S \in \mI, \\
0 & \text{if } S\notin \mI ,
\end{cases}
\label{eq:Fourier_filter}
\end{align}
where we use $|\kappa - S| \le S_{\max} - S_{\min} < M$ from definition~\eqref{Def:parameter_S_max_min} and $M= S_{\max}- S_{\min} + 1$. 
Equivalently, with $z_t=e^{2\pi i t/M}$,
\begin{align}
\sum_{\kappa \in \mI} \sum_{t=0}^{M-1} z_t^{-\kappa}  z_t^{ S}
= \begin{cases} M & S \in \mI, \\
0 & S\notin \mI .
\end{cases}
\end{align}
Apply~\eqref{eq:Fourier_filter} with $S=\sum_{j\in[r]} c_j s_j$. For every $\vec s\in\mS$, the indicator of the constrained subset is
\begin{align}
\mathbbm{1}_{\mG}(\vec s)
&= \frac{1}{M}
\sum_{\kappa \in \mI} \sum_{t=0}^{M-1} z_t^{-\kappa}  z_t^{\sum_{j\in[r]} c_j s_j}
= \frac{1}{M}\sum_{\kappa \in \mI} \sum_{t=0}^{M-1} z_t^{-\kappa} \prod_{j\in [r]} z_t^{ c_j s_j}.
\label{eq:summation_rewrite_indicator}
\end{align}
Multiplying this identity by the corresponding series term and summing over $\vec s\in\mS$ yields
\begin{align}
\sum_{\vec{s} \in \mG}  \mu_{\vec{s}}
   x_1^{s_1} \cdots x_r^{s_r} 
&=\frac{1}{M}
    \sum_{\kappa \in \mI} \sum_{t=0}^{M-1} z_t^{-\kappa}
    \sum_{\vec{s}\in \mS}  \mu_{\vec{s}}\ \prod_{j\in [r]} 
    \bigl( z_t^{ c_j} x_j \bigr)^{s_j} \notag \\
&= \sum_{t=0}^{M-1} \br{ \frac{1}{M} \sum_{\kappa \in \mI} z_t^{-\kappa}}
    \sum_{\vec{s}\in \mS}  \mu_{\vec{s}}\ \prod_{j\in [r]} 
    \bigl( z_t^{ c_j} x_j \bigr)^{s_j} .
\label{eq:summation_rewrite_single}
\end{align}
This is of the target form~\eqref{eq:efficient_decomp_constraint} with $z_{w,j}=z_t^{ c_j}$ and $C_{\mG,\vec z_w}=\frac{1}{M}\sum_{\kappa\in\mI} z_t^{-\kappa}$.

For $\bar q\ge2$ constraints, multiply the individual Fourier-filter identities
and sum over their separate phase indices:
\begin{align}
\sum_{\vec{s} \in \mG}  \mu_{\vec{s}}
   x_1^{s_1} \cdots x_r^{s_r} 
&= \sum_{t_1=0}^{M-1}\cdots\sum_{t_{\bar q}=0}^{M-1}
\brr{ \frac{1}{M^{\bar q}}\prod_{q=1}^{\bar q}\br{\sum_{\kappa_q\in \mI^{(q)}} z_{t_q}^{-\kappa_q}} }
\br{e^{\sum_{j\in[r]} \prod_{q=1}^{\bar q} z_{t_q}^{ c_j^{(q)}} x_j }}_{\mS} .
\label{eq:summation_rewrite_multi}
\end{align}
Since $\bigl|\sum_{\kappa_q\in \mI^{(q)}} z_{t_q}^{-\kappa_q}\bigr|\le |\mI^{(q)}|\le M$ for each $q$, we have $|C_{\mG,\vec z_w}|\le 1$. The exchanges with the partial exponential are justified by absolute convergence. This proves the lemma. $\square$

\subsection{Decomposition of the set $\mS_Q$} \label{Decomp_the_set_mS_Q}

We introduce a binary label
\begin{align}
\vec{\sigma}_t=(\sigma_1,\dots,\sigma_t)\in\{0,1\}^t
\end{align}
and use it to define a hierarchical partition of the index set $[r]=\{1,\dots,r\}$.
The purpose of this construction is to partition the set $\mathcal S_Q$ according to a binary decision process, and to identify, for each $\vec s\in\mathcal S_Q$, an index $j\in[r]$ such that $s_j$ is controlled by the total sum $\sum_{j=1}^r s_j$.

First, set $\mathcal N_{\emptyset}:=[r]$ and recursively decompose the set $[r]$ into blocks
\begin{align}
\label{decomp_r_sigma_t}
[r]=\bigsqcup_{\vec{\sigma}_t\in\{0,1\}^t}\mathcal N_{\vec{\sigma}_t},
\end{align}
where the blocks satisfy
\begin{align}
\mathcal N_{\vec{\sigma}_{t-1}}
=
\mathcal N_{\vec{\sigma}_{t-1}\oplus 0}
\sqcup
\mathcal N_{\vec{\sigma}_{t-1}\oplus 1}.
\end{align}
Here $\oplus$ denotes the concatenation of a bit to the binary string.

More explicitly, suppose that
\begin{align}
\mathcal N_{\vec{\sigma}_{t-1}}=[x+1,x+l]
\end{align}
is an interval of length $l$. We then define its two children by
\begin{align}
\label{mN_definition_oplus}
\mathcal N_{\vec{\sigma}_{t-1}\oplus 0}
=
[x+1,\,x+\lceil l/2\rceil],
\qquad
\mathcal N_{\vec{\sigma}_{t-1}\oplus 1}
=
[x+\lceil l/2\rceil+1,\,x+l].
\end{align}
Thus, at each step, a block is divided into its left and right halves, up to rounding. If $\mathcal N_{\vec\sigma_t}=\emptyset$, both of its children are defined to be empty.

Let
\begin{align}
t_0:=\lceil \log_2 r\rceil.
\end{align}
After $t_0$ steps, every nonempty block contains exactly one element:
\begin{align}
|\mathcal N_{\vec{\sigma}_{t_0}}|=1
\qquad
\text{for every nonempty } \mathcal N_{\vec{\sigma}_{t_0}}.
\end{align}
Some blocks may be empty, but for every $j\in[r]$, there exists a unique path
\begin{align}
\vec{\sigma}_{t_0}\in\{0,1\}^{t_0}
\end{align}
such that
\begin{align}
\mathcal N_{\vec{\sigma}_{t_0}}=\{j\} .
\end{align}
In the following, we denote $\vec{\sigma}_{t_0}$ by $\vec{\sigma}$ for simplicity:
\begin{align}
\vec{\sigma}_{t_0}=\vec{\sigma}.
\end{align}

 \begin{figure*}[tt]
\centering
\includegraphics[clip, scale=0.65]{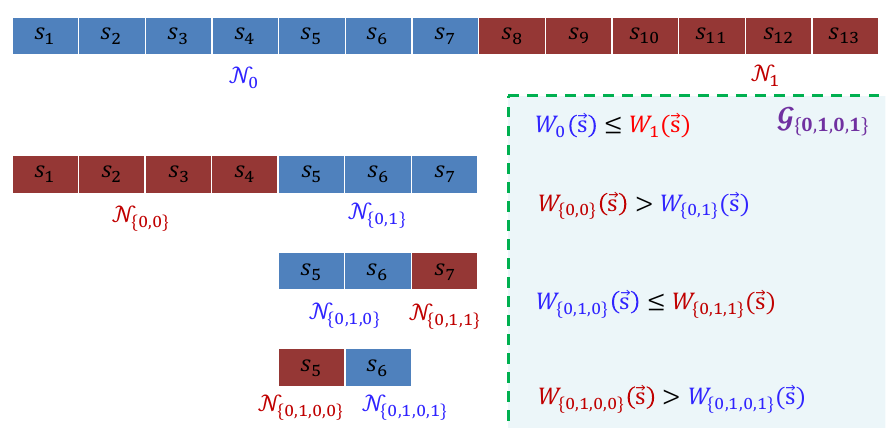}
\caption{Illustration of the hierarchical partition of the index set $[r]$.}
\label{fig:layer_decomp}
\end{figure*}

Next, for each block $\mathcal N_{\vec{\sigma}_t}$, we define the partial sum
\begin{align}
\label{Eq_W_sigma_t_vec_s}
W_{\vec{\sigma}_t}(\vec s)
:=
\sum_{j\in\mathcal N_{\vec{\sigma}_t}} s_j .
\end{align}
Using these partial sums, we recursively partition $\mathcal S_Q$.
We begin with
\begin{align}
\label{mathcal_G_emptyset_def}
\mathcal G_{\emptyset}
:= \mS_Q , 
\end{align}
and for each $\vec{\sigma}_t$ such that $|\mN_{\vec{\sigma}_t}|\ge 2$, define
\begin{align}
\label{mathcal_G_def_oplus_0}
\mathcal G_{\vec{\sigma}_t\oplus 0}
:=
\left\{
\vec s\in \mG_{\vec{\sigma}_t}
\;\middle|\;
W_{\vec{\sigma}_t\oplus 0}(\vec s)
\le
W_{\vec{\sigma}_t\oplus 1}(\vec s)
\right\},
\end{align}
and
\begin{align}
\label{mathcal_G_def_oplus_1}
\mathcal G_{\vec{\sigma}_t\oplus 1}
:=
\left\{
\vec s\in\mG_{\vec{\sigma}_t}
\;\middle|\;
W_{\vec{\sigma}_t\oplus 0}(\vec s)
>
W_{\vec{\sigma}_t\oplus 1}(\vec s)
\right\} . 
\end{align}
For $|\mN_{\vec{\sigma}_t}|=1$, Eq.~\eqref{mN_definition_oplus} gives
$\mN_{\vec{\sigma}_t\oplus1}=\emptyset$, so we define
\begin{align}
\label{mathcal_G_def_oplus_1_mN=1}
\mathcal G_{\vec{\sigma}_t\oplus 0}
:=\mG_{\vec{\sigma}_t}, \quad \mathcal G_{\vec{\sigma}_t\oplus 1}
:=\emptyset, \for |\mN_{\vec{\sigma}_t}| = 1.
\end{align}
If $|\mN_{\vec{\sigma}_t}|=0$, both child sets are defined to be empty.

By construction, these two sets are disjoint and satisfy
\begin{align}
\mathcal G_{\vec{\sigma}_t}
=
\mathcal G_{\vec{\sigma}_t\oplus 0}
\sqcup
\mathcal G_{\vec{\sigma}_t\oplus 1}.
\end{align}
Therefore, iterating this decomposition yields a partition
\begin{align}
\mathcal S_Q
=
\bigsqcup_{\vec{\sigma}\in\{0,1\}^{t_0}}
\mathcal G_{\vec{\sigma}},
\end{align}
and hence
\begin{align}
\label{sum_in_S_Q}
\sum_{\vec s\in\mathcal S_Q}
=
\sum_{\vec{\sigma}\in\{0,1\}^{t_0}}
\sum_{\vec s\in\mathcal G_{\vec{\sigma}}},
\end{align}

The meaning of this partition is the following.
For each $\vec s\in\mathcal S_Q$, at each genuine split of a block containing
at least two indices, we choose the child block with the smaller partial sum,
choosing the left child in case of a tie.
Indeed, if
\begin{align}
A=W_{\vec{\sigma}_t\oplus 0}(\vec s),
\qquad
B=W_{\vec{\sigma}_t\oplus 1}(\vec s),
\end{align}
then the selected child satisfies
\begin{align}
\min\{A,B\}\le \frac{A+B}{2}
=
\frac{W_{\vec{\sigma}_t}(\vec s)}{2}.
\end{align}
Hence, along the path determined by $\vec s$, the partial sum is at most half its previous value
after each genuine split.  Once a singleton is reached, the remaining
padding steps leave its partial sum unchanged.

Now let $\vec{\sigma}$ be the terminal path associated with $\vec s$, and write
\begin{align}
\mathcal N_{\vec{\sigma}}=\{j_{\vec{\sigma}}\} \qquad
\text{for every nonempty } \mG_{\vec{\sigma}}.
\end{align}
Since the block at depth $t_0$ consists of a single index, we have
\begin{align}
\label{W_vec_sigma_vec_s}
W_{\vec{\sigma}}(\vec s)=s_{j_{\vec{\sigma}}}.
\end{align}

Let $d_{\vec\sigma}$ be the first depth at which the selected block becomes a singleton. The balanced splitting rule implies
$d_{\vec\sigma}\ge\lfloor\log_2 r\rfloor$: after fewer splits, even the smallest descendant block has cardinality at least two. Applying the factor-$(1/2)$ estimate at each of the $d_{\vec\sigma}$ genuine splits gives
\begin{align}
s_{j_{\vec{\sigma}}}
\le
2^{-d_{\vec\sigma}}\sum_{j=1}^r s_j
\le
2^{-\lfloor\log_2 r\rfloor}\sum_{j=1}^r s_j
\le
\frac{2}{r}\sum_{j=1}^r s_j.
\end{align}

Finally, since $\vec s\in\mathcal S_Q$ implies
\begin{align}
\sum_{j=1}^r s_j\le Q,
\end{align}
we conclude that
\begin{align}
\label{s_j_vec/sigma_upp}
s_{j_{\vec{\sigma}}}
\le
\frac{2Q}{r}.
\end{align}

Thus, for every $\vec s\in\mathcal S_Q$, the binary selection procedure
produces an index $j\in[r]$ with $s_j\le2Q/r$.

\subsection{Completing the proof of Theorem~\ref{thm:exp_decomposition_main}}\label{sec:completing the proof of main technique}

We now combine the decomposition~\eqref{sum_in_S_Q} with Eq.~\eqref{e^mathcal_X_mS_Q}, which gives 
\begin{align}
\label{e^mathcal_X_mS_Q_decomp}
\br{ e^{\mathcal{X}} }_{\mS_Q} 
= \sum_{\vec{\sigma}\in\{0,1\}^{t_0}}
\sum_{\vec s\in\mathcal G_{\vec{\sigma}}} \mu_{\vec{s}}\  x_1^{s_1} \cdots x_r^{s_r} . 
\end{align}
For every node on a terminal path whose block contains at least two indices, set
$\Delta_{\vec\sigma_t}(\vec s):=W_{\vec\sigma_t\oplus0}(\vec s)-W_{\vec\sigma_t\oplus1}(\vec s)$.
The branch bit $0$ imposes $\Delta_{\vec\sigma_t}(\vec s)\in[-Q,0]$, whereas the branch bit $1$ imposes $\Delta_{\vec\sigma_t}(\vec s)\in[1,Q]$, because all quantities are integers. Once a singleton is reached, the remaining levels are padded by the dummy constraint $0\in\{0\}$. Terminal paths with $\mathcal N_{\vec\sigma}=\emptyset$ have $\mathcal G_{\vec\sigma}=\emptyset$ and are omitted. Thus each remaining $\mathcal G_{\vec\sigma}$ is described by at most $t_0=\ceil{\log_2 r}$ linear constraints, padded to exactly $t_0$, and every constraint value lies in $[-Q,Q]$. In particular,
\begin{align}
W_{\vec{\sigma}_t\oplus 0}(\vec s) - W_{\vec{\sigma}_t\oplus 1}(\vec s)
= \sum_{j\in\mathcal N_{\vec{\sigma}_t\oplus 0}} s_j - \sum_{j\in\mathcal N_{\vec{\sigma}_t\oplus 1}} s_j  \in [-Q,Q].
\end{align}
Comparing with definition~\eqref{eq:def-G-multi} and $M=S_{\max}-S_{\min} +1$, we can set
\begin{align}
M = 2Q+ 1.
\end{align}

Therefore, by using the phase-reweighting lemma (Lemma~\ref{Hugo's lemma}) with $\mS=\mS_Q$, we decompose 
\begin{align}
\sum_{\vec s\in\mathcal G_{\vec{\sigma}}} \mu_{\vec{s}}\  x_1^{s_1} \cdots x_r^{s_r} 
=  \sum_{w=1}^{M^{\ceil{\log_2(r)}}} C_{\mG_{\vec{\sigma}}, \vec z_{(w,\vec \sigma)}}  \br{e^{\mX_{\vec{z}_{(w,\vec \sigma)}}}}_{\mS_Q}.
\end{align}
For every $\vec s\in\mS_Q$, the Fourier filter gives the coefficient identity
\[
\sum_{w=1}^{M^{\ceil{\log_2(r)}}}
C_{\mG_{\vec\sigma},\vec z_{(w,\vec\sigma)}}
\prod_{j=1}^r z_{(w,\vec\sigma),j}^{s_j}
=\mathbbm{1}_{\mG_{\vec\sigma}}(\vec s).
\]
Consequently, terms outside $\mG_{\vec\sigma}$ cancel only after the phase sum is taken. Since every surviving index satisfies Eq.~\eqref{s_j_vec/sigma_upp}, the combined filtered sum may then be restricted to $s_{j_0}\le\floor{2Q/r}$, where $j_0=j_{\vec\sigma}$. We therefore expand $ \sum_{w=1}^{M^{\ceil{\log_2(r)}}} C_{\mG_{\vec{\sigma}}, \vec z_{(w,\vec \sigma)}} \br{e^{\mX_{\vec{z}_{(w,\vec\sigma)}}}}_{\mS_Q}$ as follows:
\begin{align}
& \sum_{w=1}^{M^{\ceil{\log_2(r)}}} C_{\mG_{\vec{\sigma}}, \vec z_{(w,\vec \sigma)}} \sum_{s_{j_0}=0}^{\floor{2Q/r}}  \sum_{\vec s_{\neq j_0} : \vec{s} \in\mS_{Q}} \mu_{\vec{s}}\  (z_{(w,\vec \sigma),j_0}x_{j_0})^{s_{j_0}} \prod_{j\neq j_0} \br{z_{(w,\vec \sigma),j} x_j}^{s_j}   \notag \\
&= \sum_{w=1}^{M^{\ceil{\log_2(r)}}} C_{\mG_{\vec{\sigma}}, \vec z_{(w,\vec \sigma)}}  \sum_{s_{j_0}=0}^{\floor{2Q/r}} \frac{(z_{(w,\vec \sigma),j_0}x_{j_0})^{s_{j_0}} }{s_{j_0}!} \sum_{\vec s_{\neq j_0} : \vec{s} \in\mS_{Q}}  \mu_{\vec{s}_{\neq j_0}}  \prod_{j\neq j_0} \br{z_{(w,\vec \sigma),j} x_j}^{s_j}  
\notag \\
&= \sum_{w=1}^{M^{\ceil{\log_2(r)}}} C_{\mG_{\vec{\sigma}}, \vec z_{(w,\vec \sigma)}}
\sum_{s_{j_0}=0}^{\floor{2Q/r}}
\frac{(z_{(w,\vec \sigma),j_0}x_{j_0})^{s_{j_0}}}{s_{j_0}!}
 e^{\mX_{\vec{z}_{(w,\vec \sigma)}, \neq j_0}} \notag\\
&\quad - \sum_{w=1}^{M^{\ceil{\log_2(r)}}} C_{\mG_{\vec{\sigma}}, \vec z_{(w,\vec \sigma)}}
\sum_{s_{j_0}=0}^{\floor{2Q/r}}
\frac{(z_{(w,\vec \sigma),j_0}x_{j_0})^{s_{j_0}}}{s_{j_0}!}
\sum_{\vec s_{\neq j_0} : \vec{s} \notin\mS_{Q}}
\mu_{\vec{s}_{\neq j_0}}
\prod_{j\neq j_0} \br{z_{(w,\vec \sigma),j} x_j}^{s_j}.
\end{align}

Moreover, we have 
\begin{align}
&\abs{\sum_{w=1}^{M^{\ceil{\log_2(r)}}} C_{\mG_{\vec{\sigma}}, \vec z_{(w,\vec \sigma)}} \sum_{s_{j_0}=0}^{\floor{2Q/r}}  \frac{(z_{(w,\vec \sigma),j_0}x_{j_0})^{s_{j_0}}}{s_{j_0}!}  \sum_{\vec s_{\neq j_0} : \vec{s} \notin\mS_{Q}}  \mu_{\vec{s}_{\neq j_0}}  \prod_{j\neq j_0} \br{z_{(w,\vec \sigma),j} x_j}^{s_j}  }  \notag \\
&\le \sum_{w=1}^{M^{\ceil{\log_2(r)}}}  \sum_{s_{j_0}=0}^{\floor{2Q/r}}  \frac{|x_{j_0}|^{s_{j_0}}}{s_{j_0}!}  \sum_{\vec s_{\neq j_0} : \vec{s} \notin\mS_{Q}}  \mu_{\vec{s}_{\neq j_0}}  \prod_{j\neq j_0}|x_j|^{s_j}   \notag \\
&\le M^{\ceil{\log_2(r)}} \sum_{\vec s : \vec{s} \notin\mS_{Q}}  \mu_{\vec{s}}  \prod_{j=1}^r|x_j|^{s_j} \le M^{\ceil{\log_2(r)}}  \delta_Q ,
\end{align}
where we use $\abs{C_{\mG_{\vec{\sigma}}, \vec z_{(w,\vec \sigma)}}}\le 1$.
Each terminal path $\vec\sigma$ with $\mG_{\vec\sigma}\ne\emptyset$
corresponds to a unique index $j_{\vec\sigma}$.
Therefore, we can approximate Eq.~\eqref{e^mathcal_X_mS_Q_decomp} by 
\begin{align}
\label{approx_construction_mG_s_j}
\sum_{j=1}^r \sum_{s_j=0}^{\floor{2Q/r}}   \sum_{w=1}^{M^{\ceil{\log_2(r)}}} C_{\mG_{\vec{\sigma}_j}, \vec z_{(w,\vec \sigma_j)}}  \frac{(z_{(w,\vec \sigma_j),j} x_j)^{s_j}}{s_j!} 
e^{\mX_{\vec{z}_{(w,\vec \sigma_j)}, \neq j}}
\end{align}
with error at most 
\begin{align}
\label{approx_construction_mG_s_j_error}
\sum_{j=1}^r M^{\ceil{\log_2(r)}}  \delta_Q \le r M^{\ceil{\log_2(r)}}  \delta_Q ,
\end{align}
where we define $\vec{\sigma}_j$ as the unique path in $\{0,1\}^{t_0}$ which gives $\mN_{\vec{\sigma}_j}=\{j\}$.
Taking $\mZ_j$ to be the indexed family $\{\vec z_{(w,\vec\sigma_j)}\}_w$ and setting
$C_j(\vec z_{(w,\vec\sigma_j)}):=C_{\mG_{\vec\sigma_j},\vec z_{(w,\vec\sigma_j)}}$ gives the phase families and coefficients in Theorem~\ref{thm:exp_decomposition_main}.

Finally, the triangle inequality gives 
\begin{align}
&\abs{ e^{\mathcal{X}}  -  \sum_{j=1}^r \sum_{s_j=0}^{\floor{2Q/r}}   \sum_{w=1}^{M^{\ceil{\log_2(r)}}} C_{\mG_{\vec{\sigma}_j}, \vec z_{(w,\vec \sigma_j)}}  \frac{(z_{(w,\vec \sigma_j),j} x_j)^{s_j}}{s_j!} 
e^{\mX_{\vec{z}_{(w,\vec \sigma_j)}, \neq j}}}  \notag \\
&\le  \abs{ e^{\mathcal{X}}  -   \br{ e^{\mathcal{X}} }_{\mS_Q} }
 +  \abs{ \br{ e^{\mathcal{X}} }_{\mS_Q}  -  \sum_{j=1}^r \sum_{s_j=0}^{\floor{2Q/r}}   \sum_{w=1}^{M^{\ceil{\log_2(r)}}} C_{\mG_{\vec{\sigma}_j}, \vec z_{(w,\vec \sigma_j)}}  \frac{(z_{(w,\vec \sigma_j),j} x_j)^{s_j}}{s_j!} 
e^{\mX_{\vec{z}_{(w,\vec \sigma_j)}, \neq j}}} \notag \\ 
&\le \bigl(1+rM^{\ceil{\log_2(r)}}\bigr)\delta_Q
\le  2 r M^{\ceil{\log_2(r)}}   \delta_Q ,
\end{align}
where the first term is the original Taylor tail and the second is the sum of the completion errors in Eq.~\eqref{approx_construction_mG_s_j_error}. This proves the error bound~\eqref{thm:exp_decomposition_main_error} for $|\delta(\vec{x})|$ in Eq.~\eqref{thm:exp_decomposition_main_approx}. 
This completes the proof. $\square$

\subsection{Proof of Corollary~\ref{corol:exp_decomposition_main}}

We prove Corollary~\ref{corol:exp_decomposition_main} by applying the same
argument with the replacement  
\begin{align}
\{ s_j \}_{j=1}^r \longrightarrow  \{ S_{u_p}(\vec{s}) \}_{p=1}^{\tilde r}
\end{align}
We show how to decompose $\tilde{\mS}_Q$ in analogy with Sec.~\ref{Decomp_the_set_mS_Q}. 
First, we replace Eq.~\eqref{decomp_r_sigma_t} by
\begin{align}
\label{decomp_tilde_r_sigma_t}
[\tilde r]=\bigsqcup_{\vec{\sigma}_t\in\{0,1\}^t}\mathcal N_{\vec{\sigma}_t},
\end{align}
Analogously to Eq.~\eqref{Eq_W_sigma_t_vec_s}, define
\begin{align}
\label{tilde:Eq_W_sigma_t_vec_s}
\tilde{W}_{\vec{\sigma}_t}(\vec s)
:=
\sum_{p \in\mathcal N_{\vec{\sigma}_t}} S_{u_p}(\vec{s}).
\end{align}

Equations~\eqref{mathcal_G_emptyset_def}--\eqref{mathcal_G_def_oplus_1} are replaced by 
\begin{align}
\label{mathcal_G_emptyset_def_tilde}
\tilde{\mathcal G}_{\emptyset}
:= \tilde{\mS}_Q , 
\end{align}
and, for each $\vec\sigma_t$ with $|\mN_{\vec\sigma_t}|\ge2$, by
\begin{align}
\label{mathcal_G_def_oplus_0_tilde}
\tilde{\mathcal G}_{\vec{\sigma}_t\oplus 0}
:=
\left\{
\vec s\in \tilde{\mG}_{\vec{\sigma}_t}
\;\middle|\;
\tilde{W}_{\vec{\sigma}_t\oplus 0}(\vec s)
\le
\tilde{W}_{\vec{\sigma}_t\oplus 1}(\vec s)
\right\},
\quad 
\tilde{\mathcal G}_{\vec{\sigma}_t\oplus 1}
:=
\left\{
\vec s\in \tilde{\mG}_{\vec{\sigma}_t}
\;\middle|\;
\tilde{W}_{\vec{\sigma}_t\oplus 0}(\vec s)
>
\tilde{W}_{\vec{\sigma}_t\oplus 1}(\vec s)
\right\}. 
\end{align}
We adopt the same definition as Eq.~\eqref{mathcal_G_def_oplus_1_mN=1} for $|\mN_{\vec{\sigma}_t}| = 1$, and the same empty-node convention for $|\mN_{\vec{\sigma}_t}|=0$. 
At depth $\lceil\log_2\tilde r\rceil$, write
$\mathcal N_{\vec\sigma}=\{p_{\vec\sigma}\}$ for each nonempty terminal block,
and let $\vec\sigma_p$ denote the unique terminal path with block $\{p\}$.
The analogues of Eq.~\eqref{W_vec_sigma_vec_s} and inequality~\eqref{s_j_vec/sigma_upp} are  
\begin{align}
\label{tilde_W_vec_sigma_vec_s}
\tilde{W}_{\vec{\sigma}}(\vec s)=S_{u_{p_{\vec{\sigma}}}}(\vec{s}), 
\end{align}
and 
\begin{align}
\label{tilde_s_j_vec/sigma_upp}
S_{u_{p_{\vec{\sigma}}}}(\vec{s})
\le
\floor{\frac{2Q}{\tilde r}}.
\end{align}

For each terminal path $\vec\sigma_p$, every active comparison is a linear constraint in $\vec s$. Since all $S_{u_p}(\vec s)$ are nonnegative and their total is at most $Q$ on $\tilde{\mS}_Q$, every comparison value lies in $[-Q,Q]$. As in the proof of Theorem~\ref{thm:exp_decomposition_main}, singleton levels are padded by dummy constraints. Hence Lemma~\ref{Hugo's lemma} applies with $M=2Q+1$ and at most $\ceil{\log_2\tilde r}$ nontrivial constraints.

The Fourier coefficient identity first restricts the combined phase sum to
the corresponding set $\tilde{\mG}_{\vec\sigma_p}$ and therefore to
$S_{u_p}(\vec s)\le\floor{2Q/\tilde r}$.  For each phase entry, we now extend
the sum from $\tilde{\mS}_Q\cap\{\vec s:S_{u_p}(\vec s)\le\floor{2Q/\tilde r}\}$
to all nonnegative integer multi-indices satisfying
$S_{u_p}(\vec s)\le\floor{2Q/\tilde r}$.  This removes the extra restrictions
that $\tilde{\mS}_Q$ can impose on the individual variables.
After this extension, the multinomial identity
\[
\sum_{\substack{(s_j)_{j\in u_p}\in\mathbb Z_{\ge0}^{u_p}\\
                  \sum_{j\in u_p}s_j=s}}
\prod_{j\in u_p}\frac{(z_jx_j)^{s_j}}{s_j!}
=\frac{(\mX_{\vec z,u_p})^s}{s!}
\]
regroups the variables in $u_p$ for each integer $s\ge0$, with the standard
convention that the $s=0$ term equals $1$ when $u_p=\emptyset$.  Summing the
complementary variables gives the full exponential, yielding the approximation
\begin{align}
\sum_{p=1}^{\tilde r} \sum_{s=0}^{\floor{2Q/ \tilde{r}}}   \sum_{w=1}^{M^{\ceil{\log_2(\tilde{r})}}} C_{\tilde{\mG}_{\vec{\sigma}_p}, \vec z_{(w,\vec \sigma_p)}}
 \frac{(\mX_{\vec{z}_{(w,\vec \sigma_p)},u_p})^{s}}{s!}
 e^{\mX_{\vec{z}_{(w,\vec \sigma_p)}, u_p^\co}}.
\end{align}
More explicitly, for fixed $p$ and $w$, the added completion term, before
multiplication by its Fourier coefficient, is exactly the subseries below,
where $\vec z=\vec z_{(w,\vec\sigma_p)}$:
\[
\sum_{\substack{(s_j)_{j\in u_p}\in\mathbb Z_{\ge0}^{u_p},\ 
                  \sum_{j\in u_p}s_j\le\floor{2Q/\tilde r},\\
                  (s_j)_{j\notin u_p}\in\mathbb Z_{\ge0}^{u_p^\co},\ 
                  \vec s\notin\tilde{\mS}_Q}}
\mu_{\vec s}\prod_{j=1}^r(z_jx_j)^{s_j}.
\]
Its absolute value is bounded by $\tilde\delta_Q$.  Multiplication by the
Fourier coefficient does not increase this bound, since its absolute value is
at most one.  Summing over the terminal paths and phase entries, the total
completion error satisfies
\begin{align}
\label{approx_construction_mG_s_j_error_corol}
\sum_{p=1}^{\tilde r} M^{\ceil{\log_2(\tilde r)}}   \tilde \delta_Q
\le \tilde r M^{\ceil{\log_2(\tilde r)}}  \tilde \delta_Q .
\end{align}
The common range used in this application of Lemma~\ref{Hugo's lemma} is justified by
\begin{align}
\tilde{W}_{\vec{\sigma}_t\oplus 0}(\vec s) - \tilde{W}_{\vec{\sigma}_t\oplus 1}(\vec s)  =
\sum_{p \in\mathcal N_{\vec{\sigma}_t\oplus 0}} S_{u_p}(\vec{s}) - \sum_{p \in\mathcal N_{\vec{\sigma}_t\oplus 1}} S_{u_p}(\vec{s}) \in [-Q,Q].
\end{align}
Taking $\tilde{\mZ}_p$ to be the resulting indexed phase family and attaching the displayed Fourier coefficients defines $\tilde C_p$ in Corollary~\ref{corol:exp_decomposition_main}.
Adding the original tail $|e^{\mathcal X}-\br{e^{\mathcal X}}_{\tilde{\mS}_Q}|\le\tilde\delta_Q$ yields
\[
|\tilde\delta(\vec x)|
\le \bigl(1+\tilde rM^{\ceil{\log_2\tilde r}}\bigr)\tilde\delta_Q
\le 2\tilde r(2Q+1)^{\ceil{\log_2\tilde r}}\tilde\delta_Q.
\]
This proves Corollary~\ref{corol:exp_decomposition_main}.
\section{Differential formalism for quantum dynamics}\label{sec:diff-form-SR}

In this section, we show that Theorem~\ref{thm:exp_decomposition_main} provides a powerful mechanism to control the Schmidt rank of operators generated from a local Hamiltonian by filtering the relevant differential monomials.

\subsection{Parameterization and differential extractor}\label{subsec:param-diff}
Let us parameterize the Hamiltonian by independent scalars attached to the blocks $H_{p,p'}$:
\begin{align}
\label{parameterize_Hamilaonitn_lamnda}
&H(\vec\lambda)  := V_0 + \sum_{0\le p < p' \le \ell} \lambda_{p,p'}  H_{p,p'} ,
\notag \\
&V_0 := \sum_{0\le p \le \ell}  H_{p,p} ,\quad  
\vec\lambda=\{\lambda_{p,p'}\}_{0\le p < p' \le \ell} .
\end{align}
Indices range over $\{0,1,\ldots,\ell\}$, hence the number of parameters is $N=\frac{\ell(\ell+1)}{2}$.  
Define the ``total derivative'' operator
\begin{align}
\mathcal{D}  :=  \sum_{0\le p < p' \le \ell} \frac{\partial}{\partial \lambda_{p,p'}} .
\end{align}

Let $\Phi$ be a function or functional for which the operator-valued map
$F(\vec\lambda):=\Phi[H(\vec\lambda)]$ has a norm-absolutely convergent
Taylor expansion on a neighborhood of the closed unit polydisc.
For such $F$, $\mD$ generates simultaneous translation in all coordinates;
whenever the directional Taylor series about $\vec\lambda$ converges in norm at
$\vec\lambda+\vec{\mathbf{1}}$,
\begin{align}
\bigl(e^{\mD} F\bigr)(\vec\lambda)  =  F(\vec\lambda+\vec{\mathbf{1}}), 
\quad \vec{\mathbf{1}}:=\{ 1 \}_{0\le p < p'\le\ell}.
\end{align}
Evaluating the translation identity at the origin gives
\begin{align}
e^{\mathcal{D}}  \Phi\bigl[H(\vec\lambda)\bigr]\Big|_{\vec\lambda=\vec 0}
 =  \Phi\bigl[H(\vec{\mathbf{1}})\bigr]
 =  \Phi[H] .
\label{eq:shift-identity}
\end{align}
Expanding the exponential in \eqref{eq:shift-identity} yields
\begin{align}
\Phi[H]
 =  \sum_{m=0}^\infty \frac{\mD^m}{m!} \Phi\bigl[H(\vec\lambda)\bigr]\Big|_{\vec\lambda=\vec 0}.
\label{eq:diff-expansion}
\end{align}

We relabel the pair index $(p,p')$ by a single index $j$.
Fix a bijection $j\leftrightarrow(p_j,p'_j)$ for $j\in[N]$ ($N=\frac{\ell(\ell+1)}{2}$) and write
$\lambda_j:=\lambda_{p_j,p'_j}$, $\partial_{\lambda_j}:=\partial/\partial\lambda_{p_j,p'_j}$.
Here $\mD^m=(\sum_{j=1}^N \partial_{\lambda_j})^m$ expands multinomially into mixed partials.  
Writing the multi-index $\vec s=(s_1,\ldots,s_N)$,
we obtain
\begin{align}
e^{\mD}
 =\sum_{\vec s\in\mathbb Z_{\ge0}^N} \mu_{\vec{s}} 
\partial_{\lambda_1}^{s_1}\cdots \partial_{\lambda_N}^{s_N},
\quad
N=\frac{\ell(\ell+1)}{2}, 
\label{eq:Dm-multinomial}
\end{align}
where we use Eq.~\eqref{eq:multinomial-expansion//0}. 
Thus, we obtain the correspondence 
\begin{align}
x_j\leftrightarrow \partial_{\lambda_j}, \quad 
\mX \leftrightarrow \mD 
\end{align}
with the variables in Theorem~\ref{thm:exp_decomposition_main}.

For a family of Hamiltonians labeled by an additional variable $x$, we similarly define
\begin{align}
\label{parameterize_Hamilaonitn_lamnda_x_dependent-initial}
&H^{(x)}(\vec\lambda)  := V^{(x)}_0 + \sum_{0\le p < p' \le \ell} \lambda_{p,p'}  H^{(x)}_{p,p'} ,
\notag \\
&V^{(x)}_0 := \sum_{0\le p \le \ell}  H^{(x)}_{p,p}.
\end{align}
The extension to functionals of this family is given in
Sec.~\ref{sec:practical-generalization}.

%


\subsection{Cut-count constraints}\label{sec:Cut-count constraints}

Recall the parametrization
\begin{align}
 H(\vec\lambda)
 =V_0+\sum_{j=1}^{N}\lambda_j H_{p_j,p'_j},
 \qquad
 \mD=\sum_{j=1}^{N}\partial_{\lambda_j},
 \qquad
 V_0=\sum_{q=0}^{\ell}H_{q,q},
 \label{eq:H-lambda-cut-count}
\end{align}
where $0\le p_j<p'_j\le\ell$ and $N=\ell(\ell+1)/2$.
We assume that $\Phi[H(\vec\lambda)]$ has a norm-absolutely convergent
Taylor expansion on a neighborhood of the closed unit polydisc.

For the cut $(p-1,p)$, set
\begin{align}
 u_p
 :=\{j\in[N]\mid p_j<p\le p'_j\},
 \qquad p=1,\ldots,\ell,
 \label{eq:up-def-revised}
\end{align}
and, for $\vec z=(z_1,\ldots,z_N)\in\mathbb C_1^N$, write
\begin{align}
 H_{\vec z,p}
 &:=\sum_{j\in u_p}z_jH_{p_j,p'_j},
 \notag\\
 H_{\vec z,<p}
 &:=\sum_{q=0}^{p-1}H_{q,q}
   +\sum_{j:\,p'_j<p}z_jH_{p_j,p'_j},
 \notag\\
 H_{\vec z,>p}
 &:=\sum_{q=p}^{\ell}H_{q,q}
   +\sum_{j:\,p_j\ge p}z_jH_{p_j,p'_j}.
 \label{def_H_vec_p_<p_>p}
\end{align}
For a multi-index $\vec s\in\mathbb Z_{\ge0}^N$, define its cut weight by
\begin{align}
 w(\vec s)
 :=\sum_{j=1}^{N}(p'_j-p_j)s_j,
 \label{eq:cut-weight-short}
\end{align}
and define the norm-absolute cut tail
\begin{align}
 \widetilde\delta_Q(\Phi)
 :=
 \sum_{\substack{\vec s\in\mathbb Z_{\ge0}^N\\w(\vec s)>Q}}
 \frac{1}{s_1!\cdots s_N!}
 \left\|
 \left.
 \partial_{\lambda_1}^{s_1}\cdots\partial_{\lambda_N}^{s_N}
 \Phi[H(\vec\lambda)]
 \right|_{\vec\lambda=\vec 0}
 \right\|.
 \label{eq:absolute-cut-tail-revised}
\end{align}
Finally, abbreviate
\begin{align}
 \Phi_{p,\vec z}^{(s)}
 :=\frac{1}{s!}
 \left.
 \left(\frac{d}{d\lambda}\right)^s
 \Phi\!\left[
 H_{\vec z,<p}+H_{\vec z,>p}+\lambda H_{\vec z,p}
 \right]
 \right|_{\lambda=0}.
 \label{eq:Phi-p-z-s}
\end{align}

\begin{prop}[Cut-count decomposition]\label{Prop:buffered_schmidt_est}
For every $Q\in\mathbb Z_{\ge0}$, there exist finite indexed families
$\widetilde{\mZ}_p$ of phase vectors in $\mathbb C_1^N$ and coefficients
$\widetilde C_p(\vec z)$ such that
\begin{align}
 \Phi[H]
 =\sum_{p=1}^{\ell}
   \sum_{s=0}^{\lfloor 2Q/\ell\rfloor}
   \sum_{\vec z\in\widetilde{\mZ}_p}
   \widetilde C_p(\vec z)\,
   \Phi_{p,\vec z}^{(s)}
   +\delta \Phi_Q,
 \label{application_corol:exp_decomposition_main_equation_fin}
\end{align}
where
\begin{align}
 \abs{\widetilde{\mZ}_p}
 &\le (2Q+1)^{\ceil{\log_2\ell}},
\qquad
 \abs{\widetilde C_p(\vec z)}
 \le 1,
 \qquad 
 \norm{\delta \Phi_Q}
 \le
 2\ell(2Q+1)^{\ceil{\log_2\ell}}
 \widetilde\delta_Q(\Phi).
 \label{eq:cut-decomposition-bounds}
\end{align}
\end{prop}

\begin{remark}[Physical meaning]
\begin{itemize}
 \item $u_p$ is the set of interaction blocks crossing the cut
 $(p-1,p)$, and $H_{\vec z,p}$ is their phase-twisted sum.

 \item $\Phi_{p,\vec z}^{(s)}$ is the order-$s$ response obtained by
 switching on the cut-crossing operator $H_{\vec z,p}$ around a background
 containing only terms that stay on one side of the cut.

 \item $H_{\vec z,<p}$ and $H_{\vec z,>p}$ contain the terms supported
 entirely on the left and right of the cut, respectively.  In particular,
 \begin{align}
  H_{\vec z,<p}+H_{\vec z,>p}
  =V_0+\sum_{j\notin u_p}z_jH_{p_j,p'_j}
 \end{align}
 contains no interaction crossing $(p-1,p)$.  The notation $>p$ refers
 to the right side of the cut and therefore includes the layer $L_p$.

 \item The cut weight $w(\vec s)$ counts the total number of crossed cuts,
 with multiplicity, in the Taylor monomial indexed by $\vec s$.
 A length-$r$ interaction insertion therefore contributes $r$ to this
 weight.

 \item $\widetilde\delta_Q(\Phi)$ is the norm-absolute weight of
 all Taylor monomials whose total cut cost exceeds $Q$.  It controls
 the final remainder through Eq.~\eqref{eq:cut-decomposition-bounds}; no
 further decomposition of that remainder is required.

 \item The phases $z_j$ implement a discrete Fourier filter.  The resulting
 phase-twisted Hamiltonians are generally non-Hermitian and are only
 algebraic auxiliary operators.
\end{itemize}

The proposition refines truncation by total Taylor order using geometric
cost: before completion of the noncrossing sums, every retained monomial
has total geometric cost at most $Q$ and is assigned to a cut across which
at most $2Q/\ell$ interaction insertions remain explicit.  Completing these
sums incorporates all noncrossing terms into the background
$H_{\vec z,<p}+H_{\vec z,>p}$.  The additional monomials with cost above $Q$
are accounted for in the remainder bound.  Thus, in applications to dynamics,
the cut-crossing interaction appears explicitly only through derivatives
of order $O(Q/\ell)$.  Long-range insertions are
penalized automatically because an interaction of length $r$ consumes
$r$ units of the cut budget.  A separate dynamical estimate is still
needed to show that $\widetilde\delta_Q(\Phi)$ is small.
\end{remark}

\subsection{Proof of Proposition~\ref{Prop:buffered_schmidt_est}}

Set
\begin{align}
 &F(\vec\lambda):=\Phi[H(\vec\lambda)],\qquad 
 \mD_{\vec z,u}:=\sum_{j\in u}z_j\partial_{\lambda_j},
 \notag\\
 &\partial^{\vec s}:=\partial_{\lambda_1}^{s_1}\cdots
 \partial_{\lambda_N}^{s_N}, \qquad 
 \mu_{\vec s}:=\frac{1}{s_1!\cdots s_N!}.
 \label{eq:F-Dzu-revised}
\end{align}
Norm-absolute convergence gives
\begin{align}
 \Phi[H]
 =\left.e^{\mD}F(\vec\lambda)\right|_{\vec\lambda=\vec 0}
 =\sum_{\vec s\in\mathbb Z_{\ge0}^N}
 \mu_{\vec s}
 \left.\partial^{\vec s}F(\vec\lambda)
 \right|_{\vec\lambda=\vec 0}.
 \label{eq:full-Taylor-cut-proof}
\end{align}

For $p\in[\ell]$, define the crossing number
\begin{align}
 S_{u_p}(\vec s):=\sum_{j\in u_p}s_j.
 \label{C/p_vec_s}
\end{align}
Since $H_{p_j,p'_j}$ crosses precisely the cuts
$p_j+1,\ldots,p'_j$, we have
\begin{align}
 \sum_{p=1}^{\ell}S_{u_p}(\vec s)
 =\sum_{j=1}^{N}(p'_j-p_j)s_j
 =w(\vec s).
 \label{eq:cut-count-identity-proof}
\end{align}
Hence
\begin{align}
 \widetilde{\mathcal S}^{\rm cut}_Q
 :=\left\{\vec s\in\mathbb Z_{\ge0}^N\,\middle|\,
 \sum_{p=1}^{\ell}S_{u_p}(\vec s)\le Q\right\}
 =\{\vec s\in\mathbb Z_{\ge0}^N\mid w(\vec s)\le Q\},
 \label{Def:check_mG_cut}
\end{align}
which is exactly the constraint set $\widetilde{\mathcal S}_Q$ in
Corollary~\ref{corol:exp_decomposition_main}, with the family of subsets
$u_1,\ldots,u_\ell$.

Apply that corollary under the correspondence
$x_j\leftrightarrow\partial_{\lambda_j}$.  The derivatives commute, and
the proof of the corollary is a coefficientwise linear recombination of
the Taylor series, so the same identity holds for the operator-valued
function $F$.  We obtain
\begin{align}
 \Phi[H]
 ={}&\sum_{p=1}^{\ell}
 \sum_{s=0}^{\lfloor2Q/\ell\rfloor}
 \sum_{\vec z\in\widetilde{\mZ}_p}
 \widetilde C_p(\vec z)
 \frac{(\mD_{\vec z,u_p})^s}{s!}
 e^{\mD_{\vec z,[N]\setminus u_p}}
 F(\vec\lambda)
 \Big|_{\vec\lambda=\vec 0}
 +\delta \Phi_Q,
 \label{application_corol:exp_decomposition_main_equation}
\end{align}
with
\begin{align}
 \abs{\widetilde{\mZ}_p}
 \le (2Q+1)^{\ceil{\log_2\ell}},
 \qquad
 \abs{\widetilde C_p(\vec z)}\le1.
 \label{eq:phase-set-proof}
\end{align}

The remainder estimate follows directly from the norm-valued analogue of
the scalar tail $\widetilde\delta_Q$ used in
Corollary~\ref{corol:exp_decomposition_main}.  Indeed, that proof uses only scalar phase
coefficients of modulus at most one and the triangle inequality.
Replacing each scalar Taylor monomial by the operator-valued coefficient
$\mu_{\vec s}\partial^{\vec s}F(\vec 0)$ therefore gives
\begin{align}
 \norm{\delta \Phi_Q}
 &\le
 2\ell(2Q+1)^{\ceil{\log_2\ell}}
 \sum_{\vec s\notin\widetilde{\mathcal S}^{\rm cut}_Q}
 \mu_{\vec s}\,
 \norm{\partial^{\vec s}F(\vec 0)}
 \notag\\
 &=
 2\ell(2Q+1)^{\ceil{\log_2\ell}}
 \widetilde\delta_Q(\Phi).
 \label{eq:phase-decomposition-error-revised}
\end{align}
No explicit representation of $\delta \Phi_Q$ is needed.

It remains to rewrite each differential branch in Hamiltonian form.
For fixed $p$ and $\vec z$, introduce two scalar parameters
\begin{align}
 G_{p,\vec z}(\lambda,\nu)
 :=\Phi\!\left[
 V_0
 +\lambda\sum_{j\in u_p}z_jH_{p_j,p'_j}
 +\nu\sum_{j\notin u_p}z_jH_{p_j,p'_j}
 \right].
 \label{eq:G-lambda-nu}
\end{align}
By the chain rule, for all $s,q\in\mathbb Z_{\ge0}$,
\begin{align}
 \left.
 \partial_\lambda^s\partial_\nu^q
 G_{p,\vec z}(\lambda,\nu)
 \right|_{\lambda=\nu=0}
 =
 \left.
 (\mD_{\vec z,u_p})^s
 (\mD_{\vec z,[N]\setminus u_p})^q
 F(\vec\lambda)
 \right|_{\vec\lambda=\vec0}.
 \label{eq:two-directional-derivatives}
\end{align}
Consequently, norm-absolute convergence and
$e^{\partial_\nu}G(\lambda,0)=G(\lambda,1)$ give
\begin{align}
 &\left.
 (\mD_{\vec z,u_p})^s
 e^{\mD_{\vec z,[N]\setminus u_p}}
 F(\vec\lambda)
 \right|_{\vec\lambda=\vec0}
 \notag\\
 &\qquad=
 \left.
 \partial_\lambda^s e^{\partial_\nu}
 G_{p,\vec z}(\lambda,\nu)
 \right|_{\lambda=\nu=0}
 =
 \left.
 \partial_\lambda^sG_{p,\vec z}(\lambda,1)
 \right|_{\lambda=0}.
 \label{eq:shift-two-parameters}
\end{align}
Since
\begin{align}
 j\notin u_p
 \quad\Longleftrightarrow\quad
 p'_j<p\ \text{ or }\ p_j\ge p,
 \label{eq:complement-up-revised}
\end{align}
we have
\begin{align}
 G_{p,\vec z}(\lambda,1)
 =\Phi\!\left[
 H_{\vec z,<p}+H_{\vec z,>p}+\lambda H_{\vec z,p}
 \right].
 \label{eq:G-H-left-right}
\end{align}
Substitution into
Eq.~\eqref{application_corol:exp_decomposition_main_equation} gives
Eq.~\eqref{application_corol:exp_decomposition_main_equation_fin} and
completes the proof. $\square$

\subsection{Upper bound on $\widetilde\delta_Q(\Phi)$}

\begin{lemma}[Length-profile majorant for the absolute Taylor tail]
\label{lem:length-profile-majorant}
Let
\[
F(\vec\lambda):=\Phi[H(\vec\lambda)],
\qquad
r_j:=p'_j-p_j\in[\ell].
\]
Assume that there exist nonnegative numbers
$\{a_j\}_{j=1}^N$ and $\{b_s\}_{s\ge0}$ such that
\begin{align}
\left\|
\partial_{\lambda_{j_1}}\cdots
\partial_{\lambda_{j_s}}
F(\vec 0)
\right\|
\le
b_s\,a_{j_1}\cdots a_{j_s}
\label{eq:coefficientwise-majorant}
\end{align}
for every $s\ge0$ and every
$j_1,\ldots,j_s\in[N]$.

Then
\begin{align}
\widetilde\delta_Q(\Phi)
\le
\sum_{m=0}^{\infty}\frac{b_m}{m!}
\sum_{\substack{
r_1,\ldots,r_m\in[\ell]\\
r_1+\cdots+r_m>Q
}}
A_{r_1}\cdots A_{r_m},
\label{eq:length-sequence-majorant}
\end{align}
where the auxiliary length profile $\{A_r\}_{r=1}^\ell$ is defined by 
\begin{align}
A_r:=\sum_{j:\,r_j=r}a_j,
\qquad r=1,\ldots,\ell.
\label{eq:absolute-length-strength}
\end{align}
\end{lemma}

\subsubsection{Proof}
Recall the definition of $\widetilde\delta_Q(\Phi)$: 
\begin{align}
 \widetilde\delta_Q(\Phi)
 :=
 \sum_{\substack{\vec s\in\mathbb Z_{\ge0}^N\\w(\vec s)>Q}}
 \frac{1}{s_1!\cdots s_N!}
 \left\|
 \left.
 \partial_{\lambda_1}^{s_1}\cdots\partial_{\lambda_N}^{s_N}
F(\vec\lambda)
 \right|_{\vec\lambda=\vec 0}
 \right\|,
 \label{eq:absolute-cut-tail-revised:reshow}
\end{align}
with $w(\vec s)=\sum_{j=1}^Nr_js_j$, and set
$|\vec s|_1:=\sum_{j=1}^Ns_j$.
Here we have replaced $\Phi[H(\vec\lambda)]$ by $F(\vec\lambda)$.
Using assumption~\eqref{eq:coefficientwise-majorant}, we obtain 
\begin{align}
 \widetilde\delta_Q(\Phi)
&\le \sum_{\substack{\vec s\in\mathbb Z_{\ge0}^N\\w(\vec s)>Q}}
 \frac{b_{|\vec{s}|_1}}{s_1!\cdots s_N!} \prod_{j=1}^{N} a_j^{s_j}  \notag \\
 &=\sum_{m=0}^{\infty}\frac{b_m}{m!}
\sum_{\substack{
j_1,\ldots,j_m\in[N]\\
r_{j_1}+\cdots+r_{j_m}>Q
}}
a_{j_1}\cdots a_{j_m}, 
 \label{eq:absolute-cut-tail-revised:reshow_upp}
\end{align}
Summing the labels independently within each
length sector yields
\begin{align}
\sum_{\substack{
j_1,\ldots,j_m\in[N]\\
r_{j_k}=r_k\ {\rm for\ all}\ k
}}
a_{j_1}\cdots a_{j_m}
=
A_{r_1}\cdots A_{r_m} ,
\end{align}
which reduces Eq.~\eqref{eq:absolute-cut-tail-revised:reshow_upp} to Eq.~\eqref{eq:length-sequence-majorant}. $\square$

\subsection{Practical generalization}
\label{sec:practical-generalization}

The preceding decomposition also applies when the Hamiltonian depends on an
additional variable $x$.  Let $x\in\mathcal X$ and consider the family
\begin{align}
\label{parameterize_Hamilaonitn_lamnda_x_dependent}
 H^{(x)}(\vec\lambda)
 &:= V^{(x)}_0
 + \sum_{0\le p<p'\le\ell}
 \lambda_{p,p'}H^{(x)}_{p,p'},
 \notag\\
 V^{(x)}_0
 &:= \sum_{p=0}^{\ell}H^{(x)}_{p,p},
 \qquad
 \vec\lambda
 =\{\lambda_{p,p'}\}_{0\le p<p'\le\ell}.
\end{align}
The same parameters $\lambda_{p,p'}$ are used for all $x$.  We write
$H^{(\cdot)}(\vec\lambda)$ for the full family
$x\mapsto H^{(x)}(\vec\lambda)$ and set
\begin{align}
 H^{(\cdot)}:=H^{(\cdot)}(\vec 1).
\end{align}
Let $\Phi$ be an operator-valued functional of this family, and assume that
\begin{align}
 F(\vec\lambda)
 :=\Phi\!\left[H^{(\cdot)}(\vec\lambda)\right]
\end{align}
has a norm-absolutely convergent Taylor expansion under the same conditions as
in Proposition~\ref{Prop:buffered_schmidt_est}.

Using the relabeling $j\leftrightarrow(p_j,p'_j)$ and the sets $u_p$ from
Eq.~\eqref{eq:up-def-revised}, define the phase-twisted families pointwise in
$x$ by
\begin{align}
 H^{(x)}_{\vec z,p}
 &:=\sum_{j\in u_p}z_jH^{(x)}_{p_j,p'_j},
 \notag\\
 H^{(x)}_{\vec z,<p}
 &:=\sum_{q=0}^{p-1}H^{(x)}_{q,q}
 +\sum_{j:\,p'_j<p}z_jH^{(x)}_{p_j,p'_j},
 \notag\\
 H^{(x)}_{\vec z,>p}
 &:=\sum_{q=p}^{\ell}H^{(x)}_{q,q}
 +\sum_{j:\,p_j\ge p}z_jH^{(x)}_{p_j,p'_j}.
\label{eq:x-dependent-phase-blocks}
\end{align}
We use $H^{(\cdot)}_{\vec z,p}$,
$H^{(\cdot)}_{\vec z,<p}$, and
$H^{(\cdot)}_{\vec z,>p}$ for the corresponding $x$-dependent families.
For later use, set
\begin{align}
 \widetilde\delta_Q^{\rm cut}
 \!\left(\Phi;H^{(\cdot)}\right)
 &:=
 \sum_{\substack{\vec s\in\mathbb Z_{\ge0}^N\\w(\vec s)>Q}}
 \frac{1}{s_1!\cdots s_N!}
 \left\|
 \left.
 \partial_{\lambda_1}^{s_1}\cdots\partial_{\lambda_N}^{s_N}
 \Phi\!\left[H^{(\cdot)}(\vec\lambda)\right]
 \right|_{\vec\lambda=\vec0}
 \right\|,
\label{eq:x-dependent-absolute-tail}
\end{align}
where $w(\vec s)$ is the cut weight in
Eq.~\eqref{eq:cut-weight-short}.  We also abbreviate
\begin{align}
 \Phi_{p,\vec z}^{(s)}\!\left[H^{(\cdot)}\right]
 &:=\frac{1}{s!}
 \left.
 \left(\frac{d}{d\lambda}\right)^s
 \Phi\!\left[
 H^{(\cdot)}_{\vec z,<p}
 +H^{(\cdot)}_{\vec z,>p}
 +\lambda H^{(\cdot)}_{\vec z,p}
 \right]
 \right|_{\lambda=0}.
\label{eq:x-dependent-branch}
\end{align}

\begin{corol}[Cut-count decomposition for an $x$-dependent Hamiltonian]
\label{cor:cut-count-x-dependent}
For every $Q\in\mathbb Z_{\ge0}$, the same indexed phase families
$\widetilde{\mZ}_p$ and coefficients $\widetilde C_p(\vec z)$ as in
Proposition~\ref{Prop:buffered_schmidt_est} give
\begin{align}
 \Phi\!\left[H^{(\cdot)}\right]
 ={}&\sum_{p=1}^{\ell}
 \sum_{s=0}^{\lfloor2Q/\ell\rfloor}
 \sum_{\vec z\in\widetilde{\mZ}_p}
 \widetilde C_p(\vec z)\,
 \Phi_{p,\vec z}^{(s)}\!\left[H^{(\cdot)}\right]
 +\delta_Q\Phi,
\label{eq:x-dependent-cut-decomposition}
\end{align}
where
\begin{align}
 &\abs{\widetilde{\mZ}_p}
 \le(2Q+1)^{\ceil{\log_2\ell}}, \qquad
 \abs{\widetilde C_p(\vec z)}
 \le1,
 \notag\\
 &\norm{\delta_Q\Phi}
 \le
 2\ell(2Q+1)^{\ceil{\log_2\ell}}
 \widetilde\delta_Q^{\rm cut}
 \!\left(\Phi;H^{(\cdot)}\right).
\label{eq:x-dependent-cut-bounds}
\end{align}
In particular, the phase families and coefficients are independent of $x$.
\end{corol}

\textit{Proof.}
Define
$F(\vec\lambda):=\Phi[H^{(\cdot)}(\vec\lambda)]$.
The family $H^{(\cdot)}(\vec\lambda)$ has the same affine dependence on
$\vec\lambda$ as $H(\vec\lambda)$ in Eq.~\eqref{eq:H-lambda-cut-count};
the additional variable $x$ appears only in the operator coefficients
$H^{(x)}_{p,p'}$.  Hence the differential
identities, the cut-count constraint, and the phase-reweighting argument in
Proposition~\ref{Prop:buffered_schmidt_est} apply to $F$ without change.
Moreover, Eq.~\eqref{eq:x-dependent-phase-blocks} gives, pointwise in $x$,
\begin{align}
 H^{(x)}_{\vec z,<p}+H^{(x)}_{\vec z,>p}
 =V_0^{(x)}+\sum_{j\notin u_p}z_jH^{(x)}_{p_j,p'_j}.
\end{align}
Thus each differential branch has the form
\eqref{eq:x-dependent-branch}, while the remainder obeys the bound in
Eq.~\eqref{eq:x-dependent-cut-bounds}.  This proves
Eqs.~\eqref{eq:x-dependent-cut-decomposition} and
\eqref{eq:x-dependent-cut-bounds}.
$\square$

\begin{remark}
The variable $x$ may represent time, imaginary time, a driving parameter, or
any other external label.  In the dynamical application one takes
$x=\tau$ and, for example,
\begin{align}
 \Phi\!\left[H^{(\cdot)}\right]
 =\mathcal T\exp\!\left[-i\int_0^tH^{(\tau)}\,d\tau\right].
\end{align}
The derivatives with respect to $\lambda_{p,p'}$ then insert the entire
$x$-dependent interaction channel $H^{(x)}_{p,p'}$, whereas each phase $z_j$ is fixed throughout the $x$-dependence.  Thus every term in the
corollary is a single differential response associated with an $x$-dependent
phase-twisted functional evaluation.  These auxiliary Hamiltonians
need not be Hermitian.  If $\Phi$ acts pointwise rather than on the full
family, one simply fixes $x$ and replaces $H^{(\cdot)}$ by $H^{(x)}$
throughout the corollary.
\end{remark}

\subsection{Simple exercise}
\label{sec:simple-exercise-Hm}

We now apply the cut-count formalism to the polynomial $\Phi[A]=A^m$.
This example separates three situations which behave rather differently:
(i) a single finite-range Hamiltonian raised to a power,
(ii) a single exponentially decaying Hamiltonian treated up to a norm error,
and (iii) a product of different Hamiltonians.
The repeated use of the same Hamiltonian is essential in the first two cases.
Throughout this subsection we take $m\in\mathbb Z_{\ge1}$; the case $m=0$ is immediate.

For later use, define the cut-weight truncation of $H^m$ by
\begin{align}
 [H^m]_Q
 :=
 \sum_{\substack{\vec s\in\mathbb Z_{\ge0}^N\\ w(\vec s)\le Q}}
 \mu_{\vec s}\,
 \left.
 \partial_{\lambda_1}^{s_1}\cdots\partial_{\lambda_N}^{s_N}
 \bigl[H(\vec\lambda)\bigr]^m
 \right|_{\vec\lambda=\vec 0}.
 \label{eq:def-cut-weight-truncation-Hm}
\end{align}
For $\Phi[A]=A^m$, we use the shorthand
$\widetilde\delta_Q^{\rm cut}(H^m):=\widetilde\delta_Q(\Phi)$,
with $F(\vec\lambda)=[H(\vec\lambda)]^m$ in
Eq.~\eqref{eq:absolute-cut-tail-revised}.
Thus,
\begin{align}
 \norm{H^m-[H^m]_Q}
 \le \widetilde\delta_Q^{\rm cut}(H^m).
 \label{eq:cut-weight-truncation-Hm-error}
\end{align}
No new physical truncation is hidden in this definition: $[H^m]_Q$ is simply the sum of the Taylor monomials whose total number of crossed cuts, counted with multiplicity, is at most $Q$.

\subsubsection{Finite-range interactions}

Assume first that only nearest-neighbor layers interact, so that
\begin{align}
 H=\sum_{p=0}^{\ell}H_{p,p}
   +\sum_{p=1}^{\ell}H_{p-1,p}.
\end{align}
Let
\begin{align}
 \mR:=\max_{p\in[\ell]}\mathrm{SR}_p(H_{p-1,p}).
 \label{eq:def-R-finite-range}
\end{align}
Since every off-diagonal insertion has interaction length one and the polynomial degree is $m$, every Taylor monomial satisfies $w(\vec s)\le m$. Hence $[H^m]_m=H^m$ exactly. This is the setting in which Refs.~\cite{PhysRevB.85.195145,arad2013area} derive the conventional bound quoted below.

\begin{lemma}[Finite-range power]
\label{lem:finite-range-Hm-SR}
For every cut $(p_0-1,p_0)$ and every $\ell\ge1$,
\begin{align}
 \mathrm{SR}_{p_0}(H^m)
 \le{}&
 \ell\left(\left\lfloor\frac{2m}{\ell}\right\rfloor+1\right)
 (2m+1)^{\lceil\log_2\ell\rceil}
 d_0^{2\ell|\partial\Lambda|}
 \notag\\
 &\times
 \max_{0\le s\le\min\{m,\lfloor2m/\ell\rfloor\}}
 \left\{
 \binom{m+s+1}{2s+1}\mR^s
 \right\}.
 \label{eq:finite-range-Hm-rank-explicit}
\end{align}
In particular, if $\ell\ge2$, then for fixed $k$ and $d_0$, a $k$-local Hamiltonian satisfies
\begin{align}
 \mathrm{SR}_{p_0}(H^m)
 \le
 \exp\left[
 O_{k,d_0}\left(
 \ell|\partial\Lambda|
 +\frac{m}{\ell}\log(m|\partial\Lambda|+2)
 +\log(m+2)\log(\ell+1)
 \right)
 \right].
 \label{eq:finite-range-Hm-rank-simple}
\end{align}
Suppressing the phase-filter overhead displayed explicitly above, we can write this schematically as
\begin{align}
 \mathrm{SR}_{p_0}(H^m)
 \le
 (m|\partial\Lambda|)^{O(m/\ell)}
 e^{O(\ell|\partial\Lambda|)}.
 \label{eq:finite-range-Hm-rank-schematic}
\end{align}
\end{lemma}

\subsubsection{Proof}
Apply Proposition~\ref{Prop:buffered_schmidt_est} to $\Phi[A]=A^m$ with $Q=m$. The cut tail vanishes, and the order of the explicit cut-crossing derivative is bounded by $s\le 2m/\ell$.
For a fixed selected cut $p$ and phase vector $\vec z$, let
$\zeta_p$ be the phase attached to the unique pair $(p-1,p)$ and write
\begin{align}
 A:=H_{\vec z,<p}+H_{\vec z,>p},
 \qquad
 B:=H_{\vec z,p}=\zeta_pH_{p-1,p},
 \qquad |\zeta_p|=1.
\end{align}
Across the cut $(p-1,p)$, the background contains no crossing term and therefore has the form
\begin{align}
 A=A_{<p}\otimes\hat 1_{\ge p}
   +\hat 1_{<p}\otimes A_{\ge p}.
 \label{eq:A-left-right-split}
\end{align}
Consequently,
\begin{align}
 \mathrm{SR}_p(A^q)\le q+1.
 \label{eq:SR-Aq}
\end{align}
Moreover, the normalized derivative has the exact noncommutative word expansion
\begin{align}
 \frac{1}{s!}
 \left.\left(\frac{d}{d\lambda}\right)^s
 (A+\lambda B)^m\right|_{\lambda=0}
 =
 \sum_{\substack{m_0+\cdots+m_s=m-s\\m_j\ge0}}
 A^{m_0}BA^{m_1}B\cdots BA^{m_s}.
 \label{eq:derivative-word-expansion-Hm}
\end{align}
Using submultiplicativity of the operator Schmidt rank together with Eqs.~\eqref{eq:SR-Aq} and \eqref{eq:def-R-finite-range},
\begin{align}
 \mathrm{SR}_p\left[
 \frac{1}{s!}
 \left.\left(\frac{d}{d\lambda}\right)^s
 (A+\lambda B)^m\right|_{\lambda=0}
 \right]
 &\le
 \mR^s
 \sum_{m_0+\cdots+m_s=m-s}
 \prod_{j=0}^{s}(m_j+1)
 \notag\\
 &=
 \mR^s\binom{m+s+1}{2s+1}.
 \label{eq:branch-rank-Hm}
\end{align}
The last identity follows by extracting the coefficient of $x^{m-s}$ from
$(1-x)^{-2(s+1)}$.

Proposition~\ref{Prop:buffered_schmidt_est} uses at most
$(2m+1)^{\lceil\log_2\ell\rceil}$ phase vectors for each $p$ and $s$.
The operator-basis expansion underlying Eq.~\eqref{trivial_Schmidt_ineq}
applies in either direction between two cuts.  Hence moving the Schmidt-rank
estimate from the selected cut $p$ to the target cut $p_0$ costs at most
$d_0^{2\ell|\partial\Lambda|}$.
Summing over $p$, $s$, and the phase vectors gives Eq.~\eqref{eq:finite-range-Hm-rank-explicit}.

Assume now that $\ell\ge2$.  For each $p\in[\ell]$, let $S_p$ be a layer of minimum cardinality in $\{L_{p-1},L_p\}$.  Every local term contributing to $H_{p-1,p}$ intersects both layers.  Expanding each such term in a local operator basis on its support in $S_p$, and collecting equal basis elements into the operator on the opposite side of the cut, gives
\begin{align}
 \mR
 \le
 \max_{p\in[\ell]}
 \sum_{q=1}^{k}\binom{|S_p|}{q}d_0^{2q}
 \le
 \bigl(2d_0^2|\partial\Lambda|\bigr)^k.
 \label{eq:R-finite-k-local}
\end{align}
Indeed, every adjacent pair contains at least one interior layer when $\ell\ge2$, and hence $|S_p|\le|\partial\Lambda|$.  Combining this with the standard binomial estimate gives Eq.~\eqref{eq:finite-range-Hm-rank-simple}.  When $\ell=1$, the explicit estimate \eqref{eq:finite-range-Hm-rank-explicit} remains valid with $\mR$ left explicit; a boundary-size-only specialization requires first replacing the two aggregate endpoint regions by a decomposition containing an explicit intermediate layer.

Assuming that the ordered partition admits an integer $\ell\ge2$ within a constant factor of the value below, optimizing the two leading terms in Eq.~\eqref{eq:finite-range-Hm-rank-simple} gives
\begin{align}
 \ell
 \asymp
 \sqrt{\frac{m\log(m|\partial\Lambda|+2)}{|\partial\Lambda|}},
 \label{eq:optimal-ell-finite-Hm}
\end{align}
and hence
\begin{align}
 \mathrm{SR}_{p_0}(H^m)
 \le
 \exp\left[
 O_{k,d_0}\left(
 \sqrt{m|\partial\Lambda|\log(m|\partial\Lambda|+2)}
 +\log^2(m+2)
 \right)
 \right].
 \label{eq:optimized-finite-Hm}
\end{align}
If this choice of $\ell$ is unavailable, one instead uses
Eq.~\eqref{eq:finite-range-Hm-rank-simple} at an admissible endpoint.
For the optimizing choice above, the present phase-filtering construction improves, up to lower-order logarithmic terms, the conventional form quoted in Refs.~\cite{PhysRevB.85.195145,arad2013area},
$(m|\partial\Lambda|)^{O(\sqrt{m|\partial\Lambda|})}$, to
$\exp[O(\sqrt{m|\partial\Lambda|\log(m|\partial\Lambda|)})]$.

\subsubsection{Exponentially decaying interactions}

The exact Schmidt rank of $H^m$ cannot in general be subexponential in $m$, even when $H$ is two-local, has bounded norm, and has exponentially decaying interactions. Indeed, place $m$ disjoint qubit pairs $(L_j,R_j)$ across a fixed cut, with the $j$th pair at distance proportional to $j$, and set
\begin{align}
 H:=\sum_{j=1}^{m}e^{-\mu j}X_{L_j}\otimes X_{R_j}.
 \label{eq:exp-decay-Hm-counterexample-H}
\end{align}
The summands commute and square to the identity. Therefore,
\begin{align}
 H^m
 =
 \sum_{\substack{S\subseteq[m]\\ |S|\equiv m\ ({\rm mod}\ 2)}}
 c_S\,
 X_S^{L}\otimes X_S^{R},
 \qquad c_S>0,
 \label{eq:exp-decay-Hm-counterexample-expansion}
\end{align}
where $X_S^{L}:=\prod_{j\in S}X_{L_j}$ and similarly for $X_S^R$.
The Pauli strings are linearly independent, and hence
\begin{align}
 \mathrm{SR}(H^m)=2^{m-1}.
 \label{eq:exp-decay-Hm-counterexample-rank}
\end{align}
Thus a useful statement for exponentially decaying interactions must allow a norm error.

For the remainder of this subsection, take
$a_j:=\norm{H_{p_j,p'_j}}$ in Lemma~\ref{lem:length-profile-majorant}.
Then its auxiliary profile satisfies $A_r=K_r$, where $K_r$ is the
Hamiltonian layer strength defined in Eq.~\eqref{eq:setup-H-length-strengths}.
We assume that these length strengths satisfy
\begin{align}
 K_r\le \kappa\,\ell|\partial\Lambda|e^{-\mu r},
 \qquad r=1,\ldots,\ell,
 \label{eq:Kr-exponential-decay}
\end{align}
for constants $\kappa,\mu>0$ independent of $m$, $Q$, and $\ell$.

\begin{corol}[Approximate power for exponentially decaying interactions]
\label{cor:exp-decay-Hm-approx}
For every $Q\in\mathbb Z_{\ge0}$, the cut-weight truncation $[H^m]_Q$ satisfies, for every $0<\eta<\mu$,
\begin{align}
 \norm{H^m-[H^m]_Q}
 \le
 e^{-\eta Q}
 \left[
 \norm{V_0}
 +\sum_{r=1}^{\ell}e^{\eta r}K_r
 \right]^m
 \le
 e^{-\eta Q}
 \left[
 \norm{V_0}
 +\frac{\kappa\ell|\partial\Lambda|}{e^{\mu-\eta}-1}
 \right]^m.
 \label{eq:exp-decay-Hm-error}
\end{align}

To state the Schmidt-rank bound, let
\begin{align}
 \mR_Q
 :=
 \max_{p\in[\ell]}
 \max_{|z_j|=1}
 \mathrm{SR}_p\left(
 \sum_{\substack{j\in u_p\\p'_j-p_j\le Q}}
 z_jH_{p_j,p'_j}
 \right).
 \label{eq:def-RQ}
\end{align}
Then $[H^m]_Q$ admits an exact decomposition satisfying
\begin{align}
 \mathrm{SR}_{p_0}([H^m]_Q)
 \le{}&
 (mQ+1)\,
 \ell\left(\left\lfloor\frac{2Q}{\ell}\right\rfloor+1\right)
 (2Q+1)^{\lceil\log_2\ell\rceil}
 d_0^{2\ell|\partial\Lambda|}
 \notag\\
 &\times
 \max_{0\le s\le\min\{m,\lfloor2Q/\ell\rfloor\}}
 \left\{
 \binom{m+s+1}{2s+1}\mR_Q^s
 \right\}.
 \label{eq:exp-decay-Hm-rank-explicit}
\end{align}
If the Hamiltonian is $k$-local and either $Q<\ell$, or the layer decomposition has first been modified so that, for every selected cut, the union of the supports of the retained crossing terms contains $O(Q|\partial\Lambda|)$ sites, then
\begin{align}
 \mR_Q\le \bigl(Cd_0^2Q|\partial\Lambda|\bigr)^k.
 \label{eq:RQ-k-local}
\end{align}
When $Q\ge\ell$, this conclusion does not hold for the unmodified decomposition, because $H_{0,\ell}$ is included and may have a Schmidt rank that diverges with the endpoint volumes.  One must first use an independently justified edge truncation, or apply the entire cut-count construction to an extended layer decomposition with length $\widetilde\ell>Q$.  In the latter case, every occurrence of $\ell$ in Eqs.~\eqref{eq:exp-decay-Hm-rank-explicit}--\eqref{eq:exp-decay-Hm-rank-simple} refers to $\widetilde\ell$.
If an edge truncation changes the Hamiltonian, its contribution to the
power-approximation error must be added separately.

Under either of the preceding conditions, and after imposing $Q\ge m$,
\begin{align}
 \mathrm{SR}_{p_0}([H^m]_Q)
 \le
 \exp\left[
 O_{k,d_0}\left(
 \ell|\partial\Lambda|
 +\frac{Q}{\ell}\log(Q|\partial\Lambda|+2)
 +\log(mQ+2)\log(\ell+1)
 \right)
 \right].
 \label{eq:exp-decay-Hm-rank-simple}
\end{align}
Thus the finite-range estimate is recovered with the geometric degree $m$ replaced by the effective cut budget $Q$.
\end{corol}

\subsubsection{Proof}
For $F(\vec\lambda)=[H(\vec\lambda)]^m$ and an ordered list $j_1,\ldots,j_s$, each derivative inserts the operators
$H_{p_{j_1},p'_{j_1}},\ldots,H_{p_{j_s},p'_{j_s}}$ into distinct positions of a word of length $m$. Hence, for $s\le m$,
\begin{align}
 \norm{\partial_{\lambda_{j_1}}\cdots\partial_{\lambda_{j_s}}F(\vec0)}
 \le
 \frac{m!}{(m-s)!}\,
 \norm{V_0}^{m-s}
 \prod_{a=1}^{s}\norm{H_{p_{j_a},p'_{j_a}}},
 \label{eq:Hm-ordered-derivative-bound}
\end{align}
and the derivative vanishes for $s>m$.
Applying Lemma~\ref{lem:length-profile-majorant} with
$a_j=\norm{H_{p_j,p'_j}}$,
$b_s=m!\norm{V_0}^{m-s}/(m-s)!$ for $0\le s\le m$, and
$b_s=0$ for $s>m$, gives
\begin{align}
 \widetilde\delta_Q^{\rm cut}(H^m)
 \le
 \sum_{s=0}^{m}\binom{m}{s}\norm{V_0}^{m-s}
 \sum_{\substack{r_1,\ldots,r_s\in[\ell]\\r_1+\cdots+r_s>Q}}
 K_{r_1}\cdots K_{r_s}.
 \label{eq:Hm-tail-before-Chernoff}
\end{align}
For $0<\eta<\mu$,
\begin{align}
 \mathbf 1_{\{r_1+\cdots+r_s>Q\}}
 \le e^{-\eta Q}e^{\eta(r_1+\cdots+r_s)}.
\end{align}
Therefore,
\begin{align}
 \widetilde\delta_Q^{\rm cut}(H^m)
 &\le
 e^{-\eta Q}
 \sum_{s=0}^{m}\binom{m}{s}\norm{V_0}^{m-s}
 \left(\sum_{r=1}^{\ell}e^{\eta r}K_r\right)^s
 \notag\\
 &=
 e^{-\eta Q}
 \left[
 \norm{V_0}+\sum_{r=1}^{\ell}e^{\eta r}K_r
 \right]^m.
 \label{eq:Hm-tail-Chernoff}
\end{align}
Equation~\eqref{eq:Kr-exponential-decay} yields the second inequality in Eq.~\eqref{eq:exp-decay-Hm-error}.

It remains to control the Schmidt rank of the retained polynomial.
Interactions with $p'_j-p_j>Q$ do not occur in Eq.~\eqref{eq:def-cut-weight-truncation-Hm}.
After removing those variables, every Taylor monomial has total degree at most
$m$ and weight in $[0,mQ]$.  A roots-of-unity filter with
$M=mQ+1$ therefore isolates the integer weights in $[0,Q]$ exactly; its
coefficients have modulus at most one.  The corresponding weight phase is
absorbed into the phase vector $\vec z$.
The binary cut partition and the phase-reweighting argument used in Proposition~\ref{Prop:buffered_schmidt_est} then assign every retained monomial to a cut $p$ for which the number of explicit crossing insertions is at most $2Q/\ell$.
For each resulting phase branch, Eq.~\eqref{eq:derivative-word-expansion-Hm} applies with the crossing operator having Schmidt rank at most $\mR_Q$, while the noncrossing background again has the left-right form \eqref{eq:A-left-right-split}.
Thus Eq.~\eqref{eq:branch-rank-Hm} holds with $\mR$ replaced by $\mR_Q$.
The total-weight filter is retained throughout, so the completion terms
with weight above $Q$ cancel exactly after summing its phases.
Counting the additional weight-filter phases, the cut phases, the values of $p$ and $s$, and finally moving from $p$ to $p_0$, gives Eq.~\eqref{eq:exp-decay-Hm-rank-explicit}.
To prove the $k$-local estimate, denote the crossing operator in Eq.~\eqref{eq:def-RQ} by
\begin{align*}
 B_{p,\vec z}^{[Q]}
 :=
 \sum_{\substack{j\in u_p\\p'_j-p_j\le Q}}
 z_jH_{p_j,p'_j}.
\end{align*}
If $Q<\ell$, no retained block connects both aggregate endpoint regions.  Split the blocks into the three classes $p_j=0$, $p'_j=\ell$, and $1\le p_j<p'_j\le\ell-1$.  For the first class, expand every local term on its right support; for the second, expand on its left support; and for the third, expand on either side of the selected cut.  In each case the chosen support is contained in at most $Q$ explicit layers and hence has at most $Q|\partial\Lambda|$ sites.  Counting the local operator-basis elements and using subadditivity gives Eq.~\eqref{eq:RQ-k-local}, with the constant absorbing the three classes.

For the support-based version used after a geometric modification, let $e(Z)$ denote the unique off-diagonal block index containing $h_Z$, and define
\begin{align*}
 u_{p,Q}&:=\{j\in u_p:p'_j-p_j\le Q\},\\
 \Lambda_{p,Q}^{\times}&:=\bigcup_{Z:\,e(Z)\in u_{p,Q}}Z,
 \qquad
 n_Q^{\times}:=\max_{p\in[\ell]}|\Lambda_{p,Q}^{\times}|.
\end{align*}
A direct local-basis expansion gives
\begin{align*}
 \mR_Q
 \le
 \sum_{q=1}^{k}\binom{n_Q^{\times}}{q}d_0^{2q}
 \le
 \bigl(Cd_0^2n_Q^{\times}\bigr)^k.
\end{align*}
Thus $n_Q^{\times}\le C_{\rm geo}Q|\partial\Lambda|$ also implies Eq.~\eqref{eq:RQ-k-local}.  If $Q\ge\ell$ without either modification, $H_{0,\ell}$ invalidates both arguments.

For an absolute error $\epsilon$, it is sufficient to choose, for example,
\begin{align}
 Q
 \ge
 \max\left\{
 m,
 \frac{1}{\eta}
 \left[
 m\log\left(
 1+\norm{V_0}
 +\frac{\kappa\ell|\partial\Lambda|}{e^{\mu-\eta}-1}
 \right)
 +\log\frac{1}{\epsilon}
 \right]
 \right\}.
 \label{eq:choice-Q-Hm-exp-decay}
\end{align}
For fixed admissible $Q$, assume in addition that the effective-support condition used in Eq.~\eqref{eq:RQ-k-local} remains valid for the admissible values of $\ell$ entering the optimization (for example, after an edge truncation).  If the only available modification is a virtual extension with $\widetilde\ell>Q$, the square-root choice below is not generally admissible and one must optimize over the extended length instead.  Under the stated effective-support condition, and provided that the choice below is admissible up to integer rounding, optimizing the two leading terms in Eq.~\eqref{eq:exp-decay-Hm-rank-simple} gives
\begin{align}
 \ell
 \asymp
 \sqrt{\frac{Q\log(Q|\partial\Lambda|+2)}{|\partial\Lambda|}},
\end{align}
and therefore, up to lower-order logarithmic terms,
\begin{align}
 \mathrm{SR}_{p_0}([H^m]_Q)
 \le
 \exp\left[
 O_{k,d_0}\left(
 \sqrt{Q|\partial\Lambda|\log(Q|\partial\Lambda|+2)}
 \right)
 \right].
 \label{eq:optimized-exp-decay-Hm}
\end{align}
For a prescribed error $\epsilon$, the final choices of $Q$ and $\ell$
must also satisfy Eq.~\eqref{eq:choice-Q-Hm-exp-decay} simultaneously,
since its right-hand side depends on $\ell$ and $\norm{V_0}$.
The optimized bound is subexponential in $m$ in regimes where
$Q|\partial\Lambda|\log(Q|\partial\Lambda|+2)=o(m^2)$.
The exact-rank obstruction in Eq.~\eqref{eq:exp-decay-Hm-counterexample-rank} shows that the norm approximation is essential. To the best of our knowledge, the resulting subexponential Schmidt-rank approximation for powers of exponentially decaying Hamiltonians, under the conditions just stated, is a new consequence of the present cut-count formalism.

\subsubsection{Products of different Hamiltonians}

The preceding compression relies on the repeated power of one and the same Hamiltonian. For a product
$H_1H_2\cdots H_m$, the differential formalism remains algebraically valid, but a subexponential Schmidt-rank bound is impossible without additional structure.
Indeed, take $m$ qubits on each side of a cut and define
\begin{align}
 H_i:=X_{i,L}\otimes\hat 1_R+\hat 1_L\otimes X_{i,R},
 \qquad i=1,\ldots,m.
 \label{eq:different-H-counterexample}
\end{align}
Each $H_i$ has operator Schmidt rank two and contains no interaction across the cut. Nevertheless,
\begin{align}
 H_1H_2\cdots H_m
 =
 \sum_{S\subseteq[m]}
 X_S^L\otimes X_{S^{\rm c}}^R,
 \label{eq:different-H-product-expansion}
\end{align}
so that
\begin{align}
 \mathrm{SR}(H_1H_2\cdots H_m)=2^m.
 \label{eq:different-H-product-rank}
\end{align}
Equivalently, on the product state $|0\rangle_L^{\otimes m}|0\rangle_R^{\otimes m}$,
\begin{align}
 H_1\cdots H_m
 |0\rangle_L^{\otimes m}|0\rangle_R^{\otimes m}
 =
 \bigotimes_{i=1}^{m}
 \bigl(|1\rangle_{i,L}|0\rangle_{i,R}
      +|0\rangle_{i,L}|1\rangle_{i,R}\bigr),
 \label{eq:different-H-Bell-pairs}
\end{align}
which is an unnormalized product of $m$ Bell pairs and has state Schmidt rank $2^m$.

For $H^m$, each noncrossing segment is a power $A^q$ of one fixed operator
$A=A_{<p}\otimes\hat 1+\hat 1\otimes A_{\ge p}$ and has at most
$q+1$ left-right sectors, as used in Eq.~\eqref{eq:branch-rank-Hm}. In contrast, the product of different Hamiltonians records an independent left/right choice at every step, producing $2^m$ distinct sectors. This is the precise point at which the Schmidt-rank argument for a repeated power ceases to apply.

\section{Buffered Schmidt rank}
\label{sec:buffered-schmidt-rank}

The cut-count decomposition in the preceding section naturally produces a
sum of operators for which the cut used to control the Schmidt rank depends
on the summand.  Converting every summand to one prescribed cut is always
possible, but generally incurs a Hilbert-space-dimension factor associated
with the region between the two cuts.  We introduce the \emph{buffered
Schmidt rank} (BSR) in order to retain the native cut of each summand.
Algebraically, this is a partition-rank-type quantity in which the admissible
bipartitions are restricted by the geometry of the buffer; for the general
notion of partition rank, see Ref.~\cite{Naslund2020}.

Throughout this section, we use the ordered decomposition introduced in
Sec.~\ref{sec:setup}.  We adopt the convention $L_{[a,b]}=\varnothing$ when $a>b$, and set
\begin{align}
 A:=L_0,\qquad B:=L_{[1,\ell-1]},\qquad C:=L_\ell .
\end{align}
For each $p\in[\ell]$, define
\begin{align}
 B_{<p}:=L_{[1,p-1]},\qquad B_{\ge p}:=L_{[p,\ell-1]}.
\end{align}
The $p$th admissible cut is therefore
\begin{align}
 AB_{<p}:B_{\ge p}C
 =L_{[0,p-1]}:L_{[p,\ell]}.
\end{align}
Unless stated otherwise, the term BSR below refers to this ordered family of
buffer cuts.

\subsection{Definitions for operators, superoperators, and channels}

\begin{definition}[Buffered Schmidt rank of an operator]
\label{def:operator-bsr}
For an operator $O\in\mathcal B(\mathcal H_{ABC})$, its buffered Schmidt rank
through $B$ is
\begin{align}
 \operatorname{BSR}_{A:C\mid B}(O)
 :=\min\Biggl\{
 \sum_{p=1}^{\ell}r_p\ \Bigg|\ 
 O=\sum_{p=1}^{\ell}\sum_{a=1}^{r_p}
 O^{(p,a)}_{<p}\otimes O^{(p,a)}_{\ge p}
 \Biggr\},
 \label{eq:def-operator-bsr}
\end{align}
where $r_p\in\mathbb Z_{\ge0}$,
$O^{(p,a)}_{<p}\in\mathcal B(\mathcal H_{L_{[0,p-1]}})$, and
$O^{(p,a)}_{\ge p}\in\mathcal B(\mathcal H_{L_{[p,\ell]}})$.
We set $\operatorname{BSR}(0)=0$.
\end{definition}

Equivalently, the BSR is the minimum number $D$ of terms in a decomposition of the form
\begin{align}
 O=\sum_{j=1}^{D}
 O^{(j)}_{AB_j}\otimes O^{(j)}_{\overline B_jC},
 \qquad
 B_j\sqcup\overline B_j=B,
 \label{eq:generic-bsr-decomposition}
\end{align}
where the bipartition $B=B_j\sqcup\overline B_j$ may depend on $j$ and is
required to belong to the prescribed family of admissible cuts.  In the
ordered setting above, $B_j=B_{<p_j}$ and
$\overline B_j=B_{\ge p_j}$ for some $p_j\in[\ell]$.  More generally, one may
replace the ordered family by any specified family $\mathfrak C_B$ of
bipartitions of $B$; enlarging $\mathfrak C_B$ can only decrease the BSR.
For a family consisting of a single cut $p$, Definition~\ref{def:operator-bsr}
reduces to the usual operator Schmidt rank $\operatorname{SR}_p$.

The same definition applies to vectors by replacing the operator factors in
Eq.~\eqref{eq:def-operator-bsr} by state vectors.  In the present work,
however, the primary object is an operator such as a polynomial in $H$ or a
time-evolution operator.

\begin{definition}[Liouville BSR of a superoperator]
\label{def:liouville-bsr}
Let
\begin{align}
 \mathcal M:
 \mathcal B(\mathcal H_{ABC})\longrightarrow
 \mathcal B(\mathcal H_{ABC})
\end{align}
be a linear map.  Its Liouville buffered Schmidt rank is
\begin{align}
 \operatorname{BSR}_{\mathrm{Liouv}}(\mathcal M)
 :=\min\Biggl\{
 \sum_{p=1}^{\ell}r_p\ \Bigg|\ 
 \mathcal M
 =\sum_{p=1}^{\ell}\sum_{a=1}^{r_p}
 \mathcal M^{(p,a)}_{<p}\otimes
 \mathcal M^{(p,a)}_{\ge p}
 \Biggr\},
 \label{eq:def-liouville-bsr}
\end{align}
where the two factors are arbitrary linear maps on the operator spaces of
$L_{[0,p-1]}$ and $L_{[p,\ell]}$, respectively.
\end{definition}

Let $J(\mathcal M)$ denote the Choi operator of $\mathcal M$, and group the
input and output copies of each physical site together.  Since the Choi
isomorphism respects tensor products, Definition~\ref{def:liouville-bsr} is
equivalently expressed as
\begin{align}
 \operatorname{BSR}_{\mathrm{Liouv}}(\mathcal M)
 =\operatorname{BSR}_{AA':CC'\mid BB'}\!\left[J(\mathcal M)\right],
 \label{eq:choi-bsr-equivalence}
\end{align}
where $A'B'C'$ is the input copy.  For a quantum channel $\mathcal N$, we
refer to $\operatorname{BSR}_{\mathrm{Liouv}}(\mathcal N)$ as its
\emph{algebraic channel BSR}.  The factors in
Eq.~\eqref{eq:def-liouville-bsr} are not required to be completely positive
or trace preserving.  Requiring every summand to be a product of completely
positive maps would define a different and substantially stronger notion,
which is generally not adapted to coherent unitary dynamics.

For operational questions, it is also useful to record a Kraus-level
extension of the operator definition.

\begin{definition}[Kraus buffered Schmidt rank]
\label{def:kraus-bsr}
For a quantum channel
\begin{align}
 \mathcal N(X)=\sum_{\mu}K_\mu X K_\mu^\dagger,
\end{align}
we define
\begin{align}
 \operatorname{BSR}_{\mathrm{Kraus}}(\mathcal N)
 :=\min_{\{K_\mu\}}
 \max_\mu \operatorname{BSR}_{A:C\mid B}(K_\mu),
 \label{eq:def-kraus-bsr}
\end{align}
where the minimum is over finite Kraus representations of $\mathcal N$.
\end{definition}

The Liouville and Kraus versions are inequivalent in general.  For example,
for a family consisting of a single cut, a unitary $U$ with
$\operatorname{SR}(U)=r$ has Kraus BSR $r$, whereas the Choi operator
$|U\rangle\!\rangle\langle\!\langle U|$ has operator Schmidt rank $r^2$.
More generally, a BSR decomposition of an operator $U$ does not automatically
give a Liouville decomposition of $X\mapsto UXU^\dagger$ with the same rank,
because the two operator factors appearing on the ket and bra sides may
select different cuts.  On the other hand, for a unitary channel
$\mathcal U(X)=UXU^\dagger$, every Kraus operator is proportional to $U$, and
hence
\begin{align}
 \operatorname{BSR}_{\mathrm{Kraus}}(\mathcal U)
 =\operatorname{BSR}(U).
 \label{eq:unitary-kraus-bsr}
\end{align}

\subsection{Elementary mathematical properties}

For $p,q\in[\ell]$, define the cut-conversion factor
\begin{align}
 \chi_{p,q}
 :=D_{L_{[\min\{p,q\},\max\{p,q\}-1]}}^2,
 \qquad D_\varnothing:=1,
 \label{eq:cut-conversion-factor}
\end{align}
and set $\chi_q:=\max_{p\in[\ell]}\chi_{p,q}$.  The square in
Eq.~\eqref{eq:cut-conversion-factor} is the dimension of the corresponding
operator space.

\begin{prop}[Basic properties of the BSR]
\label{prop:basic-bsr-properties}
For arbitrary operators $O,O_1,O_2\in\mathcal B(\mathcal H_{ABC})$, the
following statements hold.
\begin{enumerate}
 \item For every admissible cut $p$,
 \begin{align}
  \operatorname{BSR}(O)\le \operatorname{SR}_p(O).
  \label{eq:bsr-below-fixed-sr}
 \end{align}

 \item The BSR is subadditive under addition:
 \begin{align}
  \operatorname{BSR}(O_1+O_2)
  \le \operatorname{BSR}(O_1)+\operatorname{BSR}(O_2).
  \label{eq:bsr-subadditivity}
 \end{align}

 \item Tracing out the buffer gives
 \begin{align}
  \operatorname{SR}_{A:C}\!\left(\operatorname{tr}_B O\right)
  \le \operatorname{BSR}_{A:C\mid B}(O).
  \label{eq:bsr-partial-trace}
 \end{align}

 \item Endpoint-local linear maps do not increase the BSR.  Namely, for
 arbitrary linear maps $\mathcal A$ on $A$ and $\mathcal C$ on $C$,
 \begin{align}
  \operatorname{BSR}\!\left[
  (\mathcal A\otimes\operatorname{id}_B\otimes\mathcal C)(O)
  \right]
  \le \operatorname{BSR}(O).
  \label{eq:bsr-endpoint-monotonicity}
 \end{align}

 \item For every target cut $q$,
 \begin{align}
  \operatorname{SR}_q(O)
  \le \chi_q\operatorname{BSR}(O).
  \label{eq:fixed-sr-from-bsr}
 \end{align}
 More precisely, any decomposition in
 Eq.~\eqref{eq:def-operator-bsr} satisfies
 \begin{align}
  \operatorname{SR}_q(O)
  \le \sum_{p=1}^{\ell}r_p\chi_{p,q}.
  \label{eq:refined-fixed-sr-from-bsr}
 \end{align}
\end{enumerate}
\end{prop}

\textit{Proof.}
Equation~\eqref{eq:bsr-below-fixed-sr} follows by restricting every summand
to the same cut $p$, and Eq.~\eqref{eq:bsr-subadditivity} follows by
concatenating two BSR decompositions.  For a single summand in
Eq.~\eqref{eq:generic-bsr-decomposition},
\begin{align}
 \operatorname{tr}_B\!\left(
 O^{(j)}_{AB_j}\otimes O^{(j)}_{\overline B_jC}
 \right)
 =\operatorname{tr}_{B_j}\!\left(O^{(j)}_{AB_j}\right)
 \otimes
 \operatorname{tr}_{\overline B_j}\!\left(
 O^{(j)}_{\overline B_jC}
 \right),
\end{align}
which proves Eq.~\eqref{eq:bsr-partial-trace}.  Endpoint-local maps can be
absorbed into the corresponding factors, proving
Eq.~\eqref{eq:bsr-endpoint-monotonicity}.

Finally, set
$M_{p,q}:=L_{[\min\{p,q\},\max\{p,q\}-1]}$.
If $p<q$, expand the right factor of a product operator across cut $p$ in an
operator basis on $M_{p,q}$; if $p>q$, expand the left factor instead.  In
either case at most $D_{M_{p,q}}^2=\chi_{p,q}$ product terms across cut $q$
are required.  Summing the resulting ranks proves
Eqs.~\eqref{eq:fixed-sr-from-bsr} and
\eqref{eq:refined-fixed-sr-from-bsr}. $\square$

For a uniform upper bound $d_0$ on the on-site dimensions, the coarse estimate
\begin{align}
 \chi_q\le D_B^2\le d_0^{2|B|}
 \le d_0^{2\ell|\partial\Lambda|}
 \label{eq:coarse-bsr-to-sr}
\end{align}
shows that converting a BSR bound into a fixed-cut Schmidt-rank bound may
cost an exponential factor in the buffer volume.

Unlike the ordinary Schmidt rank across one fixed cut, the BSR is not in
general submultiplicative under operator multiplication.  For example,
consider the simplest case in which the buffer $B$ consists of a single
block, so that the admissible cuts are $A:BC$ and $AB:C$.  Let $A,B,C$ each
contain an $r$-dimensional subspace and define
$P_{X,s}:=|s\rangle\langle s|_X$.  The two operators
\begin{align}
 O_L&:=\left(\sum_{s=1}^{r}P_{A,s}\otimes P_{B,s}\right)
 \otimes\mathbf 1_C,\\
 O_R&:=\mathbf 1_A\otimes
 \left(\sum_{s=1}^{r}P_{B,s}\otimes P_{C,s}\right)
\end{align}
obey $\operatorname{BSR}(O_L)=\operatorname{BSR}(O_R)=1$, using the cuts
$AB:C$ and $A:BC$, respectively.  However,
\begin{align}
 O_LO_R=\sum_{s=1}^{r}P_{A,s}\otimes P_{B,s}\otimes P_{C,s}
 \label{eq:bsr-product-counterexample}
\end{align}
has Schmidt rank $r$ across both cuts and therefore has BSR strictly larger
than one for $r\ge2$.  Thus the BSR should be viewed as a one-shot
factorization complexity of an operator, rather than as a multiplicative
resource under arbitrary composition.

The restriction-to-one-cut, subadditivity, and endpoint-local monotonicity
statements have direct Liouville analogues.  In the fixed-cut conversion of
a superoperator, the corresponding factor is
$D_{L_{[\min\{p,q\},\max\{p,q\}-1]}}^4$, because the local Liouville
space has dimension $D_X^2$.  For the Kraus BSR, pre- and post-processing by endpoint-local
quantum channels also cannot increase the rank, since their Kraus operators
can be multiplied into the left and right factors of each $K_\mu$.

\subsection{Low-BSR and fixed-cut low-rank approximations}

For $D\in\mathbb Z_{\ge0}$ and a norm $\|\cdot\|$, define the optimal errors
\begin{align}
 \varepsilon_D^{\mathrm{BSR}}(O)
 &:=\inf_{\widetilde O:\,\operatorname{BSR}(\widetilde O)\le D}
 \|O-\widetilde O\|,
 \label{eq:def-bsr-approx-error}\\
 \varepsilon_{D,p}^{\mathrm{SR}}(O)
 &:=\inf_{\widetilde O:\,\operatorname{SR}_p(\widetilde O)\le D}
 \|O-\widetilde O\|.
 \label{eq:def-sr-approx-error}
\end{align}
For operator dynamics we use the operator norm, while for a direct
approximation of quantum channels the natural analogue is the diamond norm.

The two approximation notions satisfy
\begin{align}
 \varepsilon_D^{\mathrm{BSR}}(O)
 &\le \min_{p\in[\ell]}\varepsilon_{D,p}^{\mathrm{SR}}(O),
 \label{eq:bsr-approx-better-than-one-cut}\\
 \varepsilon_{\chi_qD,q}^{\mathrm{SR}}(O)
 &\le \varepsilon_D^{\mathrm{BSR}}(O)
 \qquad(q\in[\ell]).
 \label{eq:bsr-approx-to-one-cut}
\end{align}
The first inequality follows from
Eq.~\eqref{eq:bsr-below-fixed-sr}; the second follows from
Eq.~\eqref{eq:fixed-sr-from-bsr}.  Hence a low-BSR approximation is weaker
than a low-rank approximation across a prescribed cut, but it can be
converted into the latter at the dimension cost $\chi_q$.

The main utility of the BSR for the differential formalism is summarized by
the following immediate observation.

\begin{prop}[Native-cut compression]
\label{prop:native-cut-compression}
Suppose that
\begin{align}
 O=\widetilde O+\Delta,
 \qquad
 \widetilde O=\sum_{p=1}^{\ell}\sum_{\nu=1}^{M_p}O_{p,\nu},
 \qquad
 \|\Delta\|\le\varepsilon,
 \label{eq:native-cut-decomposition}
\end{align}
and that
\begin{align}
 \operatorname{SR}_p(O_{p,\nu})\le r_{p,\nu}.
\end{align}
Then
\begin{align}
 \operatorname{BSR}(\widetilde O)
 \le \sum_{p=1}^{\ell}\sum_{\nu=1}^{M_p}r_{p,\nu},
 \label{eq:native-cut-bsr-bound}
\end{align}
and therefore
\begin{align}
 \varepsilon_D^{\mathrm{BSR}}(O)\le\varepsilon,
 \qquad
 D:=\sum_{p,\nu}r_{p,\nu}.
 \label{eq:native-cut-bsr-approx}
\end{align}
For a fixed target cut $q$, the corresponding bound is instead
\begin{align}
 \operatorname{SR}_q(\widetilde O)
 \le\sum_{p,\nu}\chi_{p,q}r_{p,\nu}.
 \label{eq:native-cut-fixed-rank}
\end{align}
\end{prop}

In particular, a cut-count approximation of the form obtained in the
preceding section,
\begin{align}
 \widetilde\Phi_Q
 =\sum_{p,s,\boldsymbol z}
 C_{p,s}(\boldsymbol z)\,
 \Phi_{p,\boldsymbol z}^{(s)},
 \label{eq:differential-bsr-form}
\end{align}
obeys the direct estimate
\begin{align}
 \operatorname{BSR}(\widetilde\Phi_Q)
 \le\sum_{p,s,\boldsymbol z}
 \operatorname{SR}_p\!\left(\Phi_{p,\boldsymbol z}^{(s)}\right).
 \label{eq:differential-bsr-bound}
\end{align}
Thus every differential branch can be kept at the cut selected by the
cut-count decomposition.  Converting these branchwise bounds to a fixed
target cut using Proposition~\ref{prop:basic-bsr-properties} introduces
the factors $\chi_{p,q}$.  This distinction is precisely the gain captured
by a low-BSR approximation.

A low-BSR approximation does not, by itself, imply a uniformly small
Schmidt rank across all cuts, a small matrix-product-operator bond dimension,
or efficient tensor-network contraction.  Such conclusions require either
a direct fixed-cut estimate or additional control of the conversion factors
$\chi_{p,q}$.

\subsection{A minimal operational interpretation}

We next state an information-theoretic consequence that follows directly
from the definition.  It concerns coherent quantum transmission from $A$ to
$C$ when the receiver has access only to $C$ and the buffer is initialized
without prior entanglement.
The final fidelity estimate is the usual bound on the overlap of a state
of bounded Schmidt number with a maximally entangled state
\cite{TerhalHorodecki2000}; the BSR-specific step is the bound on the
matrix rank of each induced Kraus operator proved below.

\begin{prop}[Endpoint quantum-transfer bound]
\label{prop:bsr-transfer-bound}
Let $\mathcal G_{ABC}$ be a quantum channel satisfying
\begin{align}
 \operatorname{BSR}_{\mathrm{Kraus}}(\mathcal G_{ABC})\le D.
\end{align}
Initialize $BC$ in the layer-product pure state
$|0\rangle_{BC}=(\bigotimes_{r=1}^{\ell-1}|0_r\rangle_{L_r})\otimes
|0\rangle_C$, with the empty-product convention when $\ell=1$, and define
the induced channel
\begin{align}
 \mathcal N_{A\to C}(\rho_A)
 :=\operatorname{tr}_{AB}\!\left[
 \mathcal G_{ABC}\!\left(
 \rho_A\otimes|0\rangle\langle0|_{BC}
 \right)
 \right].
 \label{eq:induced-endpoint-channel}
\end{align}
Then $\mathcal N_{A\to C}$ admits a Kraus representation
$\{M_\eta\}_\eta$ such that
\begin{align}
 \operatorname{rank}(M_\eta)\le D
 \qquad\text{for every }\eta.
 \label{eq:induced-kraus-rank}
\end{align}
Consequently, for every isometric encoder of a $K$-dimensional quantum
system into $A$ and every decoding channel from $C$ to a $K$-dimensional
output system, the entanglement fidelity satisfies
\begin{align}
 F_{\mathrm{ent}}^{(K)}
 \le \min\left\{1,\frac{D}{K}\right\}.
 \label{eq:exact-bsr-entanglement-fidelity}
\end{align}
Here $F_{\mathrm{ent}}^{(K)}$ is the overlap with the normalized maximally
entangled state $|\Phi_K\rangle=K^{-1/2}\sum_{j=1}^K|j\rangle_R|j\rangle$
after sending the second subsystem through the chosen encoder, induced
channel, and decoder.  The same bound therefore holds after optimizing over
such encoders and decoders.
In particular, for $Q\in\mathbb Z_{\ge0}$, perfect transmission of $Q$
qubits requires
\begin{align}
 D\ge 2^Q.
 \label{eq:perfect-transfer-requires-bsr}
\end{align}
\end{prop}

\textit{Proof.}
Choose a Kraus representation $\{K_\mu\}_\mu$ of $\mathcal G_{ABC}$ such
that $\operatorname{BSR}(K_\mu)\le D$ for every $\mu$, and choose a
layer-product basis
$|e\rangle_{AB}=|e_A\rangle_A\otimes
\bigotimes_{r=1}^{\ell-1}|e_r\rangle_{L_r}$
of the discarded output systems.  The induced
Kraus operators from $A$ to $C$ are
\begin{align}
 M_{\mu,e}:={}_{AB}\!\langle e|K_\mu|0\rangle_{BC}.
 \label{eq:induced-kraus-operator}
\end{align}
For each buffered product term in $K_\mu$, contraction of the left factor
with the $AB_{<p}$ part of $\langle e|$ and $|0\rangle$ produces a bra on
$A$, whereas contraction of the right factor produces a ket on $C$.
Therefore,
\begin{align}
 M_{\mu,e}
 =\sum_{j=1}^{D_\mu}|c_{\mu,e,j}\rangle_C
 \langle a_{\mu,e,j}|_A,
 \qquad D_\mu\le D,
 \label{eq:induced-rank-one-sum}
\end{align}
which proves Eq.~\eqref{eq:induced-kraus-rank}.

Let $E:\mathbb C^K\to\mathcal H_A$ be the encoder and let
$\{D_\nu\}_\nu$ be Kraus operators of the decoder.  The resulting channel on
the $K$-dimensional system has Kraus operators
$N_{\mu,e,\nu}:=D_\nu M_{\mu,e}E$, each of matrix rank at most $D$.  Therefore,
\begin{align*}
 F_{\mathrm{ent}}^{(K)}
 &=\frac{1}{K^2}\sum_{\mu,e,\nu}
   \left|\operatorname{tr}N_{\mu,e,\nu}\right|^2\\
 &\le\frac{D}{K^2}\sum_{\mu,e,\nu}
   \operatorname{tr}\!\left(N_{\mu,e,\nu}^\dagger
   N_{\mu,e,\nu}\right)
 =\frac{D}{K}.
\end{align*}
Here we used
$|\operatorname{tr}N|^2\le\operatorname{rank}(N)
\operatorname{tr}(N^\dagger N)$ and trace preservation in the last equality.
This proves Eq.~\eqref{eq:exact-bsr-entanglement-fidelity}. $\square$

For a unitary channel $\mathcal U(X)=UXU^\dagger$, the hypothesis in
Proposition~\ref{prop:bsr-transfer-bound} reduces to
$\operatorname{BSR}(U)\le D$ by Eq.~\eqref{eq:unitary-kraus-bsr}.  The same
argument remains useful when $U$ is only approximated by a low-BSR operator,
even if the approximating operator is not itself unitary.

\begin{corol}[Approximate low-BSR evolution]
\label{cor:approximate-bsr-transfer}
Let $U$ be unitary and suppose that an operator $\widetilde U$ satisfies
\begin{align}
 \|U-\widetilde U\|\le\varepsilon,
 \qquad
 \operatorname{BSR}(\widetilde U)\le D.
 \label{eq:approximate-bsr-unitary}
\end{align}
Under the same initialization and endpoint-only encoding/decoding setup as
in Proposition~\ref{prop:bsr-transfer-bound},
\begin{align}
 F_{\mathrm{ent}}^{(K)}(U)
 \le
 \min\left\{
 1,
 \left[(1+\varepsilon)\sqrt{\frac{D}{K}}+\varepsilon\right]^2
 \right\}.
 \label{eq:approximate-bsr-fidelity}
\end{align}
For $Q\in\mathbb Z_{\ge0}$ and $K=2^Q$, the exact case
$\varepsilon=0$ gives
\begin{align}
 F_{\mathrm{ent}}^{(2^Q)}(U)\le D\,2^{-Q}.
 \label{eq:qubit-transfer-bsr-bound}
\end{align}
\end{corol}

\textit{Proof.}
Use the same isometric encoder and an isometric dilation of the decoder
for both evolutions, retaining all discarded output systems.  Let $P_K$ be the
projector
$P_K:=|\Phi_K\rangle\langle\Phi_K|\otimes\mathbf 1_{\mathrm{env}}$,
where the identity acts on all environment systems.  Denote the exact and
approximate purified output vectors by $|\Psi\rangle$ and
$|\widetilde\Psi\rangle$, respectively; they differ in norm by at most
$\varepsilon$.  Moreover, the bound on individual Kraus-operator matrix ranks
in Proposition~\ref{prop:bsr-transfer-bound}, applied to
the generally unnormalized map produced by $\widetilde U$, gives
\begin{align}
 \|P_K|\widetilde\Psi\rangle\|
 \le \sqrt{\frac{D}{K}}\,
 \||\widetilde\Psi\rangle\|
 \le (1+\varepsilon)\sqrt{\frac{D}{K}}.
\end{align}
Indeed, the square of the left-hand side is the maximally entangled overlap
of the generally unnormalized reference--output operator obtained from the
normalized input $|\Phi_K\rangle$.  The trace of this operator is
$\||\widetilde\Psi\rangle\|^2$.  The second inequality uses
$\|\widetilde U\|\le\|U\|+\varepsilon=1+\varepsilon$.
The triangle inequality then proves
Eq.~\eqref{eq:approximate-bsr-fidelity}. $\square$

Thus, if for every $D\in\mathbb N$ an approximant $U_{t,D}$ satisfies
\begin{align}
 \|U_t-U_{t,D}\|\le\varepsilon_D(t),
 \qquad
 \operatorname{BSR}(U_{t,D})\le D,
\end{align}
then
\begin{align}
 F_{\mathrm{ent}}^{(2^Q)}(U_t)
 \le\inf_{D\in\mathbb N}
 \min\left\{
 1,
 \left[
 (1+\varepsilon_D(t))\sqrt{D\,2^{-Q}}
 +\varepsilon_D(t)
 \right]^2
 \right\}.
 \label{eq:dynamical-bsr-transfer-bound}
\end{align}
In the zero-error limit, the factor $2^{-Q}$ is recovered as the $D=1$
baseline, while an approximation with BSR at most $D$ gives the baseline
$\min\{1,D\,2^{-Q}\}$.

We emphasize the scope of this interpretation.  The BSR decomposition is an
algebraic sum and its summands are not probabilities or physically
accessible branches.  Proposition~\ref{prop:bsr-transfer-bound} controls
coherent quantum transmission to a receiver with access only to $C$; it does
not by itself give an upper bound on classical capacity, on the full mutual
information $I(A:C)$, or on protocols in which the buffer is measured and
its outcome is communicated to the receiver.  Pre-existing entanglement in
the buffer can also act as an additional resource.  Stronger operational
statements therefore require extra assumptions, such as a specified
instrument decomposition or a model-dependent conserved quantity.

\section{Low-BSR approximation of quantum dynamics (time-independent cases)}
\label{sec:low-rank-dynamics}

We first treat a time-independent Hamiltonian and its real-time evolution.
The extension to a time-independent Liouvillian is given in
Sec.~\ref{subsec:static-liouvillian-dynamics}.  Throughout this section,
$t\ge 0$.

\subsection{Time-independent unitary dynamics}
\label{subsec:static-unitary-dynamics}

Let
\begin{align}
    H=\sum_{Z\subseteq\Lambda:\,|Z|\le k}h_Z
\end{align}
be a time-independent, $k$-local, $g$-extensive Hamiltonian satisfying
\begin{align}
    \sum_{Z:\,Z\ni i}\lVert h_Z\rVert\le g
    \qquad (i\in\Lambda).
\end{align}
Since $H=H^\dagger$, replacing each $h_Z$ by
$(h_Z+h_Z^\dagger)/2$ if necessary preserves $H$, locality, and all
norm bounds.  We therefore take the fixed local decomposition to be
termwise Hermitian.  In particular, the blocks $H_{a,b}$ below and $V_0$
are Hermitian.  Let
\begin{align}
    H=\sum_{0\le a\le b\le\ell}H_{a,b},
    \qquad
    V_0:=\sum_{a=0}^{\ell}H_{a,a}
\end{align}
be the layer decomposition introduced above.  To avoid an arbitrary
single-index enumeration of the off-diagonal blocks, set
\begin{align}
    \mathcal E_\ell
    :=\{(a,b):0\le a<b\le\ell\}.
\end{align}
For $e=(a,b)\in\mathcal E_\ell$, write
\begin{align}
    H_e:=H_{a,b},
    \qquad
    r_e:=b-a.
\end{align}
We parameterize the off-diagonal blocks by
\begin{align}
    H(\vec{\lambda})
    :=V_0+\sum_{e\in\mathcal E_\ell}\lambda_eH_e,
    \qquad
    \vec{\lambda}=(\lambda_e)_{e\in\mathcal E_\ell}.
\label{eq:static-unitary-parameterization}
\end{align}
For a multi-index
$\vec{s}=(s_e)_{e\in\mathcal E_\ell}\in
\mathbb Z_{\ge0}^{\mathcal E_\ell}$, define
\begin{align}
    \vec{s}!
    &:=\prod_{e\in\mathcal E_\ell}s_e!,
    &
    \partial_{\vec{\lambda}}^{\vec{s}}
    &:=\prod_{e\in\mathcal E_\ell}\partial_{\lambda_e}^{s_e},
    &
    w(\vec{s})
    &:=\sum_{e\in\mathcal E_\ell}r_es_e.
\end{align}
Here $w(\vec{s})$ is the total number of crossed layer cuts,
counted with multiplicity.

For $r\in[\ell]$, define the total strength of the length-$r$ blocks by
\begin{align}
    K_r
    :=\sum_{e\in\mathcal E_\ell:\,r_e=r}\lVert H_e\rVert,
    \qquad
    K_\eta:=\sum_{r=1}^{\ell}e^{\eta r}K_r
    \quad (\eta>0).
\label{eq:static-unitary-length-strength}
\end{align}
We also use the two combinatorial quantities
\begin{align}
    N_{\mathrm{ph}}(Q)
    &:=\ell(2Q+1)^{\lceil\log_2\ell\rceil},
    &
    s_Q&:=\left\lfloor\frac{2Q}{\ell}\right\rfloor.
\label{eq:static-unitary-phase-prefactor}
\end{align}
Thus $N_{\mathrm{ph}}(Q)$ bounds the total number of pairs consisting
of a selected cut and a phase vector.

\subsubsection{High-level overview}
\label{subsubsec:static-unitary-overview}

We first explain the proof at the level of scaling.  The purpose of this
overview is to make the roles of the two truncation parameters transparent:
$m_\epsilon$ is the usual Taylor--Duhamel order, whereas $Q$ is a geometric
budget that counts how many layer cuts are crossed by all interaction
insertions in a monomial.  We begin with finite-range interactions and
suppress constants depending on $g$, $k$, the local Hilbert-space dimension,
and $|\partial\Lambda|$.  For $0<\epsilon<1$, set
\begin{align}
    L_\epsilon:=\log(1/\epsilon).
\end{align}

\paragraph{1. Taylor order and geometric cut budget.}
Expand the evolution in the off-diagonal interaction blocks $H_e$, keeping
$V_0$ unexpanded.  For a Taylor monomial labelled by
$\vec{s}=(s_e)_e$, define
\begin{align}
    |\vec{s}|_1:=\sum_e s_e,
    \qquad
    w(\vec{s}):=\sum_e r_es_e.
\label{eq:overview-order-and-cut-weight}
\end{align}
Here $|\vec{s}|_1$ is the number of off-diagonal interaction insertions,
whereas $w(\vec{s})$ is their total cut weight.  An insertion of a block
$H_e=H_{a,b}$ crosses the $r_e=b-a$ layer cuts between $L_a$ and $L_b$ and
therefore consumes $r_e$ units of the budget.  Equivalently,
\begin{align}
    w(\vec{s})=\sum_{p=1}^{\ell}C_p(\vec{s}),
\end{align}
where $C_p(\vec{s})$ is the number of insertions in the monomial that cross
the cut $(p-1,p)$.  For example, one insertion of $H_{0,2}$ and two
insertions of $H_{1,2}$ have total cut weight $2+2\times1=4$.

The Taylor--Duhamel tail is small once the total order is truncated at
\begin{align}
    m_\epsilon=O(t\ell+L_\epsilon).
\label{eq:overview-mepsilon}
\end{align}
If the interaction range in layer units is at most $R_0$, then every
monomial of order at most $m_\epsilon$ satisfies
\begin{align}
    w(\vec{s})\le R_0m_\epsilon.
\end{align}
Thus one may take
\begin{align}
    Q=O(m_\epsilon).
\label{eq:overview-Q-mepsilon}
\end{align}
The cut-weight truncation is not identical to a total-order truncation:
it may also retain monomials with $|\vec{s}|_1>m_\epsilon$ when their
geometric cost is small.  Equation~\eqref{eq:overview-Q-mepsilon} means
that every monomial retained by the ordinary order-$m_\epsilon$
truncation fits inside a cut budget of the same scale.
After coarse graining so that only nearest-neighbor layers interact,
$R_0=1$ and one can simply take $Q=m_\epsilon$: every off-diagonal insertion
adds exactly one to the cut weight.  The rigorous error bound contains the
additional prefactor $N_{\mathrm{ph}}(Q)$ from the phase filter, so the exact
choice of $Q$ also absorbs an additive $O(\!\log N_{\mathrm{ph}}(Q))$
correction.  This additive logarithmic term is displayed explicitly in the
rigorous parameter choice below.  In the order-level notation
$Q\asymp m_\epsilon$, we simply regard it as part of the accuracy budget.

\paragraph{2. Reduction to a small number of explicit crossings.}
For every retained monomial, $w(\vec{s})\le Q$ and hence
\begin{align}
    \sum_{p=1}^{\ell}C_p(\vec{s})\le Q.
\end{align}
The cut-count reduction assigns the monomial to a cut $p$ for which
\begin{align}
    C_p(\vec{s})\le \frac{2Q}{\ell}.
\label{eq:overview-small-crossing-count}
\end{align}
The phase filter then resums all noncrossing insertions and rewrites the
retained evolution as a sum of differential branches plus a controlled
remainder from the added terms of cut weight greater than $Q$.
To state one branch, let
\begin{align}
    u_p&:=\{e=(a,b):a<p\le b\},
    &
    u_{p,Q}&:=\{e\in u_p:r_e\le Q\},
\end{align}
and, for a phase vector $\vec{z}=(z_e)_e$ with $|z_e|=1$, define
\begin{align}
    A_{p,\vec{z}}
    &:=V_0+\sum_{e\notin u_p}z_eH_e,
    &
    B_{p,\vec{z}}^{[Q]}
    &:=\sum_{e\in u_{p,Q}}z_eH_e.
\label{eq:overview-A-B}
\end{align}
The first operator contains no term crossing $(p-1,p)$, while the second is
the phase-twisted crossing interaction retained by the cut budget.  The
corresponding normalized branch is
\begin{align}
    U_{p,\vec{z}}^{(s,Q)}(t)
    :=
    \frac{1}{s!}
    \left.
    \left(\frac{d}{dx}\right)^s
    e^{-it(A_{p,\vec{z}}+xB_{p,\vec{z}}^{[Q]})}
    \right|_{x=0},
\label{eq:overview-differential-branch}
\end{align}
with
\begin{align}
    0\le s\le s_Q:=\left\lfloor\frac{2Q}{\ell}\right\rfloor
    =O\!\left(t+\frac{L_\epsilon}{\ell}\right).
\label{eq:overview-sQ}
\end{align}
The total number of cut--phase pairs is at most
\begin{align}
    N_{\mathrm{ph}}(Q)
    =\ell(2Q+1)^{\lceil\log_2\ell\rceil}
    =Q^{O(\log\ell)}.
\label{eq:overview-phase-count}
\end{align}
Because each branch is estimated at its own selected cut, the native-cut
property of the buffered Schmidt rank gives schematically
\begin{align}
    \operatorname{BSR}(\widetilde U_t)
    \lesssim
    N_{\mathrm{ph}}(Q)(s_Q+1)
    \max_{p,\vec{z},\,s\le s_Q}
    \operatorname{SR}_p
    \bigl(\widetilde U_{p,\vec{z}}^{(s,Q)}(t)\bigr).
\label{eq:overview-native-cut}
\end{align}

\paragraph{3. Short-time factorization of one differential branch.}
Choose
\begin{align}
    M:=\max\{1,\lceil4gkt\rceil\},
    \qquad
    t_0:=\frac{t}{M},
\label{eq:overview-short-time-choice}
\end{align}
so that $2gkt_0\le1/2$.  Uniformly for all $t\ge0$,
$M=O(1+t)$; for $t\gtrsim(gk)^{-1}$ this is simply $M=O(t)$, and the
additive one only handles the short-time regime.

Fix $p$ and $\vec{z}$ and abbreviate
$A:=A_{p,\vec{z}}$ and $B:=B_{p,\vec{z}}^{[Q]}$.  Define the normalized
order-$r$ coefficient of one short-time factor by
\begin{align}
    E_r(t_0)
    :=
    \frac{1}{r!}
    \left.
    \left(\frac{d}{dx}\right)^r
    e^{-it_0(A+xB)}
    \right|_{x=0}.
\label{eq:overview-Er-definition}
\end{align}
Since $e^{-it(A+xB)}=[e^{-it_0(A+xB)}]^M$, the normalized derivative obeys
the exact identity
\begin{align}
    U_{p,\vec{z}}^{(s,Q)}(t)
    =
    \sum_{\substack{r_1+\cdots+r_M=s\\r_j\ge0}}
    E_{r_M}(t_0)\cdots E_{r_1}(t_0).
\label{eq:overview-weak-compositions}
\end{align}
There are
\begin{align}
    \binom{M+s-1}{s}\le2^{M+s}
\label{eq:overview-composition-number}
\end{align}
weak compositions.  No multinomial coefficient appears in
Eq.~\eqref{eq:overview-weak-compositions} because both the long-time and
short-time derivatives are normalized by the corresponding factorials.

\paragraph{4. Local commutator expansion on one short interval.}
The interaction-picture representation of Eq.~\eqref{eq:overview-Er-definition}
is
\begin{align}
    E_r(t_0)
    =(-i)^r e^{-it_0A}
    \int_{0\le\tau_r\le\cdots\le\tau_1\le t_0}
    B_A(\tau_1)\cdots B_A(\tau_r)
    \,d\tau_1\cdots d\tau_r,
\label{eq:overview-interaction-picture}
\end{align}
where
\begin{align}
    B_A(\tau)
    &:=e^{i\tau A}Be^{-i\tau A}
    =\sum_{q=0}^{\infty}
    \frac{(i\tau)^q}{q!}\operatorname{ad}_A^q(B),
    &
    \operatorname{ad}_A(X)&:=[A,X].
\label{eq:overview-commutator-series}
\end{align}
The standard support-growth estimate for nested commutators
\cite[Lemma~3]{KUWAHARA201696}, using locality and $g$-extensivity, implies
\begin{align}
    \frac{\tau^q}{q!}
    \bigl\lVert\operatorname{ad}_A^q(B)\bigr\rVert
    \le
    (2gk\tau)^q\lVert B\rVert_{\mathrm{loc}}.
\label{eq:overview-commutator-tail}
\end{align}
Because $2gkt_0\le1/2$, the terms decay geometrically as $2^{-q}$.
Truncating the series at $q=n$ therefore produces an error of order
$2^{-n}$.  At fixed $t$ and $\ell$, this gives the familiar precision
scaling $n=O(L_\epsilon)$.  When the norm growth and the total number of
phase-filtered branches are also included, the finite-range proof requires
only
\begin{align}
    n=O\bigl(m_\epsilon+\log N_{\mathrm{ph}}(Q)\bigr),
\label{eq:overview-n-scaling}
\end{align}
so $n$ remains of the same scale as $m_\epsilon$, up to the already noted
logarithmic phase-filter correction.

\paragraph{5. Schmidt rank of one branch.}
Across the selected cut,
\begin{align}
    A=A_L\otimes\mathbf 1+\mathbf 1\otimes A_R.
\end{align}
Set
\begin{align}
    R_Q:=\max_{p,\vec{z}}
    \operatorname{SR}_p(B_{p,\vec{z}}^{[Q]}).
\end{align}
For each commutator order,
\begin{align}
    \operatorname{SR}_p\!\left(\operatorname{ad}_A^q(B)\right)
    \le(q+1)R_Q.
\end{align}
After regrouping the truncated series by the number of left commutators,
the whole polynomial $\sum_{q=0}^{n}(i\tau)^q
\operatorname{ad}_A^q(B)/q!$ has Schmidt rank at most
$(n+1)R_Q$.  The left factors in this regrouping are independent of the
integration variables, so the time integrals can be absorbed into the
right factors.  Consequently,
\begin{align}
    \operatorname{SR}_p(\widetilde E_r(t_0))
    &\le[(n+1)R_Q]^r,
    \\
    \operatorname{SR}_p
    \bigl(\widetilde U_{p,\vec{z}}^{(s,Q)}(t)\bigr)
    &\le
    \binom{M+s-1}{s}[(n+1)R_Q]^s.
\label{eq:overview-branch-rank}
\end{align}
For finite-range interactions and a bounded one-dimensional boundary,
$R_Q=O(1)$.  Since $M$, $n$, and $s_Q$ are at most polynomial in
$m_\epsilon$, Eq.~\eqref{eq:overview-branch-rank} gives the mnemonic
\begin{align}
    \operatorname{SR}_p
    \bigl(\widetilde U_{p,\vec{z}}^{(s,Q)}(t)\bigr)
    \le m_\epsilon^{O(s)}.
\label{eq:overview-branch-mnemonic}
\end{align}

\paragraph{6. Final order of magnitude.}
Combining the phase count, the bound $s\le s_Q$, and
Eq.~\eqref{eq:overview-branch-rank} yields
\begin{align}
    \operatorname{BSR}(\widetilde U_t)
    \le
    m_\epsilon^{O(\log\ell+s_Q)}
    =
    m_\epsilon^{O\left(\log\ell+t+L_\epsilon/\ell\right)},
    \qquad
    m_\epsilon=O(t\ell+L_\epsilon),
\label{eq:overview-final-scaling}
\end{align}
where the phase-filter logarithms are suppressed.  In the
accuracy-dominated regime $L_\epsilon\gtrsim t\ell$, one has
$m_\epsilon\asymp L_\epsilon$, and therefore a single differential branch
may be remembered as
\begin{align}
    \operatorname{SR}_p
    \bigl(\widetilde U_{p,\vec{z}}^{(s,Q)}(t)\bigr)
    \lesssim
    [\log(1/\epsilon)]^{O\left(t+\log(1/\epsilon)/\ell\right)}.
\label{eq:overview-accuracy-dominated}
\end{align}
The full BSR has the additional phase-filter exponent $O(\log\ell)$.
Thus the base $\log(1/\epsilon)$ is the correct mnemonic when the precision
term dominates; for general $t$, the uniform base is
$m_\epsilon\asymp t\ell+\log(1/\epsilon)$.

\paragraph{Exponentially decaying interactions.}
For genuinely long-range interactions, $w(\vec{s})$ is no longer bounded by
a constant times $|\vec{s}|_1$.  Instead, exponential decay gives a direct
cut-tail estimate of the form
\begin{align}
    \delta_Q^{\mathrm{cut}}(t)
    \lesssim e^{-\eta Q+tK_\eta}.
\end{align}
After including the phase-filter prefactor, one can still choose
\begin{align}
    Q=O\bigl(t\ell+L_\epsilon+\log N_{\mathrm{ph}}(Q)\bigr),
\end{align}
so $Q$ remains proportional to the finite-range scale $m_\epsilon$ up to
logarithmic corrections.  The explicit derivative order is still
$s_Q=O(Q/\ell)$.  The main change is the crossing rank:
$R_Q=O(1)$ is replaced by a quantity polynomial in $Q$.  For one-dimensional
two-body interactions one typically has $R_Q=O(Q)$, while for a general
$k$-local interaction the safe estimate is
\begin{align}
    R_Q\le(Cd_0^2Q|\partial\Lambda|)^k
\end{align}
under the geometric assumption stated below.  The local coefficient norm of
the crossing interaction can likewise grow polynomially with $Q$ (typically
$O(Q)$ in the same one-dimensional setting), so the commutator cutoff $n$
may also acquire a polynomial dependence on $Q$.  Both effects change only
the polynomial base of the rank bound for each branch, not its essential
exponent $O(Q/\ell)$.

For
\begin{align}
    F(\vec{\lambda}):=e^{-itH(\vec{\lambda})},
\end{align}
define the norm-absolute cut tail by
\begin{align}
    \delta_Q^{\mathrm{cut}}(t)
    :=
    \sum_{\substack{\vec{s}\in
    \mathbb Z_{\ge0}^{\mathcal E_\ell}\\
    w(\vec{s})>Q}}
    \frac{1}{\vec{s}!}
    \left\lVert
    \left.
    \partial_{\vec{\lambda}}^{\vec{s}}
    F(\vec{\lambda})
    \right|_{\vec{\lambda}=\vec{0}}
    \right\rVert.
\label{eq:static-unitary-cut-tail-definition}
\end{align}

\begin{prop}[Cut-weight tail for static unitary dynamics]
\label{prop:static-unitary-cut-tail}
For every $Q\in\mathbb Z_{\ge0}$ and every $\eta>0$,
\begin{align}
    \delta_Q^{\mathrm{cut}}(t)
    \le
    \exp\bigl[-\eta Q+tK_\eta\bigr].
\label{eq:static-unitary-cut-tail-bound}
\end{align}
\end{prop}

\textit{Proof.}
For an ordered list $e_1,\ldots,e_m\in\mathcal E_\ell$, the Duhamel
expansion gives
\begin{align}
&
\left.
\partial_{\lambda_{e_1}}\cdots\partial_{\lambda_{e_m}}
 e^{-itH(\vec{\lambda})}
\right|_{\vec{\lambda}=\vec{0}}
\nonumber\\
&\quad=
(-i)^m
\sum_{\pi\in S_m}
\int_{0\le\tau_m\le\cdots\le\tau_1\le t}
 e^{-i(t-\tau_1)V_0}
 H_{e_{\pi(1)}}
 e^{-i(\tau_1-\tau_2)V_0}
 \cdots
 H_{e_{\pi(m)}}
 e^{-i\tau_mV_0}
 \,d\tau_1\cdots d\tau_m.
\label{eq:static-unitary-mixed-derivative}
\end{align}
Since $V_0$ is Hermitian, all exponential factors in
Eq.~\eqref{eq:static-unitary-mixed-derivative} have norm one.  The sum over
$m!$ permutations cancels the factor $1/m!$ in the volume $t^m/m!$ of
the ordered simplex, and hence
\begin{align}
    \left\lVert
    \left.
    \partial_{\lambda_{e_1}}\cdots\partial_{\lambda_{e_m}}
    e^{-itH(\vec{\lambda})}
    \right|_{\vec{\lambda}=\vec{0}}
    \right\rVert
    \le
    t^m\prod_{a=1}^{m}\lVert H_{e_a}\rVert.
\end{align}
Therefore,
\begin{align}
    \delta_Q^{\mathrm{cut}}(t)
    &\le
    \sum_{m=0}^{\infty}\frac{t^m}{m!}
    \sum_{\substack{r_1,\ldots,r_m\in[\ell]\\
    r_1+\cdots+r_m>Q}}
    K_{r_1}\cdots K_{r_m}
    \nonumber\\
    &\le
    e^{-\eta Q}
    \sum_{m=0}^{\infty}\frac{1}{m!}
    \left(t\sum_{r=1}^{\ell}e^{\eta r}K_r\right)^m,
\end{align}
which proves Eq.~\eqref{eq:static-unitary-cut-tail-bound}.
\hfill\(\square\)

For the cut $(p-1,p)$, define
\begin{align}
    u_p
    &:=\{e=(a,b)\in\mathcal E_\ell:a<p\le b\},
    \\
    u_{p,Q}
    &:=\{e\in u_p:r_e\le Q\}.
\label{eq:static-unitary-crossing-sets}
\end{align}
For a phase vector
\begin{align}
    \vec{z}=(z_e)_{e\in\mathcal E_\ell},
    \qquad |z_e|=1,
\end{align}
define
\begin{align}
    A_{p,\vec{z}}
    &:=V_0+\sum_{e\notin u_p}z_eH_e,
    \\
    B_{p,\vec{z}}^{[Q]}
    &:=\sum_{e\in u_{p,Q}}z_eH_e.
\label{eq:static-unitary-A-B}
\end{align}
No term in $A_{p,\vec{z}}$ crosses the cut $(p-1,p)$.
Consequently, there exist operators $A^L_{p,\vec{z}}$ and
$A^R_{p,\vec{z}}$ such that
\begin{align}
    A_{p,\vec{z}}
    =A^L_{p,\vec{z}}\otimes\mathbf 1
    +\mathbf 1\otimes A^R_{p,\vec{z}}.
\label{eq:static-unitary-A-factorization}
\end{align}

\begin{prop}[$Q$-localized cut-count decomposition]
\label{prop:static-unitary-Q-localized}
For every $Q\in\mathbb Z_{\ge0}$, there exist finite
indexed families of phase vectors
\begin{align}
    \mathcal Z_{p,Q}
    \subseteq
    \{\vec{z}:|z_e|=1\ \text{for every }e\in\mathcal E_\ell\}
\end{align}
and coefficients $c_{p,Q}(\vec{z})$ such that
\begin{align}
    e^{-itH}
    =
    \sum_{p=1}^{\ell}
    \sum_{s=0}^{s_Q}
    \sum_{\vec{z}\in\mathcal Z_{p,Q}}
    c_{p,Q}(\vec{z})
    U_{p,\vec{z}}^{(s,Q)}(t)
    +\Delta_Q^{\mathrm{cut}},
\label{eq:static-unitary-Q-localized-decomposition}
\end{align}
where
\begin{align}
    U_{p,\vec{z}}^{(s,Q)}(t)
    :=
    \frac{1}{s!}
    \left.
    \left(\frac{d}{dx}\right)^s
    e^{-it(A_{p,\vec{z}}+xB_{p,\vec{z}}^{[Q]})}
    \right|_{x=0}.
\label{eq:static-unitary-differential-branch}
\end{align}
Moreover,
\begin{align}
    |\mathcal Z_{p,Q}|
    &\le(2Q+1)^{\lceil\log_2\ell\rceil},
    &
    |c_{p,Q}(\vec{z})|&\le1,
\end{align}
and, for every $\eta>0$,
\begin{align}
    \lVert\Delta_Q^{\mathrm{cut}}\rVert
    \le
    2N_{\mathrm{ph}}(Q)
    \exp\bigl[-\eta Q+tK_\eta\bigr].
\label{eq:static-unitary-Q-localized-error}
\end{align}
\end{prop}

\textit{Proof.}
Apply the coefficientwise proof of the cut-count decomposition to the
retained set
\begin{align}
    \mathcal S_Q
    :=\{\vec{s}:w(\vec{s})\le Q\}.
\end{align}
If $\vec{s}\in\mathcal S_Q$ and $s_e\ge1$, then
$r_e\le w(\vec{s})\le Q$.  Hence, in every retained monomial,
the variables crossing the selected cut belong to $u_{p,Q}$, and the
explicit crossing derivative may be written with
$B_{p,\vec{z}}^{[Q]}$.

Completing each filtered Taylor sum to the full branch in
Eq.~\eqref{eq:static-unitary-differential-branch} introduces only
multi-indices with cut weight larger than $Q$.  For a fixed pair
$(p,\vec{z})$, the terms with different $s$ have different total
crossing multiplicity at the selected cut.  Hence, after summing over
$s$, every additional multi-index is counted at most once, and its
norm-absolute contribution is bounded by
$\delta_Q^{\mathrm{cut}}(t)$.  There are at most
$N_{\mathrm{ph}}(Q)$ such pairs in total.  Adding the original Taylor
tail gives
\begin{align}
    \lVert\Delta_Q^{\mathrm{cut}}\rVert
    \le
    \bigl[1+N_{\mathrm{ph}}(Q)\bigr]
    \delta_Q^{\mathrm{cut}}(t)
    \le
    2N_{\mathrm{ph}}(Q)\delta_Q^{\mathrm{cut}}(t).
\end{align}
Proposition~\ref{prop:static-unitary-cut-tail} now proves
Eq.~\eqref{eq:static-unitary-Q-localized-error}.
\hfill\(\square\)

Define the total off-diagonal strength
\begin{align}
    J_{\mathrm{off}}
    :=\sum_{e\in\mathcal E_\ell}\lVert H_e\rVert.
\label{eq:static-unitary-offdiagonal-strength}
\end{align}
Under the decay assumption of the setup,
$J_{\mathrm{off}}\le \bar g_1\ell|\partial\Lambda|$.

\begin{lemma}[Stability of the phase-twisted background]
\label{lem:static-unitary-background-stability}
For every $p$, every phase vector $\vec{z}$, and every
$\tau\in\mathbb R$,
\begin{align}
    \max\left\{
    \lVert e^{-i\tau A_{p,\vec{z}}}\rVert,
    \lVert e^{i\tau A_{p,\vec{z}}}\rVert
    \right\}
    \le e^{J_{\mathrm{off}}|\tau|}.
\label{eq:static-unitary-background-stability}
\end{align}
Moreover,
\begin{align}
    \operatorname{SR}_p(e^{-i\tau A_{p,\vec{z}}})=1.
\label{eq:static-unitary-background-rank}
\end{align}
\end{lemma}

\textit{Proof.}
Write
\begin{align}
    A_{p,\vec{z}}=V_0+W_{p,\vec{z}},
    \qquad
    W_{p,\vec{z}}:=\sum_{e\notin u_p}z_eH_e.
\end{align}
Then $\lVert W_{p,\vec{z}}\rVert\le J_{\mathrm{off}}$.
Passing to the interaction picture with respect to the Hermitian
operator $V_0$ and using the Dyson expansion gives
\begin{align}
    \lVert e^{-i\tau A_{p,\vec{z}}}\rVert
    \le e^{|\tau|\lVert W_{p,\vec{z}}\rVert}
    \le e^{J_{\mathrm{off}}|\tau|}.
\end{align}
The same argument applies to $e^{i\tau A_{p,\vec{z}}}$.
Equation~\eqref{eq:static-unitary-A-factorization} gives
\begin{align}
    e^{-i\tau A_{p,\vec{z}}}
    =e^{-i\tau A^L_{p,\vec{z}}}
    \otimes e^{-i\tau A^R_{p,\vec{z}}},
\end{align}
which proves Eq.~\eqref{eq:static-unitary-background-rank}.
\hfill\(\square\)

All coefficient norms below refer to the fixed local decomposition
$H=\sum_Zh_Z$.  If
\begin{align}
    O=\sum_Zc_Zh_Z,
\end{align}
set
\begin{align}
    \lVert O\rVert_{\mathrm{loc}}
    :=\sum_Z|c_Z|\lVert h_Z\rVert.
\label{eq:static-unitary-interaction-weight}
\end{align}
Define
\begin{align}
    b_Q
    &:={}
    \max_{p\in[\ell]}
    \sup_{|z_e|=1}
    \lVert B_{p,\vec{z}}^{[Q]}\rVert_{\mathrm{loc}},
    \\
    R_Q
    &:={}
    \max_{p\in[\ell]}
    \sup_{|z_e|=1}
    \operatorname{SR}_p(B_{p,\vec{z}}^{[Q]}).
\label{eq:static-unitary-bQ-RQ}
\end{align}

\begin{lemma}[Nested-commutator estimate]
\label{lem:static-unitary-nested-commutator}
Let $A$ be a phase-twisted sub-sum of the local terms of $H$, and let
$B$ be another such sub-sum.  Then, for every
$m\in\mathbb Z_{\ge0}$,
\begin{align}
    \left\lVert\operatorname{ad}_A^m(B)\right\rVert
    \le(2gk)^m m!\,\lVert B\rVert_{\mathrm{loc}},
    \qquad
    \operatorname{ad}_A(B):=[A,B].
\label{eq:static-unitary-nested-commutator-bound}
\end{align}
\end{lemma}

\textit{Proof.}
It is enough to consider one local term $b_Y$ with $|Y|\le k$.
After $a$ commutators, every generated term is supported on at most
$k(a+1)$ sites.  By $g$-extensivity, the sum of the norms of the local
terms in $A$ intersecting such a support is at most $gk(a+1)$.
Using $\lVert[X,Y]\rVert\le2\lVert X\rVert\lVert Y\rVert$ at each
step gives
\begin{align}
    \left\lVert\operatorname{ad}_A^m(b_Y)\right\rVert
    \le
    \prod_{a=0}^{m-1}2gk(a+1)\,\lVert b_Y\rVert
    =(2gk)^m m!\,\lVert b_Y\rVert.
\end{align}
Summing with the coefficient weights in
Eq.~\eqref{eq:static-unitary-interaction-weight} proves the claim.
\hfill\(\square\)

Choose $M\in\mathbb N$ and set
\begin{align}
    t_0:=\frac{t}{M},
    \qquad
    \theta_0:=2gkt_0.
\label{eq:static-unitary-short-time-parameters}
\end{align}
We assume $\theta_0<1$.  For the remainder of the unitary argument, fix
$p$ and $\vec{z}$ and abbreviate
\begin{align}
    A:=A_{p,\vec{z}},
    \qquad
    B:=B_{p,\vec{z}}^{[Q]}.
\end{align}
Define
\begin{align}
    B_A(\tau)
    &:=e^{i\tau A}Be^{-i\tau A},
    \\
    B_{A,n}(\tau)
    &:=\sum_{m=0}^{n}
    \frac{(i\tau)^m}{m!}\operatorname{ad}_A^m(B),
\label{eq:static-unitary-commutator-polynomial}
\end{align}
and set
\begin{align}
    \bar b_Q
    :=\frac{b_Q}{1-\theta_0},
    \qquad
    \varepsilon_{Q,n}
    :=\bar b_Q\theta_0^{n+1}.
\label{eq:static-unitary-delta-beta}
\end{align}

\begin{lemma}[Short-time commutator truncation]
\label{lem:static-unitary-short-time-truncation}
For every $0\le\tau\le t_0$,
\begin{align}
    \lVert B_A(\tau)-B_{A,n}(\tau)\rVert
    &\le\varepsilon_{Q,n},
    \\
    \max\{\lVert B_A(\tau)\rVert,
    \lVert B_{A,n}(\tau)\rVert\}
    &\le\bar b_Q.
\label{eq:static-unitary-short-time-truncation}
\end{align}
\end{lemma}

\textit{Proof.}
The commutator series is
\begin{align}
    B_A(\tau)
    =\sum_{m=0}^{\infty}
    \frac{(i\tau)^m}{m!}\operatorname{ad}_A^m(B).
\end{align}
Lemma~\ref{lem:static-unitary-nested-commutator} and
$2gk\tau\le\theta_0$ give
\begin{align}
    \lVert B_A(\tau)-B_{A,n}(\tau)\rVert
    \le b_Q\sum_{m=n+1}^{\infty}\theta_0^m
    =\varepsilon_{Q,n}.
\end{align}
The same geometric-series estimate gives the second inequality.
\hfill\(\square\)

\begin{lemma}[Static commutator factorization]
\label{lem:static-unitary-commutator-factorization}
Suppose that, across the cut $(p-1,p)$,
\begin{align}
    A=A_L\otimes\mathbf 1+\mathbf 1\otimes A_R,
    \qquad
    B=\sum_{\alpha=1}^{R}B^L_\alpha\otimes B^R_\alpha
    \qquad (R\ge1).
\end{align}
Then
\begin{align}
B_{A,n}(\tau)
&=
\sum_{\alpha=1}^{R}
\sum_{a=0}^{n}
\operatorname{ad}_{A_L}^{a}(B^L_\alpha)
\otimes
\Biggl[
\frac{(i\tau)^a}{a!}
\sum_{b=0}^{n-a}
\frac{(i\tau)^b}{b!}
\operatorname{ad}_{A_R}^{b}(B^R_\alpha)
\Biggr].
\label{eq:static-unitary-commutator-factorization}
\end{align}
Consequently,
\begin{align}
    \operatorname{SR}_p(B_{A,n}(\tau))
    \le(n+1)R.
\label{eq:static-unitary-commutator-rank}
\end{align}
Moreover, for every $r\in\mathbb Z_{\ge0}$,
\begin{align}
&
\operatorname{SR}_p\Biggl(
\int_{0\le\tau_r\le\cdots\le\tau_1\le t_0}
B_{A,n}(\tau_1)\cdots B_{A,n}(\tau_r)
\,d\tau_1\cdots d\tau_r
\Biggr)
\le[(n+1)R]^r.
\label{eq:static-unitary-integrated-rank}
\end{align}
\end{lemma}

\textit{Proof.}
The superoperators
$\operatorname{ad}_{A_L}\otimes\operatorname{id}$ and
$\operatorname{id}\otimes\operatorname{ad}_{A_R}$ commute.  Hence
\begin{align}
    \operatorname{ad}_A^m(B)
    =
    \sum_{\alpha=1}^{R}
    \sum_{a=0}^{m}
    \binom{m}{a}
    \operatorname{ad}_{A_L}^{a}(B^L_\alpha)
    \otimes
    \operatorname{ad}_{A_R}^{m-a}(B^R_\alpha).
\end{align}
Substituting this identity into $B_{A,n}(\tau)$ and writing $m=a+b$
gives Eq.~\eqref{eq:static-unitary-commutator-factorization}.
For the time-ordered product, fix the labels
$(\alpha_1,a_1),\ldots,(\alpha_r,a_r)$.  The corresponding left
operator is independent of the integration variables, while the
complete time integral can be absorbed into one right operator.  There
are at most $[(n+1)R]^r$ label tuples.
\hfill\(\square\)

For $r\in\mathbb Z_{\ge0}$, define the normalized coefficient of one
short-time factor by
\begin{align}
    E_r
    :=
    \frac{1}{r!}
    \left.
    \left(\frac{d}{dx}\right)^r
    e^{-it_0(A+xB)}
    \right|_{x=0}.
\label{eq:static-unitary-short-step-coefficient}
\end{align}
The interaction-picture representation is
\begin{align}
    E_r
    =
    (-i)^r e^{-it_0A}
    \int_{0\le\tau_r\le\cdots\le\tau_1\le t_0}
    B_A(\tau_1)\cdots B_A(\tau_r)
    \,d\tau_1\cdots d\tau_r.
\label{eq:static-unitary-short-step-interaction-picture}
\end{align}
For $r\ge1$, define
\begin{align}
    \widetilde E_{r,n}
    :=
    (-i)^r e^{-it_0A}
    \int_{0\le\tau_r\le\cdots\le\tau_1\le t_0}
    B_{A,n}(\tau_1)\cdots B_{A,n}(\tau_r)
    \,d\tau_1\cdots d\tau_r,
\label{eq:static-unitary-short-step-approximant}
\end{align}
and set
\begin{align}
    \widetilde E_{0,n}:=E_0=e^{-it_0A}.
\end{align}

\begin{prop}[One-step differential approximation]
\label{prop:static-unitary-one-step}
The zeroth coefficient satisfies
\begin{align}
    \operatorname{SR}_p(\widetilde E_{0,n})=1,
\end{align}
and, for every $r\ge1$,
\begin{align}
    \operatorname{SR}_p(\widetilde E_{r,n})
    \le[(n+1)R_Q]^r.
\label{eq:static-unitary-one-step-rank}
\end{align}
Moreover, for every $r\ge0$,
\begin{align}
    \max\{\lVert E_r\rVert,
    \lVert\widetilde E_{r,n}\rVert\}
    \le
    e^{J_{\mathrm{off}}t_0}
    \frac{(t_0\bar b_Q)^r}{r!},
\label{eq:static-unitary-one-step-norm}
\end{align}
and, for $r\ge1$,
\begin{align}
    \lVert E_r-\widetilde E_{r,n}\rVert
    \le
    e^{J_{\mathrm{off}}t_0}
    \frac{t_0^r}{(r-1)!}
    \varepsilon_{Q,n}\bar b_Q^{r-1}.
\label{eq:static-unitary-one-step-error}
\end{align}
\end{prop}

\textit{Proof.}
If $B=0$, the rank and error assertions are immediate.  Otherwise, the
rank bound follows from
Lemmas~\ref{lem:static-unitary-background-stability} and
\ref{lem:static-unitary-commutator-factorization}, with
$R=\operatorname{SR}_p(B)\le R_Q$.  The norm bound
follows from Eqs.~\eqref{eq:static-unitary-background-stability} and
\eqref{eq:static-unitary-short-time-truncation}, together with the
simplex volume $t_0^r/r!$.  For the error, expand the difference of the
two $r$-fold products by a telescoping sum.  Each summand contains one
factor bounded by $\varepsilon_{Q,n}$ and $r-1$ factors bounded by
$\bar b_Q$.  Multiplication by $t_0^r/r!$ proves
Eq.~\eqref{eq:static-unitary-one-step-error}.
\hfill\(\square\)

Since the generator is time independent,
\begin{align}
    e^{-it(A+xB)}
    =\left[e^{-it_0(A+xB)}\right]^M.
\end{align}
Using normalized Taylor coefficients,
\begin{align}
    U_{p,\vec{z}}^{(s,Q)}(t)
    =
    \sum_{\substack{r_1+\cdots+r_M=s\\r_j\ge0}}
    E_{r_M}\cdots E_{r_1}.
\label{eq:static-unitary-M-step-exact}
\end{align}
Define
\begin{align}
    \widetilde U_{p,\vec{z},n}^{(s,Q)}(t)
    :=
    \sum_{\substack{r_1+\cdots+r_M=s\\r_j\ge0}}
    \widetilde E_{r_M,n}\cdots\widetilde E_{r_1,n}.
\label{eq:static-unitary-M-step-approximant}
\end{align}

\begin{prop}[Approximation of a static differential branch]
\label{prop:static-unitary-branch}
For $s=0$,
\begin{align}
    \operatorname{SR}_p
    (\widetilde U_{p,\vec{z},n}^{(0,Q)}(t))=1,
\end{align}
while, for every $s\ge1$,
\begin{align}
    \operatorname{SR}_p
    (\widetilde U_{p,\vec{z},n}^{(s,Q)}(t))
    \le
    \binom{M+s-1}{s}[(n+1)R_Q]^s.
\label{eq:static-unitary-branch-rank}
\end{align}
The approximation is exact for $s=0$.  For $s\ge1$,
\begin{align}
&
\left\lVert
U_{p,\vec{z}}^{(s,Q)}(t)
-\widetilde U_{p,\vec{z},n}^{(s,Q)}(t)
\right\rVert
\le
 e^{J_{\mathrm{off}}t}
 t\varepsilon_{Q,n}
 \frac{(t\bar b_Q)^{s-1}}{(s-1)!}.
\label{eq:static-unitary-branch-error}
\end{align}
\end{prop}

\textit{Proof.}
For each weak composition $r_1+\cdots+r_M=s$, fixed-cut
submultiplicativity and Proposition~\ref{prop:static-unitary-one-step}
give the rank $[(n+1)R_Q]^s$ when $s\ge1$.  There are
$\binom{M+s-1}{s}$ weak compositions.

For the error estimate, define
\begin{align}
    f(x)
    &:=\sum_{r=0}^{\infty}
    e^{J_{\mathrm{off}}t_0}
    \frac{(t_0\bar b_Qx)^r}{r!}
    =e^{J_{\mathrm{off}}t_0}e^{t_0\bar b_Qx},
    \\
    g(x)
    &:=\sum_{r=1}^{\infty}
    e^{J_{\mathrm{off}}t_0}
    \frac{t_0^r\varepsilon_{Q,n}\bar b_Q^{r-1}}
    {(r-1)!}x^r
    =e^{J_{\mathrm{off}}t_0}t_0\varepsilon_{Q,n}x
    e^{t_0\bar b_Qx}.
\end{align}
A telescoping expansion of the $M$-fold product shows that the
order-$s$ error is bounded by the coefficient of $x^s$ in
$Mg(x)f(x)^{M-1}$.  Since $Mt_0=t$,
\begin{align}
    Mg(x)f(x)^{M-1}
    =e^{J_{\mathrm{off}}t}t\varepsilon_{Q,n}x
    e^{t\bar b_Qx}.
\end{align}
Extracting the coefficient of $x^s$ proves
Eq.~\eqref{eq:static-unitary-branch-error}.
\hfill\(\square\)

For $m\in\mathbb Z_{\ge0}$ and $x\ge0$, define the truncated
exponential
\begin{align}
    \exp_m(x):=\sum_{q=0}^{m}\frac{x^q}{q!},
    \qquad
    \exp_{-1}(x):=0.
\end{align}
Define the final approximant
\begin{align}
    \widetilde U_{t,Q,n}
    :=
    \sum_{p=1}^{\ell}
    \sum_{s=0}^{s_Q}
    \sum_{\vec{z}\in\mathcal Z_{p,Q}}
    c_{p,Q}(\vec{z})
    \widetilde U_{p,\vec{z},n}^{(s,Q)}(t).
\label{eq:static-unitary-final-approximant}
\end{align}

\begin{theorem}[Low-BSR approximation of time-independent unitary dynamics]
\label{thm:static-unitary-low-BSR}
Let $Q,n\in\mathbb Z_{\ge0}$ and $M\in\mathbb N$, and assume
$\theta_0=2gkt/M<1$.  Then, for every $\eta>0$,
\begin{align}
&
\lVert e^{-itH}-\widetilde U_{t,Q,n}\rVert
\nonumber\\
&\quad\le
2N_{\mathrm{ph}}(Q)
\exp[-\eta Q+tK_\eta]
\nonumber\\
&\qquad+
N_{\mathrm{ph}}(Q)
 e^{J_{\mathrm{off}}t}
 t\varepsilon_{Q,n}
 \exp_{s_Q-1}(t\bar b_Q).
\label{eq:static-unitary-final-error}
\end{align}
Moreover,
\begin{align}
    \operatorname{BSR}(\widetilde U_{t,Q,n})
    &\le
    N_{\mathrm{ph}}(Q)
    \left[
    1+\sum_{s=1}^{s_Q}
    \binom{M+s-1}{s}[(n+1)R_Q]^s
    \right]
\label{eq:static-unitary-final-BSR-exact}
\\
    &\le
    N_{\mathrm{ph}}(Q)(s_Q+1)
    \left[(M+s_Q)(n+1)\max\{1,R_Q\}\right]^{s_Q}.
\label{eq:static-unitary-final-BSR-simple}
\end{align}
\end{theorem}

\textit{Proof.}
The first term in Eq.~\eqref{eq:static-unitary-final-error} is the
cut-count remainder from
Proposition~\ref{prop:static-unitary-Q-localized}.  For the branch
approximation error, use $|c_{p,Q}(\vec{z})|\le1$ and sum
Eq.~\eqref{eq:static-unitary-branch-error} over $p$,
$\vec{z}$, and $1\le s\le s_Q$.  This gives the second term.

Each branch is controlled at the cut selected by the cut-count
decomposition.  The native-cut compression property of the BSR and
Eq.~\eqref{eq:static-unitary-branch-rank} give
Eq.~\eqref{eq:static-unitary-final-BSR-exact}.  Finally,
$\binom{M+s-1}{s}\le(M+s)^s$ gives
Eq.~\eqref{eq:static-unitary-final-BSR-simple}.
\hfill\(\square\)

\begin{corol}[A convenient short-time choice]
\label{corol:static-unitary-parameter-choice}
Choose
\begin{align}
    M:=\max\{1,\lceil4gkt\rceil\}.
\end{align}
Then $\theta_0\le1/2$, $\bar b_Q\le2b_Q$, and
$\varepsilon_{Q,n}\le b_Q2^{-n}$.  Hence, for every $\eta>0$,
\begin{align}
&
\lVert e^{-itH}-\widetilde U_{t,Q,n}\rVert
\nonumber\\
&\quad\le
2N_{\mathrm{ph}}(Q)e^{-\eta Q+tK_\eta}
+
N_{\mathrm{ph}}(Q)e^{J_{\mathrm{off}}t}
 t b_Q2^{-n}\exp_{s_Q-1}(2tb_Q).
\label{eq:static-unitary-error-rho-half}
\end{align}
When $t>0$, $s_Q\ge1$, $b_Q>0$, and
$\epsilon_{\mathrm{br}}>0$, the branch error is at most
$\epsilon_{\mathrm{br}}$ if
\begin{align}
    n\ge
    \max\left\{0,
    \left\lceil
    \log_2\left[
    \frac{
    N_{\mathrm{ph}}(Q)e^{J_{\mathrm{off}}t}
    t b_Q\exp_{s_Q-1}(2tb_Q)
    }{\epsilon_{\mathrm{br}}}
    \right]
    \right\rceil
    \right\}.
\label{eq:static-unitary-choice-n}
\end{align}
If $t=0$, $s_Q=0$, or $b_Q=0$, one may take $n=0$.
\end{corol}

\textit{Proof.}
Substitute $\theta_0\le1/2$ into
Theorem~\ref{thm:static-unitary-low-BSR}.
\hfill\(\square\)

Suppose now that
\begin{align}
    K_r\le\kappa\ell|\partial\Lambda|e^{-\mu r}
    \qquad (r\in[\ell])
\label{eq:static-unitary-exponential-decay}
\end{align}
for constants $\kappa,\mu>0$.  Then, for every $0<\eta<\mu$,
\begin{align}
    K_\eta
    \le
    \frac{\kappa\ell|\partial\Lambda|}
    {e^{\mu-\eta}-1}.
\label{eq:static-unitary-Keta-exponential}
\end{align}
Thus, for $\epsilon_{\mathrm{cut}}>0$, the cut-count contribution is
at most $\epsilon_{\mathrm{cut}}$ whenever the integer $Q$ satisfies
the implicit inequality
\begin{align}
    Q\ge\frac{1}{\eta}\left[
    \frac{\kappa t\ell|\partial\Lambda|}
    {e^{\mu-\eta}-1}
    +\log\left(\frac{2N_{\mathrm{ph}}(Q)}
    {\epsilon_{\mathrm{cut}}}\right)
    \right].
\label{eq:static-unitary-choice-Q}
\end{align}

For geometric bounds on $b_Q$ and $R_Q$, let $e(Z)$ denote the unique
off-diagonal block index containing an off-diagonal local term $h_Z$,
and define
\begin{align}
    \Lambda_{p,Q}^{\times}
    &:={}
    \bigcup_{Z:\,e(Z)\in u_{p,Q}}Z,
    \\
    n_Q^{\times}
    &:=\max_{p\in[\ell]}|\Lambda_{p,Q}^{\times}|.
\label{eq:static-unitary-Q-crossing-neighborhood}
\end{align}
Then
\begin{align}
    b_Q&\le gn_Q^{\times},
    \\
    R_Q
    &\le
    \sum_{q=1}^{k}\binom{n_Q^{\times}}{q}d_0^{2q}
    \le(Cd_0^2n_Q^{\times})^k,
\label{eq:static-unitary-geometric-bounds}
\end{align}
where $C$ depends only on $k$.  If the geometry, an edge truncation,
or a virtual extension of the layer decomposition ensures
\begin{align}
    n_Q^{\times}\le C_{\mathrm{geo}}Q|\partial\Lambda|,
\end{align}
then
\begin{align}
    b_Q&\le C_{\mathrm{geo}}gQ|\partial\Lambda|,
    \\
    R_Q&\le(Cd_0^2Q|\partial\Lambda|)^k.
\label{eq:static-unitary-Q-geometric-bounds}
\end{align}
The last geometric input is separate from the algebraic condition
$r_e\le Q$; in particular, it need not hold automatically when
$Q\ge\ell$ or when an endpoint layer is macroscopically large.

\subsection{Time-independent Liouvillian dynamics}
\label{subsec:static-liouvillian-dynamics}

We now work directly in Liouville space.  Let
\begin{align}
    \mathcal L
    =\sum_{Z\subseteq\Lambda:\,|Z|\le k}\mathcal L_Z
\label{eq:static-liouvillian-local-decomposition}
\end{align}
be a time-independent local Liouvillian.  We use the diamond norm and
assume
\begin{align}
    \sum_{Z:\,Z\ni i}\lVert\mathcal L_Z\rVert_\diamond
    \le g_{\mathrm L}
    \qquad (i\in\Lambda).
\label{eq:static-liouvillian-extensivity}
\end{align}
Decompose the generator into layer blocks exactly as for the
Hamiltonian,
\begin{align}
    \mathcal L
    =\sum_{0\le a\le b\le\ell}\mathcal L_{a,b},
    \qquad
    \mathcal L_0:=\sum_{a=0}^{\ell}\mathcal L_{a,a}.
\label{eq:static-liouvillian-block-decomposition}
\end{align}
We assume that $e^{t\mathcal L_0}$ is completely positive and trace
preserving for every $t\ge0$.  In particular,
\begin{align}
    \lVert e^{t\mathcal L_0}\rVert_\diamond=1.
\label{eq:static-liouvillian-diagonal-contraction}
\end{align}
This holds, for example, when $\mathcal L_0$ is a Lindblad generator
obtained by collecting local Lindblad terms supported within
individual layers.

For $e=(a,b)\in\mathcal E_\ell$, write
\begin{align}
    \mathcal L_e:=\mathcal L_{a,b},
    \qquad
    r_e:=b-a,
\end{align}
and introduce the affine family
\begin{align}
    \mathcal L(\vec{\lambda})
    :=\mathcal L_0+
    \sum_{e\in\mathcal E_\ell}\lambda_e\mathcal L_e.
\label{eq:static-liouvillian-parameterization}
\end{align}
Define
\begin{align}
    K_r^{(\mathrm L)}
    &:=\sum_{e:\,r_e=r}\lVert\mathcal L_e\rVert_\diamond,
    \\
    K_\eta^{(\mathrm L)}
    &:=\sum_{r=1}^{\ell}e^{\eta r}K_r^{(\mathrm L)},
    \\
    J_{\mathrm{off}}^{(\mathrm L)}
    &:=\sum_{e\in\mathcal E_\ell}
    \lVert\mathcal L_e\rVert_\diamond.
\label{eq:static-liouvillian-strengths}
\end{align}
For the selected cut $p$ and phase vector $\vec{z}$, define
\begin{align}
    \mathcal A_{p,\vec{z}}
    &:=\mathcal L_0+\sum_{e\notin u_p}z_e\mathcal L_e,
    \\
    \mathcal B_{p,\vec{z}}^{[Q]}
    &:=\sum_{e\in u_{p,Q}}z_e\mathcal L_e.
\label{eq:static-liouvillian-A-B}
\end{align}
Across the cut $(p-1,p)$,
\begin{align}
    \mathcal A_{p,\vec{z}}
    =\mathcal A^L_{p,\vec{z}}\otimes\operatorname{id}
    +\operatorname{id}\otimes\mathcal A^R_{p,\vec{z}}.
\label{eq:static-liouvillian-A-factorization}
\end{align}

For a superoperator $\mathcal M$, define its fixed-cut Liouville
Schmidt rank by
\begin{align}
\operatorname{SR}^{\mathrm{Liouv}}_p(\mathcal M)
:=\min\left\{
R:\mathcal M=\sum_{\alpha=1}^{R}
\mathcal M^L_\alpha\otimes\mathcal M^R_\alpha
\right\}.
\label{eq:fixed-cut-liouville-rank}
\end{align}
Relative to the fixed local decomposition of $\mathcal L$, set
\begin{align}
    \left\lVert\sum_Zc_Z\mathcal L_Z
    \right\rVert_{\mathrm{loc},\diamond}
    :=\sum_Z|c_Z|\lVert\mathcal L_Z\rVert_\diamond.
\end{align}
Define
\begin{align}
    b_Q^{(\mathrm L)}
    &:={}
    \max_{p\in[\ell]}
    \sup_{|z_e|=1}
    \lVert\mathcal B_{p,\vec{z}}^{[Q]}
    \rVert_{\mathrm{loc},\diamond},
    \\
    R_Q^{(\mathrm L)}
    &:={}
    \max_{p\in[\ell]}
    \sup_{|z_e|=1}
    \operatorname{SR}^{\mathrm{Liouv}}_p
    (\mathcal B_{p,\vec{z}}^{[Q]}).
\label{eq:static-liouvillian-bQ-RQ}
\end{align}

For superoperators, write
\begin{align}
    [\mathcal A,\mathcal B]_\circ
    :=\mathcal A\circ\mathcal B-\mathcal B\circ\mathcal A,
    \qquad
    \operatorname{ad}^{\circ}_{\mathcal A}(\mathcal B)
    :=[\mathcal A,\mathcal B]_\circ.
\label{eq:composition-commutator}
\end{align}

The following is the composition-commutator version of the local
support-counting argument used for Hamiltonians
\cite[Lemma~3]{KUWAHARA201696}; we give the superoperator proof explicitly.

\begin{lemma}[Local composition-commutator bounds]
\label{lem:static-liouvillian-commutator}
Let $\mathcal A$ be a phase-twisted sub-sum of the local terms of
$\mathcal L$, and let $\mathcal B$ be another such sub-sum.  Then,
for every $m\in\mathbb Z_{\ge0}$,
\begin{align}
    \left\lVert
    (\operatorname{ad}^{\circ}_{\mathcal A})^m(\mathcal B)
    \right\rVert_\diamond
    \le
    (2g_{\mathrm L}k)^m m!\,
    \lVert\mathcal B\rVert_{\mathrm{loc},\diamond}.
\label{eq:static-liouvillian-commutator-norm}
\end{align}
For $n\in\mathbb Z_{\ge0}$ and $\tau\in\mathbb R$, set
\begin{align}
    \mathcal B_{\mathcal A,n}(\tau)
    :=\sum_{m=0}^{n}
    \frac{(-\tau)^m}{m!}
    (\operatorname{ad}^{\circ}_{\mathcal A})^m(\mathcal B).
\label{eq:static-liouvillian-commutator-polynomial}
\end{align}
If, in addition, across the cut $(p-1,p)$,
\begin{align}
    \mathcal A
    &=\mathcal A^L\otimes\operatorname{id}
      +\operatorname{id}\otimes\mathcal A^R,
    &
    \mathcal B
    &=\sum_{\alpha=1}^{R}
    \mathcal B^L_\alpha\otimes\mathcal B^R_\alpha,
    \qquad R\ge1,
\end{align}
then
\begin{align}
    \operatorname{SR}^{\mathrm{Liouv}}_p
    (\mathcal B_{\mathcal A,n}(\tau))
    \le(n+1)R.
\label{eq:static-liouvillian-commutator-rank}
\end{align}
Moreover, for every $r\in\mathbb Z_{\ge0}$ and every $T\ge0$,
\begin{align}
&
\operatorname{SR}^{\mathrm{Liouv}}_p\Biggl(
\int_{0\le\tau_r\le\cdots\le\tau_1\le T}
\mathcal B_{\mathcal A,n}(\tau_1)\circ\cdots\circ
\mathcal B_{\mathcal A,n}(\tau_r)
\,d\tau_1\cdots d\tau_r
\Biggr)
\le[(n+1)R]^r.
\label{eq:static-liouvillian-integrated-rank}
\end{align}
\end{lemma}

\textit{Proof.}
Local superoperators on disjoint regions commute under composition,
and
\begin{align}
    \lVert[\mathcal X,\mathcal Y]_\circ\rVert_\diamond
    \le2\lVert\mathcal X\rVert_\diamond
    \lVert\mathcal Y\rVert_\diamond.
\end{align}
The support-growth argument used in
Lemma~\ref{lem:static-unitary-nested-commutator} therefore gives
Eq.~\eqref{eq:static-liouvillian-commutator-norm}.  Under the additional
factorization assumption,
\begin{align}
    \operatorname{ad}^{\circ}_{\mathcal A}
    =
    \operatorname{ad}^{\circ}_{\mathcal A^L}
    \otimes\operatorname{id}
    +\operatorname{id}\otimes
    \operatorname{ad}^{\circ}_{\mathcal A^R}.
\end{align}
The two terms commute.  Hence the same binomial regrouping as in
Lemma~\ref{lem:static-unitary-commutator-factorization} gives the
one-sided factorization
\begin{align}
\mathcal B_{\mathcal A,n}(\tau)
&=
\sum_{\alpha=1}^{R}\sum_{a=0}^{n}
(\operatorname{ad}^{\circ}_{\mathcal A^L})^a
(\mathcal B^L_\alpha)
\otimes
\Biggl[
\frac{(-\tau)^a}{a!}
\sum_{b=0}^{n-a}\frac{(-\tau)^b}{b!}
(\operatorname{ad}^{\circ}_{\mathcal A^R})^b
(\mathcal B^R_\alpha)
\Biggr].
\label{eq:static-liouvillian-commutator-factorization}
\end{align}
This proves Eq.~\eqref{eq:static-liouvillian-commutator-rank}.  For an
$r$-fold ordered product, fix the $r$ labels $(\alpha,a)$ in this
factorization.  The resulting left superoperator is independent of all
integration variables, so the complete time integral can be absorbed into
one right superoperator.  Counting the at most $[(n+1)R]^r$ label tuples
proves Eq.~\eqref{eq:static-liouvillian-integrated-rank}.
\hfill\(\square\)

\begin{lemma}[Stability of a phase-twisted Liouvillian background]
\label{lem:static-liouvillian-background-stability}
For every $\tau\ge0$,
\begin{align}
    \lVert e^{\tau\mathcal A_{p,\vec{z}}}\rVert_\diamond
    \le e^{J_{\mathrm{off}}^{(\mathrm L)}\tau}.
\label{eq:static-liouvillian-background-stability}
\end{align}
Moreover,
\begin{align}
    \operatorname{SR}^{\mathrm{Liouv}}_p
    (e^{\tau\mathcal A_{p,\vec{z}}})=1.
\end{align}
\end{lemma}

\textit{Proof.}
Write
\begin{align}
    \mathcal A_{p,\vec{z}}
    =\mathcal L_0+\mathcal W_{p,\vec{z}},
    \qquad
    \lVert\mathcal W_{p,\vec{z}}\rVert_\diamond
    \le J_{\mathrm{off}}^{(\mathrm L)}.
\end{align}
The Duhamel formula, the contraction
Eq.~\eqref{eq:static-liouvillian-diagonal-contraction}, and Gronwall's
inequality give Eq.~\eqref{eq:static-liouvillian-background-stability}.
The rank-one statement follows by exponentiating
Eq.~\eqref{eq:static-liouvillian-A-factorization}.
\hfill\(\square\)

Choose $M\in\mathbb N$, set $t_0=t/M$, and define
\begin{align}
    \theta_{0,\mathrm L}
    &:=2g_{\mathrm L}kt_0,
    \\
    \bar b_Q^{(\mathrm L)}
    &:=\frac{b_Q^{(\mathrm L)}}{1-\theta_{0,\mathrm L}},
    \\
    \varepsilon_{Q,n}^{(\mathrm L)}
    &:=\bar b_Q^{(\mathrm L)}\theta_{0,\mathrm L}^{n+1}.
\label{eq:static-liouvillian-short-time-parameters}
\end{align}
We assume $\theta_{0,\mathrm L}<1$.  For fixed
$\mathcal A=\mathcal A_{p,\vec{z}}$ and
$\mathcal B=\mathcal B_{p,\vec{z}}^{[Q]}$, define the
algebraic interaction-picture map below.  Neither $e^{\tau\mathcal A}$
nor $e^{-\tau\mathcal A}$ is required to be a quantum channel.
\begin{align}
    \mathcal B_{\mathcal A}(\tau)
    :=e^{-\tau\mathcal A}\circ\mathcal B\circ e^{\tau\mathcal A}.
\end{align}
We use the polynomial $\mathcal B_{\mathcal A,n}(\tau)$ defined in
Eq.~\eqref{eq:static-liouvillian-commutator-polynomial}.
Lemma~\ref{lem:static-liouvillian-commutator} gives, for
$0\le\tau\le t_0$,
\begin{align}
    \lVert\mathcal B_{\mathcal A}(\tau)
    -\mathcal B_{\mathcal A,n}(\tau)\rVert_\diamond
    &\le\varepsilon_{Q,n}^{(\mathrm L)},
    \\
    \max\{\lVert\mathcal B_{\mathcal A}(\tau)\rVert_\diamond,
    \lVert\mathcal B_{\mathcal A,n}(\tau)\rVert_\diamond\}
    &\le\bar b_Q^{(\mathrm L)}.
\label{eq:static-liouvillian-short-time-truncation}
\end{align}

Define the normalized Liouville differential branch by
\begin{align}
    \mathcal G_{p,\vec{z}}^{(s,Q)}(t)
    :=\frac{1}{s!}
    \left.
    \left(\frac{d}{dx}\right)^s
    e^{t(\mathcal A_{p,\vec{z}}
    +x\mathcal B_{p,\vec{z}}^{[Q]})}
    \right|_{x=0}.
\label{eq:static-liouvillian-differential-branch}
\end{align}
We define an approximant
$\widetilde{\mathcal G}_{p,\vec{z},n}^{(s,Q)}(t)$ by replacing every
$\mathcal B_{\mathcal A}(\tau)$ in each short-time factor by
$\mathcal B_{\mathcal A,n}(\tau)$, composing $M$ short-time factors,
and extracting the order-$s$ Taylor coefficient in $x$.

\begin{prop}[Approximation of a static Liouvillian branch]
\label{prop:static-liouvillian-branch}
For $s=0$,
\begin{align}
    \operatorname{SR}^{\mathrm{Liouv}}_p
    (\widetilde{\mathcal G}_{p,\vec{z},n}^{(0,Q)}(t))=1,
\end{align}
while, for every $s\ge1$,
\begin{align}
    \operatorname{SR}^{\mathrm{Liouv}}_p
    (\widetilde{\mathcal G}_{p,\vec{z},n}^{(s,Q)}(t))
    \le
    \binom{M+s-1}{s}
    [(n+1)R_Q^{(\mathrm L)}]^s.
\label{eq:static-liouvillian-branch-rank}
\end{align}
The approximation is exact for $s=0$.  For $s\ge1$,
\begin{align}
&
\left\lVert
\mathcal G_{p,\vec{z}}^{(s,Q)}(t)
-\widetilde{\mathcal G}_{p,\vec{z},n}^{(s,Q)}(t)
\right\rVert_\diamond
\nonumber\\
&\qquad\le
 e^{J_{\mathrm{off}}^{(\mathrm L)}t}
 t\varepsilon_{Q,n}^{(\mathrm L)}
 \frac{(t\bar b_Q^{(\mathrm L)})^{s-1}}{(s-1)!}.
\label{eq:static-liouvillian-branch-error}
\end{align}
\end{prop}

\textit{Proof.}
Use the interaction-picture identity
\begin{align}
&
\frac{1}{r!}
\left.
\left(\frac{d}{dx}\right)^r
 e^{t_0(\mathcal A+x\mathcal B)}
\right|_{x=0}
\nonumber\\
&\quad=
 e^{t_0\mathcal A}\circ
 \int_{0\le\tau_r\le\cdots\le\tau_1\le t_0}
 \mathcal B_{\mathcal A}(\tau_1)\circ\cdots\circ
 \mathcal B_{\mathcal A}(\tau_r)
 \,d\tau_1\cdots d\tau_r.
\end{align}
If $\mathcal B=0$, the assertions are immediate.  Otherwise, the
fixed-cut Liouville rank is submultiplicative under composition.  The
one-sided factorization and integrated-rank bound in
Lemma~\ref{lem:static-liouvillian-commutator}, together with
Lemma~\ref{lem:static-liouvillian-background-stability}, therefore
reproduce the unitary one-step estimates with $\lVert\cdot\rVert$
replaced by $\lVert\cdot\rVert_\diamond$ and the constants replaced by
their Liouvillian counterparts.  The
weak-composition count and generating-function argument in
Proposition~\ref{prop:static-unitary-branch} are unchanged.
\hfill\(\square\)

The cut-count decomposition for
$e^{t\mathcal L(\vec{\lambda})}$ uses the same phase families and
coefficients as in Proposition~\ref{prop:static-unitary-Q-localized}.
Indeed, the mixed derivatives at $\vec{\lambda}=\vec{0}$
are bounded in diamond norm by
$t^m\prod_a\lVert\mathcal L_{e_a}\rVert_\diamond$, because every
intermediate factor $e^{\tau\mathcal L_0}$ has diamond norm one.
Consequently, the Liouvillian cut tail is bounded by
\begin{align}
    \exp[-\eta Q+tK_\eta^{(\mathrm L)}].
\end{align}
Repeating the completion argument in the proof of
Proposition~\ref{prop:static-unitary-Q-localized} therefore gives
\begin{align}
    e^{t\mathcal L}
    &={}
    \sum_{p=1}^{\ell}
    \sum_{s=0}^{s_Q}
    \sum_{\vec{z}\in\mathcal Z_{p,Q}}
    c_{p,Q}(\vec{z})
    \mathcal G_{p,\vec{z}}^{(s,Q)}(t)
    +\Delta_{Q,\mathrm L}^{\mathrm{cut}},
\label{eq:static-liouvillian-Q-localized-decomposition}
\\
    \lVert\Delta_{Q,\mathrm L}^{\mathrm{cut}}\rVert_\diamond
    &\le
    2N_{\mathrm{ph}}(Q)
    \exp[-\eta Q+tK_\eta^{(\mathrm L)}].
\label{eq:static-liouvillian-Q-localized-error}
\end{align}
Define
\begin{align}
    \widetilde{\mathcal G}_{t,Q,n}
    :=
    \sum_{p=1}^{\ell}
    \sum_{s=0}^{s_Q}
    \sum_{\vec{z}\in\mathcal Z_{p,Q}}
    c_{p,Q}(\vec{z})
    \widetilde{\mathcal G}_{p,\vec{z},n}^{(s,Q)}(t).
\label{eq:static-liouvillian-final-approximant}
\end{align}

\begin{theorem}[Low Liouville-BSR approximation for static dynamics]
\label{thm:static-liouvillian-low-BSR}
Let $Q,n\in\mathbb Z_{\ge0}$ and $M\in\mathbb N$, and assume
$\theta_{0,\mathrm L}=2g_{\mathrm L}kt/M<1$.  Then, for every
$\eta>0$,
\begin{align}
&
\left\lVert e^{t\mathcal L}
-\widetilde{\mathcal G}_{t,Q,n}\right\rVert_\diamond
\nonumber\\
&\quad\le
2N_{\mathrm{ph}}(Q)
\exp[-\eta Q+tK_\eta^{(\mathrm L)}]
\nonumber\\
&\qquad+
N_{\mathrm{ph}}(Q)
 e^{J_{\mathrm{off}}^{(\mathrm L)}t}
 t\varepsilon_{Q,n}^{(\mathrm L)}
 \exp_{s_Q-1}(t\bar b_Q^{(\mathrm L)}).
\label{eq:static-liouvillian-final-error}
\end{align}
Moreover,
\begin{align}
    \operatorname{BSR}_{\mathrm{Liouv}}
    (\widetilde{\mathcal G}_{t,Q,n})
    &\le
    N_{\mathrm{ph}}(Q)
    \left[
    1+\sum_{s=1}^{s_Q}
    \binom{M+s-1}{s}
    [(n+1)R_Q^{(\mathrm L)}]^s
    \right]
\label{eq:static-liouvillian-final-BSR-exact}
\\
    &\le
    N_{\mathrm{ph}}(Q)(s_Q+1)
    \left[
    (M+s_Q)(n+1)\max\{1,R_Q^{(\mathrm L)}\}
    \right]^{s_Q}.
\label{eq:static-liouvillian-final-BSR-simple}
\end{align}
\end{theorem}

\textit{Proof.}
Combine Eq.~\eqref{eq:static-liouvillian-Q-localized-error} with
Proposition~\ref{prop:static-liouvillian-branch}, and use the native-cut
compression property for the Liouville BSR.
\hfill\(\square\)

\begin{corol}[Convenient Liouvillian parameter choice]
Choose
\begin{align}
    M:=\max\{1,\lceil4g_{\mathrm L}kt\rceil\}.
\end{align}
Then $\theta_{0,\mathrm L}\le1/2$,
$\bar b_Q^{(\mathrm L)}\le2b_Q^{(\mathrm L)}$, and
$\varepsilon_{Q,n}^{(\mathrm L)}\le b_Q^{(\mathrm L)}2^{-n}$.
Hence the branch-error term in
Eq.~\eqref{eq:static-liouvillian-final-error} is bounded by
\begin{align}
    N_{\mathrm{ph}}(Q)e^{J_{\mathrm{off}}^{(\mathrm L)}t}
    t b_Q^{(\mathrm L)}2^{-n}
    \exp_{s_Q-1}(2tb_Q^{(\mathrm L)}).
\end{align}
For fixed $t$, $Q$, and the other model parameters, the required
commutator order grows logarithmically with the inverse target branch error.
\end{corol}

\textit{Proof.}
Substitute $\theta_{0,\mathrm L}\le1/2$ into
Theorem~\ref{thm:static-liouvillian-low-BSR}.
\hfill\(\square\)

If the effective $Q$-crossing neighborhood contains at most
$n_Q^{\times}$ sites, then
\begin{align}
    b_Q^{(\mathrm L)}&\le g_{\mathrm L}n_Q^{\times},
    \\
    R_Q^{(\mathrm L)}
    &\le
    \sum_{q=1}^{k}\binom{n_Q^{\times}}{q}d_0^{4q}
    \le(Cd_0^4n_Q^{\times})^k.
\label{eq:static-liouvillian-geometric-rank}
\end{align}
The power $d_0^4$ is the dimension of the operator space on one local
Liouville space.

Theorem~\ref{thm:static-liouvillian-low-BSR} controls the Liouville
BSR of the approximant $\widetilde{\mathcal G}_{t,Q,n}$ to the channel
$e^{t\mathcal L}$.  The approximant is not guaranteed to be completely
positive and trace preserving.  The theorem does not by itself give the
same bound for the Kraus BSR, since Liouville and Kraus decompositions
are inequivalent in general.

\begin{corol}[Hamiltonian Liouvillian]
\label{corol:hamiltonian-liouvillian}
Let
\begin{align}
    \mathcal L_H(X):=-i[H,X].
\end{align}
Then the assumptions of
Theorem~\ref{thm:static-liouvillian-low-BSR} hold with
\begin{align}
    g_{\mathrm L}&\le2g,
    &
    J_{\mathrm{off}}^{(\mathrm L)}&\le2J_{\mathrm{off}},
    &
    K_\eta^{(\mathrm L)}&\le2K_\eta,
    \\
    b_Q^{(\mathrm L)}&\le2b_Q,
    &
    R_Q^{(\mathrm L)}&\le2R_Q.
\label{eq:hamiltonian-liouvillian-replacements}
\end{align}
Consequently, the theorem gives a direct low-Liouville-BSR
approximation of the unitary channel
$X\mapsto e^{-itH}Xe^{itH}$.
\end{corol}

\textit{Proof.}
For every operator $O$,
\begin{align}
    \lVert-i[O,\,\cdot\,]\rVert_\diamond\le2\lVert O\rVert.
\end{align}
This gives the first four estimates in
Eq.~\eqref{eq:hamiltonian-liouvillian-replacements}.  If
\begin{align}
    B=\sum_{\alpha=1}^{R}B^L_\alpha\otimes B^R_\alpha,
\end{align}
then
\begin{align}
    -i[B,\,\cdot\,]
    =-i\sum_{\alpha=1}^{R}
    \left(
    \mathcal L_{B^L_\alpha}\otimes\mathcal L_{B^R_\alpha}
    -
    \mathcal R_{B^L_\alpha}\otimes\mathcal R_{B^R_\alpha}
    \right),
\end{align}
where $\mathcal L_O(X):=OX$ and $\mathcal R_O(X):=XO$.  Hence its
fixed-cut Liouville Schmidt rank is at most $2R$.
\hfill\(\square\)

\section{Low-BSR approximation of quantum dynamics (time-dependent cases)}
\subsection{Time-dependent unitary dynamics}
\label{subsec:time-dependent-unitary-dynamics}

We next consider a time-dependent Hamiltonian
\begin{align}
    H^{(\tau)}
    =\sum_{0\le a\le b\le\ell}H_{a,b}^{(\tau)},
    \qquad 0\le \tau\le t,
\label{eq:time-dependent-layer-decomposition}
\end{align}
where, uniformly in $\tau$, the instantaneous Hamiltonian is $k$-local,
$g$-extensive, and satisfies the same spatial-decay assumptions as in the
static case.  We use a termwise Hermitian local decomposition, as in
the static case, and assume that each map
$\tau\mapsto h_Z^{(\tau)}$ is strongly measurable.  For
$e=(a,b)\in\mathcal E_\ell$, write
\begin{align}
    H_e^{(\tau)}:=H_{a,b}^{(\tau)},
    \qquad
    r_e:=b-a,
    \qquad
    V_0^{(\tau)}:=\sum_{a=0}^{\ell}H_{a,a}^{(\tau)},
\end{align}
and introduce the affine family
\begin{align}
    H^{(\tau)}(\vec\lambda)
    :=V_0^{(\tau)}+
    \sum_{e\in\mathcal E_\ell}\lambda_eH_e^{(\tau)}.
\label{eq:time-dependent-parameterization}
\end{align}
The corresponding propagator is
\begin{align}
    U_{x\to y}(\vec\lambda)
    :=\mathcal T\exp\left[
    -i\int_x^y H^{(\tau)}(\vec\lambda)\,d\tau
    \right].
\label{eq:time-dependent-propagator}
\end{align}
The parameters $\lambda_e$ multiply entire time-dependent interaction
channels $\tau\mapsto H_e^{(\tau)}$; in particular, the same parameter is
used at every time.  Since the lattice is finite, the uniform local bounds
make the Dyson series norm-absolutely convergent on every bounded parameter
polydisc.  Hence $U_{x\to y}(\vec\lambda)$ is well defined and entire in
the finitely many parameters $\lambda_e$.  The case $t=0$ is trivial;
below we take $t>0$.  Whenever an accuracy parameter is used, we assume
$0<\epsilon<1$.

The general unitary result below requires no additional temporal
regularity or independently assumed localization estimate.
Lemma~\ref{lem:td-transported-localization} derives the required
short-time localization from the diameter-tail bound of
Sec.~\ref{sec:setup}.  The geometric specializations and the regimes
used to simplify the final bound are identified separately below.
The additional temporal conditions in
Sec.~\ref{subsec:td-static-type-compression} concern the stronger
static-type precision dependence.

\subsubsection{High-level overview}
\label{subsubsec:time-dependent-unitary-overview}

We first explain the extension from the time-independent argument at the
level of scaling.  The cut-count reduction itself is essentially unchanged;
the new difficulty is that, after the reduction, the interaction inserted at
different times may point in unrelated operator directions.  Throughout this
overview we suppress constants depending on the locality, the local Hilbert-
space dimension, the interaction-decay profile, and the bounded boundary
size.  We also suppress the phase-filter logarithms and polynomial factors in
the cut budget.  Set
\begin{align}
    L_\epsilon:=\log(1/\epsilon).
\end{align}

\paragraph{1. The geometric reduction is unchanged.}
The parameters $\lambda_e$ multiply the entire time-dependent channels
$\tau\mapsto H_e^{(\tau)}$.  Hence the same cut-weight decomposition applies
to the time-ordered propagator.  The cut tail has the scaling
\begin{align}
    \left\|\Delta_{Q,\mathrm{td}}^{\mathrm{cut}}\right\|
    \lesssim
    N_{\mathrm{ph}}(Q)
    \exp\bigl[-\eta Q+O(t\ell)\bigr],
\label{eq:td-overview-cut-tail}
\end{align}
so, up to the phase-filter correction, it is enough to choose
\begin{align}
    Q=O(t\ell+L_\epsilon).
\label{eq:td-overview-Q}
\end{align}
Every retained term is then assigned to a cut $p$ across which only
\begin{align}
    s\le s_Q:=\left\lfloor\frac{2Q}{\ell}\right\rfloor
    =O\left(t+\frac{L_\epsilon}{\ell}\right)
\label{eq:td-overview-sQ}
\end{align}
interaction insertions remain explicit.  All other channels are resummed into
a background evolution that factorizes across the selected cut.  Therefore,
as in the static case,
\begin{align}
    \operatorname{BSR}(\widetilde U_t)
    \lesssim
    N_{\mathrm{ph}}(Q)(s_Q+1)
    \max_{p,\vec z,\,s\le s_Q}
    \operatorname{SR}_p
    \bigl(\widetilde U_{p,\vec z}^{(s,Q)}(t)\bigr).
\label{eq:td-overview-native-cut}
\end{align}
Thus the only genuinely new task is to estimate one time-dependent
differential branch at its native cut.

\paragraph{2. Why the static polynomial argument does not extend.}
For a time-independent short-range Hamiltonian, HHKL localization reduces the
problem to a buffer of width $O(t+L_\epsilon)$.  A Taylor approximation of the
central unitary then requires degree
\begin{align}
    m=O\bigl(t^2+(t+1)L_\epsilon\bigr),
\end{align}
and the decisive static input is that all coefficients are repeated powers of
one fixed Hamiltonian,
\begin{align}
    H^m,
    \qquad
    \operatorname{SR}(H^m)\le m^{O(\sqrt m)}.
\end{align}
For a general drive, the corresponding Dyson coefficients are instead
\begin{align}
    H^{(\tau_1)}H^{(\tau_2)}\cdots H^{(\tau_m)}.
\end{align}
Products of different Hamiltonians can have Schmidt rank $2^m$ even when
every factor has the noncrossing form
$H_L^{(\tau_j)}\otimes\mathbf 1+\mathbf 1\otimes H_R^{(\tau_j)}$.
Hence the cancellations
present in the full time-ordered exponential are destroyed if the Dyson words
are bounded separately.  The resulting wordwise estimate can be as bad as
$\exp[O(t^2+(t+1)L_\epsilon)]$, whereas simply retaining the whole
Lieb--Robinson buffer already gives the stronger baseline
\begin{align}
    \operatorname{SR}(\widetilde U_t)
    \le \exp[O(t+L_\epsilon)]
\end{align}
for a one-dimensional cut.

The differential formalism removes most of this loss because it leaves only
$s_Q=O(t+L_\epsilon/\ell)$ explicit crossing insertions.  It does not,
however, by itself solve the time-integration problem.  A family of rank-one
operators can visit arbitrarily many linearly independent Schmidt directions
inside an arbitrarily short interval, and its integral can have rank equal to
the number of visited directions.  Thus
\begin{align}
    \text{small rank at every fixed time}
    \quad\not\Longrightarrow\quad
    \text{small rank after time integration}.
\end{align}
A direct discretization with $N_{\mathrm{time}}$ time nodes would produce a
factor $N_{\mathrm{time}}^s$.  Even if one assumes
$N_{\mathrm{time}}=\operatorname{poly}(1/\epsilon)$, this gives
\begin{align}
    \log\operatorname{SR}
    \lesssim
    s_Q L_\epsilon
    =O\left(tL_\epsilon+\frac{L_\epsilon^2}{\ell}\right),
\label{eq:td-overview-naive-discretization}
\end{align}
which is quasi-polynomial in $1/\epsilon$ and can be worse than the trivial
Lieb--Robinson estimate.  For completely arbitrary time dependence, there is
not even a uniform quadrature bound of this type.

\paragraph{3. Replace time discretization by a common operator basis.}
Divide the evolution into $M=O(1+t)$ intervals of length at most a fixed
$t_c=O(1)$.  The normalized order-$s$ branch has the exact form
\begin{align}
    U_{p,\vec z}^{(s,Q)}(t)
    =
    \sum_{r_1+\cdots+r_M=s}
    E_{M,r_M}\cdots E_{1,r_1},
\label{eq:td-overview-short-factorization}
\end{align}
where $E_{j,r}$ is an $r$-fold ordered integral of the dressed crossing
interaction on the $j$th short interval.  If the dressed interaction on one
interval can be written, to the required accuracy, using $W$ fixed product
operators across the cut, then the corresponding $r$-fold integral has
Schmidt rank at most $W^r$: the time integrals modify only scalar
coefficients and create no new operator directions.

The crossing channel itself may depend arbitrarily on time, but it can first
be expanded in a fixed local product basis,
\begin{align}
    B_{p,\vec z}^{[Q],(\tau)}
    =\sum_{\alpha=1}^{R_Q}
    b_\alpha(\tau)
    V_\alpha^L\otimes V_\alpha^R.
\label{eq:td-overview-crossing-fixed-seeds}
\end{align}
The scalar functions $b_\alpha(\tau)$ do not increase the number of operator
directions.  It is therefore enough to study one fixed local seed $V$ under
the noncrossing time-dependent background, with the short interval
translated to start at zero,
\begin{align}
    V(A_0,x)=W_{0\to x}^{-1}VW_{0\to x},
    \qquad
    \int_0^{t_c}V(A_0,x)\,dx.
\label{eq:td-overview-integrated-seed}
\end{align}
Here the inverse may be replaced by the adjoint for a Hermitian background.
Because $W_{0\to x}$ factorizes across the selected cut, the earlier part
of the same short interval changes the product operators but not their number;
its possible nonunitary norm amplification is controlled by the transported
short-time estimate proved below.

\paragraph{4. One integrated insertion costs only
 $\boldsymbol{e^{\widetilde O(\sqrt j)}}$ directions at accuracy
 $\boldsymbol{e^{-j}}$.}
Subdivide a short interval into smaller windows of width $\delta$.  On each
window, expand only the short evolution in time-ordered nested commutators.
If
\begin{align}
    a:=\log\frac{1}{C\delta},
\end{align}
then truncating the commutator expansion at order $n$ gives an error of order
\begin{align}
    (C\delta)^{n+1}=e^{-a(n+1)}.
\end{align}
Thus an accuracy $e^{-j}$ requires only
\begin{align}
    a(n+1)\gtrsim j.
\label{eq:td-overview-an-constraint}
\end{align}
A Lieb--Robinson truncation confines the relevant commutators to a buffer of
radius $O(j)$.  For fixed locality $k$, the number of local operator channels
available in this buffer is polynomial in $Q+j$.  The number $W_j$ of fixed
product operators needed for accuracy $e^{-j}$ therefore satisfies,
schematically,
\begin{align}
    \log W_j
    \lesssim
    \log R_Q+a+n\log(Q+j+2).
\label{eq:td-overview-Wj-general}
\end{align}
The simple symmetric choice $a\asymp n\asymp\sqrt j$ gives
\begin{align}
    W_j
    \le
    R_Q\exp\left[
    O\bigl(\sqrt j\,\log(Q+j+2)\bigr)
    \right].
\label{eq:td-overview-Wj-symmetric}
\end{align}
Optimizing $a$ and $n$ independently improves this to
\begin{align}
    W_j
    \le
    R_Q\exp\left[
    O\bigl(\sqrt{j\log(Q+j+2)}\bigr)
    \right]
    =\exp\bigl[\widetilde O(\sqrt j)\bigr].
\label{eq:td-overview-Wj-optimized}
\end{align}
For a fixed finite set of driving channels, the logarithm in
Eq.~\eqref{eq:td-overview-Wj-optimized} is absent and one obtains
$W_j\le R_Qe^{O(\sqrt j)}$.  At the same accuracy, a direct Riemann sum with mesh size of order
$e^{-j}$ would use $O(te^j)$ time nodes and therefore up to
$O(tR_Qe^j)$ operator directions already for one integral.  Replacing this
exponential dependence on $j$ by $e^{\widetilde O(\sqrt j)}$ is the key
gain of the short-window expansion.

\paragraph{5. Combining several time integrals: decompose by accuracy level.}
Approximating each insertion independently is not enough, because the errors
of products add rather than multiply.  Instead, take differences between
successive accuracy levels.  This gives an exact expansion of the dressed
interaction of the form
\begin{align}
    V(A_0,x)
    &=\sum_{j=0}^{\infty}\Delta V_j(x),
    \nonumber\\
    \|\Delta V_j(x)\|
    &\lesssim e^{-cj},
    \qquad
    \#\{\text{operator directions in }\Delta V_j\}
    \lesssim W_j.
\label{eq:td-overview-shells}
\end{align}
For an $s$-fold integral, a term labelled by
$(j_1,\ldots,j_s)$ then has norm of order
\begin{align}
    e^{-c(j_1+\cdots+j_s)}
\end{align}
and uses at most $\prod_{a=1}^{s}W_{j_a}$ product operators.  Truncating the
sum to
\begin{align}
    j_1+\cdots+j_s\le J
\end{align}
produces a shell tail of order $e^{-cJ+O(s)}$.  The ordered-simplex factors
and the sum over $s\le s_Q$ add at most an $O(t)$ contribution to the
required budget when the crossing strength is bounded.  Thus, if the norm of the auxiliary phase-twisted background grows at most
as $e^{O(t)}$, one may take
\begin{align}
    J=O(L_\epsilon+t).
\label{eq:td-overview-J}
\end{align}
A possible norm growth of the phase-twisted non-Hermitian background is
added to this budget; with the crude estimate used below this gives
$J=O(L_\epsilon+\ell t)$, which reduces again to $J=O(L_\epsilon)$ whenever
$L_\epsilon\gtrsim\ell t$.

The largest rank under the constraint $\sum_a j_a\le J$ is controlled by
\begin{align}
    \sum_{a=1}^{s}\sqrt{j_a}
    \le\sqrt{sJ}.
\label{eq:td-overview-CS}
\end{align}
Including the weak compositions among the $M$ short intervals and the number
of shell allocations changes only the same exponential scale.  Consequently,
for one differential branch,
\begin{align}
    \operatorname{SR}_p
    \bigl(\widetilde U_{p,\vec z}^{(s,Q)}(t)\bigr)
    \le
    \exp\left[
    \widetilde O\bigl(s+\sqrt{sJ}\bigr)
    \right].
\label{eq:td-overview-one-branch-final}
\end{align}
This is the driven analogue of the static estimate
$m_\epsilon^{O(s)}$.  The square-root term is precisely the cost of allowing
new local operator directions to appear at different times.

\paragraph{6. Final order of magnitude and comparison with the static
 case.}
The general scaling obtained by combining the cut budget and the shell
budget is
\begin{align}
    \log\operatorname{BSR}(\widetilde U_t)
    \le
    \widetilde O\left[
    s_Q+\sqrt{s_Q(J+s_Q)}
    \right],
    \qquad
    s_Q=O\left(t+\frac{L_\epsilon}{\ell}\right).
\label{eq:td-overview-general-final}
\end{align}
If the crossing strength is bounded and the norm of the auxiliary
phase-twisted background grows at most as $e^{O(t)}$, then
$J=O(L_\epsilon+t)$.  More generally, under the crude $e^{O(\ell t)}$
norm-growth estimate, the same simplified expression follows in the
accuracy-dominated regime $L_\epsilon\gtrsim\ell t$.  In either case,
\begin{align}
    \boxed{
    \operatorname{BSR}(\widetilde U_t)
    \le
    \exp\left[
    \widetilde O\left(
    t+\sqrt{
    \left(t+\frac{L_\epsilon}{\ell}\right)L_\epsilon
    }
    \right)
    \right]
    }
\label{eq:td-overview-final-BSR}
\end{align}
for an operator-norm error at most $\epsilon$.  The simple symmetric
one-fold construction makes the hidden precision factor explicit as
\begin{align}
    \exp\left[
    O\left(
    \sqrt{
    \left(t+\frac{L_\epsilon}{\ell}\right)L_\epsilon
    }
    \log(L_\epsilon+t+\ell+2)
    \right)
    \right],
\end{align}
whereas the optimized construction places this logarithm under the square
root.

Writing $D$ for the final BSR, the accuracy-dominated regime gives
\begin{align}
    \log D
    =\widetilde O\left(\frac{L_\epsilon}{\sqrt\ell}\right),
\end{align}
and hence
\begin{align}
    \boxed{
    \varepsilon_D^{\mathrm{BSR}}(U_{0\to t})
    \le
    \exp\left[
    -\widetilde\Omega\bigl(\sqrt\ell\,\log D\bigr)
    +O(\ell t)
    \right].
    }
\label{eq:td-overview-error-vs-BSR}
\end{align}
For comparison, the time-independent argument gives
\begin{align}
    \varepsilon_D^{\mathrm{BSR,static}}(e^{-itH})
    \le
    \exp\left[
    -\widetilde\Omega\bigl(\ell\log D\bigr)
    +O(\ell t)
    \right].
\label{eq:td-overview-static-comparison}
\end{align}
Thus the algebraic cut-count gain is unchanged by the drive, but arbitrary
time dependence weakens the accuracy exponent from $\ell$ to $\sqrt\ell$.
The purpose of the operator-direction and shell arguments is to limit this
loss to a square root, instead of paying the quasi-polynomial penalty of a
direct time discretization.

\paragraph{Exponentially decaying interactions.}
Nothing in the preceding mechanism requires strictly finite range.  For
exponentially decaying interactions the integrated cut tail still has the
form
\begin{align}
    \left\|\Delta_{Q,\mathrm{td}}^{\mathrm{cut}}\right\|
    \lesssim
    N_{\mathrm{ph}}(Q)e^{-\eta Q+O(t\ell)},
\end{align}
so the same choice $Q=O(t\ell+L_\epsilon)$ and hence the same explicit
crossing order $s_Q=O(t+L_\epsilon/\ell)$ remain valid.  The only change is
that the number $R_Q$ of local crossing seeds and the number of local
commutator channels inside the relevant buffer may grow polynomially with
$Q$.  Using the geometric control of the crossing neighborhood derived
below, with preliminary truncation if needed, this replaces constant factors by powers of $Q$ and hence
only changes the logarithmic factors hidden in $\widetilde O$.  The essential
square-root dependence
\begin{align}
    \log\operatorname{BSR}(\widetilde U_t)
    =\widetilde O\left(
    t+\sqrt{\left(t+\frac{L_\epsilon}{\ell}\right)L_\epsilon}
    \right)
\end{align}
is unchanged.

\subsubsection{Cut-count reduction}
\label{subsubsec:time-dependent-cut-count}

The algebraic part of the differential formalism is unchanged.  Applying
Corollary~\ref{cor:cut-count-x-dependent} to the family
$\tau\mapsto H^{(\tau)}$ gives the same cut--phase decomposition as in the
static case.  For a selected cut $p$ and a phase vector $\vec z$, define
\begin{align}
    A_{p,\vec z}^{(\tau)}
    &:=V_0^{(\tau)}+
    \sum_{e\notin u_p}z_eH_e^{(\tau)},
    \\
    B_{p,\vec z}^{[Q],(\tau)}
    &:=\sum_{e\in u_{p,Q}}z_eH_e^{(\tau)}.
\label{eq:time-dependent-A-B}
\end{align}
No term in $A_{p,\vec z}^{(\tau)}$ crosses the cut $(p-1,p)$, whereas
$B_{p,\vec z}^{[Q],(\tau)}$ contains the cut-crossing channels allowed by
the geometric budget $Q$.  The phase $z_e$ is fixed throughout the entire
time profile; it does not depend on $\tau$.

The normalized differential branch is
\begin{align}
    U_{p,\vec z}^{(s,Q)}(t)
    :=\frac{1}{s!}
    \left.
    \left(\frac{d}{d\lambda_p}\right)^s
    \mathcal T\exp\left[
    -i\int_0^t
    \left(
    A_{p,\vec z}^{(\tau)}
    +\lambda_pB_{p,\vec z}^{[Q],(\tau)}
    \right)d\tau
    \right]
    \right|_{\lambda_p=0}.
\label{eq:time-dependent-differential-branch}
\end{align}
Thus, for every $Q\in\mathbb Z_{\ge0}$,
\begin{align}
    U_{0\to t}
    =
    \sum_{p=1}^{\ell}
    \sum_{s=0}^{s_Q}
    \sum_{\vec z\in\mathcal Z_{p,Q}}
    c_{p,Q}(\vec z)
    U_{p,\vec z}^{(s,Q)}(t)
    +\Delta_{Q,\mathrm{td}}^{\mathrm{cut}},
    \qquad
    s_Q:=\left\lfloor\frac{2Q}{\ell}\right\rfloor.
\label{eq:time-dependent-cut-count-decomposition}
\end{align}
The phase sets and coefficients obey the same bounds as in the static case,
\begin{align}
    |\mathcal Z_{p,Q}|
    \le(2Q+1)^{\lceil\log_2\ell\rceil},
    \qquad
    |c_{p,Q}(\vec z)|\le1.
\end{align}

The instantaneous strengths $K_r(\tau)$ and $K_\eta(\tau)$, and the
integrated quantity $K_\eta[0,t]$, are those defined in
Sec.~\ref{sec:setup}.  The time-ordered Duhamel expansion gives the direct
analogue of the static cut-tail estimate,
\begin{align}
    \lVert\Delta_{Q,\mathrm{td}}^{\mathrm{cut}}\rVert
    \le
    2N_{\mathrm{ph}}(Q)
    \exp\left[-\eta Q+K_\eta[0,t]\right].
\label{eq:time-dependent-cut-tail}
\end{align}
Indeed, after taking norms, the sum over the orderings of the insertion
times converts the ordered simplices into the full time cube.  Hence each
interaction channel contributes only through
$\int_0^t\lVert H_e^{(\tau)}\rVert d\tau$, and the same exponential
cut-weight majorant applies.

If $K_r(\tau)\le\kappa\ell|\partial\Lambda|e^{-\mu r}$
uniformly in $\tau$, then, for every fixed $0<\eta<\mu$,
$K_\eta[0,t]=O(t\ell|\partial\Lambda|)$.  Up to the
phase-filter logarithms, one may therefore choose
\begin{align}
    Q=O\bigl(t\ell+L_\epsilon\bigr),
    \qquad
    s_Q=O\left(t+\frac{L_\epsilon}{\ell}\right),
    \qquad
    L_\epsilon:=\log(1/\epsilon).
\label{eq:time-dependent-Q-s-scaling}
\end{align}
The remaining problem is to control the Schmidt rank of the branches in
Eq.~\eqref{eq:time-dependent-differential-branch}.

\subsubsection{Review of the HHKL--polynomial route}
\label{subsubsec:time-dependent-polynomial-obstruction}

Before applying the differential formalism, it is useful to recall the
standard HHKL--polynomial argument for a time-independent short-range
Hamiltonian~\cite{HHKL2021,PhysRevX.11.011047}.  This review isolates the precise step that uses time
independence and therefore fails for a general drive.

Fix a one-dimensional cut $A:B$, and decompose a neighborhood of the cut as
\begin{align}
    \Lambda=A_1\sqcup A_0\sqcup B_0\sqcup B_1,
    \qquad
    A=A_1\sqcup A_0,
    \qquad
    B=B_0\sqcup B_1,
\end{align}
where $A_0$ and $B_0$ have width $\ell_{\mathrm{LR}}$.  Let $H_X$ denote
the Hamiltonian restricted to a region $X$.  The HHKL decomposition
\cite{HHKL2021} gives
\begin{align}
&\left\|
    e^{-itH_{AB}}
    -e^{-itH_{A_0B_0}}
     e^{it(H_{A_0}+H_{B_0})}
     e^{-it(H_A+H_B)}
\right\|
\le C e^{-\mu\ell_{\mathrm{LR}}+vt},
\label{eq:time-dependent-HHKL-localization}
\end{align}
where $C$, $\mu$, and $v$ are determined by the Lieb--Robinson bound.  It
is therefore sufficient to choose
\begin{align}
    \ell_{\mathrm{LR}}
    =O\bigl(t+L_\epsilon\bigr),
    \qquad
    L_\epsilon:=\log(1/\epsilon).
\label{eq:time-dependent-HHKL-buffer}
\end{align}
The remaining two factors in
Eq.~\eqref{eq:time-dependent-HHKL-localization} have product form across
the cut:
\begin{align}
&e^{it(H_{A_0}+H_{B_0})}e^{-it(H_A+H_B)}
\nonumber\\
&\qquad=
\left(e^{itH_{A_0}}e^{-itH_A}\right)
\otimes
\left(e^{itH_{B_0}}e^{-itH_B}\right).
\label{eq:time-dependent-HHKL-product-factor}
\end{align}
Consequently, up to the HHKL error, the operator Schmidt rank is entirely
controlled by the central unitary $e^{-itH_{A_0B_0}}$.

For a one-dimensional finite-range Hamiltonian,
\begin{align}
    \lVert H_{A_0B_0}\rVert=O(\ell_{\mathrm{LR}}).
\end{align}
We approximate the central unitary by the Taylor polynomial
\begin{align}
    P_M(H_{A_0B_0})
    :=\sum_{m=0}^{M}\frac{(-it)^m}{m!}H_{A_0B_0}^{m}.
\label{eq:time-dependent-HHKL-central-polynomial}
\end{align}
At the scaling level used in the short-range argument, the Taylor cutoff is
chosen as
\begin{align}
    M
    =O\bigl(t\lVert H_{A_0B_0}\rVert+L_\epsilon\bigr)
    =O\bigl(t\ell_{\mathrm{LR}}+L_\epsilon\bigr)
    =O\bigl(t^2+(t+1)L_\epsilon\bigr),
\label{eq:time-dependent-HHKL-degree}
\end{align}
by the standard exponential-tail estimate for the Taylor series.  The crucial input is now the subexponential Schmidt-rank bound for a
repeated power of one fixed short-range Hamiltonian
\cite{PhysRevB.85.195145,arad2013area}.  In the form used in
the standard argument,
\begin{align}
    \operatorname{SR}\!\left(H_{A_0B_0}^{m}\right)
    \le m^{O(\sqrt m)}.
\label{eq:time-dependent-HHKL-power-rank}
\end{align}
The refined cut-count estimate established above improves logarithmic
factors in Eq.~\eqref{eq:time-dependent-HHKL-power-rank}, but the repeated-
power structure is the essential point.  By subadditivity of the Schmidt
rank,
\begin{align}
    \operatorname{SR}\!\left(P_M(H_{A_0B_0})\right)
    &\le
    \sum_{m=0}^{M}
    \operatorname{SR}\!\left(H_{A_0B_0}^{m}\right)
    \nonumber\\
    &\le
    \exp\!\left[\widetilde O(\sqrt M)\right]
    =
    \exp\!\left[
    \widetilde O\!\left(
    \sqrt{t^2+(t+1)L_\epsilon}
    \right)
    \right].
\label{eq:time-dependent-HHKL-static-rank}
\end{align}
Here $\widetilde O$ hides logarithmic factors.  This is the low-rank
approximation obtained in the static short-range problem.

The HHKL localization step itself is not the obstruction for a
time-dependent Hamiltonian.  The uniform finite-range local bounds give
the required Lieb--Robinson estimate, and HHKL localization again reduces the problem to a central time-ordered propagator supported on
a buffer of width $O(t+L_\epsilon)$.  The difference appears in the
polynomial expansion of this central propagator.  Instead of powers of one
fixed Hamiltonian, the Dyson expansion contains
\begin{align}
&\mathcal T\exp\left[-i\int_0^t H_{A_0B_0}^{(\tau)}d\tau\right]
\nonumber\\
&\quad=
\sum_{m=0}^{\infty}(-i)^m
\int_{0\le\tau_m\le\cdots\le\tau_1\le t}
H_{A_0B_0}^{(\tau_1)}
H_{A_0B_0}^{(\tau_2)}\cdots
H_{A_0B_0}^{(\tau_m)}
\,d\tau_1\cdots d\tau_m.
\label{eq:time-dependent-different-H-product}
\end{align}
There is no analogue of
Eq.~\eqref{eq:time-dependent-HHKL-power-rank} for a product of arbitrary
different Hamiltonians.  Indeed, take $m$ qubits on each side of the cut
and define
\begin{align}
    K_j
    :=X_{j,L}\otimes\hat 1_R
      +\hat 1_L\otimes X_{j,R},
    \qquad j=1,\ldots,m.
\label{eq:time-dependent-HHKL-different-H-example}
\end{align}
Every $K_j$ contains no interaction across the cut and has operator Schmidt
rank two.  Nevertheless,
\begin{align}
    K_1K_2\cdots K_m
    =\sum_{S\subseteq[m]}
    X_S^L\otimes X_{S^{\mathrm c}}^R,
\label{eq:time-dependent-HHKL-different-H-expansion}
\end{align}
where $X_S^L:=\prod_{j\in S}X_{j,L}$ and similarly on the right.  The
Pauli strings in Eq.~\eqref{eq:time-dependent-HHKL-different-H-expansion}
are linearly independent, and hence
\begin{align}
    \operatorname{SR}(K_1K_2\cdots K_m)=2^m.
\label{eq:time-dependent-HHKL-different-H-rank}
\end{align}
This example is deliberately stronger than a crossing-interaction
counterexample.  The exact evolution generated by the $K_j$'s is a product
evolution across the cut, but an individual ordered word has exponentially
large Schmidt rank.  Thus the problem is not the physical entanglement
generated by the background Hamiltonian; it is the loss of the
cancellations and resummations that occur when the Dyson words are bounded
separately.

Consequently, even before discretizing the time integrals, the only generic
wordwise estimate is exponential in the Dyson order.  With the same cutoff
scale as in Eq.~\eqref{eq:time-dependent-HHKL-degree}, a direct extension of
the polynomial argument can therefore cost
\begin{align}
    \exp[O(M)]
    =\exp\left[O\bigl(t^2+(t+1)L_\epsilon\bigr)\right],
\label{eq:time-dependent-HHKL-driven-word-cost}
\end{align}
with any quadrature of the time integrals producing an additional overhead.
The subexponential gain in
Eq.~\eqref{eq:time-dependent-HHKL-static-rank} is therefore specific to the
repeated powers of a time-independent Hamiltonian.

For comparison, one can avoid the polynomial expansion altogether and
retain the entire Lieb--Robinson buffer.  Since an arbitrary operator on
$A_0\sqcup B_0$ has Schmidt rank at most the dimension of the smaller local
operator space,
\begin{align}
    \operatorname{SR}(\widetilde U_t)
    \le
    d_0^{2\min\{|A_0|,|B_0|\}}
    \le
    \exp\left[
    O\bigl((t+L_\epsilon)|\partial\Lambda|\bigr)
    \right].
\label{eq:time-dependent-trivial-LR-rank}
\end{align}
For a one-dimensional cut with bounded boundary size, this gives the robust
baseline $\exp[O(t+L_\epsilon)]$.  The naive driven polynomial estimate in
Eq.~\eqref{eq:time-dependent-HHKL-driven-word-cost} is worse than this
support-only bound.

The conclusion is that HHKL localization is useful but not sufficient for
improving the driven bound.  In the static case, the improvement comes from
compressing all Taylor coefficients into powers of one fixed central
Hamiltonian.  For a general drive, the ordered products retain an
independent operator choice at every insertion time.  Any improvement over
Eq.~\eqref{eq:time-dependent-trivial-LR-rank} must therefore exploit the
time-dependent interaction channel before estimating the Schmidt rank of
its individual Dyson words.  This is the point at which the differential
formalism and the notion of common operator bases enter.

\subsubsection{Short-time differential branches}
\label{subsubsec:time-dependent-short-time-branches}

Fix $p$, $\vec z$, and $Q$.  Choose a short-time scale $t_c>0$
small enough that
$ c_{\mathrm{loc}}t_c\le 1/4 $, where
$c_{\mathrm{loc}}=O(gk)$ is the local commutator constant used below,
and choose a partition
\begin{align}
    0=\tau_0<\tau_1<\cdots<\tau_{M_t}=t,
    \qquad
    \tau_j-\tau_{j-1}\le t_c.
\end{align}
For fixed local parameters, $t_c=O(1)$ and the partition may be chosen so
that $M_t\le 1+t/t_c\le C(1+t)$.
Let
\begin{align}
    U_j(\lambda_p)
    :=\mathcal T\exp\left[
    -i\int_{\tau_{j-1}}^{\tau_j}
    \left(
    A_{p,\vec z}^{(\tau)}
    +\lambda_pB_{p,\vec z}^{[Q],(\tau)}
    \right)d\tau
    \right].
\label{eq:time-dependent-short-factor}
\end{align}
Then
\begin{align}
    U_{0\to t}(\lambda_p)
    =U_{M_t}(\lambda_p)\cdots U_1(\lambda_p).
\end{align}
Defining the normalized short-time coefficients
\begin{align}
    E_{j,r}
    :=\frac{1}{r!}
    \left.
    \left(\frac{d}{d\lambda_p}\right)^r
    U_j(\lambda_p)
    \right|_{\lambda_p=0},
\label{eq:time-dependent-Ejr}
\end{align}
we obtain the exact product rule
\begin{align}
    U_{p,\vec z}^{(s,Q)}(t)
    =
    \sum_{\substack{r_1+\cdots+r_{M_t}=s\\r_j\ge0}}
    E_{M_t,r_{M_t}}\cdots E_{1,r_1}.
\label{eq:time-dependent-weak-compositions}
\end{align}
Therefore it is enough to understand the Schmidt rank of $E_{j,r}$ for one
short interval.

For $a\le b$, let
\begin{align}
    W_{a\to b}
    :=\mathcal T\exp\left[
    -i\int_a^b A_{p,\vec z}^{(\tau)}d\tau
    \right]
\label{eq:time-dependent-background-propagator}
\end{align}
be the noncrossing background propagator.  Since
$A_{p,\vec z}^{(\tau)}$ contains no interaction across $(p-1,p)$,
\begin{align}
    W_{a\to b}=W_{a\to b}^{L}\otimes W_{a\to b}^{R},
    \qquad
    \operatorname{SR}_p(W_{a\to b})=1.
\label{eq:time-dependent-background-factorization}
\end{align}
The phase-twisted background need not be Hermitian, so in the following we
use the inverse rather than the adjoint.  For
$x\in[\tau_{j-1},\tau_j]$, define the dressed crossing operator
\begin{align}
    V_{p,\vec z}^{[Q]}
    \bigl(A^{(\cdot)},x;\tau_{j-1}\bigr)
    :=
    W_{\tau_{j-1}\to x}^{-1}
    B_{p,\vec z}^{[Q],(x)}
    W_{\tau_{j-1}\to x}.
\label{eq:time-dependent-dressed-crossing}
\end{align}
The interaction-picture expansion gives
\begin{align}
    E_{j,r}
    =(-i)^rW_{\tau_{j-1}\to\tau_j}
    \int_{\tau_{j-1}\le x_r\le\cdots\le x_1\le\tau_j}
    \prod_{a=1}^{r}
    V_{p,\vec z}^{[Q]}
    \bigl(A^{(\cdot)},x_a;\tau_{j-1}\bigr)
    \,dx_1\cdots dx_r.
\label{eq:time-dependent-Ejr-interaction-picture}
\end{align}
For each fixed time tuple $(x_1,\ldots,x_r)$, local conjugation by
$W_{\tau_{j-1}\to x_a}^{L}\otimes
W_{\tau_{j-1}\to x_a}^{R}$ does not increase the Schmidt rank.  Hence,
with
\begin{align}
    R_Q^{\mathrm{td}}
    :=\sup_{\tau,p,\vec z}
    \operatorname{SR}_p
    \left(B_{p,\vec z}^{[Q],(\tau)}\right),
\end{align}
one has
\begin{align}
    \operatorname{SR}_p\left[
    \prod_{a=1}^{r}
    V_{p,\vec z}^{[Q]}
    \bigl(A^{(\cdot)},x_a;\tau_{j-1}\bigr)
    \right]
    \le\left(R_Q^{\mathrm{td}}\right)^r.
\label{eq:time-dependent-fixed-tuple-rank}
\end{align}
When $R_Q^{\mathrm{td}}=O(1)$ and $r\le s_Q$, this is of order
\begin{align}
    \exp\left[
    O\left(t+\frac{L_\epsilon}{\ell}\right)
    \right].
\label{eq:time-dependent-nonintegrated-rank}
\end{align}
Thus the differential formalism already controls every nonintegrated
Duhamel term.  The unresolved issue is the time integration itself.

\subsubsection{The conceptual obstruction: a continuum of Schmidt directions}
\label{subsubsec:time-dependent-continuum-obstruction}

A pointwise Schmidt-rank bound does not control the Schmidt rank after time
integration.  To see this, partition an arbitrarily short interval $I$
into $N$ disjoint subintervals $I_1,\ldots,I_N$, and choose linearly
independent operator families
$\{L_a\}_{a=1}^{N}$ and $\{R_a\}_{a=1}^{N}$.  Define
\begin{align}
    O^{(x)}:=L_a\otimes R_a
    \qquad (x\in I_a).
\label{eq:time-dependent-rank-one-family-example}
\end{align}
Then
\begin{align}
    \sup_{x\in I}\operatorname{SR}_p(O^{(x)})=1,
\end{align}
whereas
\begin{align}
    \int_I O^{(x)}dx
    =\sum_{a=1}^{N}|I_a|L_a\otimes R_a
\end{align}
has Schmidt rank $N$ whenever the two operator families are linearly
independent.  Here $N$ may be taken as large as the available operator-space
dimension, which is already exponential in the size of the retained buffer.
This is an algebraic obstruction to inferring an integrated rank bound
from pointwise rank alone; it does not assert that every such family is a
dressed orbit generated by a uniformly bounded local Hamiltonian.
The missing information is the dimension of the operator space swept out
as $x$ varies, which the locality argument below will control at the
required accuracy.

A direct discretization makes this loss explicit.  Suppose that one uses
$N_{\mathrm{time}}$ time nodes to approximate each integral in
Eq.~\eqref{eq:time-dependent-Ejr-interaction-picture}.  The resulting
$r$-fold ordered integral contains up to
$N_{\mathrm{time}}^r$ operator products.  For the resulting discretized
coefficient $\widetilde E_{j,r}$, the elementary estimate gives
\begin{align}
    \operatorname{SR}_p(\widetilde E_{j,r})
    \lesssim
    \left(N_{\mathrm{time}}R_Q^{\mathrm{td}}\right)^r.
\label{eq:time-dependent-discretized-rank}
\end{align}
Under a Lipschitz-type regularity assumption, obtaining a local integration
error $\delta$ requires, schematically,
$N_{\mathrm{time}}=\operatorname{poly}(t/\delta)$.  Under completely
arbitrary time dependence there is not even a uniform quadrature bound of
this form.  Taking $\delta$ polynomially related to the final error
$\epsilon$ and $r\le s_Q$ gives the schematic complexity
\begin{align}
    \exp\left[
    O\left(
    \left(t+\frac{L_\epsilon}{\ell}\right)
    \log\frac{t}{\epsilon}
    \right)
    \right].
\label{eq:time-dependent-naive-quasipoly}
\end{align}
In the accuracy-dominated regime this contains a term
$O(L_\epsilon^2/\ell)$ in the exponent.  It is therefore
quasi-polynomial in $1/\epsilon$ and can be worse than the direct
Lieb--Robinson estimate in Eq.~\eqref{eq:time-dependent-trivial-LR-rank}.
A useful refinement must integrate the time dependence without paying one
independent Schmidt-rank factor for every quadrature node.

\subsubsection{Common operator bases over a time interval}
\label{subsubsec:time-dependent-operator-bases}

The relevant object is not the instantaneous Schmidt rank, but a common
Schmidt frame for the full time-dependent family.

\begin{definition}[Common operator basis]
\label{def:common-operator-basis}
Let $I\subseteq[0,t]$ be a time interval and let $O^{(x)}$ be an
operator-valued function on $I$.  For the cut $(p-1,p)$, define the
operator-basis number $\mathcal M_p[O^{(\cdot)};I]$ as the smallest integer
$\mathcal M$ for which there exist time-independent operators
$O^L_1,\ldots,O^L_{\mathcal M}$ and
$O^R_1,\ldots,O^R_{\mathcal M}$, together with essentially bounded
measurable scalar functions $f_1,\ldots,f_{\mathcal M}$, such that
\begin{align}
    O^{(x)}
    =\sum_{a=1}^{\mathcal M}
    f_a(x)O^L_a\otimes O^R_a
    \qquad\text{for almost every }x\in I.
\label{eq:common-operator-basis}
\end{align}
The operators $O^L_a$ and $O^R_a$ are supported on
$L_{[0,p-1]}$ and $L_{[p,\ell]}$, respectively.  They need not be
orthogonal and need not form complete operator bases of the two Hilbert
spaces.
\end{definition}

The corresponding uniform approximate operator-basis number is
\begin{align}
    \mathcal M_{p,\delta}^{\infty}
    [O^{(\cdot)};I]
    :=
    \min_{\widetilde O^{(\cdot)}}
    \left\{
    \mathcal M_p[\widetilde O^{(\cdot)};I]:
    \operatorname*{ess\,sup}_{x\in I}
    \lVert O^{(x)}-\widetilde O^{(x)}\rVert
    \le\delta
    \right\}.
\label{eq:approximate-common-operator-basis}
\end{align}
By definition,
\begin{align}
    \operatorname*{ess\,sup}_{x\in I}\operatorname{SR}_p(O^{(x)})
    \le\mathcal M_p[O^{(\cdot)};I],
\end{align}
but the converse fails, as shown by
Eq.~\eqref{eq:time-dependent-rank-one-family-example}.  Thus
$\mathcal M_p$ measures the temporal diversity of the Schmidt directions,
not merely the entangling complexity at one instant.

\begin{lemma}[Ordered integrals from common operator bases]
\label{lem:ordered-integrals-common-bases}
Let $O_a^{(x)}$, $a=1,\ldots,r$, be measurable and essentially bounded
operator-valued functions on $I=[x_0,x_1]$.  Suppose that
\begin{align}
    O_a^{(x)}
    =\sum_{b=1}^{\mathcal M_a}
    f_{a,b}(x)O^L_{a,b}\otimes O^R_{a,b},
    \qquad f_{a,b}\in L^\infty(I).
\label{eq:ordered-integral-basis-assumption}
\end{align}
Then the ordered integral
\begin{align}
    \mathfrak O_r
    :=\int_{x_0\le y_r\le\cdots\le y_1\le x_1}
    O_1^{(y_1)}O_2^{(y_2)}\cdots O_r^{(y_r)}
    \,dy_1\cdots dy_r
\label{eq:ordered-integral-operator}
\end{align}
satisfies
\begin{align}
    \operatorname{SR}_p(\mathfrak O_r)
    \le\prod_{a=1}^{r}\mathcal M_a.
\label{eq:ordered-integral-rank-bound}
\end{align}
In particular, if all factors belong to one family with operator-basis
number at most $\mathcal M$, then
\begin{align}
    \operatorname{SR}_p(\mathfrak O_r)\le\mathcal M^r.
\end{align}
\end{lemma}

\textit{Proof.}
Substituting Eq.~\eqref{eq:ordered-integral-basis-assumption} into
Eq.~\eqref{eq:ordered-integral-operator} gives
\begin{align}
    \mathfrak O_r
    =\sum_{b_1=1}^{\mathcal M_1}\cdots
    \sum_{b_r=1}^{\mathcal M_r}
    \mu_{b_1,\ldots,b_r}
    \left(
    O^L_{1,b_1}\cdots O^L_{r,b_r}
    \right)
    \otimes
    \left(
    O^R_{1,b_1}\cdots O^R_{r,b_r}
    \right),
\end{align}
where
\begin{align}
    \mu_{b_1,\ldots,b_r}
    :=\int_{x_0\le y_r\le\cdots\le y_1\le x_1}
    \prod_{a=1}^{r}f_{a,b_a}(y_a)
    \,dy_1\cdots dy_r.
\end{align}
The time integration changes only the scalar coefficients
$\mu_{b_1,\ldots,b_r}$ and creates no additional operator directions.
There are at most $\prod_a\mathcal M_a$ product operators in the resulting
decomposition, which proves
Eq.~\eqref{eq:ordered-integral-rank-bound}.
\hfill\(\square\)

\paragraph{One-sided variant.}
For the Schmidt-rank estimate, the two-sided fixed form in
Eq.~\eqref{eq:common-operator-basis} is stronger than necessary.  It is
enough to have, for example,
\begin{align}
    O^{(x)}=\sum_{a=1}^{\mathcal M}O^L_a\otimes O^R_a(x),
\label{eq:one-sided-operator-basis}
\end{align}
with only the left operators fixed in time.  The same proof then gives an
$\mathcal M^r$ bound for an $r$-fold ordered integral.  This one-sided form
is often convenient for commutator expansions, where the time dependence
can be collected entirely on one side of the cut.

The static argument is a special case of this construction.  If
\begin{align}
    A=A_L\otimes\mathbf 1+\mathbf 1\otimes A_R,
    \qquad
    B=\sum_{\alpha=1}^{R}B^L_\alpha\otimes B^R_\alpha,
\end{align}
then the order-$n$ commutator polynomial can be written as
\begin{align}
&\sum_{q=0}^{n}\frac{(ix)^q}{q!}\operatorname{ad}_A^q(B)
\nonumber\\
&\quad=
\sum_{\alpha=1}^{R}
\sum_{a+b\le n}
\frac{(ix)^{a+b}}{a!b!}
\operatorname{ad}_{A_L}^{a}(B^L_\alpha)
\otimes
\operatorname{ad}_{A_R}^{b}(B^R_\alpha).
\label{eq:static-common-operator-basis-example}
\end{align}
Hence the two-sided operator-basis number is at most
$R(n+1)(n+2)/2$.  Regrouping the right-hand side by $(\alpha,a)$ gives
the one-sided bound $R(n+1)$.  In the static problem, therefore, the
entire time dependence is carried by the scalar monomials $x^{a+b}$, and
the commutator expansion automatically produces a small common operator
basis.  The driven problem is difficult precisely because no analogous
fixed commutator family is available for an arbitrary path
$\tau\mapsto A^{(\tau)}$.

Applied to the dressed family in
Eq.~\eqref{eq:time-dependent-dressed-crossing}, the lemma replaces the
quadrature factor $N_{\mathrm{time}}^r$ by the common operator-basis number.
For example, if on every short interval
\begin{align}
    \mathcal M_p\left[
    V_{p,\vec z}^{[Q]}(A^{(\cdot)},\cdot;\tau_{j-1});
    [\tau_{j-1},\tau_j]
    \right]
    \le\mathcal M_Q,
\end{align}
then
\begin{align}
    \operatorname{SR}_p(E_{j,r})\le\mathcal M_Q^r,
\end{align}
and Eq.~\eqref{eq:time-dependent-weak-compositions} gives
\begin{align}
    \operatorname{SR}_p
    \left(U_{p,\vec z}^{(s,Q)}(t)\right)
    \le
    \binom{M_t+s-1}{s}\mathcal M_Q^s,
\label{eq:time-dependent-branch-rank-from-bases}
\end{align}
If an approximate common basis is used instead, this rank bound applies
to the corresponding approximant, with its error controlled separately.

The operator-basis formulation therefore identifies the intermediate
target: control the number of fixed product-operator directions needed
to approximate the whole dressed crossing family.  Pointwise rank alone
does not provide this control.  The next proposition constructs a
uniform approximate common basis from spatial locality and bounded local
strength, even for an arbitrary measurable drive.  Integrating that
family then changes only scalar coefficients.  An approximation to the
single integral at one endpoint would not, by itself, suffice for the
higher ordered integrals.

\subsubsection{Integrated dressed interactions for an arbitrary drive}
\label{subsubsec:time-dependent-integrated-dressed-interaction}

We now estimate the operator-basis complexity of the integrated dressed
interaction.  The statement is formulated for a general bipartition
$A|B$ and will later be applied to the selected cut of a differential
branch.  We write $\mathcal M_{A|B}$ for the analogue of
Definition~\ref{def:common-operator-basis} with this fixed bipartition.

Let
\begin{align}
    A_0^{(\tau)}
    =A_A^{(\tau)}\otimes\mathbf 1_B
    +\mathbf 1_A\otimes A_B^{(\tau)},
    \qquad 0\le \tau\le t_c,
\label{eq:integrated-dressed-split-H0}
\end{align}
be a strongly measurable, possibly non-Hermitian, time-dependent
$k$-local, $g$-extensive generator whose instantaneous interactions
satisfy the same spatial-decay assumptions as in the setup.  We require
the absolute local bounds to hold uniformly; in particular this includes
sub-sums and unit-modulus phase reweightings of the original interaction.
Let
\begin{align}
    W_{a\to b}
    :=\mathcal T\exp\left[-i\int_a^b A_0^{(\tau)}\,d\tau\right]
    =W_{a\to b}^{A}\otimes W_{a\to b}^{B}
\label{eq:integrated-dressed-product-propagator}
\end{align}
be its finite-volume propagator, and define the associated similarity
propagator by
\begin{align*}
    \mathcal W_{a\to b}(O)
    :=W_{a\to b}^{-1}OW_{a\to b}.
\end{align*}
The propagator is invertible, and both $W_{a\to b}$ and
$\mathcal W_{a\to b}$ factorize across $A|B$.  If
$A_0^{(\tau)}$ is Hermitian, then $W_{a\to b}^{-1}=W_{a\to b}^\dagger$.
For an operator $V$ with
\begin{align}
    \operatorname{SR}_{A|B}(V)\le R_V,
\end{align}
define its dressed orbit and its time integral by
\begin{align}
    V(A_0,x)
    &:=\mathcal W_{0\to x}(V)
    =W_{0\to x}^{-1}VW_{0\to x},
    \\
    \mathcal I_t(V;A_0)
    &:=\int_0^t V(A_0,x)\,dx,
    \qquad 0\le t\le t_c.
\label{eq:integrated-dressed-definition}
\end{align}
Below, $V$ is supported on at most $k$ sites near the cut, enlarging
$k$ if necessary to bound both the interaction locality and the seed
support size.  Constants below absorb this fixed support-size bound.

\paragraph{The direct discretization bound.}
For every $x$, multiplication on the left and right by the invertible
product operators
$W_{0\to x}^{A}\otimes W_{0\to x}^{B}$ and their inverses preserves
the Schmidt rank, and hence
\begin{align}
    \operatorname{SR}_{A|B}\bigl(V(A_0,x)\bigr)
    \le R_V.
\end{align}
More generally, every $N$-node quadrature or simple-function
approximation has the elementary rank bound
\begin{align}
    \operatorname{SR}_{A|B}
    \left(
    \sum_{j=1}^{N}w_jV(A_0,x_j)
    \right)
    \le R_V N.
\label{eq:integrated-dressed-naive-rank}
\end{align}
For this fixed local seed, the bounded local strength already implies
a short-time Lipschitz estimate, even when the generator is only
measurable.  Summing the double commutator expansion below with at least
one insertion between $x$ and $y$ gives
\begin{align*}
 \|V(A_0,x)-V(A_0,y)\|
 \le \frac{c_{\mathrm{loc}}\|V\|}
 {[1-c_{\mathrm{loc}}t_c]^2}|x-y|.
\end{align*}
Thus first-order quadrature can use $N=O(1+t/\epsilon)$ nodes for fixed
$t_c$ and local data, with rank cost $O(R_V(1+t/\epsilon))$.
This observation concerns a fixed seed; arbitrary measurable crossing
coefficients need not satisfy a quadrature error bound based on their
point samples.  The window construction avoids the inverse-accuracy
rank cost without imposing temporal regularity on those coefficients.

\paragraph{Locality derived from the setup.}
Let $X_\rho$ be the $\rho$-neighborhood of $\operatorname{supp}(V)$,
and let $A_{0,\rho}^{(\tau)}$ be the restriction of
$A_0^{(\tau)}$ to the interaction terms retained in $X_\rho$.  Denote
its similarity propagator by $\mathcal W^{[\rho]}_{a\to b}$.  The
window proof uses the following transported short-time localization
estimate, which we derive below, uniformly for $0\le a\le a+y\le t_c$:
\begin{align}
    \left\lVert
    \mathcal W_{0\to a}
    \left[
    \mathcal W_{a\to a+y}(V)
    -\mathcal W^{[\rho]}_{a\to a+y}(V)
    \right]
    \right\rVert
    \le
    C_{\mathrm{LR}}\lVert V\rVert
    e^{v y-\mu\rho}.
\label{eq:integrated-dressed-LR-localization}
\end{align}
For a Hermitian background the outer map is isometric, so this is the
usual local-approximation form of the Lieb--Robinson bound
\cite{PhysRevLett.97.050401,NachtergaeleSimsYoung2019}.  For the
possibly non-Hermitian backgrounds needed here, the following short-time
argument derives the transported estimate directly from the microscopic
diameter-tail bound in Eq.~\eqref{prop2:short/range}.

\begin{lemma}[Transported short-time localization]
\label{lem:td-transported-localization}
Under the locality, extensivity, and spatial-decay hypotheses above,
Eq.~\eqref{eq:integrated-dressed-LR-localization} holds on a sufficiently
short interval, uniformly over sub-sums and unit-modulus phase
reweightings.  The short-time scale depends only on the fixed local data.
No temporal regularity beyond strong measurability is required.
\end{lemma}

\textit{Proof.}
Write $A_0^{(\tau)}=\sum_Z a_Z^{(\tau)}$.  The one-site weighted
interaction strength obeys, for sufficiently small $\mu>0$,
\begin{align}
 G_\mu
 &:=\operatorname*{ess\,sup}_{\tau}\sup_{i\in\Lambda}
       \sum_{Z\ni i}e^{\mu\diam(Z)}\|a_Z^{(\tau)}\|
 \nonumber\\
 &\le g+(e^\mu-1)
       \sum_{R\ge1}e^{\mu(R-1)}\bar J(R)<\infty.
 \label{eq:td-weighted-local-strength}
\end{align}
This follows by summing the diameter tails.  The bound tends to $g$ as
$\mu\downarrow0$, so for $g>0$ we may choose $\mu$ with $G_\mu\le2g$
and take $c_{\mathrm{loc}}=4gk\ge2kG_\mu$.  The case $g=0$ is
trivial.  Thus the short-time condition
$c_{\mathrm{loc}}t_c\le1/4$ used below is a choice of subdivision,
not an additional condition on the drive.

Expand both the outer similarity map and the difference of the two inner
maps in local nested commutators.  Terms containing $q$ outer and $r$
inner commutators have ordered-simplex volumes $a^q/q!$ and $y^r/r!$.
Terms with all inner supports contained in $X_\rho$ cancel in the
difference, so $r\ge1$.  Every surviving nonzero inner word contains
a chain of intersecting supports from $\operatorname{supp}(V)$ to
outside $X_\rho$; hence the sum of its interaction diameters exceeds
$\rho$.  Weighting each local norm by $e^{\mu\diam(Z)}$ therefore
extracts a factor $e^{-\mu\rho}$.
The support-growth estimate bounds the sum of the remaining weighted
word norms by
$c_{\mathrm{loc}}^{q+r}(q+r)!\|V\|$.
Consequently, the left-hand side of
Eq.~\eqref{eq:integrated-dressed-LR-localization} is at most
\begin{align}
 &e^{-\mu\rho}\|V\|
   \sum_{q\ge0}\sum_{r\ge1}\binom{q+r}{r}
       (c_{\mathrm{loc}}a)^q(c_{\mathrm{loc}}y)^r
 \nonumber\\
 &\quad=
 e^{-\mu\rho}\|V\|
 \frac{c_{\mathrm{loc}}y}
 {(1-c_{\mathrm{loc}}a)
  [1-c_{\mathrm{loc}}(a+y)]}
 \le \frac49 e^{-\mu\rho}\|V\|.
 \label{eq:td-transported-localization-derived}
\end{align}
All series converge absolutely since $a+y\le t_c$.
This proves the required estimate, for example with
$C_{\mathrm{LR}}=1$ and $v=0$ on these short intervals.
Only absolute local norms and support growth were used, so the argument
is uniform in the phases and does not use Hermiticity, a graph
convolution condition, or a bound on the global condition number.
\hfill\(\square\)

In one dimension, for a bounded-size initial support,
\begin{align}
    |X_\rho|\le C_X(\rho+1)|\partial A|.
\label{eq:integrated-dressed-buffer-volume}
\end{align}

Because $A_{0,\rho}^{(\tau)}$ contains no interaction across $A|B$, it
can be expanded in a fixed collection of one-sided local operators,
\begin{align}
    A_{0,\rho}^{(\tau)}
    =\sum_{\beta=1}^{\mathfrak B_\rho}
    f_\beta(\tau)h_\beta,
    \qquad
    h_\beta=h_\beta^A\otimes\mathbf 1_B
    \ \text{or}\ 
    h_\beta=\mathbf 1_A\otimes h_\beta^B.
\label{eq:integrated-dressed-local-operator-basis}
\end{align}
A complete operator basis on every subset of cardinality at most $k$
gives the crude but uniform bound
\begin{align}
    \mathfrak B_\rho
    \le
    2\sum_{r=1}^{k}\binom{|X_\rho|}{r}d_0^{2r}
    \le
    \bigl(C_Bd_0^2|X_\rho|\bigr)^k.
\label{eq:integrated-dressed-local-basis-count}
\end{align}
A scalar seed is handled exactly; otherwise we may pad this local
basis with a zero coefficient so that $\mathfrak B_\rho\ge1$.
The physical interaction decomposition is used for norm estimates,
whereas Eq.~\eqref{eq:integrated-dressed-local-operator-basis} is used
only to count time-independent operator directions.

\begin{prop}[Windowed operator-basis approximation]
\label{prop:integrated-dressed-window-basis}
Under the locality and decay hypotheses stated above, let
$0<\delta\le t_c$, $n\in\mathbb Z_{\ge0}$, and $\rho\ge1$.  With
$c_{\mathrm{loc}}=O(gk)$, suppose that
\begin{align}
    c_{\mathrm{loc}}t_c\le\frac14,
    \qquad
    \theta
    :=\frac{c_{\mathrm{loc}}\delta}
    {1-c_{\mathrm{loc}}t_c}
    \le\frac12.
\label{eq:integrated-dressed-short-time-condition}
\end{align}
Then there exists an operator family
$\widetilde V_{\delta,n,\rho}(x)$ on $[0,t]$ such that
\begin{align}
    \operatorname*{ess\,sup}_{0\le x\le t}
    \left\lVert
    V(A_0,x)-\widetilde V_{\delta,n,\rho}(x)
    \right\rVert
    \le
    \eta_{\delta,n,\rho},
\label{eq:integrated-dressed-uniform-error}
\end{align}
where
\begin{align}
    \eta_{\delta,n,\rho}
    &:={}
    C_{\mathrm{LR}}\lVert V\rVert
    e^{v\delta-\mu\rho}
    +4\lVert V\rVert\theta^{n+1}.
\label{eq:integrated-dressed-eta}
\end{align}
Moreover, the full piecewise family has a common operator basis satisfying
\begin{align}
    \mathcal M_{A|B}
    \left[
    \widetilde V_{\delta,n,\rho}^{(\cdot)};[0,t]
    \right]
    \le
    2R_V
    \left\lceil\frac{t}{\delta}\right\rceil
    (n+1)\mathfrak B_\rho^n.
\label{eq:integrated-dressed-window-basis-bound}
\end{align}
Consequently, there exists an operator
$\widetilde{\mathcal I}_t(V;A_0)$ such that
\begin{align}
    \left\lVert
    \mathcal I_t(V;A_0)-\widetilde{\mathcal I}_t(V;A_0)
    \right\rVert
    &\le t\eta_{\delta,n,\rho},
    \\
    \operatorname{SR}_{A|B}
    \bigl(\widetilde{\mathcal I}_t(V;A_0)\bigr)
    &\le
    2R_V
    \left\lceil\frac{t}{\delta}\right\rceil
    (n+1)\mathfrak B_\rho^n.
\label{eq:integrated-dressed-integral-bound}
\end{align}
\end{prop}

\textit{Proof.}
Partition $[0,t]$ into windows
$I_j=[t_j,t_{j+1}]$ of width at most $\delta$.  For
$x=t_j+y\in I_j$, the propagator composition law gives
\begin{align}
    \mathcal W_{0\to x}
    =\mathcal W_{0\to t_j}\circ
    \mathcal W_{t_j\to t_j+y}.
\label{eq:integrated-dressed-correct-composition}
\end{align}
This order is important: with the convention
$W_{0\to x}=W_{t_j\to x}W_{0\to t_j}$, the short-time map acts on the
original operator $V$, and the earlier evolution appears as the outer
map in Eq.~\eqref{eq:integrated-dressed-correct-composition}.  By
Eq.~\eqref{eq:integrated-dressed-product-propagator}, the outer map is a
product map across $A|B$ and therefore does not increase the number of
product-operator directions.

For the localized short-time propagator, the time-dependent commutator
expansion is
\begin{align}
&\mathcal W_{t_j\to t_j+y}^{[\rho]}(V)
\nonumber\\
&\quad={}
\sum_{r=0}^{\infty}i^r
\int_{0\le s_r\le\cdots\le s_1\le y}
\operatorname{ad}_{A_{0,\rho}^{(t_j+s_r)}}
\cdots
\operatorname{ad}_{A_{0,\rho}^{(t_j+s_1)}}(V)
\,ds_1\cdots ds_r.
\label{eq:integrated-dressed-time-dependent-commutator-series}
\end{align}
The same support-growth argument as in the static nested-commutator
estimate applies even though the generators at different insertion times
are unrelated and need not be Hermitian.  Uniform $k$-locality and
$g$-extensivity imply
\begin{align}
&\left\lVert
\operatorname{ad}_{A_{0,\rho}^{(\tau_r)}}
\cdots
\operatorname{ad}_{A_{0,\rho}^{(\tau_1)}}(V)
\right\rVert
\le
c_{\mathrm{loc}}^r r!\lVert V\rVert.
\label{eq:integrated-dressed-time-dependent-commutator-norm}
\end{align}
Let $P_{j,n,\rho}(y)V$ denote the truncation of
Eq.~\eqref{eq:integrated-dressed-time-dependent-commutator-series} at
$r=n$, and set, for $x=t_j+y\in I_j$,
\begin{align*}
    \widetilde V_{\delta,n,\rho}(x)
    :=\mathcal W_{0\to t_j}
    \bigl(P_{j,n,\rho}(y)V\bigr).
\end{align*}
The first part of the error, obtained by replacing the full short
propagator by its spatial restriction, is bounded by
Eq.~\eqref{eq:integrated-dressed-LR-localization}.  For the truncation
error, expand also the outer map $\mathcal W_{0\to t_j}$.  A term with
$q$ earlier commutators and $r$ short-window commutators contains
$q+r$ local commutators in total.  Since $V$ has fixed support of size at
most $k$, the support-growth estimate gives the bound
$c_{\mathrm{loc}}^{q+r}(q+r)!\lVert V\rVert$.  The two ordered
simplices have volumes $t_j^q/q!$ and $y^r/r!$, respectively.  Therefore
\begin{align}
&\left\lVert
\mathcal W_{0\to t_j}
\left[
\mathcal W_{t_j\to t_j+y}^{[\rho]}(V)
-P_{j,n,\rho}(y)V
\right]
\right\rVert
\nonumber\\
&\quad\le
\lVert V\rVert
\sum_{r=n+1}^{\infty}
\sum_{q=0}^{\infty}
\binom{q+r}{r}
(c_{\mathrm{loc}}t_j)^q
(c_{\mathrm{loc}}y)^r
\nonumber\\
&\quad=
\frac{\lVert V\rVert}{1-c_{\mathrm{loc}}t_j}
\sum_{r=n+1}^{\infty}
\left(
\frac{c_{\mathrm{loc}}y}
{1-c_{\mathrm{loc}}t_j}
\right)^r
\le 4\lVert V\rVert\theta^{n+1}.
\label{eq:integrated-dressed-commutator-tail}
\end{align}
Here we used
$\sum_{q\ge0}\binom{q+r}{r}x^q=(1-x)^{-r-1}$ and
Eq.~\eqref{eq:integrated-dressed-short-time-condition}.  This estimate
is the precise short-time stability statement needed for the
nonunitary outer similarity map; no global condition-number bound is
used.  Combining it with
Eq.~\eqref{eq:integrated-dressed-LR-localization} proves
Eqs.~\eqref{eq:integrated-dressed-uniform-error} and
\eqref{eq:integrated-dressed-eta}.

It remains to count operator directions.  Choose a Schmidt decomposition
\begin{align}
    V=\sum_{\alpha=1}^{R_V}V_\alpha^A\otimes V_\alpha^B.
\end{align}
Substituting
Eq.~\eqref{eq:integrated-dressed-local-operator-basis} into the order-$r$
term of Eq.~\eqref{eq:integrated-dressed-time-dependent-commutator-series}
produces labels
$(\alpha,\beta_1,\ldots,\beta_r)$.  Since every $h_{\beta_a}$ acts
entirely on one side of the bipartition, each labeled nested commutator is
again a single product operator across $A|B$.  All insertion-time
dependence is contained in a scalar coefficient obtained by integrating
the functions $f_{\beta_a}$.  Thus one window requires at most
\begin{align}
    R_V\sum_{r=0}^{n}\mathfrak B_\rho^r
    \le 2R_V(n+1)\mathfrak B_\rho^n
\label{eq:integrated-dressed-one-window-count}
\end{align}
fixed product operators.  Applying the product outer map
$\mathcal W_{0\to t_j}$ changes those operators but not their number.
Finally, take the union of the bases for all windows and extend the
corresponding scalar functions by zero outside their own windows.  This
gives Eq.~\eqref{eq:integrated-dressed-window-basis-bound}.  Integrating
the common-basis representation changes only the scalar coefficients, so
Eq.~\eqref{eq:integrated-dressed-integral-bound} follows.
\hfill\(\square\)

\paragraph{Why the Lieb--Robinson buffer does not cost
$\boldsymbol{d_0^{O(\rho)}}$.}
At a window boundary $t_j$, the dressed operator
$V(A_0,t_j)$ still has Schmidt rank at most $R_V$, and the
Lieb--Robinson bound also makes it quasi-local in a region of radius
$O(t_j+\rho)$.  A direct expansion of that quasi-local operator in a
complete operator basis would, however, introduce the much larger factor
$d_0^{O(\rho)}$.  The proof above avoids this loss.  The $R_V$ Schmidt
factors of $V$ are kept as fixed seed operators, and only the sequences of
$k$-local generators acting on those seeds are counted.  This is the
origin of the sparse factor $\mathfrak B_\rho^n$ rather than the full
operator-space dimension of the buffer.

\paragraph{Choice of the parameters.}
Let
\begin{align}
    L_{\epsilon,t}
    :=\log\left[
    \frac{C_*(1+t_c)\max\{1,\lVert V\rVert\}}
    {\epsilon}
    \right],
\label{eq:integrated-dressed-L-epsilon-t}
\end{align}
where $C_*$ is chosen large enough to absorb the constants in
Eq.~\eqref{eq:integrated-dressed-eta}.  For fixed $t_c$,
$L_{\epsilon,t}=L_\epsilon+O(1)$ as $\epsilon\to0$.  Choose
\begin{align}
    \rho_\epsilon=C_\rho L_{\epsilon,t}
\label{eq:integrated-dressed-rho-choice}
\end{align}
so that the first term in
Eq.~\eqref{eq:integrated-dressed-eta}, after multiplication by $t$, is
at most $\epsilon/2$.

Write
\begin{align}
    a:=\log\frac{1-c_{\mathrm{loc}}t_c}
    {c_{\mathrm{loc}}\delta},
    \qquad
    \delta=\frac{1-c_{\mathrm{loc}}t_c}
    {c_{\mathrm{loc}}}e^{-a}.
\label{eq:integrated-dressed-a-definition}
\end{align}
The commutator-tail contribution is at most $\epsilon/2$ provided
\begin{align}
    a(n+1)\ge L_{\epsilon,t}.
\label{eq:integrated-dressed-error-constraint}
\end{align}
On the other hand, Proposition~\ref{prop:integrated-dressed-window-basis}
gives, up to constants and rounding,
\begin{align}
\log\mathcal M_{A|B}
\left[
\widetilde V_{\delta,n,\rho_\epsilon}^{(\cdot)};[0,t]
\right]
\le
\log\bigl[C(1+t)R_V\bigr]
+a+n b_\epsilon+O(\log(n+2)),
\label{eq:integrated-dressed-general-complexity}
\end{align}
where
\begin{align}
    b_\epsilon
    :=\max\left\{1,\log\mathfrak B_{\rho_\epsilon}\right\}.
\end{align}
In the one-dimensional setting of
Eq.~\eqref{eq:integrated-dressed-buffer-volume},
\begin{align}
    b_\epsilon
    =O\left(
    k\log\bigl[d_0|\partial A|(L_{\epsilon,t}+1)\bigr]
    \right).
\label{eq:integrated-dressed-b-epsilon}
\end{align}

A simple symmetric choice,
\begin{align}
    a\asymp n\asymp\sqrt{L_{\epsilon,t}},
\end{align}
reproduces the direct estimate suggested by the short-window argument:
\begin{align}
    \mathcal M_{A|B}
    \le
    C(1+t)R_V
    \exp\left[
    O\left(
    k\sqrt{L_{\epsilon,t}}\log(L_{\epsilon,t}+2)
    \right)
    \right].
\label{eq:integrated-dressed-symmetric-choice}
\end{align}
Equivalently, for fixed local parameters, this is
\begin{align}
    (1+t)R_V
    \exp\left[
    O\left(
    \sqrt{\log(1/\epsilon)}\,
    \log\log(1/\epsilon)
    \right)
    \right].
\end{align}

The window width and the commutator order are independent parameters, so
one can improve this choice.  Minimizing the leading exponent
$a+n b_\epsilon$ under
Eq.~\eqref{eq:integrated-dressed-error-constraint} gives
\begin{align}
    a\asymp\sqrt{L_{\epsilon,t}b_\epsilon},
    \qquad
    n\asymp\sqrt{\frac{L_{\epsilon,t}}{b_\epsilon}}.
\label{eq:integrated-dressed-optimized-parameters}
\end{align}
With this optimized choice, denote the resulting approximants by
$\widetilde V_{\epsilon}^{(\cdot)}$ and
$\widetilde{\mathcal I}_{t,\epsilon}(V;A_0)$.  Then
\begin{align}
    \max\left\{
    \mathcal M_{A|B}
    \left[\widetilde V_{\epsilon}^{(\cdot)};[0,t]\right],
    \operatorname{SR}_{A|B}
    \bigl(\widetilde{\mathcal I}_{t,\epsilon}(V;A_0)\bigr)
    \right\}
    \le
    C(1+t)R_V
    \exp\left[
    O\left(\sqrt{L_{\epsilon,t}b_\epsilon}\right)
    \right].
\label{eq:integrated-dressed-optimized-general}
\end{align}
Hence, in one dimension,
\begin{align}
    \boxed{
    \max\left\{
    \mathcal M_{A|B}
    \left[\widetilde V_{\epsilon}^{(\cdot)};[0,t]\right],
    \operatorname{SR}_{A|B}
    \bigl(\widetilde{\mathcal I}_{t,\epsilon}(V;A_0)\bigr)
    \right\}
    \le
    C(1+t)R_V
    \exp\left[
    O\left(
    \sqrt{
    kL_{\epsilon,t}\log(L_{\epsilon,t}+2)
    }
    \right)
    \right]
    }
\label{eq:integrated-dressed-final-optimized}
\end{align}
for an absolute approximation error at most $\epsilon$.  In the usual
asymptotic notation, the precision dependence is
\begin{align}
    \exp\left[
    O\left(
    \sqrt{
    \log(1/\epsilon)\log\log(1/\epsilon)
    }
    \right)
    \right]
    =
    \exp\left[
    \widetilde O\left(\sqrt{\log(1/\epsilon)}\right)
    \right].
\label{eq:integrated-dressed-final-mnemonic}
\end{align}
This is qualitatively smaller than the direct rank
$O(tR_V/\epsilon)$ in
Eq.~\eqref{eq:integrated-dressed-naive-rank}.

\begin{remark}[Fixed driving channels]
If the localized background admits a fixed number of operator channels
independent of $\rho_\epsilon$, then $b_\epsilon=O(1)$.  In that case
Eq.~\eqref{eq:integrated-dressed-optimized-general} reduces to the sharper
form
\begin{align}
    \mathcal M_{A|B}
    \le
    C(1+t)R_V
    \exp\left[O\left(\sqrt{\log(1/\epsilon)}\right)\right].
\end{align}
For a completely arbitrary drive, the additional
$\sqrt{\log\log(1/\epsilon)}$ in
Eq.~\eqref{eq:integrated-dressed-final-mnemonic} comes from the number of
local operator directions available inside the Lieb--Robinson buffer.
\end{remark}

\begin{remark}[Nonunitary, imaginary-time, and Liouvillian evolutions]
Proposition~\ref{prop:integrated-dressed-window-basis} is formulated for
the similarity flow
$\mathcal W_{a\to b}(O)=W_{a\to b}^{-1}OW_{a\to b}$ and therefore
already covers the phase-twisted non-Hermitian backgrounds used below.
The double-Dyson estimate
Eq.~\eqref{eq:integrated-dressed-commutator-tail} shows explicitly why
the outer nonunitary map changes only the local constants when
$c_{\mathrm{loc}}t_c<1$: a $k$-local seed is amplified by at most a
short-time $O(1)$ factor in the relevant commutator expansion.  The same
weighted support-growth argument in
Lemma~\ref{lem:td-transported-localization} proves the transported
localization estimate Eq.~\eqref{eq:integrated-dressed-LR-localization}
from the same microscopic diameter-tail bound.  Hence the dependence
on $\epsilon$ is unchanged.

This statement concerns algebraic similarity branches.  It does not
assert the same bound for $W^\dagger O W$ when $W$ is nonunitary, nor for
a normalized nonlinear non-Hermitian state evolution.  Imaginary-time
conjugation is included by taking the corresponding local generator in
the similarity definition above.  In Liouville
space the proof is identical with composition commutators and the diamond
norm, provided the same microscopic diameter-tail bound holds in that
norm, as in the extension stated below; ordinary
Schmidt rank is replaced by Liouville-space Schmidt rank and the local
basis count uses $d_0^4$ instead of $d_0^2$.
\end{remark}

\subsubsection{Accuracy shells for a time-dependent crossing channel}
\label{subsubsec:time-dependent-accuracy-shells}

The preceding one-fold estimate was stated for a fixed seed operator $V$.
A differential branch contains instead the time-dependent crossing family
$B_{p,\vec z}^{[Q],(\tau)}$.  We first record the reduction from this
family to finitely many fixed seed operators.

Fix $p$, $\vec z$, and $Q$.  Since
$B_{p,\vec z}^{[Q],(\tau)}$ is $k$-local and every retained interaction
crosses the cut $(p-1,p)$ with layer length at most $Q$, it admits a
representation
\begin{align}
    B_{p,\vec z}^{[Q],(\tau)}
    =\sum_{\alpha=1}^{R_Q}
    b_{p,\vec z,\alpha}(\tau)
    V_{p,\alpha}^{L}\otimes V_{p,\alpha}^{R},
\label{eq:td-crossing-channel-basis}
\end{align}
where the product operators are independent of $\tau$ and may be chosen
with unit norm.  The coefficients are measurable.  Here $R_Q$ denotes
the size of a common local product basis for the whole time-dependent
family, rather than only the pointwise rank $R_Q^{\mathrm{td}}$.  We use
the uniform coefficient bound
\begin{align}
    b_Q
    :=\sup_{p,\vec z}
    \operatorname*{ess\,sup}_{0\le\tau\le t}
    \sum_{\alpha=1}^{R_Q}
    \left|b_{p,\vec z,\alpha}(\tau)\right|.
\label{eq:td-crossing-channel-weight}
\end{align}
One may always obtain such a representation by expanding each local term
in a fixed local operator basis.  If
$\Lambda_{p,Q}^{\times}$ is the union of the supports of the retained
crossing terms and
$n_Q^{\times}:=\max_p|\Lambda_{p,Q}^{\times}|$, then
\begin{align}
    R_Q
    \le
    \sum_{q=1}^{k}\binom{n_Q^{\times}}{q}d_0^{2q}
    \le
    \bigl(Cd_0^2n_Q^{\times}\bigr)^k.
\label{eq:td-RQ-geometric-bound}
\end{align}
The boundedness of $b_Q$ follows from the setup.  At any selected cut,
assign each crossing term to its sites on the left and sum the
one-site diameter tails over the elementary layers.  A site $r$ layers
from the right side can belong to a crossing term only if its diameter
is at least $r$.  Therefore, uniformly in $p$ and $\tau$,
\begin{align}
 \sum_{Z:\,Z\text{ crosses }p}\|h_Z^{(\tau)}\|
 \le |\partial\Lambda|\sum_{r\ge1}\bar J(r).
 \label{eq:td-crossing-local-strength-derived}
\end{align}
Expanding each term in a normalized local unitary operator basis costs
at most $d_0^{2k}$ in the sum of absolute coefficients.  The phases
have unit modulus and restricting to retained terms only decreases this
bound.  Thus
\begin{align}
 b_Q\le d_0^{2k}|\partial\Lambda|\bar g_1,
 \label{eq:td-crossing-coefficient-bound-derived}
\end{align}
independently of $Q$, the endpoint sizes, and the temporal regularity.

\paragraph{Geometric bounds without an endpoint-size assumption.}
The algebraic restriction $r_e\le Q$ alone does not bound the physical
size of an aggregate endpoint layer.  For the size-independent scaling
statements, this issue can be removed with an explicitly controlled
preliminary truncation.  Let $H^{[R],(\tau)}$ retain every term within a
single coarse layer and every off-diagonal local term with
$\diam(Z)\le R$, where $R\ge1$ is an integer.  Applying the same
layer sum to discarded terms, and then summing over the $\ell$ cuts,
gives
\begin{align}
 \sup_\tau\|H^{(\tau)}-H^{[R],(\tau)}\|
 &\le \ell|\partial\Lambda|\,T_R,
 &T_R&:=\sum_{r\ge1}\bar J(\max\{r,R+1\}).
 \label{eq:td-offdiagonal-range-truncation}
\end{align}
The exponential diameter-tail bound implies
$T_R\le C e^{-\mu_*R}$ for some $C,\mu_*>0$.
Both Hamiltonians are Hermitian, so Duhamel's formula yields
\begin{align}
 \|U_{0\to t}-U^{[R]}_{0\to t}\|
 \le t\ell|\partial\Lambda|T_R.
 \label{eq:td-offdiagonal-range-truncation-error}
\end{align}
Choose
$R=O\!\left(\log[C(1+t)\ell(1+|\partial\Lambda|)/\epsilon]\right)$
to make this error at most $\epsilon/2$, and apply the subsequent
construction to $H^{[R]}$ with error budget $\epsilon/2$.
This truncation preserves all local strength and decay bounds.

Every retained term crossing a selected cut is now supported within
$R$ elementary layers of that cut, including inside either aggregate
endpoint.  The ordered-layer geometry consequently gives
\begin{align}
 n_Q^{\times}&\le C R|\partial\Lambda|,
 &n_{Q,\rho}^{\times}
 &\le C|\partial\Lambda|(R+\rho+1).
 \label{eq:td-truncated-crossing-geometry}
\end{align}
For the accuracy-dependent choice of $Q\ge1$ below, $R=O(Q)$ after
enlarging fixed constants.  Hence the bounds with $Q$ in place of $R$
follow without an additional endpoint-size hypothesis.
The exact theorem below also applies directly to the untruncated
Hamiltonian with its actual crossing-support sizes; the preliminary
truncation is needed only to obtain the displayed size-independent
geometric specializations when those sizes are not already controlled.

For a short interval $I=[a,b]$ with $0<b-a\le t_c$, let
\begin{align}
    \mathcal V_{p,\vec z}^{(\tau;a)}
    :=W_{a\to\tau}^{-1}
    B_{p,\vec z}^{[Q],(\tau)}
    W_{a\to\tau},
    \qquad \tau\in I,
\label{eq:td-dressed-crossing-family-short-window}
\end{align}
where $W$ is the noncrossing background propagator from
Eq.~\eqref{eq:time-dependent-background-propagator}.  After translating
the interval $I$ to $[0,b-a]$, apply
Proposition~\ref{prop:integrated-dressed-window-basis} with
$A_0^{(\tau)}=A_{p,\vec z}^{(a+\tau)}$.  The coefficients of this
generator are either $1$ or unit-modulus phases, so its absolute local
bounds are uniform in $p$ and $\vec z$.
Lemma~\ref{lem:td-transported-localization} therefore gives the
transported localization estimate uniformly as well.  The background also factorizes across the selected
cut.  Applying the proposition to every fixed seed
$V_{p,\alpha}^{L}\otimes V_{p,\alpha}^{R}$, the scalar functions
$b_{p,\vec z,\alpha}(\tau)$ create no new operator directions.
Therefore the one-fold estimate extends to
Eq.~\eqref{eq:td-dressed-crossing-family-short-window} with an additional
factor $R_Q$ in the basis number and a factor $b_Q$ in the error.

To state the resulting hierarchy, let
\begin{align}
    n_{Q,\rho}^{\times}
    :=\max_p
    \left|
    \left(\Lambda_{p,Q}^{\times}\right)_{\rho}
    \right|
\end{align}
be the volume of the radius-$\rho$ neighborhood of the crossing support,
and define the local channel count
\begin{align}
    \mathfrak B_{Q,\rho}
    :=\max\left\{1,\bigl(Cd_0^2n_{Q,\rho}^{\times}\bigr)^k\right\}.
\label{eq:td-local-channel-count-Q-rho}
\end{align}
For one-dimensional layers with bounded cross section,
\begin{align}
    n_{Q,\rho}^{\times}
    \le C|\partial\Lambda|(Q+\rho+1),
\label{eq:td-buffer-volume-Q-rho}
\end{align}
whenever the crossing support has the stated size.  The preliminary
truncation above ensures this condition for the geometric specializations,
with its error included in the final accuracy budget.

\begin{corol}[Accuracy-indexed common bases]
\label{corol:td-accuracy-indexed-common-bases}
Under the locality and diameter-decay hypotheses of this subsection,
there exist constants $c_0,C_0>0$, depending only on the locality,
extensivity, and decay parameters, such that the following holds.  For
every integer $j\ge0$, every short interval $I=[a,b]$ with
$0<b-a\le t_c$, and every
$p,\vec z,Q$, there is an operator-valued family
$\widetilde{\mathcal V}_{p,\vec z}^{[j]}(\tau;a)$ satisfying
\begin{align}
    \operatorname*{ess\,sup}_{\tau\in I}
    \left\|
    \mathcal V_{p,\vec z}^{(\tau;a)}
    -\widetilde{\mathcal V}_{p,\vec z}^{[j]}(\tau;a)
    \right\|
    \le C_0b_Qe^{-c_0j}.
\label{eq:td-level-j-uniform-error}
\end{align}
Its common operator-basis number across the selected cut can be chosen as
\begin{align}
    \mathcal M_p\left[
    \widetilde{\mathcal V}_{p,\vec z}^{[j]}(\cdot;a);I
    \right]
    \le W_Q^{\mathrm{sym}}(j),
\label{eq:td-level-j-basis-symmetric}
\end{align}
where
\begin{align}
    W_Q^{\mathrm{sym}}(j)
    &:={}
    C_0R_Q
    \exp\left[
    C_0\sqrt{j+1}\,\mathfrak b_Q(j)
    \right],
\label{eq:td-WQ-symmetric-definition}
\\
    \mathfrak b_Q(j)
    &:={}
    \max\left\{
    1,
    \log\mathfrak B_{Q,\rho_j}
    \right\},
    \qquad
    \rho_j=\left\lceil C_0(j+1)\right\rceil.
\label{eq:td-bQj-definition}
\end{align}
Using the optimized choice of the short subwindow width and the
commutator cutoff from
Eq.~\eqref{eq:integrated-dressed-optimized-parameters}, the stronger
bound
\begin{align}
    \mathcal M_p\left[
    \widetilde{\mathcal V}_{p,\vec z}^{[j]}(\cdot;a);I
    \right]
    \le W_Q^{\mathrm{opt}}(j),
\label{eq:td-level-j-basis-optimized}
\end{align}
with
\begin{align}
    W_Q^{\mathrm{opt}}(j)
    &:={}
    C_0R_Q
    \exp\left[
    C_0\sqrt{(j+1)\mathfrak b_Q(j)}
    \right]
\label{eq:td-WQ-optimized-definition}
\end{align}
may be used instead.
\end{corol}

\textit{Proof.}
Apply Proposition~\ref{prop:integrated-dressed-window-basis} separately
to the $R_Q$ fixed seed operators in
Eq.~\eqref{eq:td-crossing-channel-basis}.  Choose the localization radius
so that the Lieb--Robinson error is $O(e^{-c_0j})$.  For the symmetric
choice, take the logarithmic inverse subwindow width and the commutator
cutoff both of order $\sqrt{j+1}$.  The number of subwindows contributes
$\exp[O(\sqrt{j+1})]$, whereas every commutator insertion contributes one
of at most $\mathfrak B_{Q,\rho_j}$ local channels.  This gives
Eq.~\eqref{eq:td-WQ-symmetric-definition}.  Optimizing the two independent
parameters gives Eq.~\eqref{eq:td-WQ-optimized-definition}.  Finally,
Eq.~\eqref{eq:td-crossing-channel-weight} converts the channelwise
approximation error into the factor $b_Q$ in
Eq.~\eqref{eq:td-level-j-uniform-error}.
\hfill\(\square\)

The hierarchy in Corollary~\ref{corol:td-accuracy-indexed-common-bases}
must not be inserted into a product by treating
$C_0b_Qe^{-c_0j}$ as an independently multiplicative approximation
error.  Approximation errors of different factors add under a telescoping
estimate.  The multiplicative decay needed for the multi-integral is
obtained instead by passing to differences of successive approximants.

Define the accuracy shells
\begin{align}
    \Delta\mathcal V_{p,\vec z}^{[0]}(\tau;a)
    &:={}
    \widetilde{\mathcal V}_{p,\vec z}^{[0]}(\tau;a),
\nonumber\\
    \Delta\mathcal V_{p,\vec z}^{[j]}(\tau;a)
    &:={}
    \widetilde{\mathcal V}_{p,\vec z}^{[j]}(\tau;a)
    -\widetilde{\mathcal V}_{p,\vec z}^{[j-1]}(\tau;a),
    \qquad j\ge1.
\label{eq:td-accuracy-shell-definition}
\end{align}
After changing $c_0$ and $C_0$ by constants, convergence in essential
supremum in Eq.~\eqref{eq:td-level-j-uniform-error} implies
\begin{align}
    \mathcal V_{p,\vec z}^{(\tau;a)}
    =\sum_{j=0}^{\infty}
    \Delta\mathcal V_{p,\vec z}^{[j]}(\tau;a)
\label{eq:td-shell-exact-expansion}
\end{align}
for almost every $\tau\in I$, with convergence in the essential-supremum
norm, and
\begin{align}
    \operatorname*{ess\,sup}_{\tau\in I}
    \left\|
    \Delta\mathcal V_{p,\vec z}^{[j]}(\tau;a)
    \right\|
    \le C_0b_Qe^{-c_0j}.
\label{eq:td-shell-norm-decay}
\end{align}
Moreover, by subadditivity of the common operator-basis number,
\begin{align}
    \mathcal M_p\left[
    \Delta\mathcal V_{p,\vec z}^{[j]}(\cdot;a);I
    \right]
    \le
    W_Q(j)+W_Q(j-1),
\label{eq:td-shell-basis-number}
\end{align}
where $W_Q(-1):=0$ and $W_Q$ may be either
$W_Q^{\mathrm{sym}}$ or $W_Q^{\mathrm{opt}}$.
Equation~\eqref{eq:td-shell-norm-decay}, rather than the error of an
individual approximation, is the source of the factor
$e^{-c_0(j_1+\cdots+j_s)}$ below.

\subsubsection{Shell truncation of a differential branch}
\label{subsubsec:time-dependent-shell-truncation-branch}

Return to the short-time partition used in
Eq.~\eqref{eq:time-dependent-weak-compositions}, and write $M:=M_t$.
Thus $0=\tau_0<\tau_1<\cdots<\tau_M=t$, with
$\tau_m-\tau_{m-1}\le t_c$ and $M\le C(1+t)$.  For every interval
$I_m=[\tau_{m-1},\tau_m]$, use the shell expansion
Eq.~\eqref{eq:td-shell-exact-expansion} with $a=\tau_{m-1}$.

For a weak composition $\boldsymbol r=(r_1,\ldots,r_M)$ of $s$, write
$\boldsymbol j=(j_{m,a})_{1\le m\le M,\,1\le a\le r_m}$ for the $s$
shell indices assigned to the insertions, and set
\begin{align}
    |\boldsymbol j|_1
    :=\sum_{m=1}^{M}\sum_{a=1}^{r_m}j_{m,a}.
\end{align}
Let $E_{m,r_m}(\boldsymbol j_m)$ denote the coefficient in
Eq.~\eqref{eq:time-dependent-Ejr-interaction-picture}, including its
prefactor $(-i)^{r_m}W_{\tau_{m-1}\to\tau_m}$, with the $a$th
dressed interaction replaced by
$\Delta\mathcal V_{p,\vec z}^{[j_{m,a}]}$.  The exact shell expansion of
the branch is then
\begin{align}
    U_{p,\vec z}^{(s,Q)}(t)
    =
    \sum_{\substack{r_1+\cdots+r_M=s\\r_m\ge0}}
    \sum_{\boldsymbol j\in\mathbb Z_{\ge0}^{s}}
    E_{M,r_M}(\boldsymbol j_M)\cdots
    E_{1,r_1}(\boldsymbol j_1).
\label{eq:td-branch-exact-shell-expansion}
\end{align}
For $J\in\mathbb Z_{\ge0}$, define the shell-truncated branch by
\begin{align}
    \widetilde U_{p,\vec z}^{(s,Q;J)}(t)
    :=
    \sum_{\substack{r_1+\cdots+r_M=s\\r_m\ge0}}
    \sum_{\substack{\boldsymbol j\in\mathbb Z_{\ge0}^{s}\\
                      |\boldsymbol j|_1\le J}}
    E_{M,r_M}(\boldsymbol j_M)\cdots
    E_{1,r_1}(\boldsymbol j_1).
\label{eq:td-shell-truncated-branch}
\end{align}
For $s=0$, this is understood as the exact noncrossing background branch.

The phase-twisted background is generally non-Hermitian.  We therefore
keep the following stability factor explicit:
\begin{align}
    \Gamma_{\mathrm{off}}(t)
    :=\sup_{p,\vec z}
    \log\left[
    \prod_{m=1}^{M}
    \left\|W_{\tau_{m-1}\to\tau_m}\right\|
    \right].
\label{eq:td-background-stability-factor}
\end{align}
For a Hermitian background, $\Gamma_{\mathrm{off}}(t)=0$.  More
generally, because the phases have unit modulus, one may use the crude
bound
\begin{align}
    \Gamma_{\mathrm{off}}(t)
    \le
    \int_0^t J_{\mathrm{off}}(\tau)\,d\tau.
\label{eq:td-background-stability-crude}
\end{align}
Indeed, use the unitary propagator of $V_0^{(\tau)}$ as the background
in the Dyson expansion of each $W_{a\to b}$.  Its remaining interaction
has norm at most $J_{\mathrm{off}}(\tau)$, giving the bound after
multiplication over the short intervals.  By the layer-strength estimate
in Sec.~\ref{sec:setup},
\begin{align}
 \Gamma_{\mathrm{off}}(t)
 \le \bar g_1\ell|\partial\Lambda|t.
 \label{eq:td-background-stability-derived}
\end{align}
Thus this stability bound is a consequence of the setup, not an extra
hypothesis on the drive.
Only the non-Hermitian part of the phase-twisted generator is relevant to
norm growth, so Eq.~\eqref{eq:td-background-stability-crude} may often be
improved.  The exact theorem below uses
$\Gamma_{\mathrm{off}}(t)$ and does not require such an improvement.

\begin{prop}[One differential branch]
\label{prop:td-one-branch-shell-bound}
There exist constants $c_1,C_1>0$ such that, for every
$1\le s\le s_Q$ and $J\ge0$,
\begin{align}
&\left\|
U_{p,\vec z}^{(s,Q)}(t)
-\widetilde U_{p,\vec z}^{(s,Q;J)}(t)
\right\|
\nonumber\\
&\qquad\le
\exp\left[\Gamma_{\mathrm{off}}(t)\right]
\frac{(C_1b_Qt)^s}{s!}
\exp[-c_1J+C_1s].
\label{eq:td-one-branch-shell-error}
\end{align}
Furthermore,
\begin{align}
&\operatorname{SR}_p\left(
\widetilde U_{p,\vec z}^{(s,Q;J)}(t)
\right)
\nonumber\\
&\quad\le
\binom{M+s-1}{s}
\sum_{\substack{j_1,\ldots,j_s\ge0\\j_1+\cdots+j_s\le J}}
\prod_{a=1}^{s}
\left[W_Q(j_a)+W_Q(j_a-1)\right].
\label{eq:td-one-branch-rank-exact-shell}
\end{align}
With the symmetric profile in
Eq.~\eqref{eq:td-WQ-symmetric-definition}, this implies
\begin{align}
&\operatorname{SR}_p\left(
\widetilde U_{p,\vec z}^{(s,Q;J)}(t)
\right)
\nonumber\\
&\quad\le
\binom{M+s-1}{s}
\binom{J+s}{s}
(C_1R_Q)^s
\exp\left[
C_1\sqrt{s(J+s)}\,\mathfrak b_Q(J)
\right].
\label{eq:td-one-branch-rank-symmetric}
\end{align}
With the optimized profile in
Eq.~\eqref{eq:td-WQ-optimized-definition}, one instead has
\begin{align}
&\operatorname{SR}_p\left(
\widetilde U_{p,\vec z}^{(s,Q;J)}(t)
\right)
\nonumber\\
&\quad\le
\binom{M+s-1}{s}
\binom{J+s}{s}
(C_1R_Q)^s
\exp\left[
C_1\sqrt{s(J+s)\mathfrak b_Q(J)}
\right].
\label{eq:td-one-branch-rank-optimized}
\end{align}
\end{prop}

\textit{Proof.}
For a fixed weak composition $\boldsymbol r$ and shell tuple
$\boldsymbol j$, Eq.~\eqref{eq:td-shell-norm-decay}, the ordered-simplex
volumes, and the product background give
\begin{align}
&\left\|
E_{M,r_M}(\boldsymbol j_M)\cdots
E_{1,r_1}(\boldsymbol j_1)
\right\|
\nonumber\\
&\quad\le
\exp\left[\Gamma_{\mathrm{off}}(t)\right]
(C_0b_Q)^s
e^{-c_0|\boldsymbol j|_1}
\prod_{m=1}^{M}
\frac{(\tau_m-\tau_{m-1})^{r_m}}{r_m!}.
\label{eq:td-fixed-shell-term-norm}
\end{align}
Summing the interval lengths over all weak compositions gives the exact
identity
\begin{align}
    \sum_{r_1+\cdots+r_M=s}
    \prod_{m=1}^{M}
    \frac{(\tau_m-\tau_{m-1})^{r_m}}{r_m!}
    =\frac{t^s}{s!}.
\label{eq:td-simplex-volume-composition-identity}
\end{align}
For the shell tail, choose $0<c_1<c_0$ and use
\begin{align}
&\sum_{\substack{j_1,\ldots,j_s\ge0\\j_1+\cdots+j_s>J}}
 e^{-c_0(j_1+\cdots+j_s)}
\nonumber\\
&\qquad\le
 e^{-c_1J}
 \left(
 \sum_{j=0}^{\infty}e^{-(c_0-c_1)j}
 \right)^s
 \le e^{-c_1J+C_1s}.
\label{eq:td-shell-tail-sum}
\end{align}
Equations~\eqref{eq:td-fixed-shell-term-norm}--
\eqref{eq:td-shell-tail-sum} prove
Eq.~\eqref{eq:td-one-branch-shell-error}.

For the rank, multiplication by any background factor has Schmidt rank one
across the native cut.  Lemma~\ref{lem:ordered-integrals-common-bases} and
Eq.~\eqref{eq:td-shell-basis-number} therefore give the product inside
Eq.~\eqref{eq:td-one-branch-rank-exact-shell}; the first binomial
coefficient counts weak compositions of $s$ into $M$ intervals.

There are $\binom{J+s}{s}$ shell tuples with total weight at most $J$.
For the symmetric profile, monotonicity of $\mathfrak b_Q$ and the
Cauchy--Schwarz inequality give
\begin{align}
    \sum_{a=1}^{s}
    \sqrt{j_a+1}\,\mathfrak b_Q(j_a)
    \le
    \mathfrak b_Q(J)
    \sqrt{s(J+s)}.
\label{eq:td-shell-CS-symmetric}
\end{align}
For the optimized profile,
\begin{align}
    \sum_{a=1}^{s}
    \sqrt{(j_a+1)\mathfrak b_Q(j_a)}
    \le
    \sqrt{s(J+s)\mathfrak b_Q(J)}.
\label{eq:td-shell-CS-optimized}
\end{align}
These estimates prove Eqs.~\eqref{eq:td-one-branch-rank-symmetric} and
\eqref{eq:td-one-branch-rank-optimized}.
\hfill\(\square\)

The equal distribution
$j_1\simeq\cdots\simeq j_s\simeq J/s$ is the worst case in the
exponential part of the symmetric estimate, because $\sqrt{x}$ is
concave.  This recovers the heuristic factor
\begin{align}
    \exp\left[
    O\left(\sqrt{sJ}\log(J+2)\right)
    \right]
\end{align}
when $\mathfrak b_Q(J)=O(\log(J+Q+2))$.  The shell formulation is the
rigorous reason that the decay is governed by the sum of the accuracy
indices; independent approximation errors themselves would not multiply.

\subsubsection{Low-BSR approximation of the full driven evolution}
\label{subsubsec:time-dependent-final-low-BSR}

Define the final approximant
\begin{align}
    \widetilde U_{t,Q,J}^{\mathrm{td}}
    :=
    \sum_{p=1}^{\ell}
    \sum_{s=0}^{s_Q}
    \sum_{\vec z\in\mathcal Z_{p,Q}}
    c_{p,Q}(\vec z)
    \widetilde U_{p,\vec z}^{(s,Q;J)}(t).
\label{eq:td-final-approximant}
\end{align}
For $s=0$, the exact background branch is used.  We emphasize that the
Schmidt rank of each summand is estimated at its own selected cut $p$.
The BSR is not submultiplicative under arbitrary operator products, and no
such property is used; only fixed-cut submultiplicativity inside one
branch and native-cut compression at the final summation are required.

\begin{theorem}[Low-BSR approximation of arbitrarily driven unitary dynamics]
\label{thm:time-dependent-low-BSR}
Let $H^{(\tau)}$ satisfy the local strength, microscopic spatial-decay,
and measurability hypotheses stated at the start of this subsection.
Let $Q,J\in\mathbb Z_{\ge0}$ and choose a partition into intervals of
length at most $t_c$ with $M\le C(1+t)$.
The common bases required below are supplied by
Corollary~\ref{corol:td-accuracy-indexed-common-bases}, using
Lemma~\ref{lem:td-transported-localization}; they are not an additional
assumption.  Then, for every $\eta>0$,
\begin{align}
&\left\|
U_{0\to t}-\widetilde U_{t,Q,J}^{\mathrm{td}}
\right\|
\nonumber\\
&\quad\le
2N_{\mathrm{ph}}(Q)
\exp\left[-\eta Q+K_\eta[0,t]\right]
\nonumber\\
&\qquad+
N_{\mathrm{ph}}(Q)
\exp\left[
\Gamma_{\mathrm{off}}(t)
+C_2b_Qt-c_2J
\right]
\label{eq:td-final-error}
\end{align}
for constants $c_2,C_2>0$.  Moreover,
\begin{align}
&\operatorname{BSR}\left(
\widetilde U_{t,Q,J}^{\mathrm{td}}
\right)
\nonumber\\
&\quad\le
N_{\mathrm{ph}}(Q)
\left[
1+
\sum_{s=1}^{s_Q}
\binom{M+s-1}{s}
\sum_{\substack{j_1,\ldots,j_s\ge0\\
                  j_1+\cdots+j_s\le J}}
\prod_{a=1}^{s}
\left[W_Q(j_a)+W_Q(j_a-1)\right]
\right].
\label{eq:td-final-BSR-exact}
\end{align}
In particular, the symmetric construction gives
\begin{align}
&\operatorname{BSR}\left(
\widetilde U_{t,Q,J}^{\mathrm{td}}
\right)
\nonumber\\
&\quad\le
N_{\mathrm{ph}}(Q)(s_Q+1)
\binom{M+s_Q-1}{s_Q}
\binom{J+s_Q}{s_Q}
\bigl(C_2\max\{1,R_Q\}\bigr)^{s_Q}
\nonumber\\
&\qquad\times
\exp\left[
C_2\sqrt{s_Q(J+s_Q)}\,\mathfrak b_Q(J)
\right].
\label{eq:td-final-BSR-symmetric}
\end{align}
The optimized construction replaces the last exponential by
\begin{align}
    \exp\left[
    C_2\sqrt{s_Q(J+s_Q)\mathfrak b_Q(J)}
    \right].
\label{eq:td-final-BSR-optimized-factor}
\end{align}
\end{theorem}

\textit{Proof.}
The first term in Eq.~\eqref{eq:td-final-error} is the cut-count remainder
in Eq.~\eqref{eq:time-dependent-cut-tail}.  Summing
Eq.~\eqref{eq:td-one-branch-shell-error} over $1\le s\le s_Q$ and over
all cut--phase pairs gives
\begin{align}
&N_{\mathrm{ph}}(Q)
 e^{\Gamma_{\mathrm{off}}(t)-c_1J}
 \sum_{s=1}^{s_Q}
 \frac{(C_1e^{C_1}b_Qt)^s}{s!}
\nonumber\\
&\qquad\le
N_{\mathrm{ph}}(Q)
 \exp\left[
 \Gamma_{\mathrm{off}}(t)+C_2b_Qt-c_2J
 \right],
\end{align}
which is the second term.  Equation~\eqref{eq:td-final-BSR-exact} follows
from Proposition~\ref{prop:td-one-branch-shell-bound} and native-cut
compression, Proposition~\ref{prop:native-cut-compression}.  The final two
bounds follow from Eqs.~\eqref{eq:td-one-branch-rank-symmetric} and
\eqref{eq:td-one-branch-rank-optimized}, together with subadditivity.
\hfill\(\square\)

\paragraph{Choice of the parameters.}
Set
\begin{align}
    L_\epsilon:=\log(1/\epsilon).
\end{align}
It is sufficient to choose $Q$ and $J$ so that
\begin{align}
    \eta Q
    &\ge
    K_\eta[0,t]
    +L_\epsilon
    +\log\left[4N_{\mathrm{ph}}(Q)\right],
\label{eq:td-choice-Q-final}
\\
    c_2J
    &\ge
    \Gamma_{\mathrm{off}}(t)
    +C_2b_Qt
    +L_\epsilon
    +\log\left[2N_{\mathrm{ph}}(Q)\right].
\label{eq:td-choice-J-final}
\end{align}
Then
\begin{align}
    \left\|
    U_{0\to t}-\widetilde U_{t,Q,J}^{\mathrm{td}}
    \right\|\le\epsilon.
\end{align}
Under the preceding uniform exponential-decay bound, with fixed
$0<\eta<\mu$, and bounded local parameters, these choices have the
scaling
\begin{align}
    Q
    &=O\left(
    t\ell|\partial\Lambda|
    +L_\epsilon
    +\log N_{\mathrm{ph}}(Q)
    \right),
\label{eq:td-Q-scaling-final}
\\
    J
    &=O\left(
    L_\epsilon
    +\Gamma_{\mathrm{off}}(t)
    +b_Qt
    +\log N_{\mathrm{ph}}(Q)
    \right),
\label{eq:td-J-scaling-final}
\\
    s_Q
    &=O\left(
    t|\partial\Lambda|
    +\frac{L_\epsilon}{\ell}
    +\frac{\log N_{\mathrm{ph}}(Q)}{\ell}
    \right).
\label{eq:td-sQ-scaling-final}
\end{align}
For a one-dimensional cut with bounded $|\partial\Lambda|$,
Eqs.~\eqref{eq:td-crossing-coefficient-bound-derived} and
\eqref{eq:td-background-stability-derived} give $b_Q=O(1)$ and
$\Gamma_{\mathrm{off}}(t)=O(\ell t)$.  Hence this reduces,
up to phase-filter logarithms, to
\begin{align}
    Q=O(t\ell+L_\epsilon),
    \qquad
    J=O(L_\epsilon+\ell t),
    \qquad
    s_Q=O\left(t+\frac{L_\epsilon}{\ell}\right).
\label{eq:td-main-parameter-scales}
\end{align}
If the phase-twisted background has a uniform $O(1)$ stability rate, then
$J=O(L_\epsilon+t)$; none of the precision exponents below are worsened.

The binomial coefficients satisfy
\begin{align}
    \log\binom{M+s-1}{s}
    &\le s\log\left[e\left(1+\frac{M}{s}\right)\right],
\nonumber\\
    \log\binom{J+s}{s}
    &\le C\left(s+\sqrt{sJ}\right).
\label{eq:td-binomial-log-bounds}
\end{align}
Combining these inequalities with
Theorem~\ref{thm:time-dependent-low-BSR} yields the explicit symmetric
estimate
\begin{align}
&\log\operatorname{BSR}\left(
\widetilde U_{t,Q,J}^{\mathrm{td}}
\right)
\nonumber\\
&\quad\le
\log N_{\mathrm{ph}}(Q)
+C\,s_Q\log\left[
C(M+s_Q)(J+s_Q+1)\max\{1,R_Q\}
\right]
\nonumber\\
&\qquad+
C\sqrt{s_Q(J+s_Q)}\,\mathfrak b_Q(J).
\label{eq:td-final-log-BSR-explicit-symmetric}
\end{align}
The optimized construction gives instead
\begin{align}
&\log\operatorname{BSR}\left(
\widetilde U_{t,Q,J}^{\mathrm{td}}
\right)
\nonumber\\
&\quad\le
\log N_{\mathrm{ph}}(Q)
+C\,s_Q\log\left[
C(M+s_Q)(J+s_Q+1)\max\{1,R_Q\}
\right]
\nonumber\\
&\qquad+
C\sqrt{s_Q(J+s_Q)\mathfrak b_Q(J)}.
\label{eq:td-final-log-BSR-explicit-optimized}
\end{align}

In one dimension,
Eq.~\eqref{eq:td-buffer-volume-Q-rho} implies
\begin{align}
    \mathfrak b_Q(J)
    =O\left(
    k\log\left[
    d_0|\partial\Lambda|(Q+J+2)
    \right]
    \right).
\label{eq:td-bQJ-one-dimensional}
\end{align}
Consequently, for bounded $|\partial\Lambda|$ and fixed local
parameters, the exact scaling consequence of
Eq.~\eqref{eq:td-final-log-BSR-explicit-symmetric} is
\begin{align}
&\log\operatorname{BSR}\left(
\widetilde U_{t,\epsilon}^{\mathrm{td}}
\right)
\nonumber\\
&\quad\le
\widetilde O\left[
s_Q+\sqrt{s_Q\bigl(J+s_Q\bigr)}
\right],
\qquad
s_Q=O\left(t+\frac{L_\epsilon}{\ell}\right),
\label{eq:td-final-BSR-general-mnemonic}
\end{align}
where $J$ is chosen according to
Eq.~\eqref{eq:td-choice-J-final}.  If the phase-twisted background has an
$O(1)$ stability rate, so that $J=O(L_\epsilon+t)$, or if one works in the
accuracy-dominated regime $L_\epsilon\gtrsim\ell t$ under the crude bound
$\Gamma_{\mathrm{off}}(t)=O(\ell t)$, this reduces to
\begin{align}
    \boxed{
    \operatorname{BSR}\left(
    \widetilde U_{t,\epsilon}^{\mathrm{td}}
    \right)
    \le
    \exp\left[
    \widetilde O\left(
    t+
    \sqrt{
    \left(t+\frac{L_\epsilon}{\ell}\right)L_\epsilon
    }
    \right)
    \right]
    }
\label{eq:td-final-BSR-mnemonic}
\end{align}
The corresponding explicit precision-dependent term for the symmetric
one-fold construction is
\begin{align}
    \exp\left[
    O\left(
    \sqrt{
    \left(t+\frac{L_\epsilon}{\ell}\right)L_\epsilon
    }
    \log(L_\epsilon+t+\ell+2)
    \right)
    \right],
\label{eq:td-final-BSR-symmetric-explicit-scaling}
\end{align}
whereas the optimized construction gives
\begin{align}
    \exp\left[
    O\left(
    \sqrt{
    \left(t+\frac{L_\epsilon}{\ell}\right)L_\epsilon
    \log(L_\epsilon+t+\ell+2)
    }
    \right)
    \right],
\label{eq:td-final-BSR-optimized-explicit-scaling}
\end{align}
up to the same lower-order phase-filter and channel-count factors.  Outside
the two regimes just stated, one should use the general formula
Eq.~\eqref{eq:td-final-BSR-general-mnemonic} with the actual stability
budget $J$; the theorem does not discard that dependence.

\paragraph{Error as a function of the BSR.}
To avoid confusing the cut budget $Q$ with the logarithm of the final
rank, write
\begin{align}
    D:=\operatorname{BSR}\left(
    \widetilde U_{t,\epsilon}^{\mathrm{td}}
    \right),
    \qquad
    q_D:=\log D.
\end{align}
In the accuracy-dominated regime
$L_\epsilon\gtrsim\ell t$, the mnemonic estimate
Eq.~\eqref{eq:td-final-BSR-mnemonic} gives
\begin{align}
    q_D
    =\widetilde O\left(
    \frac{L_\epsilon}{\sqrt\ell}
    \right).
\end{align}
Equivalently, there is a low-BSR approximation obeying
\begin{align}
    \boxed{
    \varepsilon_D^{\mathrm{BSR}}(U_{0\to t})
    \le
    \exp\left[
    -\widetilde\Omega\left(
    \sqrt\ell\,\log D
    \right)
    +O(\ell t)
    \right]
    }
\label{eq:td-error-vs-BSR-mnemonic}
\end{align}
where the tilde hides polylogarithmic factors.  Keeping the logarithm in
the symmetric construction gives, schematically,
\begin{align}
    \varepsilon_D^{\mathrm{BSR}}(U_{0\to t})
    \le
    \exp\left[
    -\Omega\left(
    \frac{\sqrt\ell\,\log D}
    {\log(\ell+\log D+2)}
    \right)
    +O(\ell t)
    \right],
\label{eq:td-error-vs-BSR-symmetric-explicit}
\end{align}
whereas the optimized construction replaces the denominator by the square
root of the same logarithm.

For comparison, the time-independent analysis gives, up to the analogous
polylogarithmic factors,
\begin{align}
    \varepsilon_D^{\mathrm{BSR,static}}(e^{-itH})
    \le
    \exp\left[
    -\widetilde\Omega\left(
    \ell\log D
    \right)
    +O(\ell t)
    \right].
\label{eq:td-static-comparison-error-vs-BSR}
\end{align}
Thus arbitrary time dependence changes the accuracy exponent from the
static scale $\ell\log D$ to the driven scale
$\sqrt\ell\log D$.  This loss is the quantitative price of allowing the
drive to explore new local operator directions at every time.  The shell
construction shows that the loss is only square-root in the buffer length,
rather than the full quasi-polynomial penalty produced by direct time
discretization.

\begin{remark}[Scope of the result]
The exact statements are
Eqs.~\eqref{eq:td-final-error},
\eqref{eq:td-final-BSR-exact}, and
\eqref{eq:td-final-log-BSR-explicit-symmetric}.  The compact formulas
Eqs.~\eqref{eq:td-final-BSR-mnemonic} and
\eqref{eq:td-error-vs-BSR-mnemonic} suppress the phase-filter overhead,
the polynomial channel count $R_Q$, and logarithmic factors from the local
basis in the Lieb--Robinson buffer.  The geometric bounds are either used directly or obtained by the
preliminary truncation in
Eq.~\eqref{eq:td-offdiagonal-range-truncation-error}, with the error
budget split as described there.  The transported localization and the
bound $\Gamma_{\mathrm{off}}(t)\le\bar g_1\ell|\partial\Lambda|t$
are derived from the setup.  A stronger $O(t)$ stability bound is
optional; without it, the exact theorem applies for all $t$, and the
compact precision-dominated formula is used in the stated regime
$L_\epsilon\gtrsim\ell t$.
\end{remark}

\begin{remark}[Liouvillian extension]
Consider a strongly measurable time-dependent local Liouvillian with
the diamond-norm analogue of the microscopic diameter-tail bound
Eq.~\eqref{prop2:short/range}.  As in the static Liouvillian argument,
assume that its diagonal background $\mathcal L_0^{(\tau)}$ generates
a completely positive, trace-preserving propagator.  The weighted
commutator proof of Lemma~\ref{lem:td-transported-localization} then
derives the transported localization estimate in diamond norm; it need
not be assumed independently.  A layer-strength bound alone does not
replace this microscopic decay condition.  The cut-count tail is
controlled in diamond norm exactly as in the static Liouvillian argument.
More generally, the same abstract extension holds if the transported
estimate is supplied directly instead of the microscopic decay bound.  For each phase-twisted differential branch, replace
$\operatorname{ad}$ by the composition commutator
$\operatorname{ad}^{\circ}$ and apply the double-Dyson estimate above in
diamond norm.  Lemma~\ref{lem:static-liouvillian-commutator} supplies the
required local factorial bound, and factorization of the noncrossing
background supplies the product Liouville map.

Consequently, Theorem~\ref{thm:time-dependent-low-BSR} has a direct
Liouville-space version with operator norms replaced by diamond norms,
BSR replaced by Liouville BSR, $R_Q$ replaced by the corresponding
Liouville rank, and $d_0^2$ replaced by $d_0^4$ in the local basis count.
The norm growth of the phase-twisted product background is retained in
the Liouville analogue of $\Gamma_{\mathrm{off}}(t)$.  Under these
explicit inputs, the dependence on $L_\epsilon$, $s_Q$, and the shell
budget $J$ is unchanged.
Under the microscopic decay hypothesis, the preliminary geometric
truncation is also available.  The original physical propagator is
contractive, so Duhamel's formula bounds the truncation error by
$e^{d_R}-1$, where
$d_R:=\int_0^t\|\mathcal L^{(\tau)}-
\mathcal L^{[R],(\tau)}\|_\diamond\,d\tau$.
The microscopic decay bound makes $d_R$ exponentially small in $R$.
Choosing $d_R\le\log(1+\epsilon/2)$ therefore allocates error
$\epsilon/2$ to this step without requiring the truncated generator
to be CPTP, and again gives $R=O(Q)$ for the stated parameter choice.
\end{remark}

\subsection{Time-dependent dynamics with static-type compression complexity}
\label{subsec:td-static-type-compression}

Time dependence does not by itself require the additional precision cost
of the general driven bound.  We now identify two classes for which the
static dependence is recovered: piecewise-constant driving with a
controlled number of time intervals, and driving that admits accurate
low-degree polynomial approximations in time.  No commutativity between
Hamiltonians at different times is assumed.
Appendix~\ref{subsec:renyi-free-fermions} also proves static-type
R\'enyi-entanglement scaling for arbitrarily driven one-dimensional
bilinear fermionic systems, under the spatial-locality and Gaussian
product-state hypotheses stated there, without a temporal regularity
assumption.

\paragraph{Roadmap: approximation in time versus rank counting.}
The key object is the dressed boundary family, not just its integral at
one final time.  On a short interval $I=[a,b]$, fix a cut--phase pair,
write $A^{(\tau)}=A_{p,\vec z}^{(\tau)}$ for its noncrossing background,
and let $V$ be a product seed across the selected cut $p$.  Then
\[
    F^{(x)}:=W_{a\to x}^{-1}VW_{a\to x},\qquad x\in I.
\]
Each $F^{(x)}$ has Schmidt rank one, but its factors can change with
$x$, so integrating over time can increase the rank.  We seek a
\emph{uniform} approximation to this family in a small set of fixed
product operators.  Uniformity preserves the intermediate-time
information needed in higher time-ordered integrals.  For a physical
background the inverse is the adjoint; the inverse notation also covers
the non-Hermitian phase-twisted backgrounds.

For piecewise-constant driving, the same generator is repeated within
each short interval.  Its commutator expansion supplies a common basis
there, and the subsequent count retains the number of time intervals.
For polynomially approximable driving, the construction is less direct:
we use a truncated expansion to certify approximation accuracy, but
\emph{not} to estimate Schmidt rank.  The following steps explain this
distinction before the formal statements.

First, approximate $A^{(\tau)}$ locally by a degree-$\nu$ polynomial
$A_{[\nu]}^{(\tau)}$, with propagator $W^{[\nu]}$, and consider its
\emph{exact} dressed family
\[
    F_{[\nu]}^{(x)}
    :=(W^{[\nu]}_{a\to x})^{-1}VW^{[\nu]}_{a\to x}.
\]
Local stability controls $F-F_{[\nu]}$.
Neither $W^{[\nu]}$ nor $F_{[\nu]}$ is generally a finite-degree
polynomial in time.  Only the auxiliary function $S_n(x)$ obtained by
truncating the commutator expansion of $F_{[\nu]}^{(x)}$ at order $n$
is a polynomial, with $\deg S_n\le n(\nu+1)$ and a uniformly small
remainder on $I$.  Its operator coefficients may contain many
noncommuting words; no low-rank bound on these coefficients is needed.

Next, take $P=n(\nu+1)+1$ Chebyshev points $x_v\in I$ and their cardinal
polynomials $\ell_v(x)$.  Interpolate the \emph{exact sample values},
not the truncated ones:
\[
    \widetilde F^{(x)}
    :=\sum_{v=1}^{P}\ell_v(x)F_{[\nu]}^{(x_v)}.
\]
Why does the polynomial $S_n$ justify this different approximant?
Interpolation reproduces $S_n$ exactly, so
\begin{align*}
    F_{[\nu]}^{(x)}-\widetilde F^{(x)}
    &=\bigl(F_{[\nu]}^{(x)}-S_n(x)\bigr)\\
    &\quad-\sum_{v=1}^{P}\ell_v(x)
       \bigl(F_{[\nu]}^{(x_v)}-S_n(x_v)\bigr).
\end{align*}
Thus the interpolation error is bounded by the truncation error times
$1+\sup_{x\in I}\sum_v|\ell_v(x)|$; the bound $1+2P$ suffices below.
The change from truncated values to exact samples is therefore an
\emph{error-controlled replacement}, not an exact reversal of the
truncation.

The rank estimate now uses only the sample representation.  Since
$W^{[\nu]}_{a\to x_v}$ factorizes across the cut, every
$F_{[\nu]}^{(x_v)}$ is a product operator.  All remaining $x$-dependence
is carried by the scalar polynomials $\ell_v(x)$.  Consequently,
$\widetilde F^{(x)}$ has a common product-operator basis of size at most
$P$.  The samples are kept as exact conjugates; expanding them again
into noncommutative words would discard the structure used for this
rank bound.  \emph{The truncated series certifies accuracy, whereas the
exact samples control rank.}

Finally, insert $\widetilde F^{(x)}$ in place of the dressed seed in the
original time-ordered branch.  For a scalar crossing coefficient
$c(x)$, one insertion becomes
\[
    \int_a^b c(x)F^{(x)}\,dx
    \;\simeq\;
    \sum_{v=1}^{P}
       \left[\int_a^b c(x)\ell_v(x)\,dx\right]
       F_{[\nu]}^{(x_v)}.
\]
If the uniform seed error is $\delta$, the error here is at most
$\delta\int_a^b|c(x)|\,dx$.
For $r$ insertions from such families, expansion of the interpolants
produces at most $P^r$ product operators, with all ordering constraints
remaining in scalar integrals.  Including the $R_Q$ crossing seeds
replaces $P$ by $R_QP$.  Since $r\le s_Q$, a polynomially bounded $P$
contributes only $s_Q\log P$ to the logarithmic rank cost.  Together
with a controlled interval count, this recovers the static-type buffer
tradeoff proved below.

Only the dressed seeds are replaced: the actual scalar crossing
coefficients and the outer background factors remain unchanged.
In particular, this is not a global replacement of the dynamics by
that of the polynomial Hamiltonian.  Nor does absorbing time integrals
into scalar coefficients, by itself, assert efficient computation of
those coefficients or of the sampled operators.  The statements below
concern approximation error and representation rank; construction time
requires a separate input model.

We retain the notation of
Sec.~\ref{subsec:time-dependent-unitary-dynamics}, including
$A_{p,\vec z}^{(\tau)}$, $B_{p,\vec z}^{[Q],(\tau)}$,
$W_{a\to b}$, $\mathcal V_{p,\vec z}^{(\tau;a)}$, $R_Q$, $b_Q$,
$N_{\mathrm{ph}}(Q)$, and $s_Q$.
The statements below concern operator-norm approximations and their BSR
or fixed-cut Schmidt rank.  Constants depend only on fixed local data,
unless additional driving parameters are displayed explicitly.
We may decrease the short-time scale used above so that
\begin{align}
    2k(g+1)t_c\le\frac14.
\label{eq:reg-short-time-scale}
\end{align}
This also accommodates polynomial approximants whose local strength is
at most $g+1$.  The cases $t=0$ and $b_Q=0$ are immediate; below we
write the argument for $t>0$ and $b_Q>0$.

\subsubsection{The integrated boundary interaction and a uniform-basis criterion}
\label{subsubsec:reg-integrated-boundary}

For a fixed cut $A|B$, the relevant one-insertion object is already
present in Eq.~\eqref{eq:integrated-dressed-definition}:
\begin{align}
    \mathcal I_u(V;A_0)
    =\int_0^u W_{0\to x}^{-1}VW_{0\to x}\,dx.
\label{eq:reg-essential-integral}
\end{align}
For the physical Hermitian background
$A_0^{(\tau)}=H_A^{(\tau)}+H_B^{(\tau)}$, this is precisely
$\int_0^u U_{A|B,0\to x}^{\dagger}VU_{A|B,0\to x}\,dx$.
At each time, product conjugation preserves $\operatorname{SR}_{A|B}(V)$.
The integration can nevertheless combine many different Schmidt
directions.  The static--driven distinction therefore concerns the
compression of this time-dependent operator family, not the
instantaneous rank of its members.

A low-rank approximation to the single integral at one endpoint is not,
by itself, sufficient for higher ordered integrals: it need not preserve
the intermediate-time information used in
Eq.~\eqref{eq:time-dependent-Ejr-interaction-picture}.
We use the stronger, composable condition from
Definition~\ref{def:common-operator-basis}: a uniform approximation to
the dressed family in a common product-operator basis.  For example, an
interpolant of the form
\begin{align}
    \widetilde V^{(x)}
    =\sum_{a=1}^{P}\ell_a(x)W_{0\to x_a}^{-1}VW_{0\to x_a}
\label{eq:reg-sample-frame}
\end{align}
has common operator-basis number at most
$P\operatorname{SR}_{A|B}(V)$.
Its integral is a sum of the same $P$ sampled operators, with weights
$\int\ell_a(x)\,dx$.
More importantly, Lemma~\ref{lem:ordered-integrals-common-bases}
controls every ordered product of the interpolated family.
Only the scalar coefficient functions are integrated; the number of
operator directions does not increase further.
Throughout the differential branches, the inverse in these formulas
must be retained because the phase-twisted background need not be
Hermitian.

For use in both applications, allow an arbitrary refinement
\begin{align}
    0=\tau_0<\tau_1<\cdots<\tau_M=t,
    \qquad |I_j|:=\tau_j-\tau_{j-1}\le t_c.
\label{eq:reg-refined-time-partition}
\end{align}
In particular, $M$ need not be $O(1+t)$ when the partition resolves
additional switching times.  The definition
Eq.~\eqref{eq:td-background-stability-factor} of
$\Gamma_{\mathrm{off}}(t)$ is used with this refined partition; its
upper bound Eq.~\eqref{eq:td-background-stability-crude} is unchanged.

\begin{prop}[Compression from a uniform common basis]
\label{prop:reg-uniform-basis-compression}
Suppose that, for every retained cut--phase pair and every $I_j$, there
is a family $\widetilde{\mathcal V}_{p,\vec z,j}^{(\tau)}$ satisfying
\begin{align}
    \operatorname*{ess\,sup}_{\tau\in I_j}
    \left\|\mathcal V_{p,\vec z}^{(\tau;\tau_{j-1})}
       -\widetilde{\mathcal V}_{p,\vec z,j}^{(\tau)}\right\|
    &\le b_Q\delta,
    \qquad 0<\delta\le1,
\label{eq:reg-frame-accuracy}
    \\
    \mathcal M_p[
       \widetilde{\mathcal V}_{p,\vec z,j}^{(\cdot)};I_j]
    &\le\mathcal M_Q,
    \qquad \mathcal M_Q\ge1.
\label{eq:reg-frame-size}
\end{align}
Replace the dressed factors in
Eq.~\eqref{eq:time-dependent-Ejr-interaction-picture} by these families,
keep the background factors exact, and then use the sums in
Eqs.~\eqref{eq:time-dependent-weak-compositions} and
\eqref{eq:time-dependent-cut-count-decomposition}.
The resulting operator $\widetilde U$ obeys
\begin{align}
    \|U_{0\to t}-\widetilde U\|
    &\le
    2N_{\mathrm{ph}}(Q)e^{-\eta Q+K_\eta[0,t]}
    \nonumber\\[-2pt]
    &\quad+
    N_{\mathrm{ph}}(Q)e^{\Gamma_{\mathrm{off}}(t)}
    \delta b_Qt\,e^{3b_Qt},
\label{eq:reg-compression-error}
    \\
    \operatorname{BSR}(\widetilde U)
    &\le N_{\mathrm{ph}}(Q)
       \sum_{s=0}^{s_Q}\binom{M+s-1}{s}\mathcal M_Q^s
    \nonumber\\
    &\le N_{\mathrm{ph}}(Q)(s_Q+1)
       \binom{M+s_Q-1}{s_Q}\mathcal M_Q^{s_Q}.
\label{eq:reg-compression-rank}
\end{align}
Here $\eta>0$ is the cut-tail parameter, as in
Eq.~\eqref{eq:time-dependent-cut-tail}.
\end{prop}

\textit{Proof.}
The local nested-commutator estimate gives, for a unit-norm seed
supported on at most $k$ sites and $0\le x-a\le t_c$,
\begin{align}
    \|W_{a\to x}^{-1}VW_{a\to x}\|
    \le\sum_{q\ge0}[2gk(x-a)]^q\le\frac43.
\label{eq:reg-local-orbit-norm}
\end{align}
This estimate uses absolute local interaction strengths and is valid
for the phase-twisted, non-Hermitian background.
Equations~\eqref{eq:td-crossing-channel-basis} and
\eqref{eq:td-crossing-channel-weight} therefore give
$\|\mathcal V_{p,\vec z}^{(\tau;a)}\|\le4b_Q/3$.
Both the original and the approximated dressed factors have norm at
most $3b_Q$.
A telescoping expansion of a product containing $s\ge1$ insertions
bounds its error by $s\delta b_Q(3b_Q)^{s-1}$.
For a fixed weak composition of $s$, the product of simplex volumes is
$\prod_j |I_j|^{r_j}/r_j!$.  Summing over compositions gives
\begin{align}
    \sum_{r_1+\cdots+r_M=s}
       \prod_{j=1}^{M}\frac{|I_j|^{r_j}}{r_j!}
    =\frac{t^s}{s!}.
\label{eq:reg-simplex-volumes}
\end{align}
After multiplication by the background norm factors, the total error
of one order-$s$ branch is at most
\begin{align*}
    e^{\Gamma_{\mathrm{off}}(t)}
       s\delta b_Q(3b_Q)^{s-1}\frac{t^s}{s!}.
\end{align*}
Summing over $s$ and the cut--phase pairs proves
Eq.~\eqref{eq:reg-compression-error}.
Lemma~\ref{lem:ordered-integrals-common-bases} bounds each short-time
coefficient by $\mathcal M_Q^{r_j}$.
Fixed-cut submultiplicativity, followed by the weak-composition count
and native-cut compression, proves
Eq.~\eqref{eq:reg-compression-rank}.
No submultiplicativity of BSR is used.
\hfill$\square$

Choose $Q$ so that
Eq.~\eqref{eq:td-choice-Q-final} holds and set
\begin{align}
    \delta_Q
    :=\frac{\epsilon}{2N_{\mathrm{ph}}(Q)(1+b_Qt)}
       \exp\left[-\int_0^t J_{\mathrm{off}}(\tau)\,d\tau
                    -3b_Qt\right].
\label{eq:reg-delta-Q}
\end{align}
Taking $\delta=\delta_Q$ in the proposition gives total error at most
$\epsilon$.
The logarithm of the rank is bounded explicitly by
\begin{align}
    \log\operatorname{BSR}(\widetilde U)
    &\le\log N_{\mathrm{ph}}(Q)+\log(s_Q+1)
    \nonumber\\
    &\quad+s_Q\log\left[e(M+s_Q)\mathcal M_Q\right].
\label{eq:reg-log-rank-explicit}
\end{align}
Thus the time-partition size and the common-basis size enter through
$s_Q\log M$ and $s_Q\log\mathcal M_Q$, respectively.
The small tolerance in Eq.~\eqref{eq:reg-delta-Q} accounts for the
nonunitary amplification; it must not be omitted.

\paragraph{Geometric consequences.}
Use the geometric crossing-support bounds obtained above, with the
preliminary truncation and its error budget when needed, so that
$R_Q=\operatorname{poly}(Q,|\partial\Lambda|)$.
Together with Eq.~\eqref{eq:td-crossing-coefficient-bound-derived} and
the uniform exponential layer-strength bound, these give
$b_Q=O(|\partial\Lambda|)$.  If $M$ and $\mathcal M_Q$ are polynomially bounded in
$1+t+\ell+|\partial\Lambda|+L_\epsilon$, then
Eq.~\eqref{eq:reg-log-rank-explicit} gives
\begin{align}
    \log\operatorname{BSR}(\widetilde U)
    =\widetilde O\left(
       1+|\partial\Lambda|t+\frac{L_\epsilon}{\ell}
       \right).
\label{eq:reg-static-type-BSR}
\end{align}
Converting to a prescribed cut and optimizing the admissible buffer
width, exactly as in Sec.~\ref{sec:renyi-upper-bound}, yields an
operator-norm $\epsilon$-approximant with
\begin{align}
    \log D_\epsilon
    =\widetilde O\left(
       |\partial\Lambda|(1+t)
       +\sqrt{|\partial\Lambda|L_\epsilon}\right).
\label{eq:reg-static-type-fixed-cut}
\end{align}
In one dimension with bounded $|\partial\Lambda|$, this is
$\widetilde O(1+t+\sqrt{L_\epsilon})$.
The same endpoint and admissible-buffer qualifications as in the
preceding geometric corollaries apply.
Unlike the shell estimate for an arbitrary drive, the applications
below have a basis size polynomial in $\log(1/\delta_Q)$.
Consequently the crude stability bound
$\Gamma_{\mathrm{off}}(t)=O(\ell|\partial\Lambda|t)$ affects only
logarithmic factors; no accuracy-dominated regime is needed here.

\subsubsection{Discrete driving}
\label{subsubsec:reg-discrete-driving}

Suppose that a common time partition has $N$ intervals and that the
Hamiltonian is constant on each interval:
\begin{align}
    H^{(\tau)}=H_j,
    \qquad a_{j-1}\le\tau<a_j,
    \qquad 0=a_0<\cdots<a_N=t.
\label{eq:reg-discrete-schedule}
\end{align}
The $H_j$ may be different and need not commute.
Here $N$ counts intervals in the schedule, including repeated uses of
the same Hamiltonian; it is not the number of distinct Hamiltonian
seeds.
Refining each interval to length at most $t_c$ gives
\begin{align}
    M\le N+\lceil t/t_c\rceil.
\label{eq:reg-discrete-M}
\end{align}

\begin{corol}[Piecewise-constant driving]
\label{corol:reg-discrete-driving}
Choose $Q$ and $\delta_Q$ as above, and let
\begin{align}
    n=\left\lceil\frac{\log(1/\delta_Q)}{\log4}\right\rceil.
\label{eq:reg-discrete-n}
\end{align}
For the schedule in Eq.~\eqref{eq:reg-discrete-schedule}, there is an
$\epsilon$-approximant satisfying
Eq.~\eqref{eq:reg-compression-rank} with
\begin{align}
    \mathcal M_Q=R_Q\frac{(n+1)(n+2)}{2},
    \qquad M\le N+\lceil t/t_c\rceil.
\label{eq:reg-discrete-basis}
\end{align}
In particular, $N=\operatorname{poly}(1+t+L_\epsilon)$ gives the
static-type bounds
Eqs.~\eqref{eq:reg-static-type-BSR} and
\eqref{eq:reg-static-type-fixed-cut}, using the geometric
truncation described above if needed.
\end{corol}

\textit{Proof.}
On a refined interval $I=[a,b]$, the noncrossing background $A$ is
constant.  Replace the dressed orbit of each fixed crossing seed $V$
by
\begin{align}
    \sum_{q=0}^{n}\frac{[i(\tau-a)]^q}{q!}
       \operatorname{ad}_A^q(V).
\label{eq:reg-discrete-commutator-polynomial}
\end{align}
The uniform remainder is at most $4^{-n}\|V\|$ by
Eq.~\eqref{eq:reg-short-time-scale} and the same local commutator
estimate as in Eq.~\eqref{eq:reg-local-orbit-norm}.
Equation~\eqref{eq:static-common-operator-basis-example} gives at most
$(n+1)(n+2)/2$ fixed product operators per product seed.
Summing the $R_Q$ seeds produces error at most $b_Q\delta_Q$ and the
basis size in Eq.~\eqref{eq:reg-discrete-basis}.
Apply Proposition~\ref{prop:reg-uniform-basis-compression}.
\hfill$\square$

For fixed spatial data, the schedule-dependent contribution to the
logarithmic rank is at most $s_Q\log(N+C(1+t)+s_Q)$, in addition to
the static logarithmic factors.
There is no inverse-minimum-interval-length factor in this rank bound.
In particular, arbitrarily short pulses are allowed if their number is
controlled.  The gain comes from applying the cut-count reduction to
the entire drive and distributing its $s_Q$ explicit insertions among
the intervals, rather than multiplying $N$ separately optimized
static rank bounds.
If $N$ grows faster than any polynomial in $L_\epsilon$, its dependence
must be retained.  This is consistent with the rapidly growing stage
counts in the driven lower-bound constructions.

\subsubsection{Polynomially approximable time dependence}
\label{subsubsec:reg-polynomial-driving}

A degree bound on $H^{(\tau)}$ does not make the propagator, or its
dressed boundary orbit, a polynomial of the same degree.  The useful
fact is more specific: after truncating the local commutator expansion
at order $n$, the dressed orbit is a polynomial of degree at most
$n(\nu+1)$ when the Hamiltonian has time degree $\nu$.
Interpolation of this polynomial structure produces a small common
basis without enumerating its noncommutative words.
We use $\nu$ for the time-polynomial degree to avoid confusion with the
selected cut $p$.

The approximation assumption will be imposed term by term in the
local interaction norm, rather than in the extensive norm $\|H\|$.
For a common partition $0=a_0<\cdots<a_N=t$, suppose that there are
Hermitian operator polynomials $h_{Z,j}^{[\nu]}(\tau)$, supported on
the same $Z$ as $h_Z^{(\tau)}$, such that
\begin{align}
    \deg_\tau h_{Z,j}^{[\nu]}&\le\nu,
    \\
    \sup_j\operatorname*{ess\,sup}_{\tau\in[a_{j-1},a_j]}
    \sup_{x\in\Lambda}
    \sum_{Z\ni x}
       \|h_Z^{(\tau)}-h_{Z,j}^{[\nu]}(\tau)\|
    &\le\zeta.
\label{eq:reg-local-polynomial-approximation}
\end{align}
Here $0\le\zeta\le1$, so the approximating interactions have one-site
strength at most $g+1$.
No bound on a monomial-basis coefficient norm is required.
The partition and the degree may depend on the target accuracy, but
the approximation is simultaneous for all local terms.

\begin{theorem}[Static-type compression from local polynomial approximation]
\label{thm:reg-polynomial-compression}
Choose $Q$ and $\delta_Q$ as in
Eq.~\eqref{eq:td-choice-Q-final} and
Eq.~\eqref{eq:reg-delta-Q}.
Assume Eq.~\eqref{eq:reg-local-polynomial-approximation} with
$\zeta\le\delta_Q/2$.
Let $n\ge1$ be an integer for which
\begin{align}
    P:=n(\nu+1)+1,
    \qquad (1+2P)4^{-n}\le\frac{\delta_Q}{2}.
\label{eq:reg-polynomial-P-n}
\end{align}
Such an $n$ can be chosen with
\begin{align}
    n=O\left(\log\frac{\nu+2}{\delta_Q}\right).
\label{eq:reg-polynomial-n-size}
\end{align}
Then $U_{0\to t}$ has an operator-norm $\epsilon$-approximant satisfying
Eq.~\eqref{eq:reg-compression-rank} with
\begin{align}
    \mathcal M_Q=R_QP,
    \qquad M\le N+\lceil t/t_c\rceil.
\label{eq:reg-polynomial-common-size}
\end{align}
Consequently, using the geometric bounds obtained above, the
static-type bounds hold whenever the number of approximation pieces
and the degree obey
\begin{align}
    \max\{N,\nu\}\le C\bigl(
       1+t+\log(1/\delta_Q)\bigr)^c,
\label{eq:reg-polynomial-growth-condition}
\end{align}
for constants $C,c$ independent of system size and accuracy.
The theorem also includes exact polynomial driving by taking
$\zeta=0$ and $N=1$.
\end{theorem}

\textit{Proof.}
Refine the approximation partition as in
Eq.~\eqref{eq:reg-refined-time-partition}.
Fix a refined interval $I=[a,b]$, a cut--phase pair, and one unit-norm
crossing product seed $V$ supported on at most $k$ sites.
Abbreviate the true background by $A^{(\tau)}$.
Delete the crossing terms from the polynomial approximants and apply
the same fixed phase weights as in
Eq.~\eqref{eq:time-dependent-A-B}; call the resulting noncrossing
background $A_{[\nu]}^{(\tau)}$.
It has time degree at most $\nu$, local strength at most $g+1$, and
local interaction-norm distance at most $\zeta$ from $A^{(\tau)}$.
Let $W^{[\nu]}$ be its propagator and write
\begin{align}
    F^{(x)}&:=W_{a\to x}^{-1}VW_{a\to x},
    &F_{[\nu]}^{(x)}&:=
       (W^{[\nu]}_{a\to x})^{-1}VW^{[\nu]}_{a\to x}.
\label{eq:reg-two-seed-orbits}
\end{align}

First compare the two local orbits.
For arbitrary $k$-local generators $K_1,\ldots,K_q$ with one-site
strengths $g_1,\ldots,g_q$, successive support growth gives
\begin{align}
    \|\operatorname{ad}_{K_q}\cdots
       \operatorname{ad}_{K_1}(V)\|
    \le (2k)^q q!\,g_1\cdots g_q\,\|V\|.
\label{eq:reg-mixed-local-commutator}
\end{align}
Indeed, a nonzero commutator must meet the support built so far,
which contains at most $k(q+1)$ sites after $q$ steps.
Telescoping each ordered word, with one generator replaced by the
local difference, and integrating its ordered simplex yields
\begin{align}
    \sup_{x\in I}\|F^{(x)}-F_{[\nu]}^{(x)}\|
    &\le\frac{\zeta}{g+1}
       \sum_{q\ge1}q[2k(g+1)|I|]^q\|V\|
    \nonumber\\
    &=\frac{2k|I|\zeta}
             {[1-2k(g+1)|I|]^2}\|V\|
    \le\zeta\|V\|.
\label{eq:reg-local-orbit-perturbation}
\end{align}
This is a local estimate: neither $\|H\|$ nor the number of lattice
sites appears.

Next truncate the time-ordered commutator series of $F_{[\nu]}^{(x)}$
at order $n$, and denote the result by $S_n(x)$.
An order-$q$ term is an integral of a product of $q$ degree-$\nu$
polynomials over a simplex with upper endpoint $x$.
It is therefore a polynomial in $x-a$ of degree at most $q(\nu+1)$,
irrespective of the noncommutativity of the operator coefficients.
The local estimate above also gives
\begin{align}
    \deg_x S_n&\le n(\nu+1)=P-1,
    &\sup_{x\in I}\|F_{[\nu]}^{(x)}-S_n(x)\|
       &\le4^{-n}\|V\|.
\label{eq:reg-polynomial-seed-tail}
\end{align}
We do not estimate the Schmidt ranks of the coefficients of $S_n$.
Instead, interpolate the exact family $F_{[\nu]}$ at the $P$ mapped
Chebyshev roots~\cite{Trefethen2019}
\begin{align}
    y_v=\cos\frac{(2v-1)\pi}{2P},
    \qquad x_v=a+\frac{|I|}{2}(1+y_v),
    \qquad v=1,\ldots,P,
\label{eq:reg-Chebyshev-nodes}
\end{align}
Let $\ell_v(x)$ be the cardinal polynomials of these nodes.
The elementary bound
$\sup_{x\in I}\sum_v|\ell_v(x)|\le2P$ suffices here.
For completeness, on $[-1,1]$ the cardinal polynomials have the form
\begin{align}
    \ell_v(y)=\frac1P\left[
       1+2\sum_{m=1}^{P-1}T_m(y_v)T_m(y)\right],
\label{eq:reg-cardinal-bound}
\end{align}
so $|\ell_v(y)|\le2$ follows from $|T_m(y)|\le1$.
Since interpolation reproduces $S_n$ exactly, the family
\begin{align}
    \widetilde F^{(x)}
    :=\sum_{v=1}^{P}\ell_v(x)
          (W^{[\nu]}_{a\to x_v})^{-1}V
             W^{[\nu]}_{a\to x_v}
\label{eq:reg-interpolated-seed-family}
\end{align}
satisfies
\begin{align}
    \sup_{x\in I}\|F^{(x)}-\widetilde F^{(x)}\|
    &\le\bigl[\zeta+(1+2P)4^{-n}\bigr]\|V\|
    \le\delta_Q\|V\|,
\label{eq:reg-polynomial-frame-error}
    \\
    \mathcal M_p[\widetilde F^{(\cdot)};I]
    &\le P.
\label{eq:reg-polynomial-seed-frame-size}
\end{align}
The last inequality holds because each sampled propagator factorizes
across the selected cut, so each sampled conjugate of the product
seed is again a product operator.
This is the step that avoids an exponential count of ordered words.

Finally, retain the actual scalar crossing coefficients
$b_{p,\vec z,\alpha}(x)$ in
Eq.~\eqref{eq:td-crossing-channel-basis} and replace only each dressed
seed by its interpolant.
There are at most $R_QP$ common product operators, and the total
uniform error is at most $b_Q\delta_Q$.
Proposition~\ref{prop:reg-uniform-basis-compression} proves the stated
error and rank bounds.
The size estimate for $n$ follows by substituting
$n=C\log[(\nu+2)/\delta_Q]$ into
Eq.~\eqref{eq:reg-polynomial-P-n} with a sufficiently large constant.
\hfill$\square$

The theorem does not approximate the full evolution by the evolution
of $\sum_Zh_{Z,j}^{[\nu]}$ in global operator norm.
The polynomial Hamiltonian is used only to build common frames for
local dressed seeds; the exact background factors and actual crossing
coefficients are retained.
Consequently the sufficient approximation tolerance is local and does
not carry an extensive $|\Lambda|$ factor.
This also explains why the transported-localization estimate derived
for the arbitrary-drive construction is not needed in this argument:
only the local commutator bound is used to build the frame.
Spatial decay is still needed for the cut-tail estimate and the
geometric crossing-channel count.

For the next corollary, let $E_\rho$ be the filled Bernstein ellipse
with foci $\pm1$ and boundary
$\{(z+z^{-1})/2:|z|=\rho\}$, where $\rho>1$.

\begin{corol}[Uniformly analytic local driving]
\label{corol:reg-analytic-driving}
Suppose that $[0,t]$ admits $N=O(1+t)$ intervals of length at most $t_c$.
On each interval $I_j=[a_j,b_j]$, define the rescaled local interaction
\begin{align}
    \widehat h_{Z,j}(y)
    :=h_Z^{(a_j+(b_j-a_j)(y+1)/2)},
    \qquad -1\le y\le1.
\end{align}
Assume that every $\widehat h_{Z,j}$ extends analytically to a common
Bernstein ellipse $E_\rho$, $\rho>1$, and that
\begin{align}
    \sup_j\sup_{x\in\Lambda}
       \sum_{Z\ni x}\sup_{z\in E_\rho}
          \|\widehat h_{Z,j}(z)\|
    \le g_\rho.
\label{eq:reg-analytic-local-bound}
\end{align}
Here $\rho$ and $g_\rho$ are independent of the system size and the
target accuracy.  Then the static-type bounds
Eqs.~\eqref{eq:reg-static-type-BSR} and
\eqref{eq:reg-static-type-fixed-cut} hold under the same spatial
assumptions.  Noncommuting local interactions are allowed.
\end{corol}

\textit{Proof.}
Truncate the Chebyshev expansion of each local interaction at degree
$\nu$.  The standard coefficient estimate reviewed in
Ref.~\cite[Theorems~8.1 and 8.2]{Trefethen2019}
(see also Ref.~\cite[Theorem~20]{TangTian2023CSGuide}) gives
\begin{align}
    \sup_{y\in[-1,1]}
    \|\widehat h_{Z,j}(y)-\widehat h_{Z,j}^{[\nu]}(y)\|
    \le\frac{2\rho^{-\nu}}{\rho-1}
           \sup_{z\in E_\rho}\|\widehat h_{Z,j}(z)\|.
\label{eq:reg-analytic-Chebyshev-error}
\end{align}
The same coefficient-contour proof applies directly in operator norm,
without a Hilbert-space-dimension factor.
Summing Eq.~\eqref{eq:reg-analytic-Chebyshev-error} over $Z\ni x$
shows that Eq.~\eqref{eq:reg-local-polynomial-approximation} holds with
\begin{align}
    \zeta=\frac{2g_\rho\rho^{-\nu}}{\rho-1}.
\end{align}
It is enough to take
\begin{align}
    \nu=\max\left\{0,
       \left\lceil\frac{\log[4g_\rho/((\rho-1)\delta_Q)]}
                          {\log\rho}\right\rceil\right\}
       =O(1+\log(1/\delta_Q)).
\label{eq:reg-analytic-degree}
\end{align}
Because the real-time local interactions are Hermitian, their
Chebyshev coefficients are Hermitian as well.
Theorem~\ref{thm:reg-polynomial-compression} therefore applies, with
$P=O([1+\log(1/\delta_Q)]^2)$.
\hfill$\square$

\begin{corol}[Linear ramps and bounded-frequency modulation]
\label{corol:reg-elementary-drives}
Assume the same spatial hypotheses and uniformly bounded local
strengths of the fixed Hamiltonians appearing below.
The static-type compression bounds apply to a linear ramp
\begin{align}
    H^{(\tau)}=(1-\tau/t)H_0+(\tau/t)H_1,
    \qquad 0\le\tau\le t,
\label{eq:reg-linear-ramp}
\end{align}
and to a finite collection of bounded-frequency tones
\begin{align}
    H^{(\tau)}=H_{\mathrm{dc}}
       +\sum_{a=1}^{N_{\mathrm f}}
       \bigl[\cos(\omega_a\tau)H_{a,\mathrm c}
             +\sin(\omega_a\tau)H_{a,\mathrm s}\bigr],
    \qquad |\omega_a|\le\Omega,
\label{eq:reg-finite-frequency-drive}
\end{align}
where $N_{\mathrm f}$ and $\Omega$ are fixed independently of the
system size and accuracy.  The Hamiltonians need not commute.
The conclusion also holds for any fixed-degree polynomial ramp with
uniform real-time local strength.
\end{corol}

\textit{Proof.}
The ramp is an exact degree-one polynomial, so use
Theorem~\ref{thm:reg-polynomial-compression} with $\zeta=0$ and $N=1$.
The same argument treats fixed-degree polynomial ramps.
For Eq.~\eqref{eq:reg-finite-frequency-drive}, partition time into
$O(1+t)$ intervals of length at most $t_c$ and fix any $\rho>1$.
On the associated complex-time ellipses,
$|\operatorname{Im}\tau|\le
 t_c(\rho-\rho^{-1})/4$.
The bounds $|\sin(\omega\tau)|,|\cos(\omega\tau)|
\le e^{|\omega\operatorname{Im}\tau|}$ therefore imply
Eq.~\eqref{eq:reg-analytic-local-bound}, with a constant determined
by the combined one-site strengths, $\rho$, and $\Omega$.
Apply Corollary~\ref{corol:reg-analytic-driving}.
\hfill$\square$

\paragraph{Uniformity and scope.}
The degree, interval count, and analytic constants must be controlled
at the precision required by Eq.~\eqref{eq:reg-delta-Q}.
Neither an unqualified $C^\infty$ assumption nor a fixed number of
Hamiltonian seeds guarantees this control: the scalar control
functions may still contain increasingly fine temporal structure.
If the hypotheses hold uniformly at every accuracy, the rank-to-spectrum
argument of Sec.~\ref{sec:renyi-upper-bound} gives the static
small-$\alpha$ dependence $\widetilde O(\alpha^{-1})$ at fixed time
and bounded boundary size for product initial states.
The statements here are compression bounds, not assertions that the
sampled product propagators or scalar ordered integrals are available
at unit computational cost.  A matching MPO construction-time theorem
requires an explicit input representation and the corresponding
coefficientwise construction, separately from the present rank proof.

\section{R\'enyi upper bounds from low-rank dynamics}
\label{sec:renyi-upper-bound}

This section converts the low-BSR approximations of Secs.~VI and VII into
bounds on the full Schmidt spectrum of an evolved pure state.  The argument
has two logically separate steps.  First, Proposition
\ref{prop:basic-bsr-properties} converts BSR into ordinary Schmidt rank across
a prescribed cut.  Second, the Eckart--Young characterization of the optimal
low-rank approximation~\cite{Eckart1936} converts the resulting rank--error curve into a
Schmidt-tail estimate and hence into a R\'enyi bound.
The connection between Schmidt tails, R\'enyi entropies, and MPS
approximability is developed in
Refs.~\cite{PhysRevB.73.094423,PhysRevLett.100.030504}.
For complementary constraints on instantaneous entanglement growth, see
Refs.~\cite{PhysRevLett.111.170501,PhysRevA.109.042404,KimEtAl2025SpectralSIE}.

Fix a target cut $q\in[\ell]$.  Let $U_t$ denote either $e^{-itH}$ or the
time-ordered propagator $U_{0\to t}$, and let $\ket{\psi_0}$ be a normalized
product state across that cut.  Write
\begin{align}
 \ket{\psi_t}:=U_t\ket{\psi_0},
 \qquad
 \lambda_1\ge\lambda_2\ge\cdots\ge0,
 \qquad
 \sum_{j\ge1}\lambda_j^2=1
 \label{eq:renyi-state-and-schmidt}
\end{align}
for the evolved state and its Schmidt coefficients, padded with zeros
beyond the finite Schmidt rank.  For $0<\alpha<1$, its
R\'enyi entanglement is
\begin{align}
 E_\alpha(\psi_t)
 :=\frac{1}{1-\alpha}
 \log\left(\sum_{j\ge1}\lambda_j^{2\alpha}\right).
 \label{eq:renyi-entanglement-definition}
\end{align}

\subsection{From a BSR approximation to a Schmidt tail}

\begin{prop}[BSR--SR--tail bridge]
\label{prop:bsr-sr-tail-bridge}
Suppose that an operator $\widetilde U$ satisfies
\begin{align}
 \|U_t-\widetilde U\|\le\epsilon,
 \qquad
 \operatorname{BSR}(\widetilde U)\le D.
 \label{eq:renyi-bsr-approximant}
\end{align}
Then
\begin{align}
 \operatorname{SR}_q(\widetilde U)
 \le \chi_qD,
 \label{eq:renyi-bsr-to-sr}
\end{align}
and the Schmidt coefficients of $\ket{\psi_t}$ obey
\begin{align}
 \sum_{j>\chi_qD}\lambda_j^2\le\epsilon^2.
 \label{eq:renyi-tail-from-bsr}
\end{align}
More generally, if $\ket{\psi_0}$ has Schmidt rank $r_0$ across the target
cut, then the rank cutoff in Eq.~\eqref{eq:renyi-tail-from-bsr} is
$r_0\chi_qD$.
\end{prop}

\textit{Proof.}
Equation~\eqref{eq:renyi-bsr-to-sr} is item~5 of Proposition
\ref{prop:basic-bsr-properties}.  The vector
$\ket{\widetilde\psi}:=\widetilde U\ket{\psi_0}$ has Schmidt rank at most
$\chi_qD$ and satisfies
\begin{align}
 \|\ket{\psi_t}-\ket{\widetilde\psi}\|
 \le\|U_t-\widetilde U\|\le\epsilon.
\end{align}
The Schmidt version of the Eckart--Young theorem states that the minimum
squared distance from $\ket{\psi_t}$ to a possibly unnormalized vector
of Schmidt rank at most $R$ is $\sum_{j>R}\lambda_j^2$.
Taking $R=\chi_qD$ proves
Eq.~\eqref{eq:renyi-tail-from-bsr}.  If the initial Schmidt rank is $r_0$,
the same operator produces a vector of Schmidt rank at most $r_0\chi_qD$.
\hfill$\square$

It is useful to record the pointwise consequence without introducing a new
paper-wide rank notation.  Suppose that for every $L\ge0$ we have an
approximant $\widetilde U_L$ with
\begin{align}
 \|U_t-\widetilde U_L\|\le e^{-L},
 \qquad
 R(L):=\operatorname{SR}_q(\widetilde U_L).
 \label{eq:renyi-rank-profile-definition}
\end{align}
Then Proposition~\ref{prop:bsr-sr-tail-bridge} gives
\begin{align}
 \sum_{j>R(L)}\lambda_j^2\le e^{-2L}.
 \label{eq:renyi-rank-profile-tail}
\end{align}
Consequently, with the convention that the supremum of the empty set is zero,
\begin{align}
 L_s:=\sup\{L\ge0:R(L)<s\},
 \qquad
 \lambda_s\le e^{-L_s}.
 \label{eq:renyi-generalized-inverse}
\end{align}
This formulation automatically handles small $s$: when no available
approximant has rank below $s$, Eq.~\eqref{eq:renyi-generalized-inverse} reduces
to the trivial bound $\lambda_s\le1$.

\subsection{Fixed-cut rank profiles supplied by the dynamical bounds}

We isolate the only input from Secs.~VI and VII that is needed below.  Set
\begin{align}
 \mathsf B&:=\max\{1,|\partial\Lambda|\},
 &
 \mathsf A_t&:=1+\mathsf B+\bar g_1\mathsf B t.
 \label{eq:renyi-A-B-definition}
\end{align}
The additive constants make the formulas uniform at $t=0$, for a zero
interaction, and for a bounded finite prefix of the spectrum.

\begin{lemma}[Fixed-cut rank profiles]
\label{lem:renyi-fixed-cut-rank-profiles}
Assume the one-dimensional or uniformly layered geometry and the local
strength and spatial-decay hypotheses of the corresponding dynamical
theorem.  The required crossing-neighborhood bounds are obtained, when
necessary, by the preliminary truncation in
Eq.~\eqref{eq:td-offdiagonal-range-truncation-error}, which also applies
to a static Hamiltonian.  There is a constant $C\ge1$, depending only
on the fixed local data, such that for every $L\ge0$ the following
statements hold after choosing the buffer width $\ell$.  The physical
target bipartition is kept fixed as the coarse-layer partition is varied.

For a time-independent Hamiltonian, there is a fixed-cut approximant of error
at most $e^{-L}$ and rank $R_{\rm st}(L)$ satisfying
\begin{align}
 \log R_{\rm st}(L)
 \le
 C\left(\mathsf A_t+\sqrt{\mathsf B L}\right)
 \log\left(e+\mathsf A_t+\mathsf B+L\right).
 \label{eq:renyi-static-fixed-rank-profile}
\end{align}
For an arbitrarily driven Hamiltonian satisfying the setup of Sec.~VII,
there is a fixed-cut approximant of error at most $e^{-L}$ and rank
$R_{\rm dr}(L)$ satisfying
\begin{align}
 \log R_{\rm dr}(L)
 \le
 C\left[
 \mathsf A_t+
 \max\left\{\sqrt{\mathsf A_tL},
 \mathsf B^{1/3}L^{2/3}\right\}
 \right]
 \log\left(e+\mathsf A_t+\mathsf B+L\right).
 \label{eq:renyi-driven-fixed-rank-profile}
\end{align}
The single logarithm is a conservative envelope for the explicit phase,
channel-count, and local-basis factors in the preceding theorems.
\end{lemma}

\textit{Proof.}
For $0\le L\le1$, use the approximant constructed at $L=1$.  Its error
is at most $e^{-1}\le e^{-L}$, and, after increasing the constant $C$, its
rank obeys the displayed bounds because
$\mathsf A_t\ge1+\mathsf B$ absorbs the $L=1$ terms.  It therefore
suffices to prove the estimates for $L\ge1$.

For a target cut $q$, Proposition~\ref{prop:basic-bsr-properties} and
Eq.~\eqref{eq:coarse-bsr-to-sr} add at most
\begin{align}
 \log\chi_q\le2\ell\mathsf B\log d_0
 \label{eq:renyi-cut-conversion-cost}
\end{align}
to the logarithm of the BSR.

If the preliminary spatial truncation is needed, allocate half the error
budget to it and half to the subsequent approximation.  Replacing $L$ by
$L+\log2$ only changes the constants below.  The bounds on the crossing
strength and background stability derived in Sec.~VII apply also in the
static case.  In particular, $b_Qt\le C\mathsf A_t$, while both
$tJ_{\mathrm{off}}$ in the static case and
$\Gamma_{\mathrm{off}}(t)$ in the driven case are bounded by
$C\mathsf A_t\ell$.  For a fixed admissible $\eta>0$, the corresponding
exponential-moment budget $tK_\eta$ or $K_\eta[0,t]$ has the same bound.
The cut-tail and branch-error budgets therefore admit choices
\begin{align}
 Q&=O(\mathsf A_t\ell+L),
 &s_Q&=O\left(\mathsf A_t+\frac{L}{\ell}\right),
 \label{eq:renyi-general-cut-budgets}
\end{align}
with static truncation order $n=O(\mathsf A_t\ell+L)$ or driven shell
budget $J=O(\mathsf A_t\ell+L)$, respectively.  Here the phase count is
included in the error budgets: since
\begin{align}
 \log N_{\mathrm{ph}}(Q)
 &=\log\ell+\lceil\log_2\ell\rceil\log(2Q+1)
 \nonumber\\
 &\le C\log(e+\ell)\log(e+Q),
 \label{eq:renyi-phase-overhead}
\end{align}
a sufficiently large fixed prefactor in the choice of $Q$ absorbs this
term.  For the widths chosen below, all truncation parameters and channel
counts are polynomial in $e+\mathsf A_t+\mathsf B+L$.
Consequently, the phase contribution to the logarithm of the rank is at
most
$C\mathsf B\ell\log(e+\mathsf A_t+\mathsf B+L)$, using
$\log(e+\ell)\le C\ell$ and $\mathsf B\ge1$.
It need not be bounded by a single logarithm on its own.

For the static theorem, Corollary~\ref{corol:static-unitary-parameter-choice}
and Eq.~\eqref{eq:static-unitary-final-BSR-simple} therefore give
\begin{align}
 \log R_{\rm st}(L)
 \le C\left(\mathsf B\ell+\mathsf A_t+\frac{L}{\ell}\right)
 \log(e+\mathsf A_t+\mathsf B+L).
 \label{eq:renyi-static-buffer-cost}
\end{align}
Choose
$\ell=\lceil\max\{1,\sqrt{L/\mathsf B}\}\rceil$.
Then
\begin{align}
 \mathsf B\ell+\frac{L}{\ell}
 =O\left(\mathsf B+\sqrt{\mathsf BL}\right),
\end{align}
which proves Eq.~\eqref{eq:renyi-static-fixed-rank-profile}.

For the driven theorem, use the general estimate
Eq.~\eqref{eq:td-final-log-BSR-explicit-symmetric} and choose
\begin{align}
 \ell=\left\lceil\max\left\{1,
 \min\left\{(L/\mathsf B)^{2/3},\frac{L}{\mathsf A_t}\right\}
 \right\}\right\rceil.
 \label{eq:renyi-driven-buffer-choice}
\end{align}
This ensures $\mathsf A_t\ell=O(\mathsf A_t+L)$, so the general
stability bound already gives $J=O(L+\mathsf A_t)$.
No stronger stability assumption is needed.  Since
$s_Q=O(\mathsf A_t+L/\ell)$, we obtain
\begin{align}
 s_Q+\sqrt{s_Q(J+s_Q)}
 \le C\left[
 \mathsf A_t+\sqrt{\mathsf A_tL}
 +\frac{L}{\sqrt\ell}
 \right].
\end{align}
The chosen width also satisfies
\begin{align}
 \mathsf B\ell+\frac{L}{\sqrt\ell}
 \le C\left[\mathsf B+
 \max\left\{\sqrt{\mathsf A_tL},
 \mathsf B^{1/3}L^{2/3}\right\}\right].
 \label{eq:renyi-driven-buffer-balance}
\end{align}
Indeed, $L\le\mathsf A_t$ gives $\ell=1$.  Otherwise, the choice
$\ell\asymp(L/\mathsf B)^{2/3}$ gives the second term in the maximum.
If instead $\ell\asymp L/\mathsf A_t$, the defining comparison implies
$\mathsf B^2L\le\mathsf A_t^3$, and hence
$\mathsf BL/\mathsf A_t\le\sqrt{\mathsf A_tL}$.
Together with the phase and polynomial factors bounded above, these
estimates prove Eq.~\eqref{eq:renyi-driven-fixed-rank-profile}.

Finally, if a chosen width exceeds the available number
$\bar\ell+1$ of cuts, use the exact propagator instead.  Its operator
Schmidt rank obeys
$\log\operatorname{SR}_q(U_t)\le2\bar\ell\mathsf B\log d_0$,
which is no larger than the buffer-conversion cost already bounded
above, up to a fixed constant.  This covers finite systems without
requiring an unavailable buffer width.
\hfill$\square$

\subsection{Decay of the Schmidt coefficients}

For $s\ge3$, define the section-local quantity
\begin{align}
 Y_s
 :=c_0\frac{\log s}
 {\log\left(e+\mathsf A_t+\mathsf B+\log s\right)},
 \label{eq:renyi-Ys-definition}
\end{align}
where $c_0>0$ is chosen sufficiently small in terms of the constant in
Lemma~\ref{lem:renyi-fixed-cut-rank-profiles}.  Write
$x_+:=\max\{x,0\}$.

\begin{prop}[Schmidt-coefficient envelopes]
\label{prop:renyi-schmidt-envelopes}
Under the static rank profile
Eq.~\eqref{eq:renyi-static-fixed-rank-profile},
\begin{align}
 \lambda_s
 \le
 \exp\left[-c_1
 \frac{(Y_s-\mathsf A_t)_+^2}{\mathsf B}
 \right]
 \qquad(s\ge3).
 \label{eq:renyi-static-coefficient-envelope}
\end{align}
Under the driven rank profile
Eq.~\eqref{eq:renyi-driven-fixed-rank-profile},
\begin{align}
 \lambda_s
 \le
 \max\Bigg\{
 \exp\left[-c_1
 \frac{(Y_s-\mathsf A_t)_+^2}{\mathsf A_t}
 \right],
 \exp\left[-c_1
 \frac{(Y_s-\mathsf A_t)_+^{3/2}}{\sqrt{\mathsf B}}
 \right]
 \Bigg\}
 \qquad(s\ge3),
 \label{eq:renyi-driven-coefficient-envelope}
\end{align}
for a constant $c_1>0$ depending only on the fixed local data.
\end{prop}

\textit{Proof.}
In the static case, choose
\begin{align}
 L=c_2\frac{(Y_s-\mathsf A_t)_+^2}{\mathsf B}
 \label{eq:renyi-static-inversion-choice}
\end{align}
with $c_2>0$ sufficiently small.  If $Y_s\le\mathsf A_t$, the conclusion is
the trivial bound $\lambda_s\le1$.  Otherwise,
$\mathsf A_t+\sqrt{\mathsf BL}\le Y_s$ after decreasing $c_2$, and
\begin{align}
 \log(e+\mathsf A_t+\mathsf B+L)
 \le C\log(e+\mathsf A_t+\mathsf B+\log s).
\end{align}
The definition of $Y_s$ therefore ensures
$R_{\rm st}(L)<s$.  Equation~\eqref{eq:renyi-generalized-inverse} gives
Eq.~\eqref{eq:renyi-static-coefficient-envelope}.

For the driven profile, the case $Y_s\le\mathsf A_t$ is again trivial.
For $Y_s>\mathsf A_t$, choose
\begin{align}
 L=c_2\min\left\{
 \frac{(Y_s-\mathsf A_t)_+^2}{\mathsf A_t},
 \frac{(Y_s-\mathsf A_t)_+^{3/2}}{\sqrt{\mathsf B}}
 \right\}.
 \label{eq:renyi-driven-inversion-choice}
\end{align}
Then both $\sqrt{\mathsf A_tL}$ and
$\mathsf B^{1/3}L^{2/3}$ are at most a sufficiently small multiple of
$(Y_s-\mathsf A_t)_+$.  The same logarithmic comparison yields
$R_{\rm dr}(L)<s$.  Finally,
$e^{-\min\{a,b\}}=\max\{e^{-a},e^{-b}\}$, proving
Eq.~\eqref{eq:renyi-driven-coefficient-envelope}.  \hfill$\square$

The positive parts in Eqs.~\eqref{eq:renyi-static-coefficient-envelope} and
\eqref{eq:renyi-driven-coefficient-envelope} are essential.  Below the
finite threshold at which $Y_s$ exceeds $\mathsf A_t$, the rank profile places
no nontrivial restriction on the accuracy; the corresponding finite prefix is
handled directly in the R\'enyi sum.

\subsection{R\'enyi entanglement}

\begin{theorem}[R\'enyi upper bounds]
\label{thm:renyi-upper-bounds}
Under the hypotheses of Lemma~\ref{lem:renyi-fixed-cut-rank-profiles}, there
is a constant $C\ge1$, depending only on the fixed local data, such that the
following holds for every $0<\alpha<1$.

For time-independent dynamics,
\begin{align}
 E_\alpha(\psi_t)
 \le\frac{C}{1-\alpha}\Bigg[
 1+\mathsf A_t\log(e+\mathsf A_t)
 +\frac{\mathsf B}{\alpha}
 \log^2\left(
 e+\mathsf A_t+\frac{\mathsf B}{\alpha}
 \right)
 \Bigg].
 \label{eq:renyi-static-final-bound}
\end{align}
For arbitrarily driven dynamics under the same spatial hypotheses
and the measurable-drive setup of Sec.~VII,
\begin{align}
 E_\alpha(\psi_t)
 \le\frac{C}{1-\alpha}\Bigg[
 1+\mathsf A_t\log(e+\mathsf A_t)
 &+\frac{\mathsf A_t}{\alpha}
 \log^2\left(
 e+\mathsf A_t+\frac{\mathsf A_t}{\alpha}
 \right)
 \notag\\
 &+\frac{\mathsf B}{\alpha^2}
 \log^3\left(
 e+\mathsf A_t+\frac{\mathsf B}{\alpha^2}
 \right)
 \Bigg].
 \label{eq:renyi-driven-final-bound}
\end{align}
In particular, uniformly in system size at fixed time, boundary size,
and local interaction data,
\begin{align}
 E_\alpha^{\rm static}
 &=O\left(\alpha^{-1}\log^2(1/\alpha)\right),
 &
 E_\alpha^{\rm driven}
 &=O\left(\alpha^{-2}\log^3(1/\alpha)\right)
 \qquad(\alpha\downarrow0).
 \label{eq:renyi-final-small-alpha}
\end{align}
\end{theorem}

\textit{Proof.}
We give the cutoff argument explicitly because this is where the logarithmic
powers enter.  Since $\lambda_s\le1$, the first two coefficients contribute
at most two.  For a nonincreasing positive envelope $f$, the sum is bounded,
up to an absolute constant, by $1+\int_2^\infty f(x)\,dx$.  Set $x=e^X$;
then $dx=e^X dX$.

For the static envelope, the exponent in the integral is bounded by
\begin{align}
 X-\frac{c\alpha}{\mathsf B}
 \left[
 \frac{X}{\log(e+\mathsf A_t+\mathsf B+X)}
 -C\mathsf A_t
 \right]_+^2.
 \label{eq:renyi-static-integral-exponent}
\end{align}
Choose
\begin{align}
 X_{\rm st}:=C\left[
 \mathsf A_t\log(e+\mathsf A_t)
 +\frac{\mathsf B}{\alpha}
 \log^2\left(e+\mathsf A_t+\frac{\mathsf B}{\alpha}\right)
 \right].
 \label{eq:renyi-static-cutoff}
\end{align}
For a sufficiently large $C$, every $X\ge X_{\rm st}$ satisfies both
\begin{align}
 \frac{X}{\log(e+\mathsf A_t+\mathsf B+X)}
 &\ge2C\mathsf A_t,
 \\
 \frac{\alpha X}{\mathsf B
 \log^2(e+\mathsf A_t+\mathsf B+X)}
 &\ge C.
 \label{eq:renyi-static-cutoff-conditions}
\end{align}
Hence Eq.~\eqref{eq:renyi-static-integral-exponent} is at most $-X$ for
$X\ge X_{\rm st}$.  The part below the cutoff is at most $e^{X_{\rm st}}$,
and the remaining integral is bounded by $e^{-X_{\rm st}}$.  Therefore
\begin{align}
 \sum_s\lambda_s^{2\alpha}
 \le C e^{X_{\rm st}},
\end{align}
which gives Eq.~\eqref{eq:renyi-static-final-bound}.

For the driven envelope, use $\max\{a,b\}\le a+b$ after raising to the power
$2\alpha$.  The quadratic-branch integrand is bounded by $e^{-X}$ for
\begin{align}
 X\ge X_1:=C\left[
 \mathsf A_t\log(e+\mathsf A_t)
 +\frac{\mathsf A_t}{\alpha}
 \log^2\left(e+\mathsf A_t+\frac{\mathsf A_t}{\alpha}\right)
 \right],
 \label{eq:renyi-driven-cutoff-one}
\end{align}
and the $3/2$-branch integrand is bounded by $e^{-X}$ for
\begin{align}
 X\ge X_2:=C\left[
 \mathsf A_t\log(e+\mathsf A_t)
 +\frac{\mathsf B}{\alpha^2}
 \log^3\left(e+\mathsf A_t+\frac{\mathsf B}{\alpha^2}\right)
 \right].
 \label{eq:renyi-driven-cutoff-two}
\end{align}
Indeed, the first condition is the static calculation with
$\mathsf B$ replaced by $\mathsf A_t$.  For the second branch, after the
positive part is active, domination of the Jacobian $e^X$ reduces to
\begin{align}
 \frac{\alpha\sqrt X}
 {\sqrt{\mathsf B}\,
 \log^{3/2}(e+\mathsf A_t+\mathsf B+X)}
 \ge C,
\end{align}
which is ensured by Eq.~\eqref{eq:renyi-driven-cutoff-two}.  Splitting each
integral at its cutoff gives
\begin{align}
 \sum_s\lambda_s^{2\alpha}
 \le C(e^{X_1}+e^{X_2}).
\end{align}
Taking the logarithm, using
$\log(e^{X_1}+e^{X_2})\le\log2+X_1+X_2$, and dividing by $1-\alpha$ proves
Eq.~\eqref{eq:renyi-driven-final-bound}.  \hfill$\square$

\begin{remark}[Optimality of the spectrum exponents]
The static construction of Sec.~IX gives
$E_\alpha=\Omega(\alpha^{-1})$, while the multiscale driven construction of
Sec.~X gives $E_\alpha=\Omega(\alpha^{-2})$.  These lower bounds concern
families whose system size may grow as $\alpha\downarrow0$, with time,
boundary size, and local data uniformly bounded, or their thermodynamic
limits.  Thus the powers of $\alpha$ in the system-size-uniform bounds of
Theorem~\ref{thm:renyi-upper-bounds} are optimal; the remaining gap is
polylogarithmic.  Equivalently, the associated approximate-rank lower bounds
match the square-root and $2/3$ powers obtained by inverting the static and
driven fixed-cut rank profiles.
\end{remark}

\begin{remark}[Other initial states and Liouville space]
If the initial state has Schmidt rank $r_0$, every rank cutoff above is
multiplied by $r_0$, which adds at most $\log r_0/(1-\alpha)$ to the displayed
bounds.  For operator R\'enyi entanglement, use the normalized vectorization
$D_\Lambda^{-1/2}|U_t\rangle\!\rangle$, equivalently the pure Choi state
of the unitary channel.  Applying $U_t$ to the system halves of on-site
maximally entangled pairs gives a product initial state across the spatial
cut, with the local dimensions squared.  The approximation error is at most
$D_\Lambda^{-1/2}\|U_t-\widetilde U\|_2
\le\|U_t-\widetilde U\|$, where $\|\cdot\|_2$ is the
Hilbert--Schmidt norm, so the same argument applies.
For a general mixed output of Liouvillian dynamics, however, the pure-state
Schmidt spectrum in Eq.~\eqref{eq:renyi-state-and-schmidt} is not an
entanglement measure; one must specify a mixed-state measure separately.
\end{remark}

\section{Lower bounds: entanglement generation and circuit encoding}
\label{sec:lower-static-discrete-explicit}

The lower bounds in this part have two logically different forms.  The first
is an entanglement-generation statement: a dynamical state develops a large
Schmidt tail, which directly yields lower bounds on MPS or MPO bond dimension
and on R\'enyi entanglement.  The second is a circuit-encoding statement: a
small-amplitude sector, identified by a fixed postselection, can contain the
output of a deep quantum circuit.  Keeping these two mechanisms separate is
important because a state can have large Schmidt rank but small gate
complexity, as happens for a product of Bell pairs.

Throughout the entanglement statements, $\epsilon_{\rm sq}$ denotes a squared
state-vector error and $\epsilon_{\rm op}$ denotes an operator-norm error.  In
the circuit-encoding statements, $\varepsilon$ denotes the amplitude of the
flagged branch, so its probability is $\varepsilon^2$.  The two proof chains
used below may be summarized as
\begin{align}
 \text{large Schmidt tail}
 &\Longrightarrow
 \text{MPS/MPO rank lower bound and R\'enyi lower bound},
 \\
 \text{hard flagged branch of amplitude }\varepsilon
 &\Longrightarrow
 \text{circuit lower bound at squared error }O(\varepsilon^2).
 \label{eq:two-lower-bound-chains}
\end{align}
The first implication is unconditional once a tail estimate is available.  The
second requires a lower bound for preparing the normalized branch in
the relevant postselected circuit model and a fixed flag projector of
controlled circuit cost.  It uses the stability of normalized
postselection proved below.

\subsection{Entanglement generation under a time-independent Hamiltonian}

\begin{figure}[!t]
\centering
\includegraphics[width=0.94\textwidth]{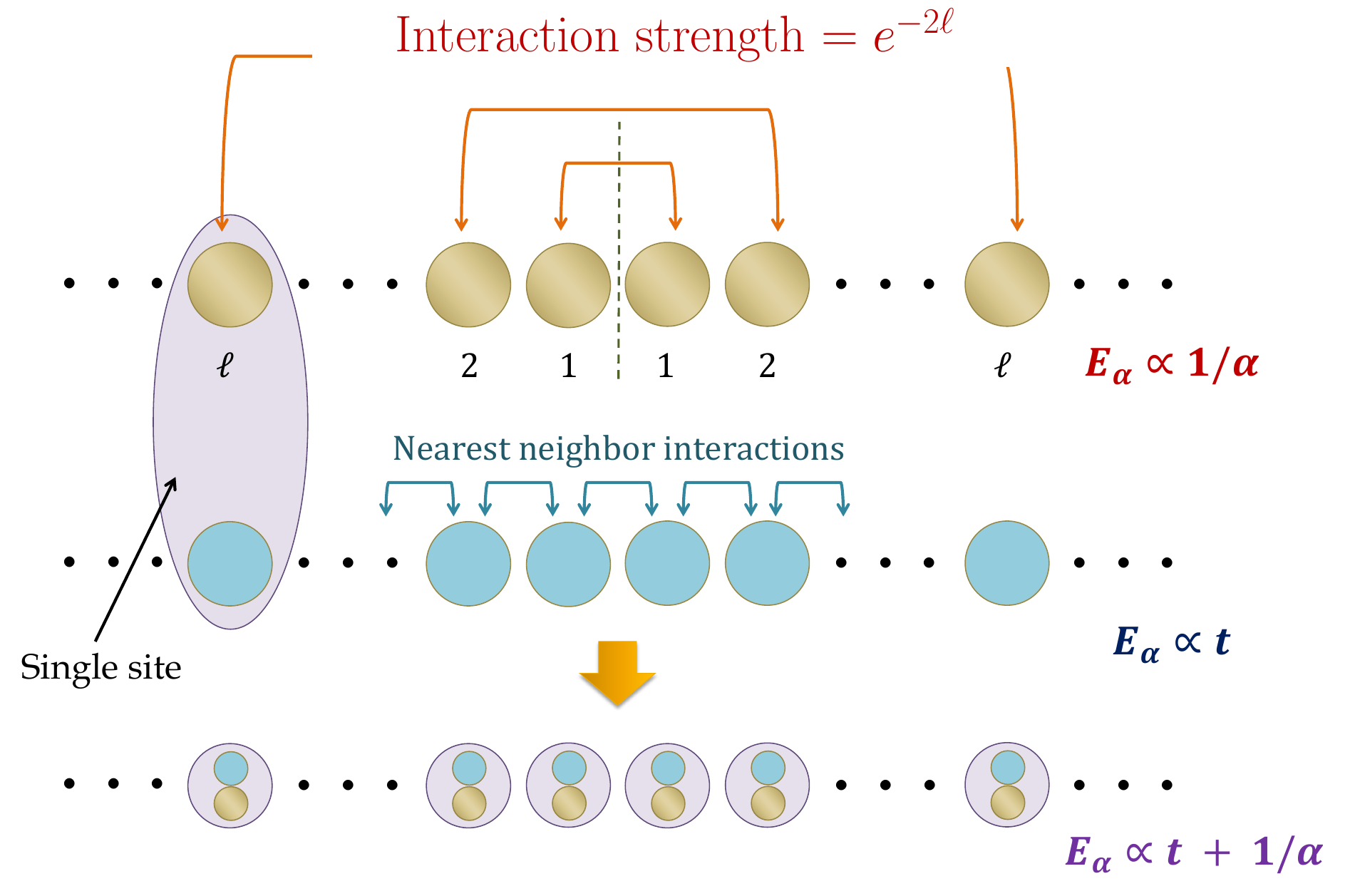}
\caption{Schematic relation between the static weak-pair lower bound and the
usual finite-range contribution.  In the upper row, the $\ell$th pair lies
$\ell$ sites from the cut on each side, so its physical separation is
$2\ell$ and its coupling is $e^{-2\ell}$.  The pairwise contributions add to
$E_\alpha=\Omega(\alpha^{-1})$, as proved in
Lemma~\ref{lem:static-weak-pairs-explicit}.  The middle row depicts, only at
the level of mechanism, a nearest-neighbor dynamics with a contribution linear
in time.  If the two dynamics act on different local tensor factors, R\'enyi
entanglement is additive, giving the bottom-row scaling
$E_\alpha=\Omega(t+\alpha^{-1})$.}
\label{fig:static-entanglement-mechanisms}
\end{figure}

Figure~\ref{fig:static-entanglement-mechanisms} isolates the weak-pair
mechanism used in this subsection and places it in the broader scaling picture.
Only its upper row is needed for Lemma~\ref{lem:static-weak-pairs-explicit};
the nearest-neighbor row is included to show how an independent linear-in-time
contribution can be combined on a separate on-site tensor factor.

\begin{lemma}[Static weak-pair construction]
\label{lem:static-weak-pairs-explicit}
Consider a one-dimensional chain separated into a left and a right half.  For
$\ell=1,2,\ldots,M$, choose a disjoint qubit pair $(L_\ell,R_\ell)$ whose two
sites are at graph distance $2\ell$, and define
\begin{align}
 H_M:=\sum_{\ell=1}^M V_\ell,
 \qquad
 V_\ell:=e^{-2\ell}
 \left(
 \ket{0,0}\bra{1,1}+\ket{1,1}\bra{0,0}
 \right)_{L_\ell R_\ell}.
 \label{eq:static-H-explicit}
\end{align}
Then $H_M$ satisfies the decay profile $\bar J(r)=e^{-r}$.  For the product
initial state
\begin{align}
 \ket{\psi_0}:=\bigotimes_{\ell=1}^M\ket{0,0}_{L_\ell R_\ell}
\end{align}
and time $t=1$, one has, for every $0<\alpha<1$,
\begin{align}
 E_\alpha\!\left(e^{-iH_M}\ket{\psi_0}\right)
 \ge \frac12\sum_{\ell=1}^M e^{-4\alpha\ell}.
 \label{eq:static-Renyi-finite-explicit}
\end{align}
In the thermodynamic limit,
\begin{align}
 E_\alpha\!\left(e^{-iH}\ket{\psi_0}\right)
 \ge \frac{1}{2(e^{4\alpha}-1)}
 \ge \frac{1}{2e^4\alpha}.
 \label{eq:static-Renyi-infinite-explicit}
\end{align}
\end{lemma}

\textit{Proof.}
The supports of the $V_\ell$ are disjoint, hence the terms commute.  With
$x_\ell=e^{-2\ell}$,
\begin{align}
 e^{-iV_\ell}\ket{0,0}
 =\cos(x_\ell)\ket{0,0}-i\sin(x_\ell)\ket{1,1}.
\end{align}
The R\'enyi entanglement is therefore additive over the pairs.  Since
$x_\ell\le e^{-2}$, it is enough to use
\begin{align}
 \log\!\left(
 |\cos x|^{2\alpha}+|\sin x|^{2\alpha}
 \right)
 \ge \frac{1-\alpha}{2}x^{2\alpha}
 \qquad
 (0\le x\le e^{-2},\ 0<\alpha<1).
\end{align}
The case $x=0$ is immediate.  For $x>0$, put $p=\sin^2x$ and
$u=p^\alpha-p$.  Since
$(1-p)^\alpha\ge1-p$ and
$1-p^{1-\alpha}\ge(1-\alpha)(1-p)$, the left-hand side is at least
$\log(1+u)\ge(\log2)u$.  For $x\le e^{-2}<1/4$,
$\sin x/x\ge1-x^2/6\ge95/96$ and $1-p\ge15/16$, and hence
\begin{align*}
 (\log2)u
 \ge
 (\log2)\left(\frac{95}{96}\right)^2\frac{15}{16}
 (1-\alpha)x^{2\alpha}
 >\frac{1-\alpha}{2}x^{2\alpha}.
\end{align*}
This proves the estimate used in
Eq.~\eqref{eq:static-Renyi-finite-explicit}.  Moreover,
$\sum_\ell\norm{V_\ell}<\infty$ and
$\sum_\ell\sin^2(x_\ell)<\infty$, so $H_M$ converges in norm and the
infinite tensor-product state is well defined.  Its Schmidt decomposition is
the infinite product of the two-level decompositions above, and its partial
R\'enyi entropies are the increasing sums of the nonnegative pair
contributions.  Monotone convergence therefore justifies taking
$M\to\infty$.  Summing the geometric series gives the first
inequality in Eq.~\eqref{eq:static-Renyi-infinite-explicit}.  Finally,
$e^{4\alpha}-1\le\alpha(e^4-1)\le\alpha e^4$ for $0<\alpha\le1$.
\hfill$\square$

The same state gives a direct approximate-rank lower bound with completely
explicit parameters.

\begin{prop}[Explicit static Schmidt-tail bound]
\label{prop:static-rank-explicit}
Let $\ket{\psi_M}:=e^{-iH_M}\ket{\psi_0}$ and let $1\le m\le M$.  Then
\begin{align}
 \inf_{\operatorname{SR}(\phi)\le2^{m-1}}
 \norm{\ket{\psi_M}-\ket{\phi}}^2
 &\ge
 2^{m-2}e^{-2m(m+1)}
 \ge e^{-5m^2}.
 \label{eq:static-tail-parametric}
\end{align}
The infimum may be taken over arbitrary, possibly unnormalized, vectors.
Consequently, if $D_{\epsilon_{\rm sq}}(\psi_M)$ denotes the least Schmidt
rank required at squared error $\epsilon_{\rm sq}$, then
\begin{align}
 \epsilon_{\rm sq}<e^{-5m^2}
 \quad\Longrightarrow\quad
 D_{\epsilon_{\rm sq}}(\psi_M)>2^{m-1}.
 \label{eq:static-rank-parametric}
\end{align}
In particular, if
$L:=\log(1/\epsilon_{\rm sq})\ge160$ and
$M\ge\lfloor\sqrt{L/10}\rfloor$, then
\begin{align}
 D_{\epsilon_{\rm sq}}(\psi_M)
 >\exp\!\left[
 \frac{\log2}{2\sqrt{10}}\sqrt{L}
 \right].
 \label{eq:static-rank-asymptotic-explicit}
\end{align}
\end{prop}

\textit{Proof.}
For a subset $S\subseteq[m]$, consider the Schmidt coefficient in which the
pairs in $S$ are excited, the pairs in $[m]\setminus S$ are unexcited, and all
pairs with $\ell>m$ are unexcited.  Its squared value is
\begin{align}
 p_S
 &=p_{\varnothing}\prod_{\ell\in S}\tan^2(e^{-2\ell}),
 \\
 p_{\varnothing}
 &=\prod_{\ell=1}^M\cos^2(e^{-2\ell}).
\end{align}
Using $\prod_j(1-a_j)\ge1-\sum_j a_j$ for $0\le a_j\le1$,
\begin{align}
 p_{\varnothing}
 \ge1-\sum_{\ell=1}^{\infty}\sin^2(e^{-2\ell})
 \ge1-\sum_{\ell=1}^{\infty}e^{-4\ell}
 >\frac12.
\end{align}
Moreover, $\tan x\ge x$ for $0\le x<\pi/2$, and hence
\begin{align}
 p_S
 \ge\frac12\exp\!\left[-4\sum_{\ell\in S}\ell\right]
 \ge\frac12e^{-2m(m+1)}.
\end{align}
There are $2^m$ such Schmidt coefficients.  By the Schmidt version of the
Eckart--Young theorem, the optimal squared error of an approximation of rank at most $r$,
even when the approximating vector is unnormalized, is the sum of all but the
$r$ largest squared Schmidt coefficients.  Thus, for $r=2^{m-1}$, at least
$2^{m-1}$ of the selected weights remain in the tail.  This gives the first
inequality in Eq.~\eqref{eq:static-tail-parametric}.  The second follows
from
\begin{align}
 2^{m-2}e^{-2m(m+1)}\ge e^{-5m^2}
 \qquad(m\ge1).
\end{align}
For the final claim, take $m=\lfloor\sqrt{L/10}\rfloor$.  Then
$5m^2<L$, while $m-1\ge\frac12\sqrt{L/10}$ for $L\ge160$.
\hfill$\square$

\paragraph{MPO consequence.}
If an open-boundary MPO $\widetilde U$ of bond dimension $D$ satisfies
\begin{align}
 \norm{\widetilde U-e^{-iH_M}}\le\epsilon_{\rm op},
\end{align}
then $\widetilde U\ket{\psi_0}$ is an MPS of bond dimension at most $D$ and
has squared state error at most $\epsilon_{\rm op}^2$.  Therefore,
\begin{align}
 \epsilon_{\rm op}<e^{-5m^2/2}
 \quad\Longrightarrow\quad
 D>2^{m-1}.
 \label{eq:static-MPO-parametric}
\end{align}
This is a representation-size lower bound.  A lower bound on the classical
construction time requires an output model to be specified separately.

\subsection{Encoding quantum circuits into a flagged error sector}
\label{subsec:flagged-circuit-encoding}

The entanglement lower bounds above measure the size of an MPS or MPO needed
to reproduce the final state.  We now introduce a complementary notion that
asks how much circuit structure can be hidden in a postselectable small-amplitude
sector.  The flag is part of the physical construction and is not chosen after
the state has been produced.

\begin{definition}[Flagged error-space circuit depth]
\label{def:flagged-error-depth}
Let $P_F$ be a fixed orthogonal projector and let $\ket{\Psi}$ be a normalized
state with
\begin{align}
 \varepsilon:=\norm{P_F\ket{\Psi}}>0,
 \qquad
 \ket{\psi_\varepsilon}:=
 \frac{P_F\ket{\Psi}}{\varepsilon}.
 \label{eq:flagged-error-branch-definition}
\end{align}
Equivalently, after an isometric identification of the two orthogonal sectors
with a data register and a logical flag, padding the smaller sector if
necessary, one may write
\begin{align}
 \ket{\Psi}
 =\sqrt{1-\varepsilon^2}\,\ket{\psi_{\rm fail}}\ket0_F
 +\varepsilon\,\ket{\psi_\varepsilon}\ket1_F.
 \label{eq:simple-flagged-decomposition}
\end{align}
The logical-flag notation is an identification of orthogonal sectors;
it does not make an arbitrary encoding or decoding a free operation.
Circuit depth is measured on the specified physical registers.
For a fixed circuit model $\mathfrak C$, including its input state, gate
set, register geometry, and ancilla allowance, and $\delta\ge0$, define
the conditional error-space depth at vector-norm accuracy $\delta$ by
\begin{align}
 \mathsf{Depth}^{\mathfrak C}_{\delta}(\Psi;P_F)
 :=\min\left\{d:\ \exists\text{ a depth-}d\text{ circuit in }\mathfrak C
 \text{ preparing }\ket{\widetilde\psi},\ 
 \norm{\ket{\widetilde\psi}-\ket{\psi_\varepsilon}}\le\delta
 \right\}.
 \label{eq:error-space-depth-definition}
\end{align}
The minimum is understood as $+\infty$ if no such circuit exists.  The branch
probability is $\varepsilon^2$.  Thus a global squared state error of order
$\delta^2\varepsilon^2$ is the natural scale for resolving the conditional
branch to vector-norm accuracy $\delta$.
\end{definition}

Two circuit models will be useful.  The first is ordinary nearest-neighbor
unitary depth.  The second allows computational-basis postselection at the end
of the circuit.  The latter is the robust resource for transferring a lower
bound from the flagged branch to an approximation of the full state.
If the flag measurement and any decoding have depth $d_F$ within the
allowed gate and ancilla model, a depth-$d$ full-state preparation followed
by this operation gives a postselected branch preparation of depth at most
$d+d_F$.  Lemma~\ref{lem:normalized-flag-stability-explicit} controls
its error.  Thus the branch lower bound must apply to a postselected model
that includes this procedure; an ordinary unitary-depth lower bound alone
does not suffice.

\begin{lemma}[Normalized flagged-branch stability]
\label{lem:normalized-flag-stability-explicit}
Let $P$ be an orthogonal projection and let $\ket{\Psi}$ and
$\ket{\widetilde\Psi}$ be normalized states.  Put
\begin{align}
 p:=\norm{P\ket{\Psi}}^2>0,
 \qquad
 \eta:=\norm{\ket{\Psi}-\ket{\widetilde\Psi}}<\sqrt p.
\end{align}
Then $P\ket{\widetilde\Psi}\ne0$ and
\begin{align}
 \norm{
 \frac{P\ket{\Psi}}{\norm{P\ket{\Psi}}}
 -
 \frac{P\ket{\widetilde\Psi}}{\norm{P\ket{\widetilde\Psi}}}
 }
 \le\frac{2\eta}{\sqrt p}.
 \label{eq:normalized-branch-stability}
\end{align}
In particular, for $0\le\delta_{\rm sq}<4$, with
$\ket{\Psi_P}:=P\ket\Psi/\norm{P\ket\Psi}$ and
$\ket{\widetilde\Psi_P}:=P\ket{\widetilde\Psi}/
\norm{P\ket{\widetilde\Psi}}$,
\begin{align}
 \norm{\ket{\Psi}-\ket{\widetilde\Psi}}^2
 \le\frac{\delta_{\rm sq} p}{4}
 \quad\Longrightarrow\quad
 \norm{\ket{\Psi_P}-\ket{\widetilde\Psi_P}}^2\le\delta_{\rm sq}.
 \label{eq:normalized-branch-transfer}
\end{align}
\end{lemma}

\textit{Proof.}
Write $x=P\ket{\Psi}$ and $y=P\ket{\widetilde\Psi}$.  Projection is
contractive, so $\norm{x-y}\le\eta$ and
$\norm y\ge\norm x-\eta>0$.  Hence
\begin{align}
 \norm{\frac{x}{\norm x}-\frac{y}{\norm y}}
 \le
 \frac{\norm{x-y}}{\norm x}
 +\frac{|\norm y-\norm x|}{\norm x}
 \le\frac{2\eta}{\sqrt p}.
 \qquad\square
\end{align}

\subsubsection{A discrete-drive embedding by spacetime postselection}

There is a particularly simple circuit-encoding construction if one does not
impose a uniform bound on the local Hilbert-space dimension.  It is the
standard tensor-network representation of a quantum circuit, interpreted as a
physical state followed by a product postselection, using the
gate-teleportation construction of Ref.~\cite{GottesmanChuang1999}.

Let
\begin{align}
 \mathcal C=U_Q\cdots U_2U_1
 \label{eq:depth-Q-circuit}
\end{align}
be an $n$-qubit nearest-neighbor brickwork circuit of depth $Q$.  Add identity
gates so that every wire has one input and one output leg at every layer.  Let
\begin{align}
 \ket{\Omega}:=\frac{\ket{00}+\ket{11}}{\sqrt2}
\end{align}
be the normalized Bell state.  For each one- or two-qubit gate $U$, prepare its
normalized Choi state by applying $U$ to the output halves of the appropriate
Bell pairs.  Connect consecutive time slices by the Bell projector
$\ket{\Omega}\bra{\Omega}$.  The elementary teleportation identity
\cite{BennettEtAl1993}
\begin{align}
 {}_{AB}\!\bra{\Omega}
 \left(\ket\phi_A\otimes\ket\Omega_{BC}\right)
 =\frac12\ket\phi_C
 \label{eq:teleportation-contraction}
\end{align}
shows that every contracted wire segment contributes a factor $1/2$ in
amplitude.  Include an additional input register in state $\ket{0^n}$
and connect it to the first time-slice inputs by the same Bell projectors.
Let $\ket{\Xi_{\mathcal C}}$ denote the tensor product of this input
state and all the Choi resource states, before postselection.

\begin{prop}[Discrete-drive circuit embedding]
\label{prop:spacetime-postselection-embedding}
For every depth-$Q$ nearest-neighbor circuit $\mathcal C$ on $n$ qubits, there
exists a finite piecewise-constant drive, a product initial state, and a product
of local Bell projectors $P_{\rm succ}$ such that
\begin{align}
 P_{\rm succ}\ket{\Xi_{\mathcal C}}
 =2^{-M}\,\mathcal C\ket{0^n}\otimes\ket{\mathrm{succ}},
 \qquad
 M=nQ.
 \label{eq:spacetime-success-branch}
\end{align}
Here $\ket{\mathrm{succ}}$ is the normalized product of Bell states
on the projected registers.  Suppressing this fixed factor, and grouping the
success sector and its orthogonal complement into a logical flag, gives
\begin{align}
 \ket{\Xi_{\mathcal C}}
 =\sqrt{1-2^{-2M}}\,\ket{\Xi_{\rm fail}}\ket0_F
 +2^{-M}\,\mathcal C\ket{0^n}\ket1_F.
 \label{eq:spacetime-flag-decomposition}
\end{align}
The success sector is detected by local Bell measurements followed by
postselection on the all-success outcome.

If each physical site is allowed to contain all $O(Q)$ input and output
time-slice registers of one circuit wire, then the construction uses local
dimension
\begin{align}
 d_{\rm loc}=2^{O(Q)}.
 \label{eq:spacetime-local-dimension}
\end{align}
All Choi resource states can then be prepared in a constant number of
macro-layers, and the Bell readout also has constant depth.  In this
constant-time packing, the norm of a
single supersite or supersite-edge Hamiltonian can scale as $O(Q)$.  If a
uniform $O(1)$ bound on that norm is imposed, the same preparation can instead
be scheduled in $O(Q)$ layers.
\end{prop}

\textit{Proof.}
The circuit tensor network has one Bell contraction for each wire at the input
boundary and one for each transition to the next circuit layer.  Thus there
are $Q$ contractions per wire and $M=nQ$ contractions in total.  Repeated use
of Eq.~\eqref{eq:teleportation-contraction} yields
Eq.~\eqref{eq:spacetime-success-branch}.  The Choi registers belonging to
different gates are disjoint.  Registers at different circuit times are also
disjoint even when they occupy the same supersite.  Starting from a
computational-basis product state, the Bell pairs and gate Choi states are
therefore prepared by a finite piecewise-constant local drive, and all
Bell-basis changes can be performed in parallel.  In the packed
implementation, $O(Q)$ bounded-norm generators may act simultaneously on one
supersite or supersite edge, giving norm $O(Q)$; scheduling them in $O(Q)$
batches keeps that norm $O(1)$.  This also proves the local-dimension and norm
statements.
\hfill$\square$

Taking $n=Q$ gives
\begin{align}
 \varepsilon_Q=2^{-Q^2},
 \qquad
 Q=\sqrt{\log_2(1/\varepsilon_Q)}.
 \label{eq:discrete-depth-error-relation}
\end{align}
Thus the error sector can contain the output of an arbitrary depth-$Q$ circuit
with
\begin{align}
 Q=\Theta\!\left(\sqrt{\log(1/\varepsilon_Q)}\right).
 \label{eq:discrete-encoding-depth}
\end{align}
This is an embedding-capacity statement.  To turn it into an intrinsic-depth
statement, one may choose, for example, a circuit whose output contains a Bell
pair between the two ends of an $n=Q$ chain.  A nearest-neighbor unitary circuit
starting from a product state requires depth $\Omega(Q)$ to create such an
end-to-end pair by the ordinary circuit light-cone argument.  This does not,
by itself, give the same lower bound against postselected circuits.


\subsubsection{Staticization when local ancillas are unrestricted}

The preceding construction is a discrete drive.  A recent localized clock
construction shows that a piecewise-continuous local Hamiltonian protocol can
be replaced by a time-independent Hamiltonian that implements the same
evolution in the same run time up to a prescribed error, while preserving the
interaction graph and introducing logarithmically many local ancilla
qubits per data qubit in the relevant size, local-term norm, run-time, and
inverse-error parameters
\cite{MooneyEtAl2025}.  For a piecewise time-independent drive, the bump-function
construction in the same work first produces a smooth Hamiltonian with exactly
the same final unitary and then applies the staticization procedure.

Apply this result to the drive of
Proposition~\ref{prop:spacetime-postselection-embedding}.  Let
$\ket{\Phi_0}$ be the normalized product of local initial clock states
and $\ket{\Xi_{\rm stat}}$ the final joint data--clock state.  The
staticization controls the joint state-vector error
\begin{align}
 \eta_{\rm stat}:=
 \norm{\ket{\Xi_{\rm stat}}
       -\ket{\Xi_{\mathcal C}}\otimes\ket{\Phi_0}}.
 \label{eq:staticization-joint-state-error}
\end{align}
Extend the success projector as $P_{\rm succ}\otimes I_{\rm clock}$.
Choosing
\begin{align}
 \eta_{\rm stat}\le\frac{\sqrt{\delta_{\rm sq}}}{2}\,\varepsilon_Q
 \label{eq:staticization-error-requirement}
\end{align}
for $0<\delta_{\rm sq}\le1$ ensures, by
Lemma~\ref{lem:normalized-flag-stability-explicit}, that the normalized joint success
branch, including the ideal clock factor, is reproduced to squared error at most $\delta_{\rm sq}$.  Moreover, the
staticized success amplitude differs from $\varepsilon_Q$ by at most
$\eta_{\rm stat}$, so it remains $\Theta(\varepsilon_Q)$ for fixed
$0<\delta_{\rm sq}\le1$.  Since $\varepsilon_Q=2^{-Q^2}$ and the clock dimension is
polynomial in the inverse implementation error, the number of ancillary clock
qubits per internal data qubit is
$O(Q^2+\log(1/\delta_{\rm sq}))$, up to logarithmic dependence on the remaining size,
norm, and run-time parameters.  An original supersite contains $O(Q)$ internal
data qubits, so the total clock overhead per supersite is
$O(Q^3+Q\log(1/\delta_{\rm sq}))$ qubits; for fixed $\delta_{\rm sq}$ this is $O(Q^3)$.
Therefore, if no uniform bound is imposed on local dimension, the same scaling
\begin{align}
 Q=\Theta\!\left(\sqrt{\log(1/\varepsilon_Q)}\right)
 \label{eq:staticized-depth-error-relation}
\end{align}
can be realized by a time-independent Hamiltonian in an enlarged Hilbert
space, to the stated conditional accuracy.

This conclusion must be interpreted with its resource assumptions visible.
The construction does not establish the same statement for a fixed on-site
dimension, nor does the constant-time packing keep the norm per original
supersite uniformly bounded.  In addition, the staticization introduces
on-site clock momentum terms whose norms grow with the clock resolution:
for clock dimension $N_c$ and run time $T$, they are of order $N_c/T$.
Thus unrestricted local ancillas are not the only resource used here;
the clock terms also need not satisfy a uniform local-strength bound.
This construction does not show how to retain fixed local dimension,
bounded local strength, one-dimensional geometry, and a comparable
flagged-sector weight simultaneously.

\subsection{Representation size versus MPO construction time}

Let $D_{\epsilon}^{\rm MPO}$ denote the minimum conventional MPO bond dimension
at accuracy $\epsilon$, and let $T_{\epsilon}^{\rm MPO}$ denote the classical
cost of producing the representation.  The lower bounds above constrain
$D_{\epsilon}^{\rm MPO}$, but no representation-independent implication
$T_{\epsilon}^{\rm MPO}=\operatorname{poly}(D_{\epsilon}^{\rm MPO})$ follows.
For an explicit standard MPO in which all virtual indices and tensor entries
are materialized, merely writing the output costs at least linear time in the
largest bond dimension.  A sparse, factored, or otherwise succinct output may
be much cheaper.  Conversely, finding the tensors or an optimal truncation may
be harder than the size of the final MPO suggests.  A construction-time theorem
must therefore specify the Hamiltonian input format, the permitted output
format, arithmetic versus bit complexity, and coefficient precision.  The
flagged-sector viewpoint is useful for reductions, but its full abstract
formalism is not needed in the main lower-bound proof.

\section{Lower bounds for arbitrarily fast driving}
\label{sec:lower-arbitrary-drive-explicit}

We now allow piecewise-constant driving with no uniform lower bound on the
stage duration.  The construction has finitely many stages at every finite
size, but its shortest stage is exponentially small in the block size.  We
again separate two consequences.  First, parallel blocks generate a large
Schmidt tail and the optimal $\log^{2/3}(1/\epsilon)$ tensor-network exponent.
Second, one programmable block can hide a state of exponentially large circuit
depth in a branch of amplitude $e^{-\Theta(b^2)}$.

\begin{figure}[!t]
\centering
\includegraphics[width=0.85\textwidth]{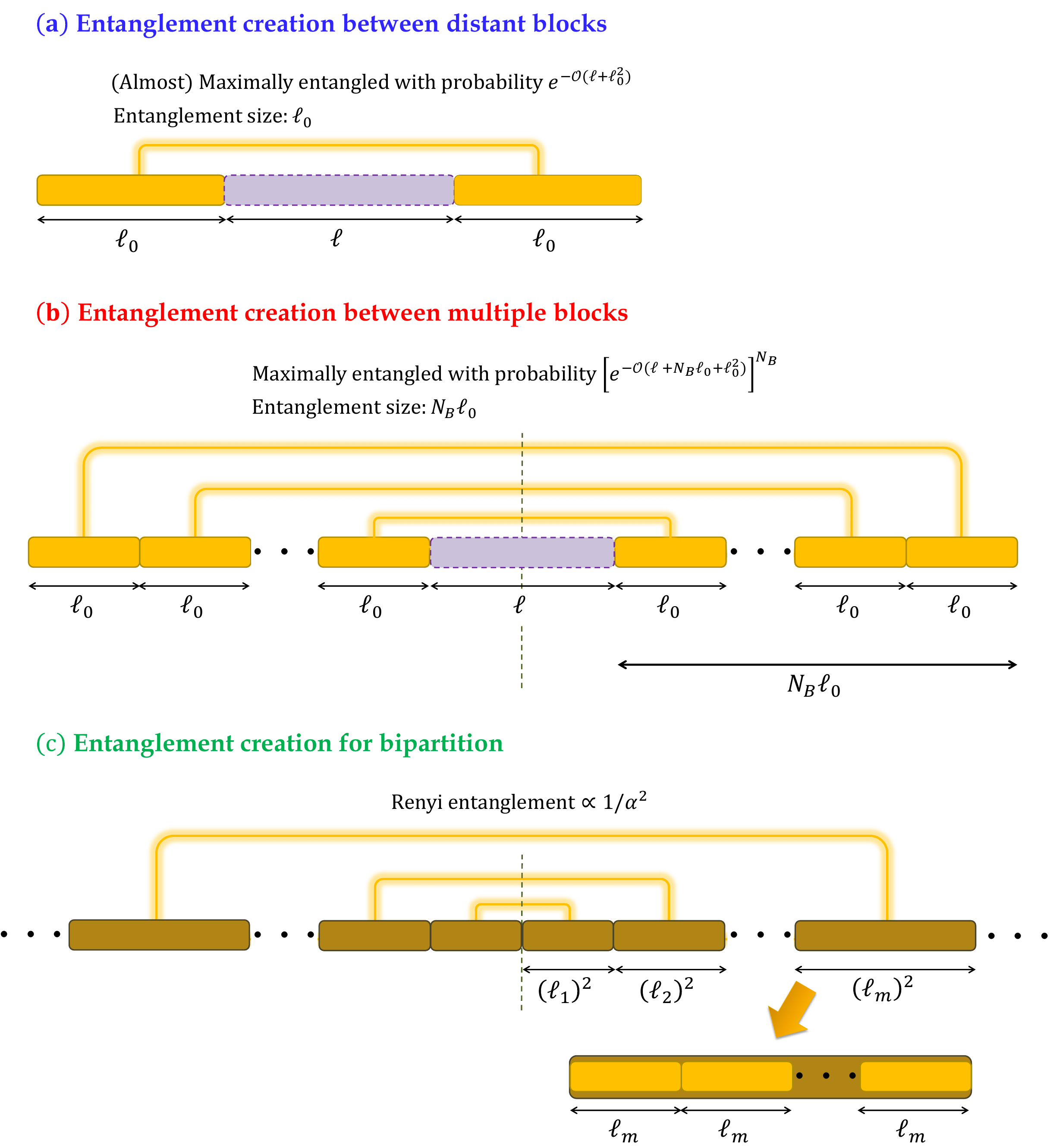}
\caption{Road map for the arbitrary-drive lower bounds.  (a) A pair of
$b$-site blocks separated by a gap $r$ is postselected onto an almost maximally
entangled state of Schmidt rank $2^b$, with probability
$e^{-O(r+b^2)}$.  (b) Running $N_B$ disjoint block-pair protocols in parallel
produces a flagged branch close to $Q=N_Bb$ Bell pairs; the total physical time and stage count are those
of one block pair, while the success probabilities multiply.  (c) Choosing
$b_m=2^m$ and allocating regions of size $Q_m=b_m^2$ on both sides of the cut
makes one scale available for every relevant $\alpha$, which yields
$E_\alpha=\Omega(\alpha^{-2})$.  Precise constants and errors are stated in
Theorem~\ref{thm:one-block-explicit}, Corollary~\ref{corol:parallel-explicit},
and Theorem~\ref{thm:uniform-renyi-explicit}.}
\label{fig:driven-lower-bound-roadmap}
\end{figure}

The geometry and the three levels of the construction are summarized in
Fig.~\ref{fig:driven-lower-bound-roadmap}.  The notation of that schematic is
related to the notation below by $\ell_0=b$, $\ell=r$,
$N_B\ell_0=Q$, and $\ell_m=b_m$.  Panel (a) is the one-block primitive,
panel (b) is its parallel repetition, and panel (c) is the dyadic multiscale
assembly used for the uniform R\'enyi lower bound.

\subsection{Entanglement generation and tensor-network lower bounds}

\subsubsection{One distant block pair}

\begin{theorem}[Distant-block protocol with explicit parameters]
\label{thm:one-block-explicit}
Let $b\ge1$ be an integer.  Let two $b$-site blocks be placed so that their
selected ancilla sites have graph distance $r\ge1$.  Each site has local
Hilbert space $\mathbb C^2\otimes\mathbb C^3$, of dimension six.  There exists
a $2$-local piecewise-constant drive satisfying the spatial decay profile
$\bar J(x)=e^{-x}$ and the uniform single-site bound
Eq.~\eqref{eq:g-extensivity} with $g=1$ at every time, a product initial state
$\ket{\psi_0}$, and product projections $P_{\rm L}$ and $P_{\rm R}$ such that
the total evolution time is smaller than one and the following statements
hold.

Define
\begin{align}
 \Delta_b&:=e^{-3b},
 \\
 c_b&:=\left|\sin(\Delta_b)\right|^b,
 \\
 \Delta_{1,b}&:=2^{-3b/2}c_b^3.
 \label{eq:one-block-parameters-explicit}
\end{align}
Then the projected amplitude obeys
\begin{align}
 a_b(r)
 &:=\norm{P_{\rm L}\otimes P_{\rm R}U_b(r)\ket{\psi_0}}
 \notag\\
 &\ge 2^{-6b-1}e^{-r-15b^2},
 \label{eq:one-block-amplitude-explicit}
\end{align}
so the success probability satisfies
\begin{align}
 p_b(r):=a_b(r)^2
 \ge2^{-12b-2}e^{-2r-30b^2}.
 \label{eq:one-block-probability-explicit}
\end{align}
Define the normalized target state
\begin{align}
 \ket{\Phi_b}
 :=2^{-b/2}\sum_{\sigma\in\{0,1\}^b}
 \ket{\sigma}_{\rm L}\ket{\sigma}_{\rm R},
\end{align}
where
$\ket{\sigma}_X:=\ket1_{{\rm anc},X}\otimes
\ket{\sigma_1,\ldots,\sigma_b}_X$ for $X\in\{{\rm L},{\rm R}\}$.
The unused qubits remain in their fixed initial states and are suppressed
in this notation.  Then
\begin{align}
 \norm{
 \frac{P_{\rm L}\otimes P_{\rm R}U_b(r)\ket{\psi_0}}{a_b(r)}
 -\ket{\Phi_b}
 }
 \le3e^{-r-3b^2}.
 \label{eq:one-block-error-explicit}
\end{align}
The protocol uses exactly
\begin{align}
 G_b:=(2b+1)2^b
 \label{eq:one-block-stage-count}
\end{align}
temporal stages if the two blocks are driven in parallel.  Its shortest stage
has duration $\Delta_{1,b}$, with
\begin{align}
 2^{-9b/2}e^{-9b^2}
 \le\Delta_{1,b}
 \le2^{-3b/2}e^{-9b^2}.
 \label{eq:one-block-min-stage}
\end{align}
\end{theorem}

\begin{figure}[!t]
\centering
\includegraphics[width=0.80\textwidth]{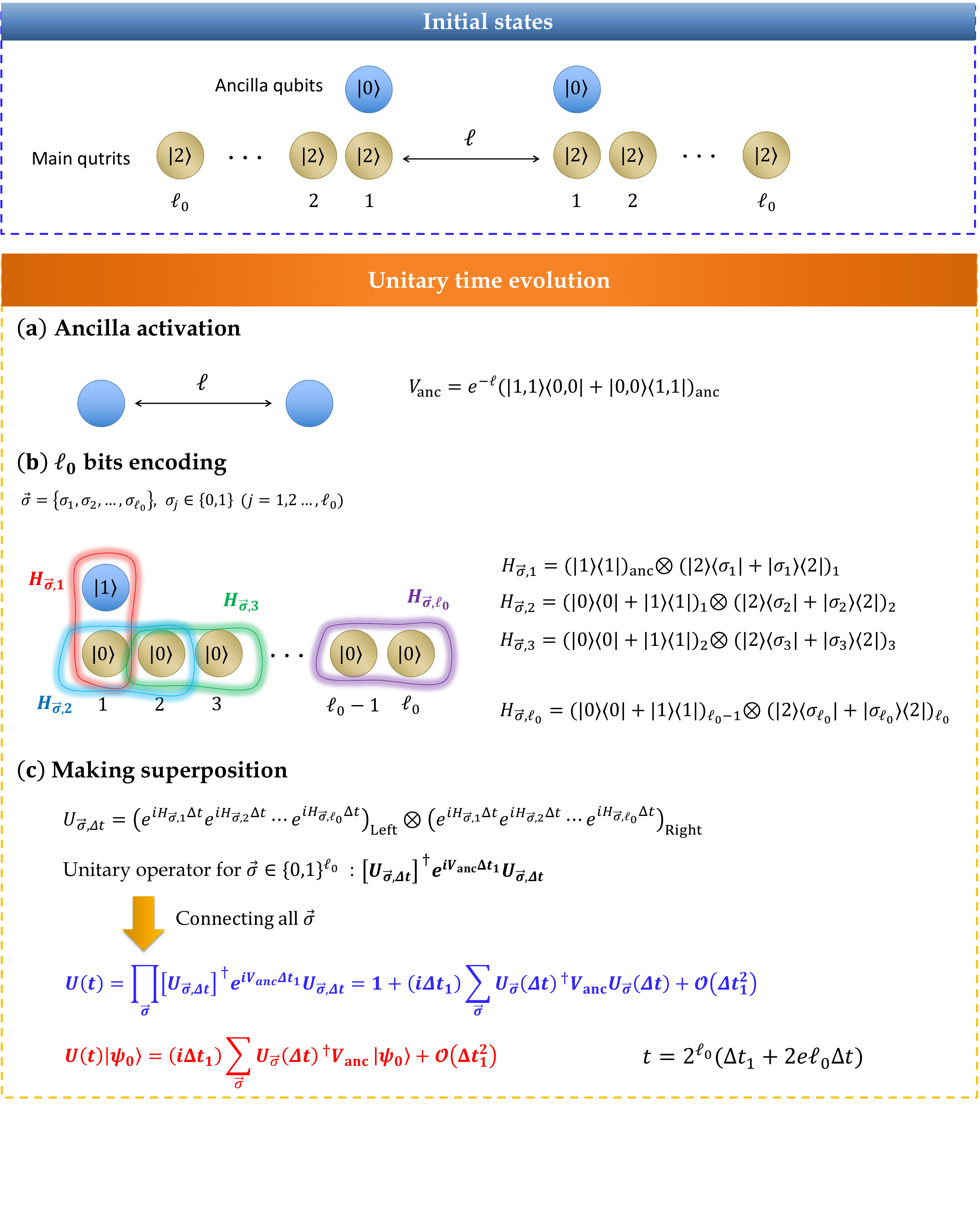}
\caption{Microscopic protocol for one distant block pair.  The initial state
contains one ancilla qubit and $b$ qutrits in each block.  (a) A weak long-range
pulse $V_{\rm anc}(r)$ jointly activates the two ancillas.  (b) Conditional
nearest-neighbor rotations encode a bit string
$\sigma\in\{0,1\}^b$ into the qutrits; the endpoint amplitude is
$|\sin\Delta_b|^b$.  (c) Conjugating the ancilla pulse by the encoding and
unencoding circuits, and multiplying the resulting factors over all
$\sigma$, makes the first-order term a coherent sum over the encoded basis.
The short duration $\Delta_{1,b}$ suppresses higher-order terms.  The displayed
notation $\ell_0,\ell,\Delta t,\Delta t_1$ corresponds respectively to
$b,r,\Delta_b,\Delta_{1,b}$ in the proof.  The parameter $\Delta t$
is a rotation angle; each encoding or unencoding stage lasts $e\Delta t$
after the $e^{-1}$ strength rescaling used in the proof.}
\label{fig:one-block-protocol-explicit}
\end{figure}

The pulse architecture behind the theorem is shown in
Fig.~\ref{fig:one-block-protocol-explicit}.  In the figure notation,
$\ell_0=b$, $\ell=r$, $\Delta t=\Delta_b$, and
$\Delta t_1=\Delta_{1,b}$.  Reading the panels from top to bottom makes the
role of the conjugation $U_{\sigma,\Delta_b}^{\dagger}
V_{\rm anc}U_{\sigma,\Delta_b}$ explicit before the quantitative estimates
are introduced.

\textit{Proof.}
On each block, only the qubit at the first site is used as an ancilla.  The
remaining qubits stay in a fixed state.  For
$\sigma=(\sigma_1,\ldots,\sigma_b)\in\{0,1\}^b$, define the unscaled encoding
operators
\begin{align}
 H_{\sigma,1}
 &:=\ket1\bra1_{\rm anc}\otimes
 \left(\ket{\sigma_1}\bra2+\ket2\bra{\sigma_1}\right)_1,
 \\
 H_{\sigma,j}
 &:=\left(\ket0\bra0+\ket1\bra1\right)_{j-1}
 \otimes
 \left(\ket{\sigma_j}\bra2+\ket2\bra{\sigma_j}\right)_j,
 \qquad j=2,\ldots,b.
 \label{eq:encoding-H-explicit}
\end{align}
For $j=1$ the operator is on-site, whereas for $j\ge2$ it is nearest-neighbor.
Under the convention that a physical Hamiltonian $K$ generates $e^{-itK}$,
the positive-sign rotations used below are implemented with the physical
Hamiltonian $-e^{-1}H_{\sigma,j}$ for time $e\Delta_b$; the inverse rotations
use the opposite sign.  Thus every implemented rotation angle has magnitude
$\Delta_b$, and the interaction norm is $e^{-1}$.  On one block define
\begin{align*}
 E_{\sigma,\Delta_b}
 :=e^{iH_{\sigma,b}\Delta_b}\cdots
 e^{iH_{\sigma,1}\Delta_b},
 \qquad
 U_{\sigma,\Delta_b}^{\dagger}
 :=E_{\sigma,\Delta_b}^{({\rm L})}\otimes
 E_{\sigma,\Delta_b}^{({\rm R})}.
\end{align*}
The initial block state is
\begin{align}
 \ket0_{\rm anc}\otimes\ket2^{\otimes b},
\end{align}
and every encoding Hamiltonian annihilates this state.  Hence both
$U_{\sigma,\Delta_b}$ and $U_{\sigma,\Delta_b}^{\dagger}$ fix
$\ket{\psi_0}$.

Set $\vartheta_b:=\pi(b+1/2)$.  The two ancillas are coupled by
\begin{align}
 V_{\rm anc}(r)
 :=e^{-r}
 \left(
 e^{i\vartheta_b}\ket{0,0}\bra{1,1}
 +e^{-i\vartheta_b}\ket{1,1}\bra{0,0}
 \right)_{\rm anc}.
 \label{eq:ancilla-coupling-explicit}
\end{align}
The fixed phase $\vartheta_b$ aligns the common phase of the projected
first-order vector with the convention for $\ket{\Phi_b}$.  The pulse
$e^{iV_{\rm anc}(r)\Delta_{1,b}}$ is implemented by the physical Hamiltonian
$-V_{\rm anc}(r)$ for time $\Delta_{1,b}$.
The full protocol is
\begin{align}
 U_b(r)
 :=\prod_{\sigma\in\{0,1\}^b}^{\longrightarrow}
 U_{\sigma,\Delta_b}^{\dagger}
 e^{iV_{\rm anc}(r)\Delta_{1,b}}
 U_{\sigma,\Delta_b}.
 \label{eq:one-block-unitary-explicit}
\end{align}
The order of the strings is arbitrary but fixed.  At every temporal stage,
the active local terms have disjoint supports, and every active term has norm
at most $e^{-\operatorname{diam}(Z)}$.  Therefore the drive is $2$-local,
satisfies $\bar J(x)=e^{-x}$, and obeys Eq.~\eqref{eq:g-extensivity} with
$g=1$.

Write $\ket{\sigma_{\le j}}$ for the state in which the ancilla equals one,
the first $j$ qutrits store $\sigma_1,\ldots,\sigma_j$, and the remaining
qutrits equal two.  Direct iteration gives
\begin{align}
 &e^{iH_{\sigma,b}\Delta_b}\cdots
 e^{iH_{\sigma,1}\Delta_b}\ket{\sigma_{\le0}}
 \notag\\
 &\quad=
 \sum_{j=0}^{b-1}(i\sin\Delta_b)^j\cos\Delta_b
 \ket{\sigma_{\le j}}
 +(i\sin\Delta_b)^b\ket{\sigma_{\le b}}.
 \label{eq:encoding-endpoint-correct-explicit}
\end{align}
In particular, the endpoint coefficient contains no additional cosine.

Set
\begin{align}
 A_\sigma:=U_{\sigma,\Delta_b}^{\dagger}
 V_{\rm anc}(r)U_{\sigma,\Delta_b},
 \qquad
 x_b:=2^be^{-r}\Delta_{1,b}.
\end{align}
Since $x_b\le2^{-b/2}e^{-r}\le1$, expand every exponential in its
norm-convergent power series.  The sum of the norms of all product terms of
total degree at least two is bounded by $e^{x_b}-1-x_b$.  Hence
\begin{align}
 \norm{
 U_b(r)-\left(\hat{\mathbf 1}+i\Delta_{1,b}\sum_\sigma A_\sigma\right)
 }
 \le e^{x_b}-1-x_b
 \le x_b^2
 =2^{2b}e^{-2r}\Delta_{1,b}^2.
 \label{eq:one-block-remainder-explicit}
\end{align}
Let $R_b$ denote the right-hand side.  On each block, take
\begin{align*}
 P_X:=\ket1\bra1_{{\rm anc},X}\otimes
 \bigotimes_{j=1}^b
 \left(\ket0\bra0+\ket1\bra1\right)_{j,X},
 \qquad X\in\{{\rm L},{\rm R}\},
\end{align*}
with the unused qubits projected onto their fixed states.  This product
projection removes all non-endpoint terms in
Eq.~\eqref{eq:encoding-endpoint-correct-explicit}.  By the choice of
$\vartheta_b$, the projected first-order vector is exactly
$A_{0,b}\ket{\Phi_b}$, where
\begin{align}
 A_{0,b}
 &=e^{-r}2^{b/2}c_b^2\Delta_{1,b}
 =e^{-r}2^{-b}c_b^5.
 \label{eq:one-block-target-norm-explicit}
\end{align}
Moreover,
\begin{align}
 \frac{R_b}{A_{0,b}}
 =e^{-r}c_b.
 \label{eq:one-block-ratio-explicit}
\end{align}
The elementary bounds $\Delta_b/2\le\sin\Delta_b\le\Delta_b$ imply
\begin{align}
 2^{-b}e^{-3b^2}\le c_b\le e^{-3b^2}.
 \label{eq:cb-two-sided-explicit}
\end{align}
Hence $e^{-r}c_b<1/2$ and
\begin{align}
 a_b(r)
 &\ge A_{0,b}-R_b
 \ge\frac12e^{-r}2^{-b}c_b^5
 \notag\\
 &\ge2^{-6b-1}e^{-r-15b^2},
\end{align}
which proves Eqs.~\eqref{eq:one-block-amplitude-explicit} and
\eqref{eq:one-block-probability-explicit}.  If $y$ is the projected output and
$x=A_{0,b}\ket{\Phi_b}$, then
\begin{align}
 \norm{\frac{y}{\norm y}-\frac{x}{\norm x}}
 \le\frac{2R_b}{A_{0,b}-R_b}
 \le\frac{2e^{-r}c_b}{1-e^{-r}c_b}
 \le3e^{-r-3b^2},
\end{align}
which gives Eq.~\eqref{eq:one-block-error-explicit}.

The total physical time is
\begin{align}
 T_b
 &=2^b\left(2eb\Delta_b+\Delta_{1,b}\right)
 \notag\\
 &\le2^{b+1}eb e^{-3b}+2^{-b/2}e^{-9b^2}
 \le4e^{-2}+2^{-1/2}e^{-9}
 <0.542<1.
 \label{eq:one-block-total-time-explicit}
\end{align}
The first sequence on the second line is maximal at $b=1$.  Finally,
Eq.~\eqref{eq:one-block-min-stage} follows by inserting
Eq.~\eqref{eq:cb-two-sided-explicit} into the definition of
$\Delta_{1,b}$.  Its upper bound is smaller than the encoding-stage duration
$e\Delta_b=e^{1-3b}$, so $\Delta_{1,b}$ is indeed the shortest stage.
\hfill$\square$

\subsubsection{Parallel generation of $Q$ Bell pairs}

This is the construction depicted in
Fig.~\ref{fig:driven-lower-bound-roadmap}(b): all block pairs are disjoint, so
they may be driven synchronously and only the branch probabilities multiply.

\begin{corol}[General parallelization bound]
\label{corol:parallel-explicit}
Let $N_B,b\ge1$ and set $Q=N_Bb$.  Place $N_B$ left blocks and $N_B$ right
blocks on the two sides of a central gap of length $r$.  The blocks may be
paired so that every paired ancilla distance is at most $r+2Q$.  Run the
protocol of Theorem~\ref{thm:one-block-explicit} on all block pairs in
parallel.  The joint flagged branch has success probability
\begin{align}
 p_{Q,N_B}(r)
 \ge
 \left[
 2^{-12b-2}
 e^{-2(r+2Q)-30b^2}
 \right]^{N_B},
 \label{eq:parallel-probability-explicit}
\end{align}
and its normalized state differs from $Q$ Bell pairs by at most
\begin{align}
 \delta_{Q,N_B}
 \le3N_Be^{-3b^2}.
 \label{eq:parallel-error-explicit}
\end{align}
The total time and the number of temporal stages are the same as for one block
pair, because all block pairs are disjoint and are driven synchronously.
\end{corol}

\textit{Proof.}
The evolution and the flag projection factorize over the $N_B$ block pairs.
Equation~\eqref{eq:parallel-probability-explicit} is therefore the product of
the one-block probability bounds, with $r$ replaced by the upper bound
$r+2Q$.  For normalized vectors,
\begin{align}
 \norm{\bigotimes_{j=1}^{N_B}\ket{\psi_j}
 -\bigotimes_{j=1}^{N_B}\ket{\phi_j}}
 \le\sum_{j=1}^{N_B}\norm{\ket{\psi_j}-\ket{\phi_j}},
\end{align}
which gives Eq.~\eqref{eq:parallel-error-explicit}.
\hfill$\square$

The choice relevant to the sharp error exponent is
\begin{align}
 N_B=b,
 \qquad
 Q=b^2,
 \qquad
 r\le Q.
 \label{eq:optimized-integer-choice-explicit}
\end{align}
Then every paired distance is at most $3b^2$, and
Eq.~\eqref{eq:parallel-probability-explicit} gives
\begin{align}
 p_{b^2,b}(r)
 &\ge
 2^{-(12b^2+2b)}e^{-36b^3}
 \ge e^{-48b^3},
 \label{eq:Q-probability-explicit}
\end{align}
where $(12b^2+2b)\log2\le12b^3$ for $b\ge1$.  Moreover,
\begin{align}
 \delta_{b^2,b}
 \le3be^{-3b^2}
 \le3e^{-3}<\frac14.
 \label{eq:Q-error-explicit}
\end{align}

\subsubsection{Schmidt-tail and approximate-rank lower bounds}

\begin{lemma}[Projection-to-tail transfer]
\label{lem:projection-tail-explicit}
Let $\ket\psi$ have Schmidt coefficients
$\lambda_1\ge\lambda_2\ge\cdots$.  Suppose a product projection
$P_{\rm L}\otimes P_{\rm R}$ has amplitude $a>0$ and the normalized projected
state is within distance $\delta$ of a state $\ket\phi$ with Schmidt
coefficients $\widetilde\lambda_1\ge\widetilde\lambda_2\ge\cdots$.  Then, for every
integer $D\ge0$,
\begin{align}
 \left(\sum_{j>D}\lambda_j^2\right)^{1/2}
 \ge
 a\left[
 \left(\sum_{j>D}\widetilde\lambda_j^2\right)^{1/2}
 -\delta
 \right].
 \label{eq:projection-tail-explicit}
\end{align}
\end{lemma}

\textit{Proof.}
The best approximation error over possibly unnormalized vectors of
Schmidt rank at most $D$ equals the Schmidt-tail norm.  A product
projection cannot increase Schmidt rank and is contractive.  Applying these
facts to the projected state, followed by the triangle inequality between the
normalized projected state and $\ket\phi$, gives
Eq.~\eqref{eq:projection-tail-explicit}.
\hfill$\square$

\begin{theorem}[Explicit $\log^{2/3}(1/\epsilon)$ lower bound]
\label{thm:rank-log-two-thirds-explicit}
For every integer $b\ge1$, there exists a finite one-dimensional driven system
with local dimension six, a product initial state, total time smaller than one,
and spatial decay profile $\bar J(x)=e^{-x}$ such that its final-state Schmidt
coefficients across the central cut obey
\begin{align}
 \sum_{j>2^{b^2-1}}\lambda_j^2
 \ge\frac15e^{-48b^3}.
 \label{eq:tail-main-explicit}
\end{align}
Let $D_{\epsilon_{\rm sq}}^{\rm MPS}$ denote the minimum open-boundary
MPS bond dimension among possibly unnormalized approximants with squared vector error
at most $\epsilon_{\rm sq}$.  Consequently,
\begin{align}
 \epsilon_{\rm sq}<\frac15e^{-48b^3}
 \quad\Longrightarrow\quad
 D_{\epsilon_{\rm sq}}^{\rm MPS}>2^{b^2-1}.
 \label{eq:rank-main-parametric-explicit}
\end{align}
The same bond-dimension lower bound holds for an open-boundary MPO approximation to the
propagator at operator-norm error
\begin{align}
 \epsilon_{\rm op}<5^{-1/2}e^{-24b^3}.
 \label{eq:MPO-main-parametric-explicit}
\end{align}
In particular, for each accuracy with $L:=\log(1/\epsilon_{\rm sq})$
and $L-\log5\ge6144$, choosing
$b=\lfloor((L-\log5)/96)^{1/3}\rfloor$ gives a system satisfying
\begin{align}
 D_{\epsilon_{\rm sq}}^{\rm MPS}
 >\exp\!\left[
 \frac{\log2}{2\,96^{2/3}}
 (L-\log5)^{2/3}
 \right].
 \label{eq:rank-main-asymptotic-explicit}
\end{align}
\end{theorem}

\textit{Proof.}
Use the choice in Eq.~\eqref{eq:optimized-integer-choice-explicit}.  The target
branch consists of $Q=b^2$ Bell pairs.  For
$D\le2^{Q-1}$, its Schmidt-tail norm is at least $1/\sqrt2$.  Combining
Eqs.~\eqref{eq:Q-probability-explicit}, \eqref{eq:Q-error-explicit}, and
Lemma~\ref{lem:projection-tail-explicit} gives
\begin{align}
 \sum_{j>D}\lambda_j^2
 &\ge
 p_{b^2,b}(r)
 \left(\frac1{\sqrt2}-\frac14\right)^2
 \notag\\
 &\ge\frac15e^{-48b^3},
\end{align}
because $(1/\sqrt2-1/4)^2>1/5$.  This proves
Eqs.~\eqref{eq:tail-main-explicit} and
\eqref{eq:rank-main-parametric-explicit}.  Applying an MPO to the product initial
state gives an MPS of no larger bond dimension, and an operator-norm error
$\epsilon_{\rm op}$ gives squared state error at most
$\epsilon_{\rm op}^2$, proving Eq.~\eqref{eq:MPO-main-parametric-explicit}.

For the final inequality, set
\begin{align}
 x:=\left(\frac{L-\log5}{96}\right)^{1/3},
 \qquad
 b:=\lfloor x\rfloor.
\end{align}
The assumption implies $x\ge4$.  Then $48b^3<L-\log5$ and
$b^2-1\ge x^2/2$.  Substitution into
Eq.~\eqref{eq:rank-main-parametric-explicit} gives
Eq.~\eqref{eq:rank-main-asymptotic-explicit}.
\hfill$\square$

\subsubsection{Uniform R\'enyi lower bound with an explicit constant}

Panel (c) of Fig.~\ref{fig:driven-lower-bound-roadmap} previews the multiscale
choice below.  Each dyadic scale is an independent tensor factor; for a given
$\alpha$, one chooses the scale $b_m\asymp\alpha^{-1}$ and discards the
others.

\begin{lemma}[Schmidt tail implies R\'enyi entanglement]
\label{lem:tail-to-renyi-explicit}
Let $\ket\psi$ be normalized with Schmidt coefficients
$\lambda_1\ge\lambda_2\ge\cdots$, let $0<\alpha<1$, and let $D\ge1$ be an
integer.  If
\begin{align}
 \sum_{j>D}\lambda_j^2\ge p_D>0,
\end{align}
then
\begin{align}
 E_\alpha(\psi)
 \ge\log D+\frac{\alpha}{1-\alpha}\log p_D
 =\frac{1}{1-\alpha}
 \log\!\left(p_D^\alpha D^{1-\alpha}\right).
 \label{eq:tail-to-renyi-explicit}
\end{align}
\end{lemma}

\textit{Proof.}
If $\lambda_D^2>p_D/D$, the first $D$ Schmidt coefficients already give the
claimed lower bound.  Otherwise, $\lambda_j^2\le p_D/D$ for $j>D$ and, because
$\alpha-1<0$,
\begin{align}
 \sum_{j>D}\lambda_j^{2\alpha}
 =\sum_{\substack{j>D\\\lambda_j>0}}
 \lambda_j^2(\lambda_j^2)^{\alpha-1}
 \ge p_D(p_D/D)^{\alpha-1}.
 \qquad\square
\end{align}

\begin{theorem}[Uniform explicit $\alpha^{-2}$ lower bound]
\label{thm:uniform-renyi-explicit}
For every $M\ge1$, there exists a finite one-dimensional driven system
with local dimension six, a product initial state, total time smaller than one,
and spatial decay profile $\bar J(x)=e^{-x}$ such that
\begin{align}
 E_\alpha(U_M\ket{\psi_{0,M}})
 \ge\frac{1}{8\cdot384^2\alpha^2}
 \label{eq:uniform-renyi-final-explicit}
\end{align}
for every
\begin{align}
 \frac{1}{192\,2^M}\le\alpha\le\frac1{384}.
 \label{eq:uniform-renyi-finite-range}
\end{align}
The constructions may be chosen compatibly in $M$.  Hence the corresponding
thermodynamic-limit state satisfies
Eq.~\eqref{eq:uniform-renyi-final-explicit} for every fixed
$0<\alpha\le1/384$.
\end{theorem}

\textit{Proof.}
For $m=1,2,\ldots,M$, set
\begin{align}
 b_m:=2^m,
 \qquad
 Q_m:=b_m^2.
\end{align}
Reserve disjoint left and right regions of size $Q_m$.  Place the scales in
increasing order away from the central cut.  The distance $r_m$ between the
$m$th pair is bounded by the total size of the preceding scales:
\begin{align}
 r_m
 \le1+2\sum_{j=1}^{m-1}Q_j
 =1+2\sum_{j=1}^{m-1}4^j
 \le4^m=Q_m.
 \label{eq:scale-distance-explicit}
\end{align}
Apply the construction of Theorem~\ref{thm:rank-log-two-thirds-explicit} to
each scale and run the disjoint drives in parallel.  The union of the finitely
many switching times is again a finite piecewise-constant schedule.  The total
time remains smaller than one, the final state factorizes over the scales, and
adding a new outer scale leaves all previously constructed tensor factors
unchanged.

Fix $\alpha$ in the interval
Eq.~\eqref{eq:uniform-renyi-finite-range}, and choose
\begin{align}
 m:=\left\lfloor
 \log_2\!\left(\frac{1}{192\alpha}\right)
 \right\rfloor,
 \qquad
 b:=b_m.
 \label{eq:dyadic-choice-explicit}
\end{align}
Then $1\le m\le M$ and
\begin{align}
 \frac{1}{384\alpha}\le b\le\frac{1}{192\alpha}.
 \label{eq:dyadic-window-explicit}
\end{align}
For the $m$th factor, use
$D=2^{b^2-1}$ and
$p_D=\frac15e^{-48b^3}$ in
Lemma~\ref{lem:tail-to-renyi-explicit}.  This gives
\begin{align}
 E_\alpha
 \ge
 (b^2-1)\log2
 -\frac{\alpha}{1-\alpha}(48b^3+\log5).
 \label{eq:renyi-before-constants-explicit}
\end{align}
Since $b\ge2$, $\alpha\le1/384$, and
$\alpha b\le1/192$,
\begin{align}
 (b^2-1)\log2&\ge\frac12b^2,
 \\
 \frac{48\alpha b}{1-\alpha}&\le\frac{96}{383},
 \\
 \frac{\alpha\log5}{1-\alpha}&\le\frac{\log5}{383}.
\end{align}
Consequently,
\begin{align}
 E_\alpha
 \ge
 \left(\frac12-\frac{96}{383}\right)b^2
 -\frac{\log5}{383}
 \ge\frac18b^2
 \ge\frac{1}{8\cdot384^2\alpha^2}.
\end{align}
R\'enyi entanglement is additive on tensor products, so the same bound holds
for the full state.

For the thermodynamic limit, let $W_m$ be the full propagator on the $m$th
scale.  Telescoping the products of unitaries in
Eq.~\eqref{eq:one-block-unitary-explicit} gives
\begin{align}
 \norm{W_m-\hat{\mathbf 1}}
 \le b_m2^{b_m}\Delta_{1,b_m}
 \le b_m2^{-b_m/2}e^{-9b_m^2}.
 \label{eq:dyadic-propagator-convergence}
\end{align}
The right-hand side is summable over $m$.  Extending each finite-size
propagator by the identity on the remaining sites therefore yields an
operator-norm limit and a normalized limiting state in the infinite tensor
product based on the initial product state.  Additivity and monotone
convergence of the nonnegative entropy contributions give the stated
lower bound in this limit.
\hfill$\square$

\subsection{Encoding an exponentially deep state into one flagged block}
\label{subsec:arbitrary-drive-error-space}

The multi-block construction above is optimized for Schmidt rank: it produces
$Q=b^2$ Bell pairs with a branch weight of order $e^{-O(b^3)}$.  The Bell-pair
branch itself, however, has only $O(Q)$ ordinary gate complexity with
unrestricted two-qubit connectivity.  A different
optimization reuses the one-block conjugation pattern of
Fig.~\ref{fig:one-block-protocol-explicit}, but assigns an independent phase
and duration to the ancilla pulse associated with each string $\sigma$.  This
programs the coefficient of every basis vector independently.  The resulting
branch has a smaller entanglement dimension than the parallel construction,
but a much stronger circuit-depth lower bound inside the flagged sector.

For a normalized coefficient vector
$a=(a_\sigma)_{\sigma\in\{0,1\}^b}$, define the encoded state
\begin{align}
 \ket{\Phi_a}
 :=\sum_{\sigma\in\{0,1\}^b}
 a_\sigma\ket{\sigma}_{\rm L}\ket{\sigma}_{\rm R},
 \qquad
 \sum_\sigma|a_\sigma|^2=1.
 \label{eq:programmable-encoded-state}
\end{align}
This is an arbitrary $b$-qubit state in the $2^b$-dimensional code subspace
spanned by
$\{\ket\sigma_{\rm L}\ket\sigma_{\rm R}\}_\sigma$.

\begin{prop}[Programmable one-block branch]
\label{prop:programmable-one-block}
Under the assumptions and parameter choice of
Theorem~\ref{thm:one-block-explicit}, the pulse associated with each
$\sigma$ can be assigned an independent phase and duration so that the
normalized projected branch is close to $\ket{\Phi_a}$ for any normalized
coefficient vector $a$.  More precisely, with
$\vartheta_b=\pi(b+1/2)$, write
$a_\sigma=|a_\sigma|e^{i\phi_\sigma}$, with $\phi_\sigma$ arbitrary when
$a_\sigma=0$, and replace the $\sigma$th ancilla pulse by
\begin{align}
 V_{\rm anc}^{(\phi_\sigma)}(r)
 :=e^{-r}\left(
 e^{i(\vartheta_b-\phi_\sigma)}\ket{0,0}\bra{1,1}
 +e^{-i(\vartheta_b-\phi_\sigma)}\ket{1,1}\bra{0,0}
 \right)
 \label{eq:phase-programmed-ancilla-pulse}
\end{align}
for the duration
\begin{align}
 \tau_\sigma:=2^{b/2}|a_\sigma|\Delta_{1,b}.
 \label{eq:amplitude-programmed-pulse-time}
\end{align}
Let $\varepsilon_b(a,r)$ denote the norm of the resulting projected branch.
Then
\begin{align}
 2^{-6b-1}e^{-r-15b^2}
 \le \varepsilon_b(a,r)
 \le 2^{1-b}e^{-r-15b^2},
 \label{eq:programmable-amplitude-two-sided}
\end{align}
and
\begin{align}
 \norm{
 \frac{P_{\rm L}\otimes P_{\rm R}U_{b,a}(r)\ket{\psi_0}}
 {\varepsilon_b(a,r)}
 -\ket{\Phi_a}}
 \le3e^{-r-3b^2}.
 \label{eq:programmable-conditional-error}
\end{align}
Zero-duration pulses may be omitted.  The number of temporal stages is
at most $(2b+1)2^b$, and the total evolution time is still smaller than one.
For general $a$, a nonzero $\tau_\sigma$ can be arbitrarily smaller than
$\Delta_{1,b}$; the minimum-stage estimate
Eq.~\eqref{eq:one-block-min-stage} is retained for equal-magnitude
coefficients $|a_\sigma|=2^{-b/2}$.
\end{prop}

\textit{Proof.}
The projected first-order contribution of the $\sigma$th conjugated pulse is
proportional to
$e^{-r}c_b^2\tau_\sigma e^{i\phi_\sigma}
\ket\sigma_{\rm L}\ket\sigma_{\rm R}$.
With Eq.~\eqref{eq:amplitude-programmed-pulse-time}, their sum is
\begin{align}
 e^{-r}c_b^2 2^{b/2}\Delta_{1,b}\ket{\Phi_a},
\end{align}
whose norm is exactly the same $A_{0,b}$ as in
Eq.~\eqref{eq:one-block-target-norm-explicit}.  Moreover,
\begin{align}
 \sum_\sigma\tau_\sigma
 =2^{b/2}\Delta_{1,b}\sum_\sigma|a_\sigma|
 \le2^b\Delta_{1,b}.
\end{align}
Hence the product-expansion remainder is no larger than in
Eq.~\eqref{eq:one-block-remainder-explicit}, and the same ratio
$R_b/A_{0,b}\le e^{-r}c_b$ applies.  The lower bound and the conditional-error
estimate therefore follow exactly as in Theorem~\ref{thm:one-block-explicit}.
For the upper bound,
$\varepsilon_b(a,r)\le A_{0,b}+R_b\le2A_{0,b}$ and
$c_b\le e^{-3b^2}$.
\hfill$\square$

For circuit-depth lower bounds it is already enough to program phases.  We
identify the qutrit code subspace $\operatorname{span}\{\ket0,\ket1\}$ with
a qubit; the two fixed ancillas add only two wires, and the unused fixed
qubits are omitted.  For a sign string
$s=(s_\sigma)_\sigma\in\{\pm1\}^{2^b}$, set
\begin{align}
 \ket{\Phi_s}:=2^{-b/2}\sum_\sigma
 s_\sigma\ket\sigma_{\rm L}\ket\sigma_{\rm R}.
 \label{eq:phase-sign-state}
\end{align}

\begin{lemma}[A phase-coded state of exponential postselected depth]
\label{lem:hard-phase-state}
Fix a finite one- and two-qubit gate set $\mathcal G$ with
$g:=|\mathcal G|$, and fix $\kappa>2$.  Consider one-dimensional
nearest-neighbor circuits of width at most $\kappa b$, starting from a fixed
computational-basis product state and allowing computational-basis
postselection at the end.  For all sufficiently large $b$, there exists a sign
string $s$ such that every such circuit whose normalized postselected output,
even after optimizing an overall phase, is within distance $1/4$ of
$\ket{\Phi_s}$ has depth at least
\begin{align}
 d_b
 \ge c_{\mathcal G,\kappa}\frac{2^b}{b},
 \qquad
 c_{\mathcal G,\kappa}
 :=\frac{\log2}{64\kappa\log(3(g+1))}.
 \label{eq:hard-phase-depth}
\end{align}
\end{lemma}

\textit{Proof.}
Put $N=2^b$.  Apply a greedy packing to sign strings modulo the equivalence
$s\sim-s$.  After selecting $s$, exclude the Hamming balls of radius less than
$N/4$ around both $s$ and $-s$.  A Hamming ball of radius $N/4$ has cardinality
at most $2^{Nh_2(1/4)}$, where
$h_2(q):=-q\log_2q-(1-q)\log_2(1-q)$.  The greedy packing therefore contains at
least $2^{N(1-h_2(1/4))-1}$ rays.  Since $1-h_2(1/4)>1/8$, this is at least
$2^{N/8}$ for all sufficiently large $N$.  Every two representatives satisfy
\begin{align*}
 \frac N4\le d_H(s,s')\le\frac{3N}{4}.
\end{align*}
Consequently,
\begin{align}
 \inf_{|\omega|=1}
 \norm{\ket{\Phi_s}-\omega\ket{\Phi_{s'}}}^2
 &=2-2\left|1-\frac{2d_H(s,s')}{N}\right|
 \ge1.
 \label{eq:phase-state-separation}
\end{align}
Thus one circuit output, even up to an overall phase, cannot lie within
distance $1/4$ of two different states in this family.

Pad circuits of depth at most $d$ to depth $d$ with identity layers.
A depth-$d$ circuit of width $w$ has at most
$(g+1)^{2wd}$ choices after overcounting the possible one- and two-site gate
locations.  Marking each wire as an output, a postselection onto zero, or a
postselection onto one contributes at most $3^w$.  Summing over
$w\le\kappa b$ and over possible output-wire orderings contributes only
an additional factor
$\exp[O_\kappa(b\log b)]$.  Hence the logarithm of the number of normalized
postselected output rays is at most
\begin{align}
 2\kappa bd\log(g+1)+\kappa b\log3+O_\kappa(b\log b).
\end{align}
Take $d=\lceil c_{\mathcal G,\kappa}2^b/b\rceil-1$, so that every
integer depth below the claimed lower bound is included.  The first term
is then smaller than
$N\log2/32$, while the remaining terms are smaller than $N\log2/32$ for all
sufficiently large $b$.  The total count is therefore strictly smaller than
$2^{N/8}\le|\mathcal S|$, where $\mathcal S$ is the packed family.
At least one state in this family therefore has no such approximation.
\hfill$\square$

\begin{theorem}[Exponential flagged error-space depth for arbitrary driving]
\label{thm:arbitrary-drive-error-depth}
Fix $\mathcal G$ and $\kappa$ as in
Lemma~\ref{lem:hard-phase-state}.  For every sufficiently large $b$, there
exists a one-dimensional piecewise-constant drive satisfying the assumptions
of Theorem~\ref{thm:one-block-explicit} and a product flag projector such that
the final state has a flagged decomposition
\begin{align}
 \ket{\Psi_b}
 =\sqrt{1-\varepsilon_b^2}\,\ket{\psi_{{\rm fail},b}}\ket0_F
 +\varepsilon_b\,\ket{\psi_{\varepsilon,b}}\ket1_F,
 \label{eq:arbitrary-drive-flagged-state}
\end{align}
with
\begin{align}
 e^{-21b^2}\le\varepsilon_b\le e^{-15b^2}.
 \label{eq:arbitrary-drive-amplitude-window}
\end{align}
Here $\mathsf{Depth}^{\rm post}$ refers to the fixed-input circuit model
of Lemma~\ref{lem:hard-phase-state}, with width at most $\kappa b$.  Moreover,
\begin{align}
 \mathsf{Depth}^{\rm post}_{1/8}(\Psi_b;P_F)
 \ge c_{\mathcal G,\kappa}\frac{2^b}{b}.
 \label{eq:arbitrary-drive-depth-b}
\end{align}
Consequently, with $L_b:=\log(1/\varepsilon_b)$,
\begin{align}
 \mathsf{Depth}^{\rm post}_{1/8}(\Psi_b;P_F)
 \ge
 c_{\mathcal G,\kappa}
 \frac{\sqrt{15}}{\sqrt{L_b}}
 \exp\left[
 \frac{\log2}{\sqrt{21}}\sqrt{L_b}
 \right].
 \label{eq:arbitrary-drive-depth-epsilon}
\end{align}
In particular, the conditional error-space depth can scale as
\begin{align}
 \exp\!\left[\Omega\!\left(
 \sqrt{\log(1/\varepsilon_b)}
 \right)\right].
 \label{eq:arbitrary-drive-depth-scaling}
\end{align}
\end{theorem}

\textit{Proof.}
Choose the hard sign string from
Lemma~\ref{lem:hard-phase-state} and apply
Proposition~\ref{prop:programmable-one-block} with $r\le b^2$ and
$a_\sigma=2^{-b/2}s_\sigma$.  All pulse durations are then
$\tau_\sigma=\Delta_{1,b}$, so the minimum-stage estimate of the
one-block primitive is preserved.  The lower
bound in Eq.~\eqref{eq:programmable-amplitude-two-sided} gives
$\varepsilon_b\ge e^{-21b^2}$, while its upper bound gives
$\varepsilon_b\le e^{-15b^2}$.  For $b$ large enough,
Eq.~\eqref{eq:programmable-conditional-error} is smaller than $1/8$.  If the
actual conditional branch admitted a postselected circuit of depth smaller
than Eq.~\eqref{eq:hard-phase-depth} at error $1/8$, the triangle inequality
would approximate the hard phase state within $1/4$, contradicting the lemma.
Finally,
\begin{align}
 b\ge\sqrt{L_b/21},
 \qquad
 \frac1b\ge\frac{\sqrt{15}}{\sqrt{L_b}},
\end{align}
which yields Eq.~\eqref{eq:arbitrary-drive-depth-epsilon}.
\hfill$\square$

\paragraph{Why the single-block construction is complementary to the Bell-pair
construction.}
The multi-block protocol gives the stronger tensor-network lower bound
$\log D_\epsilon=\Omega(\log^{2/3}(1/\epsilon))$, but its hard branch is only a
product of Bell pairs and has polynomial gate complexity in the number of
pairs.  The programmable single block has a code subspace of dimension $2^b$
with $b=\Theta(\sqrt{\log(1/\varepsilon)})$; its main conclusion is not a
stronger rank bound, but a conditional circuit-depth lower bound exponential in
$\sqrt{\log(1/\varepsilon)}$.  The two constructions optimize different
resources and should be stated separately.

\subsection{Resource comparison and the scope of the results}

Table~\ref{tab:lower-bound-resource-comparison} keeps the different resource
models visible.  In particular, the square-root logarithm in the spacetime
embedding and the $\log^{2/3}$ exponent in the arbitrary-drive rank lower bound
refer to different locality and bandwidth assumptions and should not be read
as competing statements under one common model.

\begin{table*}[!t]
\caption{Comparison of the lower-bound mechanisms in
Secs.~\ref{sec:lower-static-discrete-explicit} and
\ref{sec:lower-arbitrary-drive-explicit}.  Here $\varepsilon$ is a flagged
branch amplitude, whereas $\epsilon_{\rm sq}$ is a squared state error.  The
table separates the certified consequence from the extra resources used to
obtain it.}
\label{tab:lower-bound-resource-comparison}
\centering
\footnotesize
\setlength{\tabcolsep}{4pt}
\renewcommand{\arraystretch}{1.22}
\begin{tabular}{@{}p{0.16\textwidth}p{0.23\textwidth}p{0.27\textwidth}p{0.28\textwidth}@{}}
\toprule
Construction & Dynamical resources & Hard sector or accuracy scale & Certified consequence \\
\midrule
Static weak pairs,
Lemma~\ref{lem:static-weak-pairs-explicit} &
Time independent; qubits; $\bar J(r)=e^{-r}$; $t=1$ &
A family of $2^m$ selected Schmidt weights has total discarded mass
$e^{-O(m^2)}$ at rank $2^{m-1}$ &
$E_\alpha=\Omega(\alpha^{-1})$ and
$\log D_{\epsilon_{\rm sq}}=\Omega(\sqrt{\log(1/\epsilon_{\rm sq})})$ \\
\addlinespace
Spacetime circuit embedding,
Proposition~\ref{prop:spacetime-postselection-embedding} &
Discrete drive; $d_{\rm loc}=2^{O(Q)}$; constant macro-depth with local norm
$O(Q)$, or $O(Q)$ depth at bounded norm &
An arbitrary depth-$Q$ circuit occurs with amplitude
$\varepsilon=2^{-nQ}$; choosing $n=Q$ gives $\varepsilon=2^{-Q^2}$ &
Embedding capacity
$Q=\Theta(\sqrt{\log(1/\varepsilon)})$; the stated staticization uses
growing local ancillas and clock-term norms \\
\addlinespace
One distant block pair,
Theorem~\ref{thm:one-block-explicit} &
Local dimension six; $t<1$; $(2b+1)2^b$ stages;
$\Delta_{\min}=e^{-\Theta(b^2)}$ &
An almost maximally entangled $b$-qubit branch with
$p\ge e^{-O(r+b^2)}$ &
Primitive used by both the parallel rank construction and the programmable
hard-state construction \\
\addlinespace
Parallel block pairs,
Theorems~\ref{thm:rank-log-two-thirds-explicit} and
\ref{thm:uniform-renyi-explicit} &
The one-block drives act on disjoint regions and run synchronously;
$N_B=b$, $Q=b^2$ &
A branch of $Q$ Bell pairs with $p\ge e^{-O(b^3)}$ &
$\log D_{\epsilon_{\rm sq}}=\Omega(\log^{2/3}(1/\epsilon_{\rm sq}))$
and, after the dyadic assembly, $E_\alpha=\Omega(\alpha^{-2})$ \\
\addlinespace
Programmable one block,
Theorem~\ref{thm:arbitrary-drive-error-depth} &
Same local dimension, total time, and minimum stage duration as the
one-block primitive; string-dependent signs with equal pulse durations &
A phase-coded state of postselected depth $\Omega(2^b/b)$ occurs with
$\varepsilon=e^{-\Theta(b^2)}$ &
Conditional postselected depth
$\exp[\Omega(\sqrt{\log(1/\varepsilon)})]$ \\
\bottomrule
\end{tabular}
\end{table*}
\FloatBarrier

For the choice $Q=b^2$, the number of temporal stages is explicitly
\begin{align}
 G_Q=(2\sqrt Q+1)2^{\sqrt Q}.
 \label{eq:stage-count-Q-explicit}
\end{align}
At the state-error scale
\begin{align}
 \epsilon_{\rm sq,b}:=\frac15e^{-48b^3},
\end{align}
one has, for $b\ge1$,
\begin{align}
 G_b=(2b+1)2^b\le e^{2b}
 =\exp\!\left[
 2\left(\frac{\log(1/(5\epsilon_{\rm sq,b}))}{48}\right)^{1/3}
 \right].
 \label{eq:stage-count-error-explicit}
\end{align}
This is an achieved implementation cost of the multi-block entanglement
construction, not a lower bound on all possible implementations.  The
programmable single-block construction uses at most $(2b+1)2^b$ stages and
produces a branch whose postselected circuit depth is itself
$\Omega(2^b/b)$.  Thus its explicit stage-list size and the lower bound on intrinsic
branch depth are both exponential in $b$, whereas the Bell-pair branch
has only polynomial gate complexity in $Q=b^2$.

The temporal bandwidth, measured here by the inverse minimum stage
duration, is parametrically larger.  From
Eq.~\eqref{eq:one-block-min-stage},
\begin{align}
 \Delta_{1,b}^{-1}
 =\exp\!\left[9b^2+O(b)\right]
 =\exp\!\left[
 \Theta\!\left(\log^{2/3}(1/\epsilon_{\rm sq,b})\right)
 \right].
\end{align}
Thus the stage count and the inverse shortest switching time are distinct
resources.

Finally, the proof generates entanglement from a fixed product input.  It does
not prove uniform transfer of an arbitrary unknown state.  A genuine
information-capacity statement would require, for example, an operator-norm
approximation of a block SWAP, or an entanglement-fidelity statement including
an external reference.  Synthesizing SWAPs or generic random operators from
phase-weighted conjugates is a plausible extension, but it additionally
requires a spanning theorem, control of the coefficient norm, suppression of
higher-order terms, and input-uniform action.  These points should remain
explicit open problems rather than consequences of the present protocol.

\section{Algorithmic MPO construction for static unitary dynamics}
\label{subsec:static-unitary-algorithmic-MPO}

We now turn the differential low-rank mechanism into an explicit MPO
construction for a one-dimensional time-independent Hamiltonian.  There are
two analytic approximation steps: first, an HHKL localization of the full
evolution, and second, a coordinatewise Taylor truncation of each localized
block evolution.  The subsequent coefficientwise MPO compression is exact in
the arithmetic model used below.  In particular, no nonzero Schmidt
coefficient is discarded after the retained Taylor coefficients have been
fixed.
Earlier locality-based and polynomial tensor-network approximation
methods are given in
Refs.~\cite{PhysRevLett.97.157202,PhysRevX.11.011047,PRXQuantum.2.040331,AchuthaEtAl2025}.
The construction below implements the differential cut-count formalism
developed in this work.

Throughout this section, we specialize to a one-dimensional chain of
$n$ sites with on-site dimension $d_0$, take $t\ge0$ and
$0<\epsilon\le1$, and assume finite-range interactions.
All MPOs below have open boundaries.
After a preliminary blocking by a constant number of sites, which changes
only constants depending on the interaction range, we may assume that each
local interaction is supported on one site or on two neighboring sites.
All constants below may depend on this preliminary blocking, on $k$, on
$d_0$, and on the extensivity constant $g$, but not on $n$, $t$, or the
accuracy.

\subsubsection{Overview of the construction}

Let $L$ denote the logarithmic accuracy scale, eventually
$L\asymp\log(n/\epsilon)$.  The construction has five steps.  First, an
HHKL decomposition replaces the full evolution by a product of localized
unitaries supported on intervals of length
\begin{align}
    W=O(t+L),
\end{align}
with only $O(1)$ factors straddling any fixed microscopic cut.  Second,
every interval of length at least $\sqrt L$ is partitioned into coarse blocks
of width $b\asymp\sqrt L$, so that
\begin{align}
    \ell_X=O\left(\frac{t+L}{\sqrt L}\right).
\end{align}
Shorter intervals are handled directly by dense tensorization, whose cost is
already $\exp[O(\sqrt L)]$.  Third, after writing
$H_X=\sum_{j=1}^{\ell_X}G_j$, we retain the
multivariate Taylor coefficients with the coordinatewise occupancy bound
$m_j<M$, where
\begin{align}
    M=O(t\sqrt L+L).
\end{align}
This is a rectangular truncation, not a truncation of the total Taylor
degree.

Fourth, for a fixed retained multi-index, only $O(1)$ coarse Hamiltonians
can contribute interactions near a prescribed cut.  Hence its local cut
budget is $O(M)$.  Applying the cut-count decomposition in a window of $b$
cuts leaves only
\begin{align}
    O\left(\frac{M}{b}\right)
    =O(t+\sqrt L)
\end{align}
explicit crossing insertions.  The resulting MPO representation of that
coefficient is exact.  Finally, all $M^{\ell_X}$ retained coefficients are
generated by a dynamic program on the multi-index lattice and summed as
MPOs.  The number of coefficients and the polynomial word-counting factors
produce only logarithmic overheads, yielding
\begin{align}
    \log D_{\mathrm{loc}},\ \log T_{\mathrm{loc}}
    =\widetilde O(t+\sqrt L).
\end{align}
The bounded cut multiplicity of the HHKL product then gives the same
exponent for the global MPO bond dimension, with an additional linear
factor in $n$ in the construction time.

\subsubsection{Reduction to localized block evolutions}

We use the HHKL gluing estimate \cite{HHKL2021} as the geometric input.
Its Lieb--Robinson proof bounds the error of each gluing step by
$Ct e^{vt-\mu w_{\mathrm H}}$, with constants determined by the local
interaction bounds.  With $q$ intervals, choose an integer block length
large enough that $Cqt e^{vt-\mu w_{\mathrm H}}\le\epsilon/2$.
Such a choice is possible with
\begin{align}
    w_{\mathrm H}
    =O\left(1+t+\log\frac{q}{\epsilon}\right),
\label{eq:algorithmic-MPO-HHKL-width}
\end{align}
where the chain is partitioned into consecutive intervals $X_1,\ldots,X_q$.
The interior intervals have length $w_{\mathrm H}$; the endpoint intervals
may be shorter.  Iterating the gluing estimate gives
\begin{align}
    V_{\mathrm{HHKL}}(t)
    :=
    e^{-itH_{X_1\cup X_2}}
    e^{itH_{X_2}}
    e^{-itH_{X_2\cup X_3}}
    \cdots
    e^{itH_{X_{q-1}}}
    e^{-itH_{X_{q-1}\cup X_q}},
\label{eq:algorithmic-MPO-HHKL-product}
\end{align}
with the convention that, when $q=1$,
$V_{\mathrm{HHKL}}(t):=e^{-itH_{X_1}}$ and the localization error is zero;
when $q=2$, set
$V_{\mathrm{HHKL}}(t):=e^{-itH_{X_1\cup X_2}}$.
For $q\ge3$, Eq.~\eqref{eq:algorithmic-MPO-HHKL-product} has its literal
form.  Thus, for every $q\ge2$, the product contains $2q-3$ localized
factors.  In all cases,
\begin{align}
    \norm{e^{-itH}-V_{\mathrm{HHKL}}(t)}
    \le \frac{\epsilon}{2}.
\label{eq:algorithmic-MPO-HHKL-error}
\end{align}
We write the factors in Eq.~\eqref{eq:algorithmic-MPO-HHKL-product}
as
\begin{align}
    V_a=e^{-i\sigma_a tH_{\Omega_a}},
    \qquad
    \sigma_a\in\{+1,-1\},
\label{eq:algorithmic-MPO-HHKL-factor}
\end{align}
where every interval $\Omega_a$ has length
\begin{align}
    |\Omega_a|=O\left(1+t+\log\frac{q}{\epsilon}\right).
\label{eq:algorithmic-MPO-HHKL-support}
\end{align}
The essential geometric property is bounded cut multiplicity.  Namely, every
microscopic cut is straddled by at most
\begin{align}
    \nu_{\mathrm H}=O(1)
\label{eq:algorithmic-MPO-HHKL-overlap}
\end{align}
of the intervals $\Omega_a$.  Factors supported entirely on one side of a
cut have Schmidt rank one across that cut.  Consequently, if every $V_a$ is
approximated by an MPO of bond dimension at most $D_{\mathrm{loc}}$, then
the product of all approximating MPOs has bond dimension at most
\begin{align}
    D_{\mathrm{loc}}^{\nu_{\mathrm H}}
\label{eq:algorithmic-MPO-HHKL-bond-composition}
\end{align}
across every cut.  It is therefore enough to construct one localized block
evolution
\begin{align}
    U_X^{(\sigma)}(t):=e^{-i\sigma tH_X},
    \qquad \sigma\in\{+1,-1\},
\label{eq:algorithmic-MPO-local-target}
\end{align}
where
\begin{align}
    W:=|X|=O\left(1+t+\log\frac{q}{\epsilon}\right).
\label{eq:algorithmic-MPO-local-width}
\end{align}

\subsubsection{Coordinatewise Taylor truncation of one localized block}

Fix an integer coarse-block width $b\ge1$ and decompose
\begin{align}
    X=Y_1\cup Y_2\cup\cdots\cup Y_{\ell_X},
    \qquad
    \ell_X:=\left\lceil\frac{W}{b}\right\rceil,
    \qquad
    |Y_j|=b\ (1\le j<\ell_X),
    \qquad
    1\le |Y_{\ell_X}|\le b.
\label{eq:algorithmic-MPO-coarse-partition}
\end{align}
Assign every local interaction term in $H_X$ to the leftmost coarse block
that it intersects.  This gives an exact decomposition
\begin{align}
    H_X=\sum_{j=1}^{\ell_X}G_j,
\label{eq:algorithmic-MPO-G-decomposition}
\end{align}
with
\begin{align}
    \operatorname{supp}(G_j)
    \subseteq Y_j\cup Y_{j+1},
    \qquad
    \norm{G_j}\le gb,
\label{eq:algorithmic-MPO-G-locality}
\end{align}
where $Y_{\ell_X+1}:=\emptyset$.  Introduce independent parameters for the
coarse blocks,
\begin{align}
    H_X(\boldsymbol\lambda)
    :=\sum_{j=1}^{\ell_X}\lambda_jG_j,
    \qquad
    \boldsymbol\lambda=(\lambda_1,\ldots,\lambda_{\ell_X}),
\label{eq:algorithmic-MPO-coarse-parameterization}
\end{align}
and expand
\begin{align}
    e^{-i\sigma tH_X(\boldsymbol\lambda)}
    =
    \sum_{\boldsymbol m\in\mathbb Z_{\ge0}^{\ell_X}}
    \boldsymbol\lambda^{\boldsymbol m}
    U_{\boldsymbol m}^{(\sigma)}(t).
\label{eq:algorithmic-MPO-multivariate-expansion}
\end{align}
For
$\boldsymbol m=(m_1,\ldots,m_{\ell_X})$, let
\begin{align}
    |\boldsymbol m|_1:=\sum_{j=1}^{\ell_X}m_j,
\end{align}
and define
\begin{align}
    P_{\boldsymbol m}
    &:=[\boldsymbol\lambda^{\boldsymbol m}]
    \left(\sum_{j=1}^{\ell_X}\lambda_jG_j\right)^{|\boldsymbol m|_1},
\label{eq:algorithmic-MPO-Pm-definition}
    \\
    U_{\boldsymbol m}^{(\sigma)}(t)
    &=
    \frac{(-i\sigma t)^{|\boldsymbol m|_1}}
    {|\boldsymbol m|_1!}
    P_{\boldsymbol m}.
\label{eq:algorithmic-MPO-Um-Pm}
\end{align}
Here $[\boldsymbol\lambda^{\boldsymbol m}]$ denotes coefficient extraction.
The approximation that we use is the rectangular, or coordinatewise,
truncation
\begin{align}
    \widetilde U_{X,M}^{(\sigma)}(t)
    :=
    \sum_{\substack{\boldsymbol m\in\mathbb Z_{\ge0}^{\ell_X}\\
                     0\le m_j<M\ \forall j}}
    U_{\boldsymbol m}^{(\sigma)}(t).
\label{eq:algorithmic-MPO-rectangular-truncation}
\end{align}
The integer $M$ is an occupancy cutoff for each coarse Hamiltonian $G_j$.
It is not a cutoff on the total Taylor degree: the retained polynomial
contains degrees up to $\ell_X(M-1)$.

For $x\ge0$, define the exponential tail
\begin{align}
    \tau_M(x):=\sum_{r=M}^{\infty}\frac{x^r}{r!}.
\label{eq:algorithmic-MPO-tauM}
\end{align}

\begin{prop}[Coordinatewise Taylor tail]
\label{prop:algorithmic-MPO-coordinatewise-tail}
For every $M\in\mathbb N$,
\begin{align}
    \norm{
    e^{-i\sigma tH_X}
    -\widetilde U_{X,M}^{(\sigma)}(t)
    }
    \le
    \prod_{j=1}^{\ell_X}
    \left[1+\tau_M\!\left(t\norm{G_j}\right)\right]-1.
\label{eq:algorithmic-MPO-coordinatewise-tail-bound}
\end{align}
In particular, for $0<\delta\le1$, it is sufficient to choose
\begin{align}
    M\ge
    \max\left\{
    \left\lceil 2e\,tgb\right\rceil,
    \left\lceil\log_2\frac{4\ell_X}{\delta}\right\rceil
    \right\}
\label{eq:algorithmic-MPO-M-choice}
\end{align}
in order to ensure
\begin{align}
    \norm{
    e^{-i\sigma tH_X}
    -\widetilde U_{X,M}^{(\sigma)}(t)
    }
    \le\delta.
\label{eq:algorithmic-MPO-local-error-delta}
\end{align}
\end{prop}

\textit{Proof.}
For a nonempty subset $S\subseteq[\ell_X]$, let $T_S$ be the sum of all
Taylor coefficients in Eq.~\eqref{eq:algorithmic-MPO-multivariate-expansion}
for which $m_j\ge M$ for every $j\in S$, with no restriction on the other
coordinates.  The unrestricted variables can be resummed into the Hermitian
background
\begin{align}
    H_{\overline S}:=\sum_{j\notin S}G_j.
\end{align}
For integers $r_j\ge M$, the mixed Duhamel formula gives
\begin{align}
&
    \norm{
    \left.
    \frac{1}{\prod_{j\in S}r_j!}
    \prod_{j\in S}\partial_{\lambda_j}^{r_j}
    e^{-i\sigma t(H_{\overline S}+\sum_{j\in S}\lambda_jG_j)}
    \right|_{\boldsymbol\lambda=\boldsymbol0}
    }
\nonumber\\
&\hspace{40mm}\le
    \prod_{j\in S}
    \frac{(t\norm{G_j})^{r_j}}{r_j!}.
\label{eq:algorithmic-MPO-mixed-Duhamel-bound}
\end{align}
Hence
\begin{align}
    \norm{T_S}
    \le
    \prod_{j\in S}\tau_M\!\left(t\norm{G_j}\right).
\end{align}
Inclusion--exclusion for the union of the events $\{m_j\ge M\}$ yields
\begin{align}
&
    \norm{
    e^{-i\sigma tH_X}
    -\widetilde U_{X,M}^{(\sigma)}(t)
    }
\nonumber\\
&\quad\le
    \sum_{\emptyset\ne S\subseteq[\ell_X]}
    \prod_{j\in S}\tau_M\!\left(t\norm{G_j}\right),
\end{align}
which is Eq.~\eqref{eq:algorithmic-MPO-coordinatewise-tail-bound}.
If $M\ge2e\,tgb$, then
\begin{align}
    \tau_M(t\norm{G_j})
    \le2\left(\frac{et\norm{G_j}}{M}\right)^M
    \le2^{1-M}.
\end{align}
The second condition in Eq.~\eqref{eq:algorithmic-MPO-M-choice} therefore
gives
\begin{align}
    \sum_{j=1}^{\ell_X}\tau_M(t\norm{G_j})
    \le\frac{\delta}{2}.
\end{align}
Since $\prod_j(1+x_j)\le e^{\sum_jx_j}$ and
$e^{\delta/2}-1\le\delta$ for $0<\delta\le1$, the claim follows.
\hfill\(\square\)

\subsubsection{Exact MPO representation of the retained coefficients}

The coordinatewise constraint $m_j<M$ gives a local, rather than global,
cut budget.  This is the point that converts the low-rank existence argument
into a simultaneous fixed-cut estimate.

\begin{prop}[Exact MPO rank of one retained coefficient]
\label{prop:algorithmic-MPO-retained-coefficient}
There exist constants
$c_{\mathrm{occ}},c_{\mathrm{sh}},R_{\times},C>0$, depending only on the
local interaction structure, such that the following holds.  Let
$\boldsymbol m$ satisfy $0\le m_j<M$ for every $j$, set
\begin{align}
    r:=|\boldsymbol m|_1,
    \qquad
    Q_{\mathrm{loc}}:=\left\lceil c_{\mathrm{occ}}M\right\rceil,
    \qquad
    s_{\mathrm{loc}}
    :=\left\lfloor\frac{2Q_{\mathrm{loc}}}{b}\right\rfloor,
\label{eq:algorithmic-MPO-local-budget}
\end{align}
and let $P_{\boldsymbol m}$ be defined by
Eq.~\eqref{eq:algorithmic-MPO-Pm-definition}.  Then
$P_{\boldsymbol m}$ has an exact MPO representation with bond dimension
at most
\begin{align}
    D_{\boldsymbol m}
    &\le
    d_0^{2b}
    b(s_{\mathrm{loc}}+1)
    (2Q_{\mathrm{loc}}+1)^{\lceil\log_2 b\rceil}
    (M+1)^{c_{\mathrm{sh}}}
\nonumber\\
    &\quad\times
    \max_{0\le s\le \min\{r,s_{\mathrm{loc}}\}}
    \left\{
    \binom{r+s+1}{2s+1}R_{\times}^{s}
    \right\}.
\label{eq:algorithmic-MPO-coefficient-rank-explicit}
\end{align}
Consequently,
\begin{align}
    \log D_{\boldsymbol m}
    \le
    C\left[
    b+
    \log(r+2)+
    \frac{M}{b}\log(r+M+2)
    +\log(b+1)\log(M+2)
    \right].
\label{eq:algorithmic-MPO-coefficient-rank-log}
\end{align}
\end{prop}

\textit{Proof.}
Fix a microscopic target cut $q$ in $X$.  Place around $q$ a window
containing $b$ consecutive microscopic cuts; if $q$ is within distance $b$
of an endpoint, append empty layers so that the window still contains $b$
cuts.  Refine every coarse Hamiltonian into its microscopic local terms,
\begin{align}
    G_j=\sum_{\alpha\in\mathcal I_j}g_{j,\alpha}.
\end{align}
Split $\mathcal I_j=\mathcal I_j^{0}\sqcup\mathcal I_j^{\times}$,
where $\mathcal I_j^{\times}$ contains the terms crossing at least one cut of
the window.  Attach microscopic parameters
$\boldsymbol\mu=(\mu_{j,\alpha})_{\alpha\in\mathcal I_j^{\times}}$ only to
these terms, while retaining the coarse parameters $\boldsymbol\lambda$:
\begin{align*}
    H_X(\boldsymbol\lambda,\boldsymbol\mu)
    :=\sum_{j=1}^{\ell_X}\lambda_j
      \left(
      \sum_{\alpha\in\mathcal I_j^{0}}g_{j,\alpha}
      +\sum_{\alpha\in\mathcal I_j^{\times}}
       \mu_{j,\alpha}g_{j,\alpha}
      \right).
\end{align*}
Then
$P_{\boldsymbol m}=[\boldsymbol\lambda^{\boldsymbol m}]
H_X(\boldsymbol\lambda,\boldsymbol 1)^r$.
The phase filter below acts on $\boldsymbol\mu$; the variables
$\boldsymbol\lambda$ and the terms in $\mathcal I_j^0$ are carried by the
external family and the noncrossing background.  Only a constant number of
the coarse labels $j$ have a local term capable of crossing one of the cuts
in the window.  Moreover,
each microscopic interaction crosses only a constant number of consecutive
cuts.  Therefore, for every local-term word contributing to the fixed
coarse coefficient $P_{\boldsymbol m}$, the cut weight inside the window is
bounded by
\begin{align}
    w_{\mathrm{win}}
    \le c_{\mathrm{rng}}\sum_{j\in J_{\mathrm{win}}}m_j
    \le c_{\mathrm{occ}}M,
\end{align}
where $c_{\mathrm{rng}}=O(1)$ bounds the number of window cuts crossed by
one microscopic interaction, and $c_{\mathrm{occ}}$ also absorbs the
constant cardinality of $J_{\mathrm{win}}$.  Thus every
microscopic monomial occurring in the coefficient
$[\boldsymbol\lambda^{\boldsymbol m}]H_X(\boldsymbol\lambda)^r$ has local
cut weight at most
\begin{align}
    Q_{\mathrm{loc}}=\left\lceil c_{\mathrm{occ}}M\right\rceil.
\end{align}

Apply Corollary~\ref{cor:cut-count-x-dependent} coefficientwise to the
$b$ cuts in this window, with $\Phi[A]=A^r$, with
$\boldsymbol\lambda$ treated as the external label, and with the microscopic
parameters $\boldsymbol\mu$ as the differential variables.  The
decomposition is a linear recombination of Taylor coefficients.  We may therefore apply the coarse projection
$[\boldsymbol\lambda^{\boldsymbol m}]$ after the phase filtering.  Every
microscopic monomial contributing to this projection contains exactly $m_j$
letters with coarse label $j$, so the estimate above applies term by term.
Consequently, the projected coefficient contains no microscopic monomial of
local weight larger than $Q_{\mathrm{loc}}$, and the projected cut-tail
remainder is exactly zero.  Hence $P_{\boldsymbol m}$ is an exact sum
over a selected cut $p$ in the window, a phase vector, and an explicit
crossing order
\begin{align}
    0\le s\le s_{\mathrm{loc}}
    =\left\lfloor\frac{2Q_{\mathrm{loc}}}{b}\right\rfloor.
\end{align}
The number of cut--phase pairs is at most
\begin{align}
    b(2Q_{\mathrm{loc}}+1)^{\lceil\log_2b\rceil}.
\label{eq:algorithmic-MPO-local-phase-count}
\end{align}

For a fixed selected cut and phase vector, the noncrossing and crossing
parts of the parameterized Hamiltonian have the form
\begin{align}
    A(\boldsymbol\lambda)
    &=A_L(\boldsymbol\lambda)\otimes\hat 1
      +\hat 1\otimes A_R(\boldsymbol\lambda),
\label{eq:algorithmic-MPO-A-left-right}
    \\
    B(\boldsymbol\lambda)
    &=\sum_{\alpha=1}^{R_{\times}}
      B^L_{\alpha}(\boldsymbol\lambda)
      \otimes B^R_{\alpha}(\boldsymbol\lambda).
\label{eq:algorithmic-MPO-B-crossing-rank}
\end{align}
Here $R_{\times}=O(1)$ because the microscopic interaction range and local
Hilbert-space dimension are fixed.  The normalized order-$s$ derivative is
\begin{align}
&
    \frac{1}{s!}
    \left.
    \left(\frac{d}{dx}\right)^s
    \left[A(\boldsymbol\lambda)+xB(\boldsymbol\lambda)\right]^r
    \right|_{x=0}
\nonumber\\
&\quad=
    \sum_{r_0+\cdots+r_s=r-s}
    A(\boldsymbol\lambda)^{r_0}
    B(\boldsymbol\lambda)
    A(\boldsymbol\lambda)^{r_1}
    \cdots
    B(\boldsymbol\lambda)
    A(\boldsymbol\lambda)^{r_s}.
\label{eq:algorithmic-MPO-coefficient-word-expansion}
\end{align}
For each $j$, expand
$A^{r_j}$ according to the number of left and right factors.  Summing over
$r_0,\ldots,r_s$ and over these left--right allocations produces
\begin{align}
    \sum_{r_0+\cdots+r_s=r-s}
    \prod_{j=0}^{s}(r_j+1)
    =\binom{r+s+1}{2s+1}
\label{eq:algorithmic-MPO-word-count}
\end{align}
possible sectors.  All coarse variables whose supports lie strictly on one
side of the selected cut are assigned to that side uniquely.  Only a
constant number $c_{\mathrm{sh}}$ of coarse variables have local terms on
both sides of the cut.  For a fixed left--right sector, collect all left
operators into one polynomial and all right operators into another.  The
coefficient of a shared variable of degree $m_j$ is then a sum over only its
total left degree $a_j=0,\ldots,m_j$; no additional allocation among the
$s+1$ word segments is required.  Hence coefficient extraction requires at
most
\begin{align}
    \prod_{j\in J_{\mathrm{sh}}}(m_j+1)
    \le(M+1)^{c_{\mathrm{sh}}}
\end{align}
left--right allocations.  Equations
\eqref{eq:algorithmic-MPO-B-crossing-rank}--
\eqref{eq:algorithmic-MPO-word-count} therefore imply that one branch has
Schmidt rank at most
\begin{align}
    (M+1)^{c_{\mathrm{sh}}}
    \binom{r+s+1}{2s+1}R_{\times}^{s}
\end{align}
across the selected cut $p$.

The selected cut lies at distance at most $b$ from the target cut $q$.
Converting the Schmidt decomposition from $p$ to $q$ costs at most
$d_0^{2b}$.  Summing over the selected cut, the phase vectors, and
$0\le s\le s_{\mathrm{loc}}$ gives
Eq.~\eqref{eq:algorithmic-MPO-coefficient-rank-explicit}.  Since the bound
is uniform in the target cut $q$, sequential exact Schmidt decompositions
of the operator viewed as a vector on doubled local Hilbert spaces give an
exact MPO with the same bond-dimension bound
\cite{Oseledets2011,SCHOLLWOCK201196}.  Finally, standard binomial
estimates and $s_{\mathrm{loc}}=O(M/b)$ yield
Eq.~\eqref{eq:algorithmic-MPO-coefficient-rank-log}.
\hfill\(\square\)

The preceding proposition is the precise form of the ``zero-error'' step.
For every retained multi-index, the coefficient is compressed exactly;
only its algebraic representation is changed.  Notice also why the
coordinatewise condition is essential.  A bound on the total degree alone
would give a local cut budget proportional to that total degree, whereas
$m_j<M$ restricts the number of insertions from the constant number of
coarse Hamiltonians adjacent to any target cut.

\subsubsection{Dynamic-programming construction}

The coefficient operators can be generated without enumerating individual
noncommutative words.  Set
\begin{align}
    P_{\boldsymbol0}:=\hat 1.
\end{align}
For every nonzero multi-index $\boldsymbol m$,
Eq.~\eqref{eq:algorithmic-MPO-Pm-definition} obeys the exact recursion
\begin{align}
    P_{\boldsymbol m}
    =
    \sum_{j:\,m_j\ge1}
    P_{\boldsymbol m-\boldsymbol e_j}G_j,
\label{eq:algorithmic-MPO-recursion}
\end{align}
where $\boldsymbol e_j$ is the $j$th unit vector.  Indeed, the term indexed
by $j$ collects precisely the words whose last letter is $G_j$.  Thus the
multi-indices may be processed in increasing order of
$|\boldsymbol m|_1$.  After completing the sum for a given
$\boldsymbol m$, successive exact rank factorizations along the chain
reduce its MPO bond dimension to at most $D_{\boldsymbol m}$ by
Proposition~\ref{prop:algorithmic-MPO-retained-coefficient}
\cite{Oseledets2011,SCHOLLWOCK201196}.

This coefficient bound need not hold for a partially accumulated sum.
Let $D_\star:=\max_{\boldsymbol m}D_{\boldsymbol m}$ be a uniform
bound over the retained coefficients.  Each $G_j$ is supported on at most
$2b$ sites and has an MPO of bond dimension $d_0^{O(b)}$.  Each product
in \eqref{eq:algorithmic-MPO-recursion} therefore has bond dimension at
most $D_\star d_0^{O(b)}$, and every partial sum has bond dimension at most
\begin{align}
    \ell_X D_\star d_0^{O(b)}.
\label{eq:algorithmic-MPO-intermediate-bond}
\end{align}
These intermediate bounds are sufficient for the operation count below.
The smaller coefficient bound is used only after the full update is
complete.

The number of retained multi-indices is
\begin{align}
    N_{\mathrm{ind}}=M^{\ell_X},
\label{eq:algorithmic-MPO-number-indices}
\end{align}
and every multi-index has at most $\ell_X$ predecessors.  Hence the number
of MPO updates is at most
\begin{align}
    \ell_XM^{\ell_X}.
\label{eq:algorithmic-MPO-number-updates}
\end{align}
This corrects a possible ambiguity in an order-by-order description: the
largest total degree is $\ell_X(M-1)$, but one must keep all retained
multi-indices, not merely one MPO for each total degree.

\begin{theorem}[Algorithmic MPO approximation of one localized evolution]
\label{thm:algorithmic-MPO-local-block}
Fix $\sigma\in\{+1,-1\}$ and $0<\delta\le1$.
Let $X$ be an interval of length $W$, let $b\ge1$ be an integer, and set
$\ell_X=\lceil W/b\rceil$.  Choose $M$ as in
Eq.~\eqref{eq:algorithmic-MPO-M-choice}.  Then the operator
$\widetilde U_{X,M}^{(\sigma)}(t)$ in
Eq.~\eqref{eq:algorithmic-MPO-rectangular-truncation} is an explicitly
constructible MPO satisfying
\begin{align}
    \norm{
    e^{-i\sigma tH_X}
    -\widetilde U_{X,M}^{(\sigma)}(t)
    }
    \le\delta.
\label{eq:algorithmic-MPO-local-theorem-error}
\end{align}
Its bond dimension is at most a bound $D_X$ satisfying
\begin{align}
    \log D_X
    \le{}&
    \ell_X\log M
    +C\Bigg[
    b+\log(\ell_XM+2)
    \notag\\
    &\hspace{22mm}
    +\frac{M}{b}\log(\ell_XM+2)
    +\log(b+1)\log(M+2)
    \Bigg].
\label{eq:algorithmic-MPO-local-theorem-bond}
\end{align}
In the exact-arithmetic operation model, the recursive construction
\eqref{eq:algorithmic-MPO-recursion} requires at most
\begin{align}
    \ell_XM^{\ell_X}
    \operatorname{poly}
    \left(W,M,D_X,d_0^b\right)
\label{eq:algorithmic-MPO-local-theorem-time}
\end{align}
arithmetic operations.
\end{theorem}

\textit{Proof.}
The error estimate is Proposition
\ref{prop:algorithmic-MPO-coordinatewise-tail}.  Every retained coefficient
has an exact MPO with the uniform bond-dimension bound in
Proposition~\ref{prop:algorithmic-MPO-retained-coefficient}.  There are
$M^{\ell_X}$ coefficients.  Taking their direct sum gives the
bond-dimension bound
\begin{align}
    D_X:=M^{\ell_X}D_\star,
\end{align}
which implies Eq.~\eqref{eq:algorithmic-MPO-local-theorem-bond} because
$r=|\boldsymbol m|_1\le\ell_X(M-1)$.  The recursion
\eqref{eq:algorithmic-MPO-recursion} visits at most
$\ell_XM^{\ell_X}$ directed edges in the multi-index lattice.  MPO
multiplication, addition, and exact rank reduction are polynomial in the
physical length, the local tensor dimension, and the participating bond
dimensions.  Equation~\eqref{eq:algorithmic-MPO-intermediate-bond}
controls the partial updates, and the running sum of retained coefficients
has bond dimension at most $D_X$.  This proves
Eq.~\eqref{eq:algorithmic-MPO-local-theorem-time}.
\hfill\(\square\)

\textit{Arithmetic-model qualification.}
The preceding running-time statement is an arithmetic-complexity bound.  It
assumes exact access to the local matrix entries and counts field
operations and exact zero tests at unit cost.  Exact MPO rank reduction is implemented
by successive rank factorizations using Gaussian elimination over the
field generated by the input entries, the time parameters, and $i$;
orthonormal Schmidt vectors are not required.  The phase-filter
decomposition is used only to bound ranks and need not be evaluated by the
dynamic program.  A uniform finite-precision bit-complexity bound would
require a separate stability analysis for near-zero Schmidt coefficients;
no such condition-number claim is used here.

We now optimize the parameters at the level relevant for an HHKL block.
Let
\begin{align}
    L:=\max\left\{1,\log\frac{C_0q}{\epsilon}\right\}
\label{eq:algorithmic-MPO-L-definition}
\end{align}
for a sufficiently large constant $C_0$, and assign local error
\begin{align}
    \delta:=\frac{\epsilon}{8q}
\label{eq:algorithmic-MPO-local-error-allocation}
\end{align}
to each HHKL factor.  In the accuracy-dominated regime $t\le L$, first
consider a factor with $W<\lceil\sqrt L\rceil$.  A direct dense approximation
followed by tensorization gives $\log D_X=O(W)=O(\sqrt L)$ and the same exponential
scale for its construction time.  For the remaining factors, choose
\begin{align}
    b:=\left\lceil\sqrt L\right\rceil
    \qquad\bigl(W\ge\lceil\sqrt L\rceil\bigr).
\label{eq:algorithmic-MPO-b-choice}
\end{align}
Equations~\eqref{eq:algorithmic-MPO-local-width},
\eqref{eq:algorithmic-MPO-coarse-partition}, and
\eqref{eq:algorithmic-MPO-M-choice}, with $M$ chosen as the smallest
integer satisfying the cutoff condition, then give
\begin{align}
    \ell_X
    &=O\left(\frac{t+L}{\sqrt L}\right),
\label{eq:algorithmic-MPO-ell-scaling}
    \\
    M
    &=O\left(t\sqrt L+L\right),
\label{eq:algorithmic-MPO-M-scaling}
    \\
    \frac{M}{b}
    &=O\left(t+\sqrt L\right).
\label{eq:algorithmic-MPO-M-over-b-scaling}
\end{align}
Consequently,
\begin{align}
    \log D_X
    =\widetilde O\left(t+\sqrt L\right),
\label{eq:algorithmic-MPO-local-optimized-bond}
\end{align}
and the logarithm of the construction time has the same scaling.  Here
$\widetilde O$ suppresses factors logarithmic in $t+L$; in particular, in
the accuracy-dominated regime these are the
$\log\log(q/\epsilon)$ corrections coming from the number of retained
multi-indices and from the word-counting factor.  More explicitly, one may
use
\begin{align}
    \log D_X
    =O\left[(t+\sqrt L)\log(L+2)\right]
    \qquad (t\le L).
\label{eq:algorithmic-MPO-local-explicit-polylog}
\end{align}
In the complementary regime $t>L$, the HHKL support has length
$W=O(t)$.  A direct dense approximation of the block evolution to error at
most $\delta$, followed by tensorization of the resulting block matrix,
gives
\begin{align}
    D_X\le d_0^{O(t)}.
\label{eq:algorithmic-MPO-large-time-dense}
\end{align}
Indeed, $\norm{H_X}\le gW=O(t)$, so a dense Taylor approximation of degree
$O(t^2+L)$ is sufficient up to constant-factor changes in the cutoff.  Its
arithmetic cost is $d_0^{O(W)}\operatorname{poly}(t,L)$, which has logarithm
$O(t)$ in this regime.  Thus
Eq.~\eqref{eq:algorithmic-MPO-local-optimized-bond} and the matching
construction-time scaling hold uniformly for all $t\ge0$.

\begin{corol}[Global algorithmic MPO for static unitary dynamics]
\label{corol:algorithmic-MPO-global-static-unitary}
The finite-range assumptions above give the HHKL decomposition in
Eqs.~\eqref{eq:algorithmic-MPO-HHKL-product}--
\eqref{eq:algorithmic-MPO-HHKL-overlap}.  Let
\begin{align}
    L_{n,\epsilon}
    :=\max\left\{1,\log\frac{C_1n}{\epsilon}\right\}.
\label{eq:algorithmic-MPO-global-L}
\end{align}
Then there exists an explicitly constructible MPO
$\widetilde U_{\mathrm{MPO}}(t)$ such that
\begin{align}
    \norm{e^{-itH}-\widetilde U_{\mathrm{MPO}}(t)}
    \le\epsilon,
\label{eq:algorithmic-MPO-global-error}
\end{align}
and
\begin{align}
    D_{\mathrm{MPO}}
    \le
    \exp\left[
    \widetilde O\left(
    t+\sqrt{L_{n,\epsilon}}
    \right)
    \right].
\label{eq:algorithmic-MPO-global-bond}
\end{align}
The MPO can be constructed using
\begin{align}
    n\,
    \exp\left[
    \widetilde O\left(
    t+\sqrt{L_{n,\epsilon}}
    \right)
    \right]
\label{eq:algorithmic-MPO-global-time}
\end{align}
exact-arithmetic operations, up to polynomial factors in the displayed
parameters.  The same arithmetic-model qualification as in
Theorem~\ref{thm:algorithmic-MPO-local-block} applies.
\end{corol}

\textit{Proof.}
Approximate each HHKL factor $V_a$ by the local construction with error
$\delta=\epsilon/(8q)$.  The product
Eq.~\eqref{eq:algorithmic-MPO-HHKL-product} contains at most $2q$ factors.
Since $\norm{V_a}=1$ and $\norm{\widetilde V_a}\le1+\delta$, a
telescoping estimate gives
\begin{align}
    \norm{
    \prod_aV_a-\prod_a\widetilde V_a
    }
    \le
    2q\delta(1+\delta)^{2q-1}
    \le\frac{\epsilon}{4}e^{2q\delta}
    \le\frac{\epsilon}{4}e^{1/4}
    <\frac{\epsilon}{2}.
\end{align}
Together with
Eq.~\eqref{eq:algorithmic-MPO-HHKL-error}, this proves
Eq.~\eqref{eq:algorithmic-MPO-global-error}.  Since the $X_j$ are nonempty
consecutive intervals, $q=O(n)$ after the fixed preliminary blocking; hence
the local logarithmic scale is bounded by $L_{n,\epsilon}$ after adjusting
$C_1$.

Across a fixed microscopic cut, only $\nu_{\mathrm H}=O(1)$ localized
factors have nontrivial Schmidt rank.  Therefore
\begin{align}
    D_{\mathrm{MPO}}
    \le D_X^{\nu_{\mathrm H}},
\end{align}
which proves Eq.~\eqref{eq:algorithmic-MPO-global-bond}.  Finally, there are
$O(n)$ localized factors and each is constructed within the cost of
Theorem~\ref{thm:algorithmic-MPO-local-block}.  The product MPO can be
assembled tensorwise on the factor supports.  Each $X_j$ occurs in at
most three factors, so $\sum_a|\Omega_a|\le3n$.  At every bond only
$\nu_{\mathrm H}$ factors carry a nontrivial virtual index.  Thus this final
assembly has the same arithmetic scale, which gives
Eq.~\eqref{eq:algorithmic-MPO-global-time}.
\hfill\(\square\)

The resulting construction implements the proposed strategy in a form that
separates the three relevant scales.  The HHKL width is
$O(t+L_{n,\epsilon})$; the subblock width is
$b\asymp\sqrt{L_{n,\epsilon}}$; and the coordinatewise occupancy cutoff is
$M=O(t\sqrt{L_{n,\epsilon}}+L_{n,\epsilon})$.  The local cut-count reduction
then leaves only
\begin{align}
    O\left(\frac{M}{b}\right)
    =O\left(t+\sqrt{L_{n,\epsilon}}\right)
\end{align}
explicit crossing insertions.  The retained coefficients are represented
exactly, while the number $M^{\ell_X}$ of such coefficients contributes
only the additional logarithmic factors hidden in
Eq.~\eqref{eq:algorithmic-MPO-global-bond}.

\subsection{Extension to efficiently represented time dependence}
\label{subsubsec:alg-U-driven-extension}

This extension covers both \emph{discrete driving}, supplied as a
piecewise-constant schedule, and \emph{regular continuous driving},
including uniformly analytic local Hamiltonians with certified,
efficiently computable polynomial approximants.  The latter class includes
the bounded-frequency modulations of
Corollary~\ref{corol:reg-elementary-drives}; polynomial ramps can also be
supplied exactly.  Thus \emph{piecewise-polynomial input} below describes
the finite representation used by the algorithm, not a restriction of the
original drive to piecewise-constant dynamics.

\paragraph{Connection with Section~\ref{subsec:td-static-type-compression} and notation.}
We retain $N$ for the number of intervals in a common temporal partition,
$\nu$ for the polynomial degree in time, and $\zeta$ for the uniform local
interaction-norm error, as in
Eq.~\eqref{eq:reg-local-polynomial-approximation}.
The hatted terms $\widehat h_Z^{(x)}$ used below are the algorithmic
counterparts of the local approximants $h_{Z,j}^{[\nu]}$ in that section.
In particular, $N$ counts time intervals, not distinct Hamiltonians;
$\nu$ is unrelated to the spatial occupancy cutoff $M$ used here.
Also, $n$ here continues to denote the chain length, not the auxiliary
commutator cutoff of that section.

Section~\ref{subsec:td-static-type-compression} uses polynomial
approximation and interpolation of the dressed boundary family to obtain
a common low-rank basis.  That rank argument does not by itself supply
efficient MPOs for the exact sample operators on all spatial cuts.
We therefore do not assume an oracle for those samples.  Instead, we
retain the endpoint-time dependence of every coefficient and construct it
by a finite triangular recursion.  For an analytic input, the uniform
analytic bounds of Corollary~\ref{corol:reg-analytic-driving} must be
accompanied by an effective procedure producing the polynomial data and
their error certificate; this preprocessing cost is included below.

The two spatial scales are unchanged: an HHKL interval has length
$W=O(t+L)$, and it is partitioned into subblocks of width
$b\asymp\sqrt L$.  One occupancy register per spatial subblock is shared
by the entire time interval.  In particular, neither a new occupancy
register nor a new independently truncated MPO is introduced at each
switch of the drive.  This is what prevents an artificial factor
proportional to the number of switches in the leading accuracy exponent.

\paragraph{Finite temporal input.}
The case $t=0$ gives the identity, so take $t>0$ below.
First suppose the local interactions are supplied exactly as Hermitian
matrix polynomials on a common partition
\begin{align}
 0=a_0<a_1<\cdots<a_N=t,
 \qquad h_Z^{(a_{v-1}+\Delta_v u)}
   =\sum_{d=0}^{\nu}h_{Z,v,d}u^d,
 \quad 0\le u\le1,
 \quad \Delta_v:=a_v-a_{v-1}.
\label{eq:alg-U-temporal-input}
\end{align}
Values at the finitely many partition endpoints do not affect the
propagator.  The supports and the real-time local strength obey the same
bounds as in the static construction (with $g$ replaced by the supplied
local bound when a surrogate is used).  Discrete driving is the exact
case $\nu=0$; $N=1$ and $\nu=1$ already describe a continuously varying
linear ramp.  Local entries and endpoints are inputs to the same
exact-arithmetic model as above.  We later include the cost and error of
constructing these data from an approximately specified drive.

Apply the time-dependent version of the HHKL localization input
\eqref{eq:algorithmic-MPO-HHKL-product}--
\eqref{eq:algorithmic-MPO-HHKL-overlap}.  Its localized factors are now
propagators $U^X_{0\to t}$ or their inverses, not ordinary exponentials.
The localization uses only the uniform local-strength bound and imposes
no bound on time derivatives; this also follows from the gluing proof
\cite{HHKL2021}.  An inverse is handled by the reversed generator:
\begin{align}
 (U^X_{0\to t})^{-1}
 =\mathcal T\exp\left[-i\int_0^t -H_X^{(t-x)}\,dx\right].
\label{eq:alg-U-reverse-time}
\end{align}
This operation preserves the number and degrees of the temporal pieces.
Changing only the sign of $H_X^{(x)}$ without reversing time would in
general be incorrect.  Bounded cut multiplicity and the unitary
product-error estimate used above are unchanged.

\paragraph{A coefficient recursion that retains the integration endpoint.}
On one localized interval $X$, use the same anchored subblocks and write
\begin{align}
 H_X^{(x)}(\boldsymbol\lambda)
   =\sum_{j=1}^{\ell_X}\lambda_jG_j^{(x)},
 \qquad
 U^X_{0\to x}(\boldsymbol\lambda)
   =\sum_{\boldsymbol m\ge0}\boldsymbol\lambda^{\boldsymbol m}
      U_{\boldsymbol m}(x).
\label{eq:alg-U-driven-coefficients}
\end{align}
Coefficient extraction in the nonautonomous evolution equation gives
\begin{align}
 U_{\boldsymbol0}(x)&=\hat 1,
 &U_{\boldsymbol m}(0)&=0\quad(\boldsymbol m\ne\boldsymbol0),
 \nonumber\\
 \partial_x U_{\boldsymbol m}(x)
   &=-i\sum_{j:m_j\ge1}G_j^{(x)}
                    U_{\boldsymbol m-\boldsymbol e_j}(x),
 &U_{\boldsymbol m}(x)
   &=-i\sum_{j:m_j\ge1}\int_0^x
       G_j^{(y)}U_{\boldsymbol m-\boldsymbol e_j}(y)\,dy.
\label{eq:alg-U-Volterra-recursion}
\end{align}
Here $|\boldsymbol m|_1$ is the total insertion order.  It is not the
smaller number of explicit crossings obtained after cut-count reduction.
The recursion sums all noncommuting words without listing them.

On the $v$th temporal piece write
\begin{align}
 G_j^{(a_{v-1}+\Delta_v u)}&=\sum_{d=0}^{\nu}G_{j,v,d}u^d,
 &U_{\boldsymbol m}(a_{v-1}+\Delta_v u)
    &=\sum_{q=0}^{(\nu+1)|\boldsymbol m|_1}
          U_{\boldsymbol m,v,q}u^q.
\label{eq:alg-U-time-coefficient-table}
\end{align}
Then the following updates are exact:
\begin{align}
 U_{\boldsymbol m,v,0}&=U_{\boldsymbol m}(a_{v-1}),
 \nonumber\\
 U_{\boldsymbol m,v,q+1}
   &=-\frac{i\Delta_v}{q+1}
       \sum_{j:m_j\ge1}\sum_{d=0}^{\min\{\nu,q\}}
       G_{j,v,d}U_{\boldsymbol m-\boldsymbol e_j,v,q-d},
 \nonumber\\
 U_{\boldsymbol m}(a_v)&=\sum_q U_{\boldsymbol m,v,q}.
\label{eq:alg-U-polynomial-recursion}
\end{align}
Out-of-range coefficients are zero.  Process the temporal pieces in
chronological order and the multi-indices within each piece in increasing
total degree.  Induction in this last degree proves the polynomial degree
bound in \eqref{eq:alg-U-time-coefficient-table}.  The stored data are thus
finite even though the full propagator is not a polynomial in time.
Only the coefficients with $m_j<M$ are needed, since their predecessors
lie in the same lower set.  No factor involving the inverse shortest
piece length occurs in the arithmetic operation count.

The coordinatewise tail estimate also survives time ordering.  Put
$\gamma_j:=\int_0^t\|G_j^{(x)}\|\,dx\le gtb$.  Resumming unrestricted
coordinates into their time-dependent Hermitian background and applying
mixed Duhamel expansion, followed by inclusion--exclusion, gives
\begin{align}
 \left\|U^X_{0\to t}
       -\sum_{0\le m_j<M}U_{\boldsymbol m}(t)\right\|
 \le\prod_{j=1}^{\ell_X}[1+\tau_M(\gamma_j)]-1.
\label{eq:alg-U-driven-coordinate-tail}
\end{align}
For fixed multiplicities, the sum over the ordered shuffles of the
insertion times is bounded by the product of scalar integrals divided by
the multiplicity factorials.  This proves the same mixed-coefficient bound
as in Proposition~\ref{prop:algorithmic-MPO-coordinatewise-tail}, with
$t\|G_j\|$ replaced by $\gamma_j$.  Hence the choice
\eqref{eq:algorithmic-MPO-M-choice} remains sufficient.

\paragraph{Exact rank bounds for the intermediate time coefficients.}
The remaining issue is not the number of entries in
\eqref{eq:alg-U-time-coefficient-table}, but their MPO sizes.  We give the
coefficientwise bound explicitly, rather than infer it from the rank of
the final propagator.

\begin{prop}[Piecewise-polynomial coefficient ranks]
\label{prop:alg-U-driven-coefficient-rank}
Assume \eqref{eq:alg-U-temporal-input}.  Fix $0\le m_j<M$, set
$r=|\boldsymbol m|_1$, and use $Q_{\mathrm{loc}}$ and
$s_{\mathrm{loc}}$ from \eqref{eq:algorithmic-MPO-local-budget}.
Uniformly in $0\le x\le t$, $U_{\boldsymbol m}(x)$ has an exact MPO
with bond dimension at most
\begin{align}
 D_{\boldsymbol m}^{\mathrm{td}}
 \le{}&d_0^{2b}b(s_{\mathrm{loc}}+1)
       (2Q_{\mathrm{loc}}+1)^{\lceil\log_2b\rceil}
       (M+1)^{c_{\mathrm{sh}}}
 \nonumber\\
 &\quad\times\max_{0\le s\le\min\{r,s_{\mathrm{loc}}\}}
       \left\{N^sR_\times^s
         \binom{(\nu+1)r+s}{s}\right\}.
\label{eq:alg-U-driven-coefficient-rank}
\end{align}
The constants have the same dependence as in
Proposition~\ref{prop:algorithmic-MPO-retained-coefficient}.
Every stored time coefficient satisfies
\begin{align}
 D(U_{\boldsymbol m,v,q})
    \le[(\nu+1)r+1]D_{\boldsymbol m}^{\mathrm{td}}.
\label{eq:alg-U-stored-time-rank}
\end{align}
\end{prop}

\textit{Proof.}
Use the microscopic auxiliary variables and a window of $b$ cuts from the
proof of Proposition~\ref{prop:algorithmic-MPO-retained-coefficient}.
The local cut weight of a retained monomial is still $O(M)$: neither
insertion times nor polynomial degrees change its coarse occupancy.
Coefficientwise cut-count reduction therefore has zero remainder and
leaves $s\le s_{\mathrm{loc}}$ explicit crossing insertions.
For a selected cut, every noncrossing segment propagator factorizes into a
left and a right propagator.  Expand each crossing interaction in one
fixed local product-operator basis; finite range gives $R_\times=O(1)$,
and its scalar entries are piecewise polynomials of degree at most $\nu$.

Fix the temporal pieces containing the $s$ crossing times.  There are at
most $N^s$ choices.  On each resulting ordered domain, a coefficient of
total insertion degree $q$ in a noncrossing segment propagator is a
polynomial in its variable endpoints of total degree at most
$(\nu+1)q$.  To see this, split the segment at the fixed input breakpoints.
A $q$-fold iterated integral of degree-$\nu$ polynomial generators on one
piece has endpoint degree at most $(\nu+1)q$; composition across fixed
breakpoints preserves this degree bound.  This is an algebraic statement
about a fixed coefficient, not a commutativity assumption.

Collect all left factors and all right factors before extracting the
coarse coefficient.  Only $c_{\mathrm{sh}}=O(1)$ coarse variables occur
on both sides, so their total left--right allocations cost at most
$(M+1)^{c_{\mathrm{sh}}}$.  For one allocation, the left operator-valued
coefficient, with the scalar crossing entries included on that side, is a
polynomial in the $s$ crossing times of total degree at most
$(\nu+1)r$.  It has at most $\binom{(\nu+1)r+s}{s}$ scalar monomials.
For each such monomial, absorb its ordered integral times the right
operator-valued coefficient into one right operator.  This produces an
exact product decomposition.  No noncommutative word count is paid.
Multiplication by the cut--phase count and by the cut-shift factor
$d_0^{2b}$ proves \eqref{eq:alg-U-driven-coefficient-rank}.

Finally, on each temporal piece the curve $U_{\boldsymbol m}(x)$ is a
polynomial of degree at most $(\nu+1)r$.  Every one of its time
coefficients is a linear combination of its values at
$(\nu+1)r+1$ distinct points.  The rank bound just proved applies at
every such point, giving \eqref{eq:alg-U-stored-time-rank}.  This
interpolation is only a rank proof: the algorithm constructs the time
coefficients by \eqref{eq:alg-U-polynomial-recursion}, without evaluating
an exact large-block propagator.
\hfill\(\square\)

\begin{theorem}[Constructive localized MPO for polynomial temporal data]
\label{thm:alg-U-driven-local}
Choose $M$ as in \eqref{eq:algorithmic-MPO-M-choice}, let
$r_\star:=\ell_X(M-1)$, and define
\begin{align}
 \mathcal B_X:={}&\ell_X\log(M+1)+b
    +\log(b+1)\log(M+2)
 \nonumber\\
 &+\left(1+\frac{M}{b}\right)
       \log\!\left[2+N(\nu+1)(r_\star+M+2)\right].
\label{eq:alg-U-driven-complexity-budget}
\end{align}
The rectangularly truncated propagator is constructible by
\eqref{eq:alg-U-polynomial-recursion}, has error at most $\delta$, and
has bond dimension at most $D_X:=\exp(C\mathcal B_X)$.  All intermediate
MPOs can be kept within a constant power of this same bound.  The number
of arithmetic operations is at most
\begin{align}
 N\ell_XM^{\ell_X}
 \operatorname{poly}(W,M,\nu,r_\star,D_X,d_0^b).
\label{eq:alg-U-driven-local-cost}
\end{align}
\end{theorem}

\textit{Proof.}
The error is \eqref{eq:alg-U-driven-coordinate-tail}.  Summing the
$M^{\ell_X}$ retained coefficients and applying the preceding proposition
bounds the final rank.  There are at most $(\nu+1)r_\star+1$ time
coefficients for each multi-index and each temporal piece, and each update
uses at most $\ell_X(\nu+1)$ predecessor terms.  Each $G_{j,v,d}$ has
support of width $O(b)$ and hence an MPO of dimension $d_0^{O(b)}$.
A partially accumulated update or endpoint sum has at most a polynomial
number of these bounded-rank terms.  After the whole update, exact
rank reduction uses \eqref{eq:alg-U-stored-time-rank}; after an endpoint
sum it uses \eqref{eq:alg-U-driven-coefficient-rank}.  Thus no intermediate
requires an unbounded product of the ranks from previous temporal pieces.
The elementary binomial estimates in
\eqref{eq:alg-U-driven-coefficient-rank} yield
\eqref{eq:alg-U-driven-complexity-budget} and the claimed cost.
\hfill\(\square\)

\paragraph{Approximate input and explicit examples.}
To turn polynomial approximability into a construction, assume an input
procedure supplies the local polynomial coefficients, a common partition,
and a certificate
\begin{align}
 \operatorname*{ess\,sup}_{0\le x\le t}\sup_i
  \sum_{Z\ni i}\|h_Z^{(x)}-\widehat h_Z^{(x)}\|
       \le\zeta,
 \qquad
 \widehat h_Z^{(x)}=\widehat h_Z^{(x)\dagger},
\label{eq:alg-U-computable-surrogate}
\end{align}
with unchanged supports and local strength at most $g_\star=O(g)$.
Denote the total preprocessing cost by $T_{\mathrm{in}}$.
For $g>0$, it is enough to take
$\zeta\le\min\{g,\epsilon/[64n(1+t)]\}$; the case $g=0$ is trivial.
Then Duhamel's estimate gives
\begin{align}
 \|U^H_{0\to t}-U^{\widehat H}_{0\to t}\|
 \le\int_0^t\sum_Z\|h_Z^{(x)}-\widehat h_Z^{(x)}\|\,dx
 \le nt\zeta\le\epsilon/64.
\label{eq:alg-U-surrogate-error}
\end{align}
This global approximation is an algorithmic choice, not the local-only
replacement used in Section~\ref{subsec:td-static-type-compression}.
Its stronger requested input precision contributes only
$\log[n(1+t)/\epsilon]$ to the degree for the analytic examples.

\begin{corol}[Static-type algorithmic scaling for structured driving]
\label{corol:alg-U-driven-global}
Under the same finite-range and local-strength bounds, suppose that
either exact data \eqref{eq:alg-U-temporal-input} or the certified input
\eqref{eq:alg-U-computable-surrogate} are supplied.  Put
$L=L_{n,\epsilon}$ as in \eqref{eq:algorithmic-MPO-global-L} and
$\Theta_U:=\log[2+N(\nu+1)]$.  There is an MPO with operator-norm
error at most $\epsilon$ and bounds
\begin{align}
 \log D_{\mathrm{MPO}}
   &\le\widetilde O(t+\sqrt L)
          +O\bigl((1+t+\sqrt L)\Theta_U\bigr),
 \nonumber\\
 T_{\mathrm{construct}}
   &\le T_{\mathrm{in}}+
       nN\operatorname{poly}(1+t+L,\nu+1)
       \exp\!\left[
          \widetilde O(t+\sqrt L)
          +O\bigl((1+t+\sqrt L)\Theta_U\bigr)\right].
\label{eq:alg-U-driven-global-cost}
\end{align}
Different absolute constants may occur in the two exponents.  If $N$ and
$\nu$ are polynomial in $1+t+L$, and $T_{\mathrm{in}}$ fits the displayed
cost, the bounds reduce to the static scale
$\exp[\widetilde O(t+\sqrt{L_{n,\epsilon}})]$ for bond dimension and
$n\exp[\widetilde O(t+\sqrt{L_{n,\epsilon}})]$ for construction time.
\end{corol}

\textit{Proof.}
For $t\le L$, use $W=O(t+L)$, $b=\lceil\sqrt L\rceil$, and
the smallest admissible $M=O(t\sqrt L+L)$ in
Theorem~\ref{thm:alg-U-driven-local}, allocating
constant fractions of the error to input approximation, HHKL localization,
and the localized coefficient truncations.  The proof of
Corollary~\ref{corol:algorithmic-MPO-global-static-unitary} then applies
to the time-ordered factors and their reversals.  Short intervals
$W<b$ are treated densely.  If $t>L$, then $W=O(t)$ and a dense Dyson
truncation of total order $O(t^2+L)$ suffices.  Its coefficients are
computed piecewise by the same recursion, with a single degree variable,
in $N\operatorname{poly}(t,L,\nu)d_0^{O(W)}$ operations.  This also
fits \eqref{eq:alg-U-driven-global-cost}.
\hfill\(\square\)

\emph{Discrete driving.}  A supplied piecewise-constant schedule has
$\nu=0$ and $\zeta=0$, so no temporal approximation is required.
Its interval count $N$ remains explicit in
Eq.~\eqref{eq:alg-U-driven-global-cost}.

\emph{Analytic continuous driving.}  Uniformly analytic local Hamiltonians,
including bounded-frequency modulation as in
Corollaries~\ref{corol:reg-analytic-driving} and
\ref{corol:reg-elementary-drives}, supply polynomial-size data at the
precision in Eq.~\eqref{eq:alg-U-computable-surrogate} when local
coefficients can be evaluated efficiently and the analytic bounds are
supplied.  The original drive remains continuous; only its input
representation is replaced by polynomials.  Fixed-degree polynomial
ramps require no such approximation and are exact input as well.

Neither an abstract existence theorem for quadrature nor $C^\infty$
regularity without quantitative bounds supplies this computational input.
The operation bounds remain exact-arithmetic bounds; bit lengths,
conditioning of coefficient conversion, and numerical rank decisions are
not covered by them.

\section{Algorithmic MPO construction for time-independent Liouvillian dynamics}
\label{sec:algorithmic-MPO-static-liouvillian}

We now extend the algorithmic construction of
Section~\ref{subsec:static-unitary-algorithmic-MPO} to a one-dimensional,
time-independent Liouvillian.  Matrix-product descriptions of mixed-state
dynamics were introduced in
Refs.~\cite{PhysRevLett.93.207204,PhysRevLett.93.207205}, and locality bounds
for Markovian dynamics were established in
Refs.~\cite{Poulin2010,BarthelKliesch2012}.
The diamond-norm local-cell estimate needed here is proved below.
The main new difficulty is not the
coefficientwise MPO manipulation itself, but the analytic justification of
the coordinatewise Taylor cutoff.  A naive sequential HHKL product contains
inverse factors of the form $e^{-\Delta\mathcal L_X}$, whose diamond norm
need not admit a bound independent of the block width when treated in
isolation; moreover, differentiating an $O(n)$-depth sequential product is
unsuitable for the desired MPO complexity bound.  We therefore use an explicit two-macro-layer local-cell
realization of the short-time localization and do not attach independent
parameters to the separate occurrences of a physical subblock.  Instead, a
single parameter is shared by every HHKL occurrence of the same physical
subblock.  Setting that parameter to zero then removes the corresponding
microscopic Lindblad terms consistently from the entire HHKL product, so
that the remaining background is again the HHKL approximation of a genuine
modified Lindbladian evolution.

Throughout this section, let $\Lambda=[n]$ be a finite one-dimensional
chain with on-site Hilbert-space dimension at most a fixed $d_0$, and let
\begin{align}
    \mathcal L=\sum_{Z\subseteq\Lambda}\mathcal L_Z,
    \qquad
    |Z|\le k,
    \qquad
    \operatorname{diam}(Z)\le k,
    \qquad
    \sum_{Z\ni i}\lVert\mathcal L_Z\rVert_\diamond\le g_{\mathrm L}
    \quad(i\in\Lambda),
\label{eq:alg-L-shared-local-Liouvillian}
\end{align}
where every $\mathcal L_Z$ is taken to be a complete local Lindblad term.
This convention guarantees that an arbitrary anchored sub-sum of the local
terms is again a Lindblad generator.  We assume finite-range interactions;
a preliminary blocking by a constant number of sites is allowed as in the
static unitary construction.  All MPOs have open boundaries.  We use the
exact-arithmetic model and rank-factorization procedure specified in
Section~\ref{subsec:static-unitary-algorithmic-MPO}.

Fix a sufficiently small constant time step $\Delta>0$ and, for
notational simplicity, first write
\begin{align}
    t=R\Delta,
    \qquad R\in\mathbb N.
\label{eq:alg-L-time-slicing}
\end{align}
A final time interval shorter than $\Delta$ is treated by the same
construction.  The short-time localization used below is a \emph{two-layer
local-cell decomposition}.  This point is essential: the sequential
$O(n)$ chain of relative factors is used only as a telescoping device in the
proof of the approximation estimate, and is not the product whose Taylor
coefficients are represented as an MPO.

After the fixed preliminary blocking, choose consecutive coarse cells
$C_1,\ldots,C_M$.  Up to endpoint modifications, each cell is split into
three consecutive pieces
\begin{align}
    C_j=C_j^{\mathrm L}\sqcup C_j^{\mathrm 0}\sqcup C_j^{\mathrm R},
    \qquad
    |C_j^{\mathrm L}|,|C_j^{\mathrm 0}|,|C_j^{\mathrm R}|=\Theta(B),
\label{eq:alg-L-depth-two-cell-splitting}
\end{align}
where the left and right collars have width at least $B$.  Dummy sites with
zero generator may be appended at the endpoints, so no separate small-cell
case is needed.  Their identity factors are omitted from tensor storage.  Since $B$ is larger than the interaction range, every
microscopic term is either contained in one $C_j$ or crosses one interface
$C_j:C_{j+1}$.  Define
\begin{align}
    \mathcal K_j
    &:={}
    \sum_{Z\subseteq C_j}\mathcal L_Z,
    &
    \mathcal V_j
    &:={}
    \sum_{\substack{Z\cap C_j\ne\emptyset\\
                     Z\cap C_{j+1}\ne\emptyset}}
    \mathcal L_Z,
\label{eq:alg-L-depth-two-KV-decomposition}
\end{align}
so that
\begin{align}
    \mathcal L=\mathcal K^{(0)}+\sum_{j=1}^{M-1}\mathcal V_j,
    \qquad
    \mathcal K^{(0)}:=\sum_{j=1}^{M}\mathcal K_j.
\label{eq:alg-L-depth-two-global-decomposition}
\end{align}
For the $j$th interface, let
\begin{align}
    P_j:=C_j^{\mathrm R}\cup C_{j+1}^{\mathrm L},
    \qquad
    \mathcal A_j
    :=\sum_{Z\subseteq C_j^{\mathrm R}}\mathcal L_Z
      +\sum_{Z\subseteq C_{j+1}^{\mathrm L}}\mathcal L_Z.
\label{eq:alg-L-depth-two-patch-background}
\end{align}
The patches $P_1,\ldots,P_{M-1}$ are pairwise disjoint, because the two
collars in every coarse cell are separated by the middle region
$C_j^{\mathrm 0}$.  Introduce the base-cell and interface-correction maps
\begin{align}
    \mathcal U_{j,\Delta}
    &:={}
    e^{\Delta\mathcal K_j},
    \\
    \mathcal C_{j,\Delta}
    &:={}
    e^{-\Delta\mathcal A_j}
    \circ e^{\Delta(\mathcal A_j+\mathcal V_j)}.
\label{eq:alg-L-depth-two-local-cells}
\end{align}
The short-time approximant is
\begin{align}
    \Gamma^{(2)}_\Delta
    &:={}
    \Gamma^{\mathrm{base}}_\Delta
    \circ
    \Gamma^{\mathrm{corr}}_\Delta,
    \\
    \Gamma^{\mathrm{base}}_\Delta
    &:={}
    \mathop{\circ}_{j=1}^{M}\mathcal U_{j,\Delta},
    &
    \Gamma^{\mathrm{corr}}_\Delta
    &:={}
    \mathop{\circ}_{j=1}^{M-1}\mathcal C_{j,\Delta}.
\label{eq:alg-L-depth-two-HHKL-product}
\end{align}
Within either product in
Eq.~\eqref{eq:alg-L-depth-two-HHKL-product}, all factors have disjoint
supports and hence commute.  Thus the object used by the algorithm has
macro-depth two.  Each correction cell is itself a fixed product of two
local exponentials, but this constant internal depth does not scale with
$n$ and is kept inside one local cell throughout the Schmidt-rank
analysis.

For coefficient bookkeeping, we flatten only the constant-depth internal
structure of the local cells.  Let $\mathfrak F$ be the family of elementary
exponential occurrences
\begin{align}
    f=(0,j),\quad (1,j,-),\quad (1,j,+),
\end{align}
corresponding respectively to
$e^{\Delta\mathcal K_j}$,
$e^{-\Delta\mathcal A_j}$, and
$e^{\Delta(\mathcal A_j+\mathcal V_j)}$.  Write
\begin{align}
    \mathcal E_f=e^{\sigma_f\Delta\mathcal L_f},
    \qquad
    \sigma_f\in\{+1,-1\},
\label{eq:alg-L-HHKL-product}
\end{align}
with the order chosen consistently with
Eq.~\eqref{eq:alg-L-depth-two-HHKL-product}.  Choose once and for all an
enumeration
\begin{align}
    \mathfrak F=\{1,\ldots,N_{\mathrm f}\},
    \qquad
    \Omega_f:=\operatorname{supp}(\mathcal L_f),
    \qquad
    \Xi_f:=\text{support of the local cell containing }f,
\label{eq:alg-L-flattened-factor-enumeration}
\end{align}
compatible with the two macro-layers and with the two elementary
exponentials inside each correction cell.  Thus
$\Omega_f\subseteq\Xi_f$ and $\operatorname{diam}(\Xi_f)=O(B)$.  This flattened order is used
only for algebraic differentiation and coefficient extraction; cells in
the same macro-layer are constructed by tensorization, not by an
$O(n)$-depth MPO multiplication.

Fix an integer $1\le\ell\le B$ and partition each piece
$C_j^{\mathrm L}$, $C_j^{\mathrm 0}$, and $C_j^{\mathrm R}$ into
consecutive physical subblocks of width at most $\ell$ and at least
$\ell/2$, apart from harmless endpoint pieces, and denote their family by
$\mathfrak Y$.  Thus no physical subblock crosses a coarse-cell boundary
or one of the three-piece boundaries.  Assign every microscopic term $\mathcal L_Z$ to the leftmost physical
subblock intersecting $Z$, and denote this anchor by
$a(Z)\in\mathfrak Y$.  For
each elementary occurrence $f$, define
\begin{align}
    \mathcal L_{f,Y}
    :=\sum_{\substack{a(Z)=Y\\
                      \mathcal L_Z\text{ occurs in }\mathcal L_f}}
      \mathcal L_Z,
    \qquad
    \mathcal L_f=\sum_{Y\in\mathfrak Y}\mathcal L_{f,Y}.
\label{eq:alg-L-anchor-decomposition}
\end{align}
The same parameter is attached to every occurrence of the same anchored
microscopic term:
\begin{align}
    \mathcal L_f(\boldsymbol\lambda)
    :=\sum_{Y\in\mathfrak Y}\lambda_Y\mathcal L_{f,Y}.
\label{eq:alg-L-shared-factor-generator}
\end{align}
Equivalently, define
$\mathcal K_j(\boldsymbol\lambda)$,
$\mathcal A_j(\boldsymbol\lambda)$, and
$\mathcal V_j(\boldsymbol\lambda)$ by inserting
$\lambda_{a(Z)}$ in their microscopic sums, and set
\begin{align}
    \Gamma^{(2)}_\Delta(\boldsymbol\lambda)
    &:={}
    \left[
    \mathop{\circ}_{j=1}^{M}
    e^{\Delta\mathcal K_j(\boldsymbol\lambda)}
    \right]
    \circ
    \left[
    \mathop{\circ}_{j=1}^{M-1}
    \left
    \{
    e^{-\Delta\mathcal A_j(\boldsymbol\lambda)}
    \circ
    e^{\Delta[\mathcal A_j(\boldsymbol\lambda)
                    +\mathcal V_j(\boldsymbol\lambda)]}
    \right\}
    \right].
\label{eq:alg-L-shared-parametrization}
\end{align}
The same parameter $\lambda_Y$ is used in both macro-layers and in every
one of the $R$ time steps.  A term in the middle of a coarse cell occurs
only in the base layer; a term in an interface collar can occur once in the
base layer and twice inside one correction cell; an interface-crossing term
occurs in one correction cell.  Consequently,
\begin{align}
    \nu_Y:=\bigl|\{f:\mathcal L_{f,Y}\ne0\}\bigr|,
    \qquad
    \nu_{\mathrm{occ}}:=\max_Y\nu_Y\le3.
\label{eq:alg-L-occurrence-multiplicity}
\end{align}
For brevity, we henceforth write
$\Gamma_\Delta(\boldsymbol\lambda)$ for the depth-two object
$\Gamma^{(2)}_\Delta(\boldsymbol\lambda)$.

Set
\begin{align}
    F_R(\boldsymbol\lambda)
    :=\Gamma_\Delta(\boldsymbol\lambda)^{\circ R}
    =\sum_{\boldsymbol m\in\mathbb Z_{\ge0}^{\mathfrak Y}}
    F_{R,\boldsymbol m}\,
    \boldsymbol\lambda^{\boldsymbol m},
    \qquad
    \boldsymbol\lambda^{\boldsymbol m}
    :=\prod_{Y\in\mathfrak Y}\lambda_Y^{m_Y}.
\label{eq:alg-L-long-parameterized-product}
\end{align}
For $K\in\mathbb Z_{\ge0}$, define the shared coordinatewise truncation
\begin{align}
    \mathcal T_K^{\mathrm{sh}}F_R
    :=\sum_{\substack{\boldsymbol m\\0\le m_Y\le K\ \forall Y}}
    F_{R,\boldsymbol m}\boldsymbol\lambda^{\boldsymbol m}.
\label{eq:alg-L-shared-rectangular-truncation}
\end{align}
The algorithmic target is therefore
\begin{align}
    \left.
    \mathcal T_K^{\mathrm{sh}}
    \left[\Gamma_\Delta(\boldsymbol\lambda)^{\circ R}\right]
    \right|_{\boldsymbol\lambda=\boldsymbol1},
\label{eq:alg-L-correct-truncated-object}
\end{align}
not the unprojected power of a one-step truncated polynomial.

\subsubsection{Overview and proof architecture}

Let $\mathsf L$ denote the logarithmic accuracy scale.  Eventually we take
\begin{align}
    \mathsf L
    \asymp
    \log\frac{C n(1+g_{\mathrm L}t)}{\epsilon},
    \qquad
    B=\Theta(\mathsf L),
    \qquad
    \ell=\Theta(\sqrt{\mathsf L}).
\label{eq:alg-L-three-scales}
\end{align}
The proof is organized around the following six steps.

First, the two-macro-layer local-cell product in
Eq.~\eqref{eq:alg-L-depth-two-HHKL-product} must approximate the short-time
channel uniformly after deleting any family of anchored microscopic terms.
If $S\subseteq\mathfrak Y$ and $\lambda_Y=0$ for $Y\in S$, then all
occurrences of those terms are removed consistently.  The resulting
macro-depth-two product
$\Gamma_\Delta(\boldsymbol0_S,\boldsymbol1_{S^c})$ is therefore the local-
cell approximation of the modified Lindbladian
\begin{align}
    \mathcal L^{(S^c)}
    :=\sum_{Z:\,a(Z)\notin S}\mathcal L_Z.
\label{eq:alg-L-deleted-Liouvillian}
\end{align}
In particular, the exact background evolution
$e^{x\mathcal L^{(S^c)}}$ is completely positive and trace preserving and
has diamond norm one.

Second, the full parameterized depth-two product is factored relative to this
background.  For a fixed active family $S$ and complex parameters
$|z_Y|\le\rho$, the desired exact form is
\begin{align}
    \Gamma_\Delta(\boldsymbol z_S,\boldsymbol1_{S^c})
    =
    \Gamma_\Delta(\boldsymbol0_S,\boldsymbol1_{S^c})
    \circ
    \mathcal T\exp\left[
    \int_0^\Delta
    \widetilde{\mathcal V}_{S}(x,\boldsymbol z)\,dx
    \right].
\label{eq:alg-L-shared-cocycle-overview}
\end{align}
The generator $\widetilde{\mathcal V}_{S}$ is an algebraic local cocycle
generator, not in general a Lindblad generator.  The variable $x$ in this
factorization is a rescaled interpolation parameter between the deleted and
the deformed HHKL products; it is not the physical time.  A short-time
composition-commutator expansion, of the same type as the imaginary-time
expansion used for local Gibbs operators
\cite{KUWAHARA201696,PhysRevX.11.011047}, gives
\begin{align}
    \lVert\widetilde{\mathcal V}_{S}(x,\boldsymbol z)\rVert_\diamond
    \le
    C_\rho g_{\mathrm L}
    \sum_{Y\in S}(|Y|+c_k),
\label{eq:alg-L-local-cocycle-overview-bound}
\end{align}
where $c_k=O(k)$ accounts for the fixed interaction-range collar.  The
crucial point is that the right-hand side is independent of the main-block
width $B$ and of the diameter of the union of the active subblocks.
Together with the HHKL background estimate, this gives a fixed-radius
polydisc bound of the form
\begin{align}
    \sup_{|z_Y|\le\rho}
    \lVert
    \Gamma_\Delta(\boldsymbol z_S,\boldsymbol1_{S^c})
    \rVert_\diamond
    \le
    (1+\delta_{\mathrm H})
    \exp\left[
    C_\rho g_{\mathrm L}\Delta
    \sum_{Y\in S}(|Y|+c_k)
    \right].
\label{eq:alg-L-one-step-polydisc-overview}
\end{align}

Third, Cauchy's estimate and inclusion--exclusion convert
Eq.~\eqref{eq:alg-L-one-step-polydisc-overview} into a coordinatewise
Taylor-tail bound for the $R$-step product.  The required cutoff is
\begin{align}
    K
    =O_\rho\left(
    g_{\mathrm L}t\ell
    +\log\frac{|\mathfrak Y|}{\epsilon}
    \right)
    =O(g_{\mathrm L}t\sqrt{\mathsf L}+\mathsf L).
\label{eq:alg-L-K-overview}
\end{align}
At this step, the deleted-background argument uses the shared
physical-subblock parametrization: setting parameters to zero must remove
the same microscopic terms from every occurrence.

Fourth, within one short-time slice, the shared degree $m_Y$ is
constructed by merging the temporary degrees of the elementary
exponentials inside the local cells.  If
$m_{f,Y}$ is the degree generated inside the occurrence $f$, then
\begin{align}
    m_Y=\sum_f m_{f,Y}.
\label{eq:alg-L-degree-merging-overview}
\end{align}
For a fixed $Y$, the number of factors in this sum is at most
$\nu_{\mathrm{occ}}=O(1)$.  The merge is therefore an exact local
block-sparse operation.  It increases the arithmetic work only by a
polynomial in $K$ per active subblock and does not change the final shared
register dimension $K+1$.

Fifth, a retained coefficient has local cut budget $O(K)$.  Indeed, only a
constant number of physical subblocks can contribute microscopic terms near
a fixed cut, and each of their shared occupancies is at most $K$.  More
importantly, in each short-time slice a fixed cut meets at most one base
cell and one correction cell.  All other cells in either macro-layer are
strictly one-sided and can be absorbed before differentiating.  Applying
the cut-count decomposition in a window of $\ell$ cuts leaves only
\begin{align}
    O\left(\frac{K}{\ell}\right)
    =O(g_{\mathrm L}t+\sqrt{\mathsf L})
\label{eq:alg-L-crossing-order-overview}
\end{align}
explicit crossing insertions.  The resulting Liouville-Schmidt
representation of each retained coefficient is exact in the arithmetic
model.

Finally, the one-step polynomial is assembled cellwise: the elementary
coefficients are composed only inside each constant-depth local cell,
disjoint cells are tensorized within each macro-layer, and the two
macro-layers are then composed.  The retained degree sectors are propagated
through the $R$ time steps by the lower-set recursion
\begin{align}
    F_{r,K}
    :=\mathcal T_K^{\mathrm{sh}}
    \left[\Gamma_\Delta(\boldsymbol\lambda)^{\circ r}\right],
    \qquad
    F_{r,K}
    =\mathcal T_K^{\mathrm{sh}}
    \left[F_{r-1,K}\circ F_{1,K}\right].
\label{eq:alg-L-lower-set-overview}
\end{align}
At a fixed spatial bond, only
$O(B/\ell)=O(\sqrt{\mathsf L})$ shared degree registers are simultaneously
open.  Their block-diagonal overhead is therefore
\begin{align}
    \exp\left[
    O\left(\frac{B}{\ell}\log(K+1)\right)
    \right]
    =\exp[\widetilde O(\sqrt{\mathsf L})].
\label{eq:alg-L-register-overview-cost}
\end{align}
Combining this register cost with
Eq.~\eqref{eq:alg-L-crossing-order-overview} leads to the target bond
dimension and arithmetic construction time
\begin{align}
    D_{\mathrm{MPO}}
    &\le
    \exp\left[\widetilde O(g_{\mathrm L}t+\sqrt{\mathsf L})\right],
    \\
    T_{\mathrm{construct}}
    &\le
    n\exp\left[\widetilde O(g_{\mathrm L}t+\sqrt{\mathsf L})\right].
\label{eq:alg-L-overview-final-scaling}
\end{align}

\subsubsection{Analytic and algorithmic lemmas}

The statements below are arranged in the order in which they are used.
The first three propositions provide the analytic approximation.  The
remaining statements give the exact coefficient construction, the
fixed-cut Liouville-rank bound, and the degree-labeled MPO algorithm.

\begin{prop}[Uniform depth-two short-time approximation under anchored deletions]
\label{prop:alg-L-uniform-deleted-HHKL}
Let $\boldsymbol\mu\in\{0,1\}^{\mathfrak Y}$ be an anchored deletion
pattern and define
\begin{align}
    \mathcal L^{(\boldsymbol\mu)}
    :=\sum_Z\mu_{a(Z)}\mathcal L_Z,
    \qquad
    \Gamma_\Delta^{(\boldsymbol\mu)}
    :=\Gamma^{(2)}_\Delta(\boldsymbol\mu).
\label{eq:alg-L-mu-deleted-objects}
\end{align}
There exist constants $\Delta_0,B_0,C_{\mathrm H},\mu_{\mathrm H}>0$,
depending only on the local interaction data, such that for every
$0<\Delta\le\Delta_0$, every $B\ge B_0$, every anchored deletion pattern,
and every depth-two partition with $N_{\mathrm{int}}=M-1$ coarse-cell
interfaces,
\begin{align}
    \left\lVert
    e^{\Delta\mathcal L^{(\boldsymbol\mu)}}
    -\Gamma_\Delta^{(\boldsymbol\mu)}
    \right\rVert_\diamond
    \le
    \left[1+\eta_{\mathrm H}(B)\right]^{N_{\mathrm{int}}}-1,
    \qquad
    \eta_{\mathrm H}(B)
    \le C_{\mathrm H}e^{-\mu_{\mathrm H}B}.
\label{eq:alg-L-uniform-deleted-HHKL-resummed}
\end{align}
Consequently, whenever
$N_{\mathrm{int}}\eta_{\mathrm H}(B)\le1$,
\begin{align}
    \left\lVert
    e^{\Delta\mathcal L^{(\boldsymbol\mu)}}
    -\Gamma_\Delta^{(\boldsymbol\mu)}
    \right\rVert_\diamond
    \le
    \delta_{\mathrm H}(B),
    \qquad
    \delta_{\mathrm H}(B)
    \le
    2C_{\mathrm H}N_{\mathrm{int}}e^{-\mu_{\mathrm H}B},
\label{eq:alg-L-uniform-deleted-HHKL-bound}
\end{align}
and
\begin{align}
    \left\lVert
    \Gamma_\Delta^{(\boldsymbol\mu)}
    \right\rVert_\diamond
    \le1+\delta_{\mathrm H}(B).
\label{eq:alg-L-uniform-deleted-HHKL-norm}
\end{align}
The constants are uniform in the system size, the deletion pattern, and the
positions of the deleted subblocks.
\end{prop}

\textit{Proof.}
If $N_{\mathrm{int}}=0$, the product is exact.  We therefore assume
$N_{\mathrm{int}}\ge1$ in the proof below.
Because every deleted unit is a complete local Lindblad term,
$\mathcal L^{(\boldsymbol\mu)}$ is again a Lindblad generator and its
locality and extensivity constants do not increase.  We therefore suppress
$\boldsymbol\mu$ and prove the claim for an arbitrary generator
$\mathcal K$ satisfying the same local bounds.  Apply the depth-two
partition above and write
\begin{align}
    \mathcal K
    =\mathcal K^{(0)}+\sum_{j=1}^{N_{\mathrm{int}}}\mathcal V_j,
    \qquad
    \mathcal K^{(0)}=\sum_{a=1}^{M}\mathcal K_a,
\label{eq:alg-L-depth-two-proof-decomposition}
\end{align}
with local patch backgrounds $\mathcal A_j$ as in
Eq.~\eqref{eq:alg-L-depth-two-patch-background}.  Define the intermediate
Lindblad generators
\begin{align}
    \mathcal K^{(j)}
    :=\mathcal K^{(0)}+\sum_{b=1}^{j}\mathcal V_b,
    \qquad
    \mathcal K^{(0)}\text{ as above},
\label{eq:alg-L-depth-two-intermediate-generators}
\end{align}
and the exact and localized relative factors
\begin{align}
    \Phi_j
    &:={}
    e^{-\Delta\mathcal K^{(j-1)}}
    \circ e^{\Delta\mathcal K^{(j)}},
    \\
    \widetilde\Phi_j
    &:={}
    e^{-\Delta\mathcal A_j}
    \circ e^{\Delta(\mathcal A_j+\mathcal V_j)}.
\label{eq:alg-L-depth-two-relative-factors}
\end{align}
The exact factors telescope:
\begin{align}
    e^{\Delta\mathcal K}
    =e^{\Delta\mathcal K^{(0)}}
     \circ\Phi_1\circ\cdots\circ\Phi_{N_{\mathrm{int}}}.
\label{eq:alg-L-depth-two-exact-telescope}
\end{align}
On the other hand,
\begin{align}
    e^{\Delta\mathcal K^{(0)}}
    &=\mathop{\circ}_{a=1}^{M}e^{\Delta\mathcal K_a},
    \\
    \widetilde\Phi_1\circ\cdots\circ
    \widetilde\Phi_{N_{\mathrm{int}}}
    &=\mathop{\circ}_{j=1}^{N_{\mathrm{int}}}
      \left[
      e^{-\Delta\mathcal A_j}
      \circ e^{\Delta(\mathcal A_j+\mathcal V_j)}
      \right].
\label{eq:alg-L-depth-two-local-telescope}
\end{align}
The factors in each line have pairwise disjoint supports.  Hence the
localized product in
Eq.~\eqref{eq:alg-L-depth-two-local-telescope} is exactly the depth-two
approximant $\Gamma^{(2)}_\Delta(\mathcal K)$, independently of the order in
which the interface factors are written.

Set
\begin{align}
    \mathcal P_j^{\mathrm{ex}}
    &:=e^{\Delta\mathcal K^{(j)}},
    &
    \widetilde{\mathcal P}_j
    &:={}
    e^{\Delta\mathcal K^{(0)}}
    \circ\widetilde\Phi_1\circ\cdots\circ\widetilde\Phi_j,
\end{align}
with
$\mathcal P_0^{\mathrm{ex}}=\widetilde{\mathcal P}_0
=e^{\Delta\mathcal K^{(0)}}$, and define
\begin{align}
    \mathcal R_j
    :=\widetilde{\mathcal P}_j
      \circ(\mathcal P_j^{\mathrm{ex}})^{-1},
    \qquad
    \varepsilon_j:=\lVert\mathcal R_j-\operatorname{id}\rVert_\diamond.
\end{align}
For $j\ge1$, let
\begin{align}
    \Psi_j
    :=\mathcal P_{j-1}^{\mathrm{ex}}
      \circ\widetilde\Phi_j
      \circ\Phi_j^{-1}
      \circ(\mathcal P_{j-1}^{\mathrm{ex}})^{-1}.
\label{eq:alg-L-HHKL-local-defect}
\end{align}
Then $\mathcal R_j=\mathcal R_{j-1}\circ\Psi_j$, and therefore
\begin{align}
    \varepsilon_j
    \le(1+\eta_j)\varepsilon_{j-1}+\eta_j,
    \qquad
    \eta_j:=\lVert\Psi_j-\operatorname{id}\rVert_\diamond.
\label{eq:alg-L-relative-error-recursion}
\end{align}
It remains to bound one interface defect uniformly.

The relative-evolution identity
\begin{align}
    e^{-\Delta\mathcal A}
    \circ e^{\Delta(\mathcal A+\mathcal B)}
    =\mathcal T\exp\left[
    \int_0^\Delta
    e^{-s\operatorname{ad}^{\circ}_{\mathcal A}}(\mathcal B)
    \,ds\right]
\label{eq:alg-L-relative-evolution-identity-HHKL}
\end{align}
shows that, after conjugation by $\mathcal P_{j-1}^{\mathrm{ex}}$, the exact and localized
relative factors are generated by
\begin{align}
    \mathcal G_j(s)
    &:={}
    e^{(\Delta-s)\operatorname{ad}^{\circ}_{\mathcal K^{(j-1)}}}
    (\mathcal V_j),
    \\
    \widetilde{\mathcal G}_j(s)
    &:={}
    e^{\Delta\operatorname{ad}^{\circ}_{\mathcal K^{(j-1)}}}
    e^{-s\operatorname{ad}^{\circ}_{\mathcal A_j}}
    (\mathcal V_j).
\label{eq:alg-L-two-dressed-boundary-generators}
\end{align}
The ordered-exponential Duhamel formula gives
\begin{align}
    \eta_j
    \le
    \exp\left[
    \int_0^\Delta
    \bigl(
    \lVert\mathcal G_j(s)\rVert_\diamond
    +\lVert\widetilde{\mathcal G}_j(s)\rVert_\diamond
    \bigr)\,ds\right]
    \int_0^\Delta
    \lVert\mathcal G_j(s)-\widetilde{\mathcal G}_j(s)\rVert_\diamond
    \,ds.
\label{eq:alg-L-ordered-exponential-comparison}
\end{align}

The three-piece geometry is used now.  Every local term in
$\mathcal K^{(j-1)}-\mathcal A_j$ is at distance at least $B-c_k$ from
$\operatorname{supp}(\mathcal V_j)$.  This includes the previously added
interface term $\mathcal V_{j-1}$: it lies at the opposite boundary of the
coarse cell $C_j$, separated from the $j$th interface by the middle region
and the two collars.  Set
\begin{align}
    m_B:=\left\lfloor\frac{B-c_k}{k}\right\rfloor.
\end{align}
A connected nested composition commutator starting from $\mathcal V_j$
therefore cannot distinguish $\mathcal K^{(j-1)}$ from $\mathcal A_j$ in
total commutator order smaller than $m_B$.  More explicitly,
\begin{align}
    \mathcal G_j(s)
    &=\sum_{m=0}^{\infty}
      \frac{(\Delta-s)^m}{m!}
      (\operatorname{ad}^{\circ}_{\mathcal K^{(j-1)}})^m
      (\mathcal V_j),
    \\
    \widetilde{\mathcal G}_j(s)
    &=\sum_{a,b\ge0}
      \frac{\Delta^a(-s)^b}{a!b!}
      (\operatorname{ad}^{\circ}_{\mathcal K^{(j-1)}})^a
      (\operatorname{ad}^{\circ}_{\mathcal A_j})^b
      (\mathcal V_j).
\label{eq:alg-L-two-dressed-generator-series}
\end{align}
For $a+b<m_B$, every nonzero connected word in the second line is unchanged
if each occurrence of $\mathcal K^{(j-1)}$ is replaced by $\mathcal A_j$.
The same replacement is valid in the first line for $m<m_B$.  Summing over
$a+b=m$ and using
$\sum_{a+b=m}\Delta^a(-s)^b/(a!b!)=(\Delta-s)^m/m!$
shows coefficient by coefficient that the two series agree through total
order $m_B-1$.

The support-growth proof of
Lemma~\ref{lem:static-liouvillian-commutator} is wordwise and does not
require the generator appearing at each commutator level to be the same.
Applied to mixed words whose letters are local sub-sums of
$\mathcal K^{(j-1)}$ and $\mathcal A_j$, it implies that there are
constants $c_\partial,c_{\mathrm c}>0$, depending only on the fixed local
interaction data, such that for every connected word of length $m$ starting
from one interface generator,
\begin{align}
    \left\lVert
    \operatorname{ad}^{\circ}_{\mathcal S_m}\cdots
    \operatorname{ad}^{\circ}_{\mathcal S_1}(\mathcal V_j)
    \right\rVert_\diamond
    \le
    c_\partial g_{\mathrm L}
    (c_{\mathrm c}g_{\mathrm L}k)^m m!,
\label{eq:alg-L-mixed-composition-commutator-bound}
\end{align}
where every $\mathcal S_a$ is a local sub-sum of
$\mathcal K^{(j-1)}$ or $\mathcal A_j$.  In contrast to a norm bound on a
whole coarse cell, the prefactor in
Eq.~\eqref{eq:alg-L-mixed-composition-commutator-bound} is independent of
$B$, because the initial generator $\mathcal V_j$ is supported only in a
fixed interaction-range neighborhood of one interface.  Consequently,
for $0\le s\le\Delta$,
\begin{align}
    \lVert\mathcal G_j(s)-\widetilde{\mathcal G}_j(s)\rVert_\diamond
    &\le
    c_1g_{\mathrm L}
    \frac{(c_2g_{\mathrm L}k\Delta)^{m_B}}
         {1-c_2g_{\mathrm L}k\Delta},
    \\
    \lVert\mathcal G_j(s)\rVert_\diamond
    +\lVert\widetilde{\mathcal G}_j(s)\rVert_\diamond
    &\le
    \frac{c_3g_{\mathrm L}}
         {1-c_2g_{\mathrm L}k\Delta}.
\label{eq:alg-L-depth-two-dressed-generator-bounds}
\end{align}
Choose $\Delta_0$ so that
$c_2g_{\mathrm L}k\Delta_0\le1/4$.  Substitution into
Eq.~\eqref{eq:alg-L-ordered-exponential-comparison} yields
\begin{align}
    \eta_j
    \le
    c_4g_{\mathrm L}\Delta
    e^{c_5g_{\mathrm L}\Delta}
    (c_2g_{\mathrm L}k\Delta)^{m_B}
    \le C_{\mathrm H}e^{-\mu_{\mathrm H}B}.
\label{eq:alg-L-one-interface-HHKL-error}
\end{align}
The estimate is uniform under anchored deletions: deleting terms can only
reduce the local commutator majorant and cannot reduce the geometric number
of commutators required to reach the exterior of the interface patch.

Iterating Eq.~\eqref{eq:alg-L-relative-error-recursion} from
$\varepsilon_0=0$ gives
\begin{align}
    \varepsilon_{N_{\mathrm{int}}}
    \le
    [1+\eta_{\mathrm H}(B)]^{N_{\mathrm{int}}}-1,
    \qquad
    \eta_{\mathrm H}(B):=\sup_j\eta_j.
\label{eq:alg-L-HHKL-relative-error-resummation}
\end{align}
Since $\mathcal P_{N_{\mathrm{int}}}^{\mathrm{ex}}=e^{\Delta\mathcal K}$ is CPTP,
\begin{align}
    \lVert\Gamma^{(2)}_\Delta(\mathcal K)
           -e^{\Delta\mathcal K}\rVert_\diamond
    =\lVert(\mathcal R_{N_{\mathrm{int}}}-\operatorname{id})
           \circ\mathcal P_{N_{\mathrm{int}}}^{\mathrm{ex}}\rVert_\diamond
    \le\varepsilon_{N_{\mathrm{int}}}.
\end{align}
This proves
Eq.~\eqref{eq:alg-L-uniform-deleted-HHKL-resummed}.  The linear estimate
follows from $(1+x)^N-1\le e^{Nx}-1\le2Nx$ for $Nx\le1$, and the norm bound
follows from the triangle inequality and
$\lVert e^{\Delta\mathcal L^{(\boldsymbol\mu)}}\rVert_\diamond=1$.
\hfill\(\square\)

\textit{Remark on depth.}
The ordered list $\Phi_1,\ldots,\Phi_{N_{\mathrm{int}}}$ and its localized
counterpart appear only in the telescoping error proof.  The localized
corrections $\widetilde\Phi_j$ have pairwise disjoint supports and are
assembled as the single correction macro-layer in
Eq.~\eqref{eq:alg-L-depth-two-HHKL-product}.  In particular, no derivative
or Schmidt-rank estimate below is applied to the $O(n)$-depth telescoping
sequence.

\begin{lemma}[Adjoint bound for a short ordered product]
\label{lem:alg-L-ordered-product-adjoint}
Let
\begin{align}
    \mathcal P
    :=e^{s_M\mathcal A_M}\circ\cdots\circ e^{s_1\mathcal A_1},
    \qquad
    \mathcal A_j=\sum_{Z:\,|Z|\le k}\mathcal A_{j,Z},
\end{align}
where the scalars $s_j$ may have either sign or may be complex.  Define the
cumulative local action
\begin{align}
    \alpha(\mathcal P)
    :=
    \sup_{i\in\Lambda}
    \sum_{j=1}^{M}|s_j|
    \sum_{Z\ni i}\lVert\mathcal A_{j,Z}\rVert_\diamond.
\label{eq:alg-L-cumulative-local-action}
\end{align}
For a local sub-sum
$\mathcal B=\sum_{W:\,|W|\le k}\mathcal B_W$, set
$\lVert\mathcal B\rVert_{\mathrm{loc},\diamond}
:=\sum_W\lVert\mathcal B_W\rVert_\diamond$.
There is a constant $C_{\mathrm{ad}}=C_{\mathrm{ad}}(k)$ such that, whenever
$C_{\mathrm{ad}}\alpha(\mathcal P)<1$,
\begin{align}
    \left\lVert
    \mathcal P\circ\mathcal B\circ\mathcal P^{-1}
    \right\rVert_\diamond
    \le
    \frac{
    \lVert\mathcal B\rVert_{\mathrm{loc},\diamond}}
    {1-C_{\mathrm{ad}}\alpha(\mathcal P)}.
\label{eq:alg-L-ordered-product-adjoint-bound}
\end{align}
One may take, for example, $C_{\mathrm{ad}}=2e^2k$.
\end{lemma}

\textit{Proof.}
Using the composition commutator from
Eq.~\eqref{eq:composition-commutator}, expand
\begin{align}
    \mathcal P\circ\mathcal B\circ\mathcal P^{-1}
    =
    e^{s_M\operatorname{ad}^{\circ}_{\mathcal A_M}}
    \cdots
    e^{s_1\operatorname{ad}^{\circ}_{\mathcal A_1}}
    (\mathcal B).
\label{eq:alg-L-ordered-adjoint-expansion}
\end{align}
Introduce occurrence labels $a=(j,Z)$ and weights
\begin{align}
    w_a:=|s_j|\lVert\mathcal A_{j,Z}\rVert_\diamond.
\end{align}
Fix one initial term $\mathcal B_W$.  After expanding all exponentials and
all local sums, a total order-$m$ contribution is a nested composition
commutator of $m$ local terms.  Such a commutator vanishes unless the
intersection graph of the $m+1$ supports
\begin{align}
    W,Z_{a_1},\ldots,Z_{a_m}
\end{align}
is connected.  Moreover,
\begin{align}
    \left\lVert
    [\mathcal A_{a_m},[\cdots,[\mathcal A_{a_1},
    \mathcal B_W]_\circ\cdots]_\circ]_\circ
    \right\rVert_\diamond
    \le
    2^m\lVert\mathcal B_W\rVert_\diamond
    \prod_{r=1}^{m}\lVert\mathcal A_{a_r}\rVert_\diamond.
\label{eq:alg-L-word-commutator-bound}
\end{align}
For each multiset of occurrence labels, the total scalar coefficient in the
ordered product of exponentials is the inverse product of the multiplicity
factorials.  Equivalently, it is obtained by averaging over all $m!$
orderings with coefficient $1/m!$.  Forgetting the factor-order restriction
therefore only enlarges the absolute sum, and the norm of the total
order-$m$ contribution is at most
\begin{align}
&
    \frac{2^m}{m!}
    \sum_W\lVert\mathcal B_W\rVert_\diamond
    \sum_{a_1,\ldots,a_m}
    \mathbf 1\!\left[
    \{W,Z_{a_1},\ldots,Z_{a_m}\}\text{ is connected}
    \right]
    \prod_{r=1}^{m}w_{a_r}.
\label{eq:alg-L-connected-cluster-majorant}
\end{align}

We now use a rooted-tree bound.  A connected graph on the labeled vertices
$\{0,1,\ldots,m\}$ contains a spanning tree.  For a fixed spanning tree,
orient all edges away from the root $0$, whose support is $W$.  Once the
support of a parent vertex has been fixed, the total weight of all possible
children intersecting that support is at most
\begin{align}
    \sum_{a:\,Z_a\cap Z_{\mathrm{parent}}\ne\emptyset}w_a
    \le
    \sum_{i\in Z_{\mathrm{parent}}}
    \sum_{a:\,i\in Z_a}w_a
    \le k\alpha(\mathcal P).
\label{eq:alg-L-tree-child-sum}
\end{align}
Cayley's formula gives $(m+1)^{m-1}$ labeled trees.  Hence the expression in
Eq.~\eqref{eq:alg-L-connected-cluster-majorant} is bounded by
\begin{align}
    \frac{2^m(m+1)^{m-1}}{m!}
    [k\alpha(\mathcal P)]^m
    \lVert\mathcal B\rVert_{\mathrm{loc},\diamond}.
\end{align}
For $m\ge1$,
$(m+1)^{m-1}/m!\le e^{m+1}$, and therefore the order-$m$ term is at most
\begin{align}
    [2e^2k\alpha(\mathcal P)]^m
    \lVert\mathcal B\rVert_{\mathrm{loc},\diamond}.
\end{align}
The $m=0$ term is $\mathcal B$.  Summing the resulting geometric series
proves Eq.~\eqref{eq:alg-L-ordered-product-adjoint-bound}.
\hfill\(\square\)

\begin{prop}[Shared-block cocycle factorization and polydisc stability]
\label{prop:alg-L-shared-cocycle}
Fix $\rho>1$.  There exist $\Delta_\rho>0$ and $C_\rho>0$, depending only
on $\rho$ and the local interaction data, such that the following holds for
$0<\Delta\le\Delta_\rho$.  Assume also that $B$ lies in the linear-error
regime of Proposition~\ref{prop:alg-L-uniform-deleted-HHKL}, namely
$N_{\mathrm{int}}\eta_{\mathrm H}(B)\le1$.  For every
$S\subseteq\mathfrak Y$ and every
$\boldsymbol z_S$ satisfying $|z_Y|\le\rho$, there is an exact
factorization
\begin{align}
    \Gamma_\Delta(\boldsymbol z_S,\boldsymbol1_{S^c})
    =
    \Gamma_\Delta(\boldsymbol0_S,\boldsymbol1_{S^c})
    \circ\mathcal W_{S,\Delta}(\boldsymbol z),
\label{eq:alg-L-shared-cocycle-factorization}
\end{align}
where
\begin{align}
    \mathcal W_{S,\Delta}(\boldsymbol z)
    =\mathcal T\exp\left[
    \int_0^\Delta
    \widetilde{\mathcal V}_S(x,\boldsymbol z)\,dx
    \right]
\label{eq:alg-L-shared-cocycle-W}
\end{align}
for an algebraic cocycle generator satisfying
\begin{align}
    \sup_{0\le x\le\Delta}
    \left\lVert
    \widetilde{\mathcal V}_S(x,\boldsymbol z)
    \right\rVert_\diamond
    \le
    C_\rho g_{\mathrm L}
    \sum_{Y\in S}(|Y|+c_k).
\label{eq:alg-L-shared-cocycle-generator-bound}
\end{align}
Consequently,
\begin{align}
&
    \sup_{|z_Y|\le\rho\ (Y\in S)}
    \left\lVert
    \Gamma_\Delta(\boldsymbol z_S,\boldsymbol1_{S^c})
    \right\rVert_\diamond
\nonumber\\
&\qquad\le
    \left[1+\delta_{\mathrm H}(B)\right]
    \exp\left[
    C_\rho g_{\mathrm L}\Delta
    \sum_{Y\in S}(|Y|+c_k)
    \right].
\label{eq:alg-L-shared-one-step-polydisc}
\end{align}
The constant $C_\rho$ is independent of $B$, the system size, the number of
connected components of $S$, and the diameter of
$\bigcup_{Y\in S}Y$.
\end{prop}

\textit{Proof.}
The useful interpolation is in parameter space rather than in physical
time.  For $0\le\theta\le1$, define
\begin{align}
    \lambda_Y(\theta)
    :=
    \begin{cases}
        \theta z_Y,&Y\in S,\\
        1,&Y\notin S,
    \end{cases}
    \qquad
    \Gamma_\theta
    :=\Gamma_\Delta(\boldsymbol\lambda(\theta)).
\label{eq:alg-L-theta-interpolation}
\end{align}
Set
\begin{align}
    \mathcal B_S
    :=\Gamma_0
    =\Gamma_\Delta(\boldsymbol0_S,\boldsymbol1_{S^c}),
    \qquad
    \mathcal W_\theta
    :=\mathcal B_S^{-1}\circ\Gamma_\theta.
\label{eq:alg-L-Wtheta-definition}
\end{align}
All factors are exponentials, so the inverses in this algebraic identity
exist.  We have $\mathcal W_0=\operatorname{id}$ and
$\Gamma_\theta=\mathcal B_S\circ\mathcal W_\theta$.

Write
\begin{align}
    \mathcal X_{f,\theta}
    &:=\sigma_f\Delta
    \mathcal L_f(\boldsymbol\lambda(\theta)),
    &
    \mathcal U_{f,\theta}
    &:=e^{\mathcal X_{f,\theta}},
    \\
    \mathcal V_{f,S}(\boldsymbol z)
    &:=\sum_{Y\in S}z_Y\mathcal L_{f,Y},
    &
    \mathcal P_{>f,\theta}
    &:=\mathcal U_{N_{\mathrm f},\theta}
    \circ\cdots\circ\mathcal U_{f+1,\theta}.
\label{eq:alg-L-cocycle-factor-notation}
\end{align}
The left logarithmic derivative of the product is
\begin{align}
    \mathcal K_\theta
    &:=(\partial_\theta\Gamma_\theta)
    \circ\Gamma_\theta^{-1}
    \nonumber\\
    &=
    \sum_{f=1}^{N_{\mathrm f}}
    \mathcal P_{>f,\theta}
    \circ\mathcal J_{f,\theta}
    \circ\mathcal P_{>f,\theta}^{-1},
\label{eq:alg-L-product-logarithmic-derivative}
\end{align}
where
\begin{align}
    \mathcal J_{f,\theta}
    &:=(\partial_\theta\mathcal U_{f,\theta})
    \circ\mathcal U_{f,\theta}^{-1}
    \nonumber\\
    &=
    \int_0^1
    e^{u\mathcal X_{f,\theta}}
    \circ
    \left[\sigma_f\Delta
    \mathcal V_{f,S}(\boldsymbol z)\right]
    \circ
    e^{-u\mathcal X_{f,\theta}}\,du.
\label{eq:alg-L-single-factor-log-derivative}
\end{align}
Differentiating Eq.~\eqref{eq:alg-L-Wtheta-definition} gives the exact ODE
\begin{align}
    \partial_\theta\mathcal W_\theta
    =\mathcal V_\theta\circ\mathcal W_\theta,
    \qquad
    \mathcal V_\theta
    :=\mathcal B_S^{-1}
    \circ\mathcal K_\theta\circ\mathcal B_S.
\label{eq:alg-L-Wtheta-ODE}
\end{align}
Combining Eqs.~\eqref{eq:alg-L-product-logarithmic-derivative} and
\eqref{eq:alg-L-single-factor-log-derivative}, every summand in
$\mathcal V_\theta$ is an adjoint transform of
$\sigma_f\Delta\mathcal V_{f,S}(\boldsymbol z)$ by the ordered product
\begin{align}
    \mathcal Q_{f,u,\theta}
    :=
    \mathcal B_S^{-1}
    \circ\mathcal P_{>f,\theta}
    \circ e^{u\mathcal X_{f,\theta}}.
\label{eq:alg-L-Q-product}
\end{align}

To estimate this adjoint action, write the inverse background explicitly as
$\mathcal B_S^{-1}=\mathcal U_{1,0}^{-1}\circ\cdots\circ
\mathcal U_{N_{\mathrm f},0}^{-1}$.  Hence
$\mathcal Q_{f,u,\theta}$ is an ordered product containing at most one
background inverse occurrence and at most one deformed occurrence of each
elementary exponential, together with one fractional occurrence of the
$f$th elementary exponential.  By the
two-macro-layer cell geometry and
Eq.~\eqref{eq:alg-L-occurrence-multiplicity}, every microscopic term belongs
to only $O(1)$ elementary occurrences.  It therefore enters the absolute
aggregate interaction associated with $\mathcal Q_{f,u,\theta}$ only
$O(1)$ times.  Since
$|\lambda_Y(\theta)|\le\max\{1,\rho\}$, there is a constant
$c_{\mathrm{occ}}$ such that
\begin{align}
    \alpha(\mathcal Q_{f,u,\theta})
    \le
    c_{\mathrm{occ}}\rho g_{\mathrm L}\Delta
\label{eq:alg-L-Q-cumulative-action}
\end{align}
for all $f,u,\theta$, independently of $B$, $n$, and the number of local cells.  Choose $\Delta_\rho$ so that the right-hand side of
Eq.~\eqref{eq:alg-L-Q-cumulative-action} is at most $1/(2C_{\mathrm{ad}})$.  Then
Lemma~\ref{lem:alg-L-ordered-product-adjoint} gives
\begin{align}
    \lVert\mathcal V_\theta\rVert_\diamond
    \le
    2\Delta
    \sum_{f=1}^{N_{\mathrm f}}
    \lVert\mathcal V_{f,S}(\boldsymbol z)
    \rVert_{\mathrm{loc},\diamond}.
\label{eq:alg-L-Vtheta-first-bound}
\end{align}

It remains to sum the active local terms.  By the anchor rule, every term
anchored in $Y$ is supported in a fixed-range collar $Y^{+k}$.  Therefore,
\begin{align}
    \sum_f
    \lVert\mathcal V_{f,S}(\boldsymbol z)
    \rVert_{\mathrm{loc},\diamond}
    &\le
    \rho\nu_{\mathrm{occ}}
    \sum_{Y\in S}
    \sum_{Z:\,a(Z)=Y}
    \lVert\mathcal L_Z\rVert_\diamond
    \nonumber\\
    &\le
    C\rho g_{\mathrm L}
    \sum_{Y\in S}(|Y|+c_k).
\label{eq:alg-L-active-local-norm-sum}
\end{align}
Substituting this into
Eq.~\eqref{eq:alg-L-Vtheta-first-bound} yields
\begin{align}
    \sup_{0\le\theta\le1}
    \lVert\mathcal V_\theta\rVert_\diamond
    \le
    C_\rho g_{\mathrm L}\Delta
    \sum_{Y\in S}(|Y|+c_k).
\label{eq:alg-L-Vtheta-final-bound}
\end{align}
The solution of Eq.~\eqref{eq:alg-L-Wtheta-ODE} is
\begin{align}
    \mathcal W_1
    =\mathcal T_\theta
    \exp\left[\int_0^1\mathcal V_\theta\,d\theta\right].
\end{align}
After the change of variables $x=\Delta\theta$, define
$\widetilde{\mathcal V}_S(x,\boldsymbol z)
:=\Delta^{-1}\mathcal V_{x/\Delta}$.  This gives
Eqs.~\eqref{eq:alg-L-shared-cocycle-factorization} and
\eqref{eq:alg-L-shared-cocycle-W}, while
Eq.~\eqref{eq:alg-L-Vtheta-final-bound} gives
Eq.~\eqref{eq:alg-L-shared-cocycle-generator-bound}.
Finally, Gronwall's inequality and
Proposition~\ref{prop:alg-L-uniform-deleted-HHKL} imply
\begin{align}
    \lVert\mathcal W_{S,\Delta}(\boldsymbol z)\rVert_\diamond
    \le
    \exp\left[
    C_\rho g_{\mathrm L}\Delta
    \sum_{Y\in S}(|Y|+c_k)
    \right],
\end{align}
which proves Eq.~\eqref{eq:alg-L-shared-one-step-polydisc}.
\hfill\(\square\)

\begin{prop}[Coordinatewise Taylor cutoff for the repeated depth-two product]
\label{prop:alg-L-coordinatewise-tail}
Assume Proposition~\ref{prop:alg-L-shared-cocycle}.  Set
\begin{align}
    A_R:=\left[1+\delta_{\mathrm H}(B)\right]^R,
    \qquad
    q_K(\rho)
    :=\sum_{m=K+1}^\infty\rho^{-m}
    =\frac{\rho^{-(K+1)}}{1-\rho^{-1}}.
\label{eq:alg-L-Cauchy-quantities}
\end{align}
Then
\begin{align}
&
    \left\lVert
    F_R(\boldsymbol1)
    -\left.
    \mathcal T_K^{\mathrm{sh}}F_R(\boldsymbol\lambda)
    \right|_{\boldsymbol\lambda=\boldsymbol1}
    \right\rVert_\diamond
\nonumber\\
&\quad\le
    A_R
    \left\{
    \prod_{Y\in\mathfrak Y}
    \left[
    1+
    e^{C_\rho g_{\mathrm L}t(|Y|+c_k)}q_K(\rho)
    \right]
    -1
    \right\}.
\label{eq:alg-L-coordinatewise-tail-product}
\end{align}
In particular, if $|Y|\le\ell$ for every $Y$, then
\begin{align}
&
    \left\lVert
    F_R(\boldsymbol1)
    -\left.
    \mathcal T_K^{\mathrm{sh}}F_R
    \right|_{\boldsymbol1}
    \right\rVert_\diamond
\nonumber\\
&\quad\le
    A_R
    \left\{
    \left[
    1+e^{C_\rho g_{\mathrm L}t(\ell+c_k)}q_K(\rho)
    \right]^{|\mathfrak Y|}
    -1
    \right\}.
\label{eq:alg-L-coordinatewise-tail-uniform}
\end{align}
Hence, for $0<\epsilon_{\mathrm{Tay}}\le1$, provided $A_R=O(1)$, a
truncation error at most $\epsilon_{\mathrm{Tay}}$ is ensured by
\begin{align}
    (K+1)\log\rho
    \ge
    C_\rho g_{\mathrm L}t(\ell+c_k)
    +\log\left[
    \frac{2A_R|\mathfrak Y|}
    {(1-\rho^{-1})\epsilon_{\mathrm{Tay}}}
    \right],
\label{eq:alg-L-K-sufficient}
\end{align}
up to an absolute adjustment of the constant in the logarithm.  In
particular, a sufficient cutoff can be chosen with
\begin{align}
    K
    =O_\rho\left(
    g_{\mathrm L}t\ell
    +\log\frac{|\mathfrak Y|}{\epsilon_{\mathrm{Tay}}}
    \right).
\label{eq:alg-L-K-scaling-proposition}
\end{align}
\end{prop}

\textit{Proof.}
For a nonempty subset $S\subseteq\mathfrak Y$, let
$\mathcal R_S$ denote the sum of all Taylor coefficients of $F_R$ for
which $m_Y>K$ for every $Y\in S$, with no restriction on the coordinates
in $S^c$.  Inclusion--exclusion for the union of the events
$\{m_Y>K\}$ gives the exact identity
\begin{align}
    F_R(\boldsymbol1)
    -\left.\mathcal T_K^{\mathrm{sh}}F_R\right|_{\boldsymbol1}
    =
    \sum_{\emptyset\ne S\subseteq\mathfrak Y}
    (-1)^{|S|+1}\mathcal R_S.
\label{eq:alg-L-coordinatewise-IE}
\end{align}

Fix $S$ and define the Banach-valued analytic function
\begin{align}
    G_S(\boldsymbol z_S)
    :=F_R(\boldsymbol z_S,\boldsymbol1_{S^c}).
\end{align}
Because all variables in $S^c$ have already been set to one, the coefficient
of $\prod_{Y\in S}z_Y^{m_Y}$ in $G_S$ is precisely the sum of the
coefficients of $F_R$ with the prescribed active degrees and arbitrary
inactive degrees.  Cauchy's integral formula on the torus
$|z_Y|=\rho$ therefore gives
\begin{align}
    \lVert\mathcal R_S\rVert_\diamond
    \le
    \sup_{|z_Y|=\rho\ (Y\in S)}
    \lVert G_S(\boldsymbol z_S)\rVert_\diamond
    q_K(\rho)^{|S|}.
\label{eq:alg-L-Cauchy-tail-S}
\end{align}
By submultiplicativity of the diamond norm and
Proposition~\ref{prop:alg-L-shared-cocycle},
\begin{align}
    \sup_{|z_Y|\le\rho}
    \lVert G_S(\boldsymbol z_S)\rVert_\diamond
    &\le
    \left
    \{
    [1+\delta_{\mathrm H}(B)]
    \exp\left[
    C_\rho g_{\mathrm L}\Delta
    \sum_{Y\in S}(|Y|+c_k)
    \right]
    \right\}^R
    \nonumber\\
    &=
    A_R
    \prod_{Y\in S}
    e^{C_\rho g_{\mathrm L}t(|Y|+c_k)}.
\label{eq:alg-L-R-step-polydisc-bound}
\end{align}
Combining Eqs.~\eqref{eq:alg-L-coordinatewise-IE}--
\eqref{eq:alg-L-R-step-polydisc-bound} and summing over $S$ gives
\begin{align}
&
    \left\lVert
    F_R(\boldsymbol1)
    -\left.\mathcal T_K^{\mathrm{sh}}F_R\right|_{\boldsymbol1}
    \right\rVert_\diamond
    \nonumber\\
&\quad\le
    A_R
    \sum_{\emptyset\ne S\subseteq\mathfrak Y}
    \prod_{Y\in S}
    \left[
    e^{C_\rho g_{\mathrm L}t(|Y|+c_k)}q_K(\rho)
    \right],
\end{align}
which is exactly
Eq.~\eqref{eq:alg-L-coordinatewise-tail-product}.  The uniform-size version
follows by replacing every $|Y|$ by $\ell$.

For completeness, set
\begin{align}
    x:=e^{C_\rho g_{\mathrm L}t(\ell+c_k)}q_K(\rho),
    \qquad N_Y:=|\mathfrak Y|.
\end{align}
If $N_Yx\le1$, then
$(1+x)^{N_Y}-1\le e^{N_Yx}-1\le2N_Yx$.  Since $A_R\ge1$ and $0<\epsilon_{\mathrm{Tay}}\le1$, the condition
$x\le\epsilon_{\mathrm{Tay}}/(2A_RN_Y)$ also implies $N_Yx\le1$.
Thus this condition is sufficient and is equivalent to
Eq.~\eqref{eq:alg-L-K-sufficient}, up to an inessential absolute constant.
This also proves Eq.~\eqref{eq:alg-L-K-scaling-proposition}.
\hfill\(\square\)

\begin{prop}[Exact local-cell coefficients and shared-degree merging]
\label{prop:alg-L-exact-degree-merging}
For each elementary exponential occurrence, consider the absolutely convergent expansion
\begin{align}
    e^{\sigma_f\Delta\mathcal L_f(\boldsymbol\lambda)}
    =
    \sum_{\boldsymbol r_f\in\mathbb Z_{\ge0}^{\mathfrak Y}}
    \mathcal E_{f,\boldsymbol r_f}
    \boldsymbol\lambda^{\boldsymbol r_f}.
\label{eq:alg-L-factor-coefficient-expansion}
\end{align}
Let $|\boldsymbol r_f|_1:=\sum_Yr_{f,Y}$.  The factor coefficients can be
generated without enumerating individual noncommutative words.  Namely,
\begin{align}
    \mathcal E_{f,\boldsymbol r_f}
    =\frac{(\sigma_f\Delta)^{|\boldsymbol r_f|_1}}
    {|\boldsymbol r_f|_1!}\,
    \mathcal P_{f,\boldsymbol r_f},
    \qquad
    \mathcal P_{f,\boldsymbol0}=\operatorname{id},
\label{eq:alg-L-factor-coefficient-word-normalization}
\end{align}
where the word coefficients obey the exact recursion
\begin{align}
    \mathcal P_{f,\boldsymbol r_f}
    =\sum_{Y:\,r_{f,Y}\ge1}
    \mathcal L_{f,Y}\circ
    \mathcal P_{f,\boldsymbol r_f-\boldsymbol e_Y}.
\label{eq:alg-L-factor-coefficient-recursion}
\end{align}
If $\mathfrak Y_f:=\{Y:\mathcal L_{f,Y}\ne0\}$, all sectors with
$0\le r_{f,Y}\le K$ can therefore be generated using at most
\begin{align}
    |\mathfrak Y_f|(K+1)^{|\mathfrak Y_f|}
\label{eq:alg-L-factor-coefficient-update-count}
\end{align}
coefficient updates, followed by exact MPO rank reduction.

The coefficient of the one-step product is exactly
\begin{align}
    [\boldsymbol\lambda^{\boldsymbol m}]
    \Gamma_\Delta(\boldsymbol\lambda)
    =
    \sum_{\substack{\boldsymbol r_1,\ldots,
    \boldsymbol r_{N_{\mathrm f}}\\
    \sum_f r_{f,Y}=m_Y\ \forall Y}}
    \mathcal E_{N_{\mathrm f},\boldsymbol r_{N_{\mathrm f}}}
    \circ\cdots\circ
    \mathcal E_{1,\boldsymbol r_1}.
\label{eq:alg-L-exact-shared-coefficient}
\end{align}
Therefore every coefficient retained by
$\mathcal T_K^{\mathrm{sh}}$ can be constructed by first keeping temporary
occurrence degrees $r_{f,Y}\le K$ and then applying the exact local merge
\begin{align}
    (r_{f_1,Y},\ldots,r_{f_{\nu_Y},Y})
    \longmapsto
    m_Y=\sum_{j=1}^{\nu_Y}r_{f_j,Y},
    \qquad m_Y\le K.
\label{eq:alg-L-local-degree-merge-map}
\end{align}
For one physical subblock with $\nu_Y\ge1$, the total number of retained
occurrence tuples is
\begin{align}
    N_{\mathrm{alloc}}(Y,K)
    =\sum_{m=0}^K
    \binom{m+\nu_Y-1}{\nu_Y-1}
    =\binom{K+\nu_Y}{\nu_Y}.
\label{eq:alg-L-allocation-count}
\end{align}
If $\nu_Y=0$, we use the convention
$N_{\mathrm{alloc}}(Y,K)=1$ and omit the inactive degree register.
Thus the merge is implementable by a block-sparse local tensor with output
register dimension $K+1$ and at most
$\binom{K+\nu_Y}{\nu_Y}$ nonzero degree transitions.  If
$\nu_{\mathrm{occ}}\le3$, this is $O((K+1)^3)$ per physical subblock.
For a main block containing a family $\mathfrak Y_X$ of physical subblocks,
\begin{align}
    N_{\mathrm{merge}}(X)
    \le
    \prod_{Y\in\mathfrak Y_X}
    \binom{K+\nu_Y}{\nu_Y}
    \le
    (K+1)^{\nu_{\mathrm{occ}}|\mathfrak Y_X|}.
\label{eq:alg-L-main-block-merge-cost}
\end{align}
Since $|\mathfrak Y_X|=O(B/\ell)$,
\begin{align}
    \log N_{\mathrm{merge}}(X)
    =O\left(\frac{B}{\ell}\log(K+1)\right).
\label{eq:alg-L-main-block-merge-log-cost}
\end{align}
Moreover, the one-step polynomial
$F_{1,K}=\mathcal T_K^{\mathrm{sh}}\Gamma_\Delta$ can be assembled in the
following depth-two order.  First compose the at most two elementary
polynomials inside each correction cell and perform the shared-degree merge
inside that cell.  Next tensorize all disjoint base cells and all disjoint
correction cells separately.  Finally compose the two macro-layer MPOs and
apply the shared rectangular projection.  Thus the algorithm never forms
the flattened $O(n)$-length occurrence list as an MPO product.
\end{prop}

\textit{Proof.}
Since $\mathcal L_f(\boldsymbol\lambda)$ is linear in the variables,
only the power
$r=|\boldsymbol r_f|_1$ contributes to the coefficient
$\boldsymbol\lambda^{\boldsymbol r_f}$ in the exponential series.  Hence
\begin{align}
    \mathcal E_{f,\boldsymbol r_f}
    =
    \frac{(\sigma_f\Delta)^{|\boldsymbol r_f|_1}}
    {|\boldsymbol r_f|_1!}
    [\boldsymbol\lambda^{\boldsymbol r_f}]
    \mathcal L_f(\boldsymbol\lambda)^{\circ|\boldsymbol r_f|_1},
\end{align}
which proves
Eq.~\eqref{eq:alg-L-factor-coefficient-word-normalization}.  In every
nonempty word contributing to
$\mathcal P_{f,\boldsymbol r_f}$, the leftmost letter is
$\mathcal L_{f,Y}$ for a unique $Y$ with $r_{f,Y}\ge1$.  Removing this
letter leaves a word with multi-degree
$\boldsymbol r_f-\boldsymbol e_Y$.  Summing over the possible leftmost
letters proves the exact recursion
Eq.~\eqref{eq:alg-L-factor-coefficient-recursion}.

There are $(K+1)^{|\mathfrak Y_f|}$ retained factor sectors, and each sector
has at most $|\mathfrak Y_f|$ predecessors.  Processing the sectors in
increasing total degree therefore uses at most the number of updates stated
in Eq.~\eqref{eq:alg-L-factor-coefficient-update-count}.  This recursion
sums words algebraically and never enumerates them individually.

The coefficient formula
Eq.~\eqref{eq:alg-L-exact-shared-coefficient} is the ordinary Cauchy product
for the ordered product of the factor power series.  Since all degrees are
nonnegative, a contribution to a retained shared sector $m_Y\le K$ can only
use temporary degrees $r_{f,Y}\le K$.  For a fixed $Y$ and fixed total
$m_Y=m$, the number of weak compositions among its $\nu_Y$ occurrences is
\begin{align}
    \binom{m+\nu_Y-1}{\nu_Y-1}.
\end{align}
Summing over $0\le m\le K$ and using the hockey-stick identity gives
Eq.~\eqref{eq:alg-L-allocation-count}.  The map in
Eq.~\eqref{eq:alg-L-local-degree-merge-map} is therefore an exact sparse
transition with output dimension $K+1$.  Taking the tensor product of these
local transitions over the subblocks in a local cell proves
Eqs.~\eqref{eq:alg-L-main-block-merge-cost} and
\eqref{eq:alg-L-main-block-merge-log-cost}.

For the final assertion, apply the same Cauchy-product rule only to the two
elementary exponentials inside one correction cell.  Since cells in a fixed
macro-layer have disjoint supports, their polynomial MPOs are then combined
by tensorization rather than by sequential multiplication.  The only
global composition is between the base and correction macro-layers.  The
flattened formula
Eq.~\eqref{eq:alg-L-exact-shared-coefficient} remains an exact algebraic
description of the resulting coefficient, but it is not the contraction
order used by the algorithm.
\hfill\(\square\)

\begin{lemma}[Depth-two crossing-cell rank bound]
\label{lem:alg-L-polynomial-profile-rank}
Fix a spatial cut $p$ and consider an ordered product of $r$ short-time
maps, each of which has the depth-two form
\begin{align}
    \mathcal G_a(x,\boldsymbol\lambda)
    =\mathcal G_a^{\mathrm{base}}(x,\boldsymbol\lambda)
     \circ
     \mathcal G_a^{\mathrm{corr}}(x,\boldsymbol\lambda),
    \qquad a=1,\ldots,r.
\label{eq:alg-L-depth-two-abstract-product}
\end{align}
Assume that each macro-layer is a tensor product of disjoint local cells,
each cell contains at most $c_{\mathrm{cell}}=O(1)$ elementary
exponentials, and every cell support has diameter $O(B)$.  In an elementary
exponential that belongs to a cell meeting $p$, write the generator as
\begin{align}
    \mathcal A_j(\boldsymbol\lambda)
    +x\mathcal B_j(\boldsymbol\lambda),
\label{eq:alg-L-depth-two-elementary-crossing-generator}
\end{align}
and assume the following.
\begin{enumerate}
\item[(i)] At $x=0$, every such elementary generator factorizes across
$p$:
\begin{align}
    \mathcal A_j(\boldsymbol\lambda)
    =\mathcal A_j^L(\boldsymbol\lambda)\otimes\operatorname{id}
    +\operatorname{id}\otimes\mathcal A_j^R(\boldsymbol\lambda).
\end{align}
\item[(ii)] Across $p$, at most $c_\times=O(1)$ local cells per short-time
map can have a nonzero crossing part.  Each $\mathcal B_j$ has a
decomposition
\begin{align}
    \mathcal B_j(\boldsymbol\lambda)
    =\sum_{\alpha=1}^{R_\times}
      b_{j,\alpha}(\boldsymbol\lambda)
      \mathcal B^L_{j,\alpha}\otimes
      \mathcal B^R_{j,\alpha},
\end{align}
where $R_\times=O(1)$ and every $b_{j,\alpha}$ is a scalar monomial.
\item[(iii)] A core family
$\mathfrak Y_{\mathrm{core}}$ of $h$ shared variables contains every
variable occurring in a cell that meets $p$.  Every variable outside the
core occurs only in cells lying on one fixed side of $p$.
\item[(iv)] In the cells meeting $p$, the maps
$\mathcal A_j^L$, $\mathcal A_j^R$, and $\mathcal B_j$ are homogeneous
linear functions of the core variables.
\end{enumerate}
Then, for every multi-index $0\le m_Y\le K$ and every
$s\in\mathbb Z_{\ge0}$,
\begin{align}
&
    \operatorname{SR}^{\mathrm{Liouv}}_p
    \left(
    [\boldsymbol\lambda^{\boldsymbol m}]
    \frac{1}{s!}
    \left.\frac{d^s}{dx^s}
    \bigl[\mathcal G_r(x,\boldsymbol\lambda)\circ\cdots\circ
          \mathcal G_1(x,\boldsymbol\lambda)\bigr]
    \right|_{x=0}
    \right)
\nonumber\\
&\qquad\le
    (K+1)^{c_{\mathrm{prof}}h+c_{\mathrm{prof}}}
    \binom{c_\times c_{\mathrm{cell}}r+s-1}{s}
    R_\times^s
    \binom{hK+s}{s},
\label{eq:alg-L-polynomial-profile-rank-bound}
\end{align}
where $c_{\mathrm{prof}}=O(1)$.  The same estimate holds when each variable
outside $\mathfrak Y_{\mathrm{core}}$ is independently summed over a finite
degree set before extracting the core coefficient.  It also holds for a
prefix ending inside one macro-layer or inside one local cell, after
changing only the constants.
\end{lemma}

\textit{Proof.}
The depth-two geometry is used before any differentiation.  In each
short-time map, every cell not meeting $p$ is supported completely on one
side of the cut.  Absorb all such cells into polynomial-valued one-sided
left or right maps, while keeping the degrees of core variables unresolved.
Variables outside the core may already be summed independently over their
retained finite degree sets.  Only the at most $c_\times$ cells meeting $p$ remain relevant
for the crossing derivative.  After flattening the fixed internal depth of
those cells, the number of elementary positions at which an $x$-derivative
can act is at most
\begin{align}
    M_\times\le c_\times c_{\mathrm{cell}}r.
\label{eq:alg-L-depth-two-Mcross}
\end{align}
This is the point at which the depth-two construction removes the spatial
factor $O(n/B)$: the number of differentiated positions depends on the
number of time slices, but not on the number of cells in the chain.

For one remaining elementary exponential, define
\begin{align}
    \mathcal U_{j,q}(\boldsymbol\lambda)
    :=\frac{1}{q!}
      \left.\frac{d^q}{dx^q}
      e^{\mathcal A_j(\boldsymbol\lambda)
        +x\mathcal B_j(\boldsymbol\lambda)}
      \right|_{x=0}.
\end{align}
The normalized product rule gives an exact weak-composition sum over the
$M_\times$ positions.  Hence the number of derivative allocations is at
most
\begin{align}
    \binom{M_\times+s-1}{s}.
\label{eq:alg-L-depth-two-derivative-allocation-count}
\end{align}
For $q\ge1$, the Duhamel formula reads
\begin{align}
    \mathcal U_{j,q}
    =\int_{0\le u_q\le\cdots\le u_1\le1}
      e^{(1-u_1)\mathcal A_j}
      \circ\mathcal B_j\circ
      e^{(u_1-u_2)\mathcal A_j}
      \circ\cdots\circ
      \mathcal B_j\circ e^{u_q\mathcal A_j}
      \,d\boldsymbol u.
\label{eq:alg-L-profile-Duhamel}
\end{align}
Fix a derivative allocation and one monomial--Schmidt label for each of the
$s$ crossing insertions.  Assumption (i) factorizes every exponential
segment in Eq.~\eqref{eq:alg-L-profile-Duhamel} across $p$.  Together with
the one-sided cells already absorbed above, the full integrand therefore
has the form
\begin{align}
    \mathcal X^L(\boldsymbol\lambda,\boldsymbol u)
    \otimes
    \mathcal X^R(\boldsymbol\lambda,\boldsymbol u),
\label{eq:alg-L-profile-factorized-integrand}
\end{align}
where $\boldsymbol u$ contains exactly $s$ simplex variables.

We next extract a fixed shared coefficient.  For every core variable $Y$,
distribute the degree not already fixed by the crossing insertions among
four aggregate classes: simplex-dependent left segments,
simplex-dependent right segments, simplex-independent left factors, and
simplex-independent right factors.  There are at most
$\binom{m_Y+3}{3}\le C(K+1)^3$ choices.  Thus all aggregate degree splits
contribute at most
\begin{align}
    (K+1)^{c_{\mathrm{prof}}h+c_{\mathrm{prof}}}.
\label{eq:alg-L-profile-degree-splits}
\end{align}
All allocations inside one aggregate class are summed algebraically into a
single operator coefficient.  A variable outside the core occurs only on
one fixed side, so any finite sum over its degrees is absorbed into that
one-sided coefficient and creates no Schmidt-rank factor.

For a fixed aggregate split, the coefficient of degree $q$ in a segment
$e^{a(\boldsymbol u)\mathcal A_j(\boldsymbol\lambda)}$ carries the scalar
polynomial $a(\boldsymbol u)^q$.  By assumption (iv), the left coefficient
in Eq.~\eqref{eq:alg-L-profile-factorized-integrand} is therefore an
operator-valued polynomial of total degree at most $hK$ in the $s$ simplex
variables:
\begin{align}
    \mathcal X^L_{\mathrm{coeff}}(\boldsymbol u)
    =\sum_{|\boldsymbol\beta|_1\le hK}
      \boldsymbol u^{\boldsymbol\beta}
      \mathcal X^L_{\boldsymbol\beta}.
\label{eq:alg-L-profile-left-polynomial}
\end{align}
The number of monomials is at most
\begin{align}
    \binom{hK+s}{s}.
\label{eq:alg-L-profile-polynomial-dimension}
\end{align}
For each $\boldsymbol\beta$, absorb the simplex integral of
$\boldsymbol u^{\boldsymbol\beta}$ times the corresponding right
operator-valued coefficient into one right map.  This converts the
continuum integral exactly into a finite sum of product maps.  Finally,
there are at most $R_\times^s$ crossing monomial--Schmidt labels.
Multiplying the four counts and using
Eq.~\eqref{eq:alg-L-depth-two-Mcross} proves
Eq.~\eqref{eq:alg-L-polynomial-profile-rank-bound}.  Stopping at a prefix or
inside a local cell only reduces the number of complete cells and changes
$M_\times$ by a constant, proving the last assertion.
\hfill\(\square\)

\begin{prop}[Exact Liouville-MPO rank of retained shared-degree sectors]
\label{prop:alg-L-retained-coefficient-rank}
For $1\le r\le R$, write
\begin{align}
    F_r(\boldsymbol\lambda)
    :=\Gamma_\Delta(\boldsymbol\lambda)^{\circ r}
    =\sum_{\boldsymbol m}F_{r,\boldsymbol m}
    \boldsymbol\lambda^{\boldsymbol m}.
\end{align}
Let $0\le m_Y\le K$ for every $Y$.  There exist constants
$c_0,c_1,c_2,c_3>0$, depending only on the local interaction structure,
such that, across every microscopic cut $q$,
\begin{align}
    \operatorname{SR}^{\mathrm{Liouv}}_q
    (F_{r,\boldsymbol m})
    \le
    d_0^{c_0\ell}
    (K+1)^{c_1\log(\ell+1)+c_1B/\ell}
    \left[
    c_2(r+1)(K+1)(B+1)
    \right]^{c_3K/\ell+c_3}.
\label{eq:alg-L-retained-coefficient-rank}
\end{align}
Equivalently,
\begin{align}
    \log
    \operatorname{SR}^{\mathrm{Liouv}}_q
    (F_{r,\boldsymbol m})
    =\widetilde O\left(
    \ell+\frac{B}{\ell}+\frac{K}{\ell}
    \right),
\label{eq:alg-L-retained-coefficient-rank-scaling}
\end{align}
uniformly in $q$, $n$, $r$, and $\boldsymbol m$, with the logarithmic
factors allowed to depend on $r+K+B$.  Consequently, every retained
coefficient has an exact Liouville-MPO representation with the same maximal
bond-dimension bound.

The estimate also holds for the truncated partial sectors needed by the MPO
algorithm.  Define the local-cell occurrence hull
\begin{align}
    \mathsf H(Y)
    :=\operatorname{hull}\left(
    \bigcup_{f:\,\mathcal L_{f,Y}\ne0}\Xi_f
    \right),
\end{align}
and let $I_q$ be a window of $\ell$ microscopic cuts around $q$.  Let
$\mathfrak O_q$ contain the subblocks whose hull intersects a fixed
$O(\ell+k)$ collar of $I_q$.  Then
$|\mathfrak O_q|=O(B/\ell)$, and the same rank bound applies to
\begin{align}
    F^{(q,K)}_{r,\boldsymbol m_{\mathfrak O_q}}
    :=
    \sum_{\substack{
    \boldsymbol n\in\mathbb Z_{\ge0}^{\mathfrak O_q^c}\\
    0\le n_Y\le K\ (Y\notin\mathfrak O_q)}}
    F_{r,(\boldsymbol m_{\mathfrak O_q},\boldsymbol n)},
    \qquad
    0\le m_Y\le K\quad(Y\in\mathfrak O_q).
\label{eq:alg-L-partial-local-sector}
\end{align}
The same conclusion holds for the degree-labelled variant in which every
outside degree $n_Y$ is retained as an orthogonal classical leg placed at a
marker inside $\mathsf H(Y)$, rather than summed.  Such a leg and every
occurrence carrying it lie on one fixed side of each selected cut.  The
statement remains valid, with $r$ replaced by the number of completed time
slices, for every prefix ending after one macro-layer or inside one local
cell, and for every ordered subproduct used in the coefficient recursion
below.
\end{prop}

\textit{Proof.}
The case $K=0$ is immediate: all generators are homogeneous in
$\boldsymbol\lambda$, so every retained zero-degree coefficient, partial
sector, and degree-labelled prefix is the identity map with zero degree
labels.  We henceforth assume $K\ge1$.

Fix a microscopic target cut $q$ and choose a window $I_q$ containing
$\ell$ consecutive microscopic cuts around $q$.  If the window meets a
boundary, append empty layers as in the static unitary proof.  Introduce one
auxiliary variable $\eta_Z$ for every microscopic term, shared by all
elementary occurrences and all time steps, and set
\begin{align}
    \widehat{\mathcal L}_f
    (\boldsymbol\lambda,\boldsymbol\eta)
    :=
    \sum_{Y\in\mathfrak Y}\lambda_Y
    \sum_{\substack{a(Z)=Y\\
                    \mathcal L_Z\text{ occurs in }\mathcal L_f}}
    \eta_Z\mathcal L_Z.
\label{eq:alg-L-microscopic-auxiliary-parameters}
\end{align}
Let $\widehat F_r(\boldsymbol\lambda,\boldsymbol\eta)$ be the product
of $r$ short-time depth-two local-cell maps formed from these generators.  Then
\begin{align}
    F_{r,\boldsymbol m}
    =
    [\boldsymbol\lambda^{\boldsymbol m}]
    \widehat F_r(\boldsymbol\lambda,\boldsymbol1).
\label{eq:alg-L-coefficient-from-microscopic-functional}
\end{align}

For a microscopic multi-index $\boldsymbol s=(s_Z)_Z$, define its cut
weight in $I_q$ by
\begin{align}
    w_{I_q}(\boldsymbol s)
    :=
    \sum_Z c_{I_q}(Z)s_Z,
\end{align}
where $c_{I_q}(Z)$ is the number of cuts in $I_q$ crossed by $Z$.
Only terms anchored in a fixed-range collar of $I_q$ have positive weight.
Because the window has length $\ell$, the collar intersects only $O(1)$
physical subblocks of width $\ell$.  Every local term crosses at most
$O(k)$ cuts.  Consequently, every microscopic monomial contributing to the
fixed shared coefficient satisfies
\begin{align}
    w_{I_q}(\boldsymbol s)
    \le
    c_{\mathrm{loc}}
    \sum_{Y:\,Y\cap I_q^{+k}\ne\emptyset}m_Y
    \le c_{\mathrm{loc}}'K
    \le Q_{\mathrm{loc}},
    \qquad
    Q_{\mathrm{loc}}:=\left\lceil c_{\mathrm{loc}}'K\right\rceil.
\label{eq:alg-L-local-cut-weight-rigorous}
\end{align}
This step is independent of how the degree is distributed among elementary
occurrences and time steps.

Apply the coefficientwise cut-count decomposition of
Corollary~\ref{cor:cut-count-x-dependent} to the auxiliary variables
$\boldsymbol\eta$, using the discrete pair consisting of the time-step and
elementary-occurrence labels as the external variable, and to the $\ell$ cuts in
$I_q$.  The decomposition is a linear identity between Taylor
coefficients, so the projection
$[\boldsymbol\lambda^{\boldsymbol m}]$ may be applied afterwards.  By
Eq.~\eqref{eq:alg-L-local-cut-weight-rigorous}, the coefficient of the
cut-tail remainder vanishes exactly.  We therefore obtain an exact
representation
\begin{align}
    F_{r,\boldsymbol m}
    =
    \sum_{p\in I_q}
    \sum_{s=0}^{\lfloor2Q_{\mathrm{loc}}/\ell\rfloor}
    \sum_{\boldsymbol z\in\mathcal Z_{p,Q_{\mathrm{loc}}}}
    c_{p,Q_{\mathrm{loc}}}(\boldsymbol z)
    [\boldsymbol\lambda^{\boldsymbol m}]
    \mathcal D_{p,\boldsymbol z}^{(s)}
    (\boldsymbol\lambda),
\label{eq:alg-L-coefficient-cut-count-reduction}
\end{align}
where $\mathcal D_{p,\boldsymbol z}^{(s)}$ is the normalized order-$s$
derivative obtained by switching on the phase-twisted microscopic terms
crossing the selected cut $p$.  The number of cut-phase pairs is at most
\begin{align}
    \ell(2Q_{\mathrm{loc}}+1)^{\lceil\log_2\ell\rceil},
\label{eq:alg-L-coefficient-cut-phase-count}
\end{align}
and
\begin{align}
    s\le O(K/\ell).
\label{eq:alg-L-branch-crossing-order}
\end{align}

We now estimate one branch.  At zero crossing parameter, every elementary
exponential in a cell meeting $p$ contains only terms that stay on one side
of $p$, and hence its generator has the form required in assumption (i) of
Lemma~\ref{lem:alg-L-polynomial-profile-rank}.  In one short-time slice, at
most one base cell and one correction cell meet $p$.  After flattening the
fixed internal depth of the correction cell, at most
$c_\times c_{\mathrm{cell}}=O(1)$ elementary positions can contain a
microscopic term crossing $p$.  Across $r$ slices,
\begin{align}
    M_\times\le c_\times c_{\mathrm{cell}}r.
\label{eq:alg-L-crossing-factor-position-count}
\end{align}
All cells outside this constant-size crossing family are strictly one-sided
and are absorbed before the derivative is expanded.  The union of the
supports of the crossing cells has width $O(B)$, so the number of physical
subblocks in the corresponding core is
\begin{align}
    h\le c_{\mathrm{core}}(B/\ell+1).
\label{eq:alg-L-core-variable-count}
\end{align}
Finally, finite interaction range and fixed local dimension imply that the
phase-twisted crossing generator in one elementary position has Liouville Schmidt rank
$R_\times=O(1)$ after expanding its local terms in a product-map basis.
Lemma~\ref{lem:alg-L-polynomial-profile-rank} therefore gives
\begin{align}
&
    \operatorname{SR}^{\mathrm{Liouv}}_p
    \left(
    [\boldsymbol\lambda^{\boldsymbol m}]
    \mathcal D_{p,\boldsymbol z}^{(s)}
    (\boldsymbol\lambda)
    \right)
\nonumber\\
&\qquad\le
    (K+1)^{O(B/\ell+1)}
    \binom{c_\times c_{\mathrm{cell}}r+s-1}{s}
    R_\times^s
    \binom{c_{\mathrm{core}}K(B/\ell+1)+s}{s}
\nonumber\\
&\qquad\le
    (K+1)^{O(B/\ell+1)}
    \left[
    C(r+1)(K+1)(B+1)
    \right]^{O(s+1)}.
\label{eq:alg-L-exact-branch-rank-completed}
\end{align}
Here the last line follows from the standard binomial estimate and is also
valid for $s=0$ after adjusting $C$.

Substitute Eq.~\eqref{eq:alg-L-exact-branch-rank-completed} into
Eq.~\eqref{eq:alg-L-coefficient-cut-count-reduction}.  The phase count in
Eq.~\eqref{eq:alg-L-coefficient-cut-phase-count} contributes
$(K+1)^{O(\log(\ell+1))}$.  Converting the selected cut $p$ to the target
cut $q$ across a distance at most $\ell$ costs $d_0^{O(\ell)}$ in
Liouville space.  Together with
$Q_{\mathrm{loc}}=O(K)$ and
Eq.~\eqref{eq:alg-L-branch-crossing-order}, this proves
Eq.~\eqref{eq:alg-L-retained-coefficient-rank}.

For the partial sector in
Eq.~\eqref{eq:alg-L-partial-local-sector}, note first that the variables
outside $\mathfrak O_q$ are not summed over all degrees: the coordinatewise
cutoff $0\le n_Y\le K$ is retained.  By the definition of the
local-cell occurrence hull, however, such a variable is absent from every local-cell occurrence that can contain a microscopic term crossing a cut in
$I_q$.  It occurs only in simplex-independent one-sided cells and on one
fixed side of every selected cut.  The finite sum over its retained degrees
is therefore absorbed into a one-sided coefficient map.  The last assertion
of Lemma~\ref{lem:alg-L-polynomial-profile-rank} shows that this operation
neither changes the polynomial-profile count nor introduces an additional
Liouville Schmidt-rank factor.  Linearity also allows the cut-count identity
to be applied coefficientwise and then summed over the retained outside
degrees.  This proves the partial-sector claim.  If the degree is retained
as a classical basis vector $|n_Y\rangle$ instead, attach that vector to the
same one-sided coefficient map.  Since both the leg and all occurrences of
$Y$ are on one side of the selected cut, this label-valued version has the
same cross-cut rank bound.

The argument used only the two-macro-layer cell geometry and the number of
completed time slices.  Deleting cells, stopping after one macro-layer, or
stopping inside one correction cell can only reduce $M_\times$ and the core
family, which proves the final prefix/subproduct statement.  Since the rank bound is uniform in
the target cut, successive exact rank factorizations of the vectorized
superoperator, with input and output indices grouped sitewise, give an
exact Liouville-MPO with the same maximal bond dimension.
\hfill\(\square\)

\begin{prop}[Degree-labelled MPO recursion and arithmetic construction cost]
\label{prop:alg-L-degree-DP}
Define
\begin{align}
    F_{r,K}(\boldsymbol\lambda)
    :=\mathcal T_K^{\mathrm{sh}}
    \left[
    \Gamma_\Delta(\boldsymbol\lambda)^{\circ r}
    \right].
\label{eq:alg-L-FrK-definition}
\end{align}
Then the lower-set identity
\begin{align}
    F_{r,K}
    =\mathcal T_K^{\mathrm{sh}}
    \left[F_{r-1,K}\circ F_{1,K}\right]
\label{eq:alg-L-exact-lower-set-recursion}
\end{align}
is exact.  At the coefficient level,
\begin{align}
    F_{r,K}[\boldsymbol m]
    =
    \sum_{\substack{\boldsymbol u+\boldsymbol v=\boldsymbol m\\
    0\le u_Y,v_Y\le K}}
    F_{r-1,K}[\boldsymbol u]
    \circ F_{1,K}[\boldsymbol v].
\label{eq:alg-L-exact-degree-convolution}
\end{align}

For the constructive recursion, choose a marker site
$y_\star(Y)\in Y\cap\Lambda\subseteq\mathsf H(Y)$ for every active
physical subblock and introduce a local classical degree space
$\mathcal D_Y=\operatorname{span}\{|0\rangle,\ldots,|K\rangle\}$ at that
marker.  For $\nu_Y=0$, the degree is fixed to zero and its classical
leg is omitted in the formulas below.  Define the degree-resolved tensor
\begin{align}
    \mathbf F_{r,K}
    :=\sum_{0\le m_Y\le K}
      F_{r,K}[\boldsymbol m]
      \otimes\bigotimes_{Y\in\mathfrak Y}|m_Y\rangle_Y.
\label{eq:alg-L-degree-resolved-tensor}
\end{align}
Let
\begin{align}
    \mathsf A_K^{(Y)}:
    \mathcal D_Y\otimes\mathcal D_Y\longrightarrow\mathcal D_Y,
    \qquad
    \mathsf A_K^{(Y)}(|u\rangle\otimes|v\rangle)
    :=\begin{cases}
        |u+v\rangle,&u+v\le K,\\
        0,&u+v>K.
      \end{cases}
\label{eq:alg-L-local-degree-adder}
\end{align}
Then Eq.~\eqref{eq:alg-L-exact-degree-convolution} is implemented by
operator composition together with the product of the local adders
$\bigotimes_Y\mathsf A_K^{(Y)}$.  The desired ordinary Liouville map is
obtained only at the end by the product local contraction
\begin{align}
    \left.F_{R,K}(\boldsymbol\lambda)\right|_{\boldsymbol\lambda=\boldsymbol1}
    =
    \left(\bigotimes_{Y\in\mathfrak Y}\langle+|_Y\right)
    \mathbf F_{R,K},
    \qquad
    \langle+|_Y:=\sum_{m=0}^{K}\langle m|.
\label{eq:alg-L-final-degree-leg-contraction}
\end{align}
Thus a degree is never discarded or summed before all time slices have been
merged.

Construct $\mathbf F_{1,K}$ by the cellwise depth-two procedure in
Proposition~\ref{prop:alg-L-exact-degree-merging}: compose only inside each
local cell, tensorize the disjoint cells of each macro-layer, and then
compose the two macro-layers.  There is a local degree-labelled MPO
implementation in which, across any spatial bond, at most
\begin{align}
    N_{\mathrm{reg}}
    :=(K+1)^{c_{\mathrm{reg}}B/\ell}
\label{eq:alg-L-open-register-count}
\end{align}
degree assignments can correlate the two sides, for a constant
$c_{\mathrm{reg}}=O(1)$.  Combining the exact merge of
Proposition~\ref{prop:alg-L-exact-degree-merging}, the local adders in
Eq.~\eqref{eq:alg-L-local-degree-adder}, exact MPO rank reduction, and the
coefficient-rank bound of
Proposition~\ref{prop:alg-L-retained-coefficient-rank}, one obtains an MPO
for Eq.~\eqref{eq:alg-L-final-degree-leg-contraction} with
\begin{align}
    \log D_{R,K}
    =\widetilde O\left(
    \frac{B}{\ell}
    +\ell
    +\frac{K}{\ell}
    \right).
\label{eq:alg-L-DP-bond-bound}
\end{align}
In the exact-arithmetic operation model, the construction requires at most
\begin{align}
    nR\,
    \exp\left[
    \widetilde O\left(
    \frac{B}{\ell}
    +\ell
    +\frac{K}{\ell}
    \right)
    \right]
\label{eq:alg-L-DP-time-bound}
\end{align}
arithmetic operations, up to polynomial factors in the displayed
parameters and in the local tensor dimensions.
\end{prop}

\textit{Proof.}
The coefficient of degree $\boldsymbol m$ in a product of two formal power
series with nonnegative multi-degrees is the convolution of their
coefficients.  If $m_Y\le K$, every contributing pair satisfies
$u_Y,v_Y\le K$.  Conversely, a term that has exceeded $K$ in one coordinate
can never return to the retained rectangle under further multiplication.
This proves Eqs.~\eqref{eq:alg-L-exact-lower-set-recursion} and
\eqref{eq:alg-L-exact-degree-convolution} exactly.  By definition of the
local adder, the same identity gives
\begin{align}
    \mathbf F_{r,K}
    =
    \left(\bigotimes_Y\mathsf A_K^{(Y)}\right)
    \left(\mathbf F_{r-1,K}\circ\mathbf F_{1,K}\right),
\label{eq:alg-L-degree-resolved-recursion}
\end{align}
where composition acts on the Liouville-map factors and the tensor product
of adders acts on the two copies of the classical degree legs.  Contracting
all final legs with $\langle+|$ proves
Eq.~\eqref{eq:alg-L-final-degree-leg-contraction}.  This degree-resolved
form is important: summing a degree when its spatial occurrence hull closes
would lose the information required by later time slices.

We next describe the spatial implementation.  While constructing one time
slice, transport the temporary occurrence degrees of a physical subblock
$Y$ only through its local-cell occurrence hull $\mathsf H(Y)$, enlarged by
an $O(\ell+k)$ collar, merge them, and deposit the resulting shared degree
in the classical leg at $y_\star(Y)$.  The transport register can then be
closed, but the deposited degree leg remains available to every subsequent
time convolution through Eq.~\eqref{eq:alg-L-local-degree-adder}.  Every
enlarged hull has diameter $O(B)$.  Because the base and correction
macro-layers are each formed from disjoint cells and each physical subblock
has constant occurrence multiplicity, only $O(B/\ell)$ such hulls intersect
a fixed spatial bond.  Hence at most
$(K+1)^{c_{\mathrm{reg}}B/\ell}$ assignments can correlate its two sides.
All other degree legs and their occurrences lie entirely on one side and
are absorbed into that side of a Schmidt decomposition.

Fixing the assignments of the $O(B/\ell)$ crossing degree legs leaves the
label-valued partial sector covered by
Proposition~\ref{prop:alg-L-retained-coefficient-rank}.  Degrees attached to
hulls wholly on the left or right remain orthogonal local labels in the
corresponding one-sided factor and create no additional cross-cut rank.
Thus the Liouville Schmidt rank of $\mathbf F_{r,K}$ across any bond is at
most the product of the register count and the uniform partial-sector rank
bound.  This gives Eq.~\eqref{eq:alg-L-DP-bond-bound}.  The final contraction
by $\bigotimes_Y\langle+|_Y$ is a product of local linear maps and cannot
increase the bond dimension.

Exact rank reduction is performed sector by sector for the crossing
registers.  In the elementary coefficient recursion
\eqref{eq:alg-L-factor-coefficient-recursion}, let $D_\star$ be a uniform
bound for the completed coefficients.  Each predecessor product has bond
dimension at most $D_\star d_0^{O(\ell)}$, and a partial update contains
at most $|\mathfrak Y_f|=O(B/\ell+1)$ such terms.  Exact rank reduction
to the coefficient bound is used after the update is complete.
During the construction of $\mathbf F_{1,K}$, the
prefix/subproduct version of
Proposition~\ref{prop:alg-L-retained-coefficient-rank} applies after a base
macro-layer, inside a correction cell, and after the correction macro-layer.
During the time recursion it applies after every completed time slice.
Therefore no intermediate exact bond dimension exceeds the same exponential
scale, up to constant powers produced before rank reduction.  In
particular, the algorithm never differentiates or multiplies the $O(n)$
telescoping sequence used in the proof of
Proposition~\ref{prop:alg-L-uniform-deleted-HHKL}.

The cellwise construction of one $\mathbf F_{1,K}$ uses $O(1+n/B)$ local
cells and $O(n)$ stored local tensors, with only constant-depth composition
before rank reduction.  Each local adder has $O((K+1)^2)$ nonzero
transitions.  At each of the $R$ time steps, polynomial-MPO multiplication,
sparse local degree convolution, addition, and sector-preserving exact
rank reduction are implemented by a fixed number of sweeps over the local
tensors.  Their cost is linear in the number of stored sites and polynomial
in the local tensor dimensions and participating bond dimensions.  The
local coefficient tables and degree merges contribute the
$(K+1)^{O(B/\ell)}$ factors already included in the rank bound.
Substitution gives Eq.~\eqref{eq:alg-L-DP-time-bound}.
\hfill\(\square\)

\begin{theorem}[Global algorithmic Liouville-MPO approximation]
\label{thm:alg-L-global-target}
Let $\Lambda=[n]$ be a one-dimensional chain with uniformly bounded local
Hilbert-space dimension, and let $\mathcal L$ satisfy
Eq.~\eqref{eq:alg-L-shared-local-Liouvillian}.  Assume finite interaction
range and that every deleted unit $\mathcal L_Z$ is a complete local
Lindblad term.  Use the two-macro-layer local-cell partition in
Eqs.~\eqref{eq:alg-L-depth-two-cell-splitting}--
\eqref{eq:alg-L-depth-two-HHKL-product}; no additional bounded-depth HHKL
assumption is required.  Fix $t\ge0$ and
$0<\epsilon\le1$.  Choose a constant polydisc radius, for example
$\rho=2$.  If $g_{\mathrm L}=0$, then $\mathcal L=0$ and the identity
MPO gives the conclusion.  Assume $g_{\mathrm L}>0$.  The proofs of
Propositions~\ref{prop:alg-L-uniform-deleted-HHKL} and
\ref{prop:alg-L-shared-cocycle} permit a constant
$\vartheta_\rho>0$, depending only on the fixed locality and range data
and on $\rho$, such that the fixed time step
\begin{align}
    \Delta:=\frac{\vartheta_\rho}{g_{\mathrm L}}
    \le\min\{\Delta_0,\Delta_\rho\}
\label{eq:alg-L-global-time-step}
\end{align}
is admissible.  We use this choice throughout the construction.  Define
\begin{align}
    \mathsf L_{n,t,\epsilon}
    :=\max\left\{
    1,
    \log\frac{C_0n(1+g_{\mathrm L}t)}{\epsilon}
    \right\}.
\label{eq:alg-L-global-log-scale}
\end{align}
There are choices
\begin{align}
    B&=\Theta(\mathsf L_{n,t,\epsilon}),
    &
    \ell&=\Theta(\sqrt{\mathsf L_{n,t,\epsilon}}),
    &
    K&=O\left(
    g_{\mathrm L}t\ell+\mathsf L_{n,t,\epsilon}
    \right)
\label{eq:alg-L-global-parameter-choice}
\end{align}
for which one can explicitly construct a Liouville-MPO
$\widetilde{\mathcal E}_{t,K}$ satisfying
\begin{align}
    \left\lVert
    e^{t\mathcal L}-\widetilde{\mathcal E}_{t,K}
    \right\rVert_\diamond
    \le\epsilon.
\label{eq:alg-L-global-target-error}
\end{align}
Its bond dimension and exact-arithmetic construction cost obey
\begin{align}
    D_{\mathrm{MPO}}
    &\le
    \exp\left[
    \widetilde O\left(
    g_{\mathrm L}t+\sqrt{\mathsf L_{n,t,\epsilon}}
    \right)
    \right],
    \\
    T_{\mathrm{construct}}
    &\le
    n\exp\left[
    \widetilde O\left(
    g_{\mathrm L}t+\sqrt{\mathsf L_{n,t,\epsilon}}
    \right)
    \right].
\label{eq:alg-L-global-target-complexity}
\end{align}
The constants hidden in $O(\cdot)$ and $\widetilde O(\cdot)$ depend only
on the fixed local interaction data and the on-site dimension.
\end{theorem}

\textit{Proof.}
The cases $t=0$ and $g_{\mathrm L}=0$ are represented exactly by the
identity MPO, so assume $t>0$ and $g_{\mathrm L}>0$.  Let
\begin{align}
    R:=\left\lceil\frac{t}{\Delta}\right\rceil,
    \qquad
    \Delta_r:=
    \begin{cases}
        \Delta,&1\le r<R,\\
        t-(R-1)\Delta,&r=R,
    \end{cases}
\label{eq:alg-L-nonuniform-short-times}
\end{align}
so that $0<\Delta_r\le\Delta$ and $\sum_r\Delta_r=t$.  Denote the
corresponding parameterized depth-two short-time maps by
$\Gamma_{\Delta_r}(\boldsymbol\lambda)$ and set
\begin{align}
    \mathcal F_t(\boldsymbol\lambda)
    :=\Gamma_{\Delta_R}(\boldsymbol\lambda)
      \circ\cdots\circ
      \Gamma_{\Delta_1}(\boldsymbol\lambda).
\label{eq:alg-L-nonuniform-parameterized-product}
\end{align}
All preceding propositions remain valid for this nonuniform final time slice.
Indeed, their short-time estimates are monotone in the duration, the
polydisc exponent contains the sum $\sum_r\Delta_r=t$, and the lower-set
coefficient recursion uses only nonnegative multi-degrees.  The
prefix/subproduct statement in
Proposition~\ref{prop:alg-L-retained-coefficient-rank} already covers the
corresponding ordered product with $R$ time slices.

For each $r$, write
\begin{align}
    \mathcal E_r:=e^{\Delta_r\mathcal L},
    \qquad
    \mathcal G_r:=\Gamma_{\Delta_r}(\boldsymbol1).
\end{align}
Proposition~\ref{prop:alg-L-uniform-deleted-HHKL}, applied without a
deletion, gives uniformly in $r$
\begin{align}
    \lVert\mathcal E_r-\mathcal G_r\rVert_\diamond
    \le\delta_{\mathrm H}(B),
    \qquad
    \lVert\mathcal G_r\rVert_\diamond
    \le1+\delta_{\mathrm H}(B).
\end{align}
The telescoping identity for two ordered products and
$\lVert\mathcal E_r\rVert_\diamond=1$ imply
\begin{align}
&
    \left\lVert
    e^{t\mathcal L}-
    \mathcal G_R\circ\cdots\circ\mathcal G_1
    \right\rVert_\diamond
\nonumber\\
&\qquad\le
    R\delta_{\mathrm H}(B)
    [1+\delta_{\mathrm H}(B)]^{R-1}.
\label{eq:alg-L-global-HHKL-telescope}
\end{align}
One short-time map contains $N_{\mathrm{int}}=O(n/B)$ interfaces.  Since Eq.~\eqref{eq:alg-L-global-time-step} gives
$R=O(1+g_{\mathrm L}t)$, choosing the constant in
$B=\Theta(\mathsf L_{n,t,\epsilon})$ sufficiently large ensures
\begin{align}
    N_{\mathrm{int}}\eta_{\mathrm H}(B)&\le1,
    &
    R\delta_{\mathrm H}(B)
    [1+\delta_{\mathrm H}(B)]^{R-1}&\le\epsilon/2,
\label{eq:alg-L-global-HHKL-budget}
\end{align}
and also
\begin{align}
    A_R:=[1+\delta_{\mathrm H}(B)]^R=O(1).
\label{eq:alg-L-global-AR-bound}
\end{align}

Define the globally truncated product
\begin{align}
    \widetilde{\mathcal E}_{t,K}
    :=
    \left.
    \mathcal T_K^{\mathrm{sh}}
    \mathcal F_t(\boldsymbol\lambda)
    \right|_{\boldsymbol\lambda=\boldsymbol1}.
\label{eq:alg-L-global-truncated-product}
\end{align}
The proof of Proposition~\ref{prop:alg-L-coordinatewise-tail}, applied to
the short-time maps of durations $\Delta_r$, gives exactly the same estimate with
$R\Delta$ replaced by $\sum_r\Delta_r=t$.  With fixed $\rho=2$,
$|\mathfrak Y|=O(1+(n+B)/\ell)$ including endpoint padding, and
Eq.~\eqref{eq:alg-L-global-AR-bound}, it yields
\begin{align}
    \left\lVert
    \mathcal G_R\circ\cdots\circ\mathcal G_1
    -\widetilde{\mathcal E}_{t,K}
    \right\rVert_\diamond
    \le\epsilon/2
\label{eq:alg-L-global-Taylor-budget}
\end{align}
for
\begin{align}
    K
    =O\left(
    g_{\mathrm L}t\ell+
    \log\frac{n(1+g_{\mathrm L}t)}{\epsilon}
    \right),
\end{align}
which is Eq.~\eqref{eq:alg-L-global-parameter-choice}.  The padding
contribution to $\log|\mathfrak Y|$ is absorbed in
$O(\mathsf L_{n,t,\epsilon})$, since $B=\Theta(\mathsf L_{n,t,\epsilon})$
and $\mathsf L_{n,t,\epsilon}\ge1$.  Combining
Eqs.~\eqref{eq:alg-L-global-HHKL-budget} and
\eqref{eq:alg-L-global-Taylor-budget} proves
Eq.~\eqref{eq:alg-L-global-target-error}.

For the nonuniform durations, use the degree-labelled recursion in the
chronological order of \eqref{eq:alg-L-nonuniform-parameterized-product}:
\begin{align}
    F_{0,K}&=\operatorname{id},
    &F_{r,K}
    &=\mathcal T_K^{\mathrm{sh}}\!\left[
       \Gamma_{\Delta_r}(\boldsymbol\lambda)\circ F_{r-1,K}
       \right].
\label{eq:alg-L-nonuniform-degree-recursion}
\end{align}
The new time slice is composed on the left.  For equal durations this is
equivalent to \eqref{eq:alg-L-exact-lower-set-recursion}; for unequal
durations the local-cell approximants need not commute.
The lower-set proof and rank bounds of
Proposition~\ref{prop:alg-L-degree-DP} are unchanged.  Therefore
\begin{align}
    \log D_{\mathrm{MPO}}
    &=\widetilde O\left(
    \frac{B}{\ell}+\ell+\frac{K}{\ell}
    \right)
    \\
    &=\widetilde O\left(
    g_{\mathrm L}t+\sqrt{\mathsf L_{n,t,\epsilon}}
    \right).
\end{align}
The same substitution in the arithmetic cost bound, together with
$R=O(1+g_{\mathrm L}t)$, gives
Eq.~\eqref{eq:alg-L-global-target-complexity}.
\hfill\(\square\)

\textit{Scope of the theorem.}
All analytic and algebraic inputs used above are proved within this section
or in the earlier differential and composition-commutator formalism.  In particular, the short-time Liouvillian localization estimate is not an
additional assumption: Proposition~\ref{prop:alg-L-uniform-deleted-HHKL}
proves it uniformly under anchored deletions for the explicit two-macro-layer
local-cell product.  The $O(n)$ sequence of exact relative factors appears
only in that telescoping proof and is never used as the algorithmic product
or in a Schmidt-rank estimate.  The conclusion is an
exact-arithmetic construction of a linear-map MPO approximating the channel
in diamond norm.  The rectangular Taylor truncation is not asserted to be
completely positive or trace preserving.  Likewise, a uniform
finite-precision bit-complexity bound would require a separate stability
analysis for near-zero Schmidt coefficients and is not part of the theorem.

\subsection{Extension to piecewise-polynomial and approximable Liouvillians}
\label{subsubsec:alg-L-driven-extension}

The time-dependent extension covers both \emph{discrete Lindblad driving}
and \emph{regular continuous Lindblad driving}.  A supplied
piecewise-constant schedule is exact input.  Uniformly analytic local
generator paths, including bounded-frequency modulations that preserve
the required local Lindblad structure, are included through certified,
efficiently computable polynomial approximants.  Thus the original
generator need not be piecewise constant.

\paragraph{Connection with Section~\ref{subsec:td-static-type-compression} and notation.}
We use the same temporal approximation parameters as in
Eq.~\eqref{eq:reg-local-polynomial-approximation}: $N$ counts intervals in
a common partition, $\nu$ is the maximum polynomial degree in time, and
$\zeta$ is the local approximation error.  Here the operator norm is
replaced by the diamond norm, and the supplied polynomials are denoted by
$\widehat{\mathcal L}_Z^{(x)}$.
The exact discrete case has $\nu=0$ and $\zeta=0$; a linear ramp between
local Lindbladians has $N=1$, $\nu=1$, and $\zeta=0$.
These temporal parameters are distinct from the shared spatial cutoff
$K$; $n$ still denotes the number of sites.
For analytic input, we require the uniform analytic-continuation and
local norm bounds of Corollary~\ref{corol:reg-analytic-driving} in their
diamond-norm form, together with an effective polynomial approximation
procedure.  The required global-error tolerance and preprocessing cost
are specified below.

Three ingredients must be kept separate.  The short-time localization
and polydisc estimates use locality and a physical, deletion-stable
reference evolution.  The exact MPO rank of a retained coefficient uses
its finite polynomial dependence on the remaining crossing times.
The algorithm uses a temporal coefficient table, rather than an oracle
for low-rank dressed samples.  The last two ingredients are the
Liouville-space versions of
Section~\ref{subsubsec:alg-U-driven-extension}.  The first requires care
with inverses and with polynomial approximants that need not be physical.

Assume that $\mathcal L_Z^{(x)}$ is a complete local Lindblad term at
almost every time, with the same fixed supports, finite range, and uniform
local diamond-norm bound $g_{\mathrm L}$ as in
\eqref{eq:alg-L-shared-local-Liouvillian}.  Use the same fixed cells,
patches, physical subblocks, and anchor map $a(Z)$ throughout the evolution.
On a common partition of $[0,t]$ into $N$ pieces, suppose an input
procedure supplies local superoperator polynomials
$\widehat{\mathcal L}_Z^{(x)}$ of degree at most $\nu$, with unchanged
supports and
\begin{align}
 \operatorname*{ess\,sup}_{x}\sup_i
    \sum_{Z\ni i}\|\mathcal L_Z^{(x)}
                      -\widehat{\mathcal L}_Z^{(x)}\|_\diamond
       &\le\zeta,
 &
 \operatorname*{ess\,sup}_{x}\sup_i
    \sum_{Z\ni i}\|\widehat{\mathcal L}_Z^{(x)}\|_\diamond
       &\le g_\star=O(g_{\mathrm L}).
\label{eq:alg-L-driven-surrogate}
\end{align}
Exact piecewise-polynomial input corresponds to $\zeta=0$.
The approximate local maps are not required to be Lindblad generators.
For $g_{\mathrm L}>0$, a sufficient computational accuracy is
$\zeta\le\min\{g_{\mathrm L},\epsilon/[64n(1+t)]\}$.
We may take $g_\star=g_{\mathrm L}+\zeta>0$.  Then
\begin{align}
 d_{a,b}:=\int_a^b\sum_Z
      \|\mathcal L_Z^{(x)}-\widehat{\mathcal L}_Z^{(x)}\|_\diamond\,dx,
 \qquad d_{0,t}\le nt\zeta\le\epsilon/64.
\label{eq:alg-L-driven-perturbation-budget}
\end{align}
The cases $t=0$ and $g_{\mathrm L}=0$ are again trivial.

\paragraph{Time-ordered local cells and their stability.}
Write $\widehat T_{\mathcal A}(b,a)$ for the propagator generated by a
possibly parameterized local polynomial path $\mathcal A^{(x)}$.
For an arbitrary short interval $I_r=[t_{r-1},t_r]$ of length at most
$\Delta=\vartheta_\rho/g_\star$, replace the local cells by
\begin{align}
 \widehat{\mathcal U}_{j,r}(\boldsymbol\lambda)
   &:=\widehat T_{\mathcal K_j(\boldsymbol\lambda)}(t_r,t_{r-1}),
 \nonumber\\
 \widehat{\mathcal C}_{j,r}(\boldsymbol\lambda)
   &:=\widehat T_{\mathcal A_j(\boldsymbol\lambda)}(t_r,t_{r-1})^{-1}
        \circ\widehat T_{\mathcal A_j(\boldsymbol\lambda)
                              +\mathcal V_j(\boldsymbol\lambda)}
                         (t_r,t_{r-1}),
 \nonumber\\
 \widehat\Gamma_r(\boldsymbol\lambda)
   &:=\left[\mathop{\circ}_j\widehat{\mathcal U}_{j,r}
                       (\boldsymbol\lambda)\right]
      \circ\left[\mathop{\circ}_j\widehat{\mathcal C}_{j,r}
                       (\boldsymbol\lambda)\right].
\label{eq:alg-L-driven-local-cells}
\end{align}
The symbols $\mathcal K_j,\mathcal A_j,\mathcal V_j$ in this display
are formed from the hatted local interactions with exactly the anchor
rule above.  The same $\lambda_Y$ is used in all occurrences and at all
times.  The macro-depth remains two; a switch inside $I_r$ does not create
another spatial macro-layer.  An inverse factor is implemented by a
forward propagator with generator
$-\mathcal A^{(t_{r-1}+t_r-x)}$, not by the unreversed path
$-\mathcal A^{(x)}$.

\begin{lemma}[Nonautonomous localization and shared-block stability]
\label{lem:alg-L-driven-stability}
For fixed $\rho>1$, choose $\vartheta_\rho>0$ sufficiently small in
terms of the fixed local data.  The constants in the following bounds
are independent of the temporal polynomial degree, number of pieces,
and minimum piece length.  For every anchored deletion pattern
$\boldsymbol\mu\in\{0,1\}^{\mathfrak Y}$,
\begin{align}
 \|\widehat T_{\mathcal L^{(\boldsymbol\mu)}}(b,a)\|_\diamond
      &\le e^{d_{a,b}},
 \nonumber\\
 \|\widehat T_{\mathcal L^{(\boldsymbol\mu)}}(b,a)
               -\widehat\Gamma_r(\boldsymbol\mu)\|_\diamond
      &\le e^{d_{a,b}}\delta_{\mathrm H}(B),
 \qquad [a,b]=I_r,
\label{eq:alg-L-driven-HHKL}
\end{align}
where $\delta_{\mathrm H}(B)=O(N_{\mathrm{int}}e^{-\mu_{\mathrm H}B})$
under the same small-tail condition as in
Proposition~\ref{prop:alg-L-uniform-deleted-HHKL}.
Moreover, for $S\subseteq\mathfrak Y$,
\begin{align}
 \sup_{|z_Y|\le\rho\ (Y\in S)}
 \|\widehat\Gamma_r(\boldsymbol z_S,\boldsymbol1_{S^c})\|_\diamond
 \le e^{d_{a,b}}[1+\delta_{\mathrm H}(B)]
     \exp\left[C_\rho g_\star(b-a)
                      \sum_{Y\in S}(|Y|+c_k)\right].
\label{eq:alg-L-driven-polydisc}
\end{align}
\end{lemma}

\textit{Proof.}
The deleted unhatted generator is physical, so its propagators have
diamond norm one.  Variation of constants around this reference and
Gronwall's inequality give
\begin{align}
 \|\widehat T^{(\boldsymbol\mu)}(b,a)\|_\diamond
   &\le e^{d_{a,b}},
 &\|\widehat T^{(\boldsymbol\mu)}(b,a)
             -T^{(\boldsymbol\mu)}(b,a)\|_\diamond
   &\le e^{d_{a,b}}-1.
\label{eq:alg-L-driven-reference-stability}
\end{align}
Deleting terms only reduces the perturbation budget.  This is the only
place where the possible loss of the Lindblad form in the polynomial
surrogate affects contractivity.

For localization, repeat the relative-factor telescoping proof of
Proposition~\ref{prop:alg-L-uniform-deleted-HHKL}, replacing every
exponential by its two-parameter propagator.  The exact relative factor is
\begin{align}
 \widehat T_{\mathcal A}(b,a)^{-1}
       \circ\widehat T_{\mathcal A+\mathcal V}(b,a)
 =\mathcal T\exp\left[\int_a^b
       \widehat T_{\mathcal A}(x,a)^{-1}
       \circ\mathcal V^{(x)}\circ
              \widehat T_{\mathcal A}(x,a)\,dx\right].
\label{eq:alg-L-driven-relative-identity}
\end{align}
For one interface, let $\mathcal K^{(j-1)}$ be the previously joined
path, and let $\mathcal A_j$ be its local patch background.  Conjugation
by the endpoint propagator changes the exact and local interaction-picture
generators to
\begin{align}
 \mathcal G_j(x)
  &=\operatorname{Ad}^{\circ}_{
         \widehat T_{\mathcal K^{(j-1)}}(b,x)}(\mathcal V_j^{(x)}),
 \nonumber\\
 \widetilde{\mathcal G}_j(x)
  &=\operatorname{Ad}^{\circ}_{
       \widehat T_{\mathcal K^{(j-1)}}(b,a)
          \circ\widehat T_{\mathcal A_j}(x,a)^{-1}}
                       (\mathcal V_j^{(x)}),
 \qquad
 \operatorname{Ad}^{\circ}_{P}(V):=P\circ V\circ P^{-1}.
\label{eq:alg-L-driven-defect-generators}
\end{align}
A connected composition-commutator word of order smaller than
$m_B=\lfloor(B-c_k)/k\rfloor$ cannot reach a term outside the patch.
In such words $\mathcal K^{(j-1)}$ may be replaced by $\mathcal A_j$
at every integration time.  The two expressions then agree, using the
path identity
$\widehat T_{\mathcal A_j}(b,a)
\widehat T_{\mathcal A_j}(x,a)^{-1}
=\widehat T_{\mathcal A_j}(b,x)$.
Thus the low-order cancellation does not use time-translation invariance.
The absolute mixed-word majorant is the one in
\eqref{eq:alg-L-mixed-composition-commutator-bound}, with total local
action $O(g_\star\Delta)$.  Ordered integrations cancel the factorials
as before and give an interface defect $O(e^{-\mu_{\mathrm H}B})$.
The relative-error recursion
\eqref{eq:alg-L-relative-error-recursion} is algebraic and unchanged.
At its last step multiply by the reference bound $e^{d_{a,b}}$ from
\eqref{eq:alg-L-driven-reference-stability}.  This proves
\eqref{eq:alg-L-driven-HHKL} without ever estimating a whole-cell inverse
in isolation.

For completeness, the adjoint estimate of
Lemma~\ref{lem:alg-L-ordered-product-adjoint} extends to ordered local
paths with cumulative action
$\sup_i\int\sum_{Z\ni i}\|\mathcal A_Z(x)\|_\diamond\,dx$.
Approximate a piecewise-polynomial path by step functions, apply the
connected-word majorant of that lemma, and pass to the propagator limit.
The bound depends on this integral, not on the number of steps.
For a forward elementary factor its parameter logarithmic derivative is
\begin{align}
 (\partial_\theta\widehat T_{f,\theta}(b,a))
             \circ\widehat T_{f,\theta}(b,a)^{-1}
 =\int_a^b\operatorname{Ad}^{\circ}_{
            \widehat T_{f,\theta}(b,x)}
               (\partial_\theta\mathcal L_{f,\theta}^{(x)})\,dx.
\label{eq:alg-L-driven-parameter-derivative}
\end{align}
Use the reversed path for an inverse factor.  In the proof of
Proposition~\ref{prop:alg-L-shared-cocycle}, the adjoint products now have
cumulative local action $O(\rho g_\star\Delta)$, because the same
constant occurrence multiplicity holds.  Their active local terms sum to
$O(\rho g_\star\sum_{Y\in S}(|Y|+c_k))$.
The parameter-space Gronwall argument is therefore unchanged; its deleted
base factor has norm at most
$e^{d_{a,b}}[1+\delta_{\mathrm H}(B)]$ by
\eqref{eq:alg-L-driven-HHKL}.  This proves
\eqref{eq:alg-L-driven-polydisc}.
\hfill\(\square\)

The uniform physical time grid can be chosen with
$R=\lceil t/\Delta\rceil=O(1+g_{\mathrm L}t)$, independently of $N$.
Let $J$ be the number of pieces in its common refinement with the input
partition, so $J\le N+R$.  We do not create $J$ macro-layers: this
refinement is used only in the finite integrations within local cells.
Set
\begin{align}
 \widehat F_r(\boldsymbol\lambda)
   :=\widehat\Gamma_r(\boldsymbol\lambda)\circ\cdots
                   \circ\widehat\Gamma_1(\boldsymbol\lambda),
 \qquad
 A_R^{\mathrm{td}}
   :=e^{d_{0,t}}[1+\delta_{\mathrm H}(B)]^R.
\label{eq:alg-L-driven-product}
\end{align}
Multiplication of \eqref{eq:alg-L-driven-polydisc} and the same Cauchy
and inclusion--exclusion argument as in
Proposition~\ref{prop:alg-L-coordinatewise-tail} give
\begin{align}
 \left\|\widehat F_R(\boldsymbol1)
    -\left.\mathcal T_K^{\mathrm{sh}}\widehat F_R
                         \right|_{\boldsymbol1}\right\|_\diamond
 \le A_R^{\mathrm{td}}
   \left\{\prod_{Y\in\mathfrak Y}
       [1+e^{C_\rho g_\star t(|Y|+c_k)}q_K(\rho)]-1\right\}.
\label{eq:alg-L-driven-coordinatewise-tail}
\end{align}
Consequently $K=O(g_\star t\ell+\log(|\mathfrak Y|/\epsilon))$
remains sufficient when $A_R^{\mathrm{td}}=O(1)$.  Importantly, neither
$K$ nor the locality width must be multiplied by $N$.

\paragraph{Exact time profiles and degree-labelled construction.}
For every elementary local propagator use the temporal recurrence
\eqref{eq:alg-U-polynomial-recursion} with operator multiplication replaced
by composition and $-iG_{j,v,d}$ replaced by the corresponding
superoperator coefficient.  Explicitly, on a normalized temporal piece,
\begin{align}
 \mathcal E_{\boldsymbol m,v,q+1}
    =\frac{\Delta_v}{q+1}
       \sum_{Y:m_Y\ge1}\sum_{d=0}^{\min\{\nu,q\}}
        \widehat{\mathcal L}_{f,Y,v,d}
               \circ\mathcal E_{\boldsymbol m-\boldsymbol e_Y,v,q-d}.
\label{eq:alg-L-driven-local-recursion}
\end{align}
The zero-time coefficients are the endpoint values from the preceding
piece.  For a reverse factor the polynomial coefficients in this equation
belong to the reversed negative generator.  The endpoint-time degree is
at most $(\nu+1)|\boldsymbol m|_1$.
This recurrence computes the finite scalar time algebra exactly; it does
not discretize every integration variable on a mesh of accuracy
$\epsilon$.

The replacement of
Lemma~\ref{lem:alg-L-polynomial-profile-rank} is particularly simple.
For a fixed selected cut, keep the same core of
$h=O(B/\ell+1)$ variables.  Across all $r\le R$ physical time slices,
the number of elementary temporal pieces in cells meeting this cut is at
most $C J$.  Allocate the $s$ explicit insertions among those pieces.
After a fixed allocation, every segment coefficient is a polynomial in
its variable endpoints.  The proof of the cited lemma, with its aggregate
core-degree allocations unchanged, gives
\begin{align}
 &\operatorname{SR}^{\mathrm{Liouv}}_p
    \left([\boldsymbol\lambda^{\boldsymbol m}]
           \mathcal D_{p,\boldsymbol z}^{(s)}
           (\boldsymbol\lambda)\right)
 \nonumber\\
 &\qquad\le (K+1)^{C h+C}
      \binom{C J+s-1}{s}R_\times^s
      \binom{(\nu+1)(hK+s)+s}{s}.
\label{eq:alg-L-driven-profile-rank}
\end{align}
Indeed, after choosing the temporal pieces the left coefficient has total
crossing-time degree at most $(\nu+1)(hK+s)$, including scalar crossing
entries.  Expanding only these scalar time monomials and integrating the
right coefficient gives the last binomial.  All noncommuting operator
words of the same temporal profile are already summed.  Variables outside
the core occur only in fixed one-sided cells, so their finite degree sums,
or their local classical degree legs, do not increase this time degree.
Thus the partial-sector and prefix assertions of the original lemma
continue to hold.  The number of temporal pieces enters only through
insertions in cells meeting the cut, not through all cells in the chain.

Every retained shared monomial still has cut weight $O(K)$ in a window
of $\ell$ microscopic cuts.  Apply the zero-remainder coefficientwise
cut-count identity, followed by \eqref{eq:alg-L-driven-profile-rank},
and keep the local degree registers as in
Proposition~\ref{prop:alg-L-degree-DP}.  This gives, for some fixed
constant $C$,
\begin{align}
 \log D_{\mathrm{alg}}\le C\mathcal B_{\mathrm L},\qquad
 \mathcal B_{\mathrm L}:={}&\ell+
      [B/\ell+\log(\ell+1)+1]\log(K+2)
 \nonumber\\
 &+\left(\frac{K}{\ell}+1\right)
      \log[2+J(\nu+1)(K+1)(B+1)].
\label{eq:alg-L-driven-rank-budget}
\end{align}
This bound covers completed time slices, macro-layer prefixes, and partial
local cells.  It also covers the temporal coefficients used while building
one elementary cell, up to a polynomial factor: interpolate its
endpoint-time polynomial of degree at most
$(\nu+1)K|\mathfrak Y_f|$ from its values, exactly as in
\eqref{eq:alg-U-stored-time-rank}.  That polynomial factor is absorbed by
\eqref{eq:alg-L-driven-rank-budget}.

The shared lower-set recursion now has a different map at each physical
time step and its order must be preserved:
\begin{align}
 \widehat F_{0,K}&=\operatorname{id},
 &\widehat G_{r,K}&:=\mathcal T_K^{\mathrm{sh}}\widehat\Gamma_r,
 \nonumber\\
 \widehat F_{r,K}
    &=\mathcal T_K^{\mathrm{sh}}
          [\widehat G_{r,K}\circ\widehat F_{r-1,K}],
 &\widehat F_{r,K}[\boldsymbol m]
    &=\sum_{\boldsymbol u+\boldsymbol v=\boldsymbol m}
         \widehat G_{r,K}[\boldsymbol v]
              \circ\widehat F_{r-1,K}[\boldsymbol u].
\label{eq:alg-L-driven-lower-set-recursion}
\end{align}
All indices in this sum obey the coordinatewise cutoff.  Nonnegative
degrees prove the identity exactly.  The local adders and degree markers
of \eqref{eq:alg-L-degree-resolved-tensor} and the subsequent construction
are unchanged.  The temporal pieces within one elementary occurrence are
processed first; only its endpoint coefficient table is merged into its
cell.  Disjoint cells are tensorized, the two macro-layers are composed,
and then the shared degree registers are merged across physical time
steps.  No time degree is confused with, or substituted for, a physical
subblock occupancy.

\begin{theorem}[Algorithmic Liouville-MPO for structured driving]
\label{thm:alg-L-driven-global}
Assume \eqref{eq:alg-L-driven-surrogate} with explicitly computable
local polynomial data and total preprocessing cost $T_{\mathrm{in}}$.
Let $\mathsf L=\mathsf L_{n,t,\epsilon}$ as in
\eqref{eq:alg-L-global-log-scale}, choose
\begin{align}
 B=\Theta(\mathsf L),\qquad
 \ell=\Theta(\sqrt{\mathsf L}),\qquad
 K=O(g_\star t\ell+\mathsf L),\qquad
 R=O(1+g_{\mathrm L}t),\qquad J\le N+R,
\label{eq:alg-L-driven-parameters}
\end{align}
with sufficiently large constants in the choices of $B$ and $K$, and take
the input tolerance in \eqref{eq:alg-L-driven-perturbation-budget}.
There is an explicitly
constructible Liouville-MPO $\widetilde{\mathcal E}_{t,K}^{\mathrm{td}}$
with
\begin{align}
 \|T_{\mathcal L}(t,0)-\widetilde{\mathcal E}_{t,K}^{\mathrm{td}}
                                      \|_\diamond&\le\epsilon,
 \nonumber\\
 D_{\mathrm{MPO}}&\le\exp(C\mathcal B_{\mathrm L}),
 \nonumber\\
 T_{\mathrm{construct}}
   &\le T_{\mathrm{in}}+
       nJ\operatorname{poly}(B,K,\nu+1,1+g_{\mathrm L}t)
             \exp(C\mathcal B_{\mathrm L}).
\label{eq:alg-L-driven-global-cost}
\end{align}
If $N$ and $\nu$ are polynomial in $1+g_{\mathrm L}t+\mathsf L$,
and the preprocessing cost fits this bound, then
\begin{align}
 \log D_{\mathrm{MPO}}
      &=\widetilde O(g_{\mathrm L}t+\sqrt{\mathsf L}),
 &T_{\mathrm{construct}}
      &\le n\exp[\widetilde O(g_{\mathrm L}t+\sqrt{\mathsf L})].
\label{eq:alg-L-driven-static-scale}
\end{align}
For larger temporal descriptions their overhead remains explicit in
\eqref{eq:alg-L-driven-rank-budget} and
\eqref{eq:alg-L-driven-global-cost}.
\end{theorem}

\textit{Proof.}
Choose the constant in $B$ so that
$R\delta_{\mathrm H}(B)\le\epsilon/64$.
The preceding stability lemma and telescoping the ordered products give
\begin{align}
 \|T_{\mathcal L}(t,0)-\widehat F_R(\boldsymbol1)\|_\diamond
   \le e^{d_{0,t}}[1+\delta_{\mathrm H}(B)]^R-1
   =A_R^{\mathrm{td}}-1<\epsilon/4.
\label{eq:alg-L-driven-global-analytic-error}
\end{align}
In particular $A_R^{\mathrm{td}}=O(1)$.  The cutoff in
\eqref{eq:alg-L-driven-parameters}, with a sufficiently large constant,
therefore makes \eqref{eq:alg-L-driven-coordinatewise-tail} at most
$\epsilon/2$.  Define
$\widetilde{\mathcal E}_{t,K}^{\mathrm{td}}
:=\widehat F_{R,K}(\boldsymbol1)$; these two errors prove the first
claim.  The rank bound was established for the label-valued partial
sectors and prefixes in \eqref{eq:alg-L-driven-rank-budget}.
Thus exact rank reduction after each completed table update or local
degree merge keeps the same bound, up to constant powers for temporary
products and polynomial factors for partial sums.

There are $O(B/\ell)$ active anchors in one local cell, so its temporal
coefficient tables contain at most
$(K+1)^{O(B/\ell)}$ degree sectors, with
$O((\nu+1)K(B/\ell+1))$ time coefficients per sector and temporal piece.
The sum of the numbers of such pieces over all physical time slices is
$J\le N+R$.  With $O(n)$ local tensors per macro-layer, the recurrence
\eqref{eq:alg-L-driven-local-recursion}, constant-depth cell assembly,
and shared local adders give the operation count in
\eqref{eq:alg-L-driven-global-cost}.  Substituting
\eqref{eq:alg-L-driven-parameters} into
\eqref{eq:alg-L-driven-rank-budget} proves
\eqref{eq:alg-L-driven-static-scale}.
\hfill\(\square\)

\paragraph{Examples and scope.}
\emph{Discrete driving.}  A supplied piecewise-constant schedule of
complete local Lindblad terms is included directly, with $\nu=0$ and
$\zeta=0$.

\emph{Analytic continuous driving.}  Analytic local generator paths with
uniform analytic bounds and certified, efficiently computable polynomial
approximants are included as well.  Examples are finite bounded-frequency
modulations whenever the original real-time path is a Lindblad generator
with the required deletion property.  Linear ramps between local
Lindbladians and bounded nonnegative polynomial rates multiplying fixed
local dissipators are exact polynomial inputs, with $\zeta=0$.

Component-wise polynomial approximation need not preserve positivity of
rates or complete positivity of the surrogate; the perturbative reference
bound \eqref{eq:alg-L-driven-reference-stability} is included precisely
to handle this case.  Alternatively, a structure-preserving polynomial
input may be used.

The conclusion concerns a linear-map MPO with diamond-norm error; the
truncated output is not claimed to be CPTP.  As in the static theorem,
this is an exact-arithmetic construction, not a condition-number-independent
finite-precision algorithm.  Efficient integration of one terminal dressed
interaction alone would not prove this theorem: the time representation
must be reusable in the lower-set recursion, and its intermediate
coefficient ranks must be controlled.  Piecewise-polynomial temporal data
provide both properties explicitly.

\section{Numerical details for main-text Fig.~2}
\label{sec:fig2-numerics}

This section describes the model, rank-extraction procedure, fixed
comparison curves, and numerical checks underlying main-text Fig.~2.
The calculation probes the first-order boundary contribution discussed
in the main text, rather than the full time-evolution operator.

\subsection{Model and drive protocol}
\label{sec:fig2-numerics-model}

We consider two identical open spin-$1/2$ chains, $A$ and $B$, each
containing $b=9$ sites, for a total of $18$ spins. Sites in each chain
are indexed starting at the coupling boundary. In units $J=\hbar=1$,
the uncoupled evolution is generated by
\begin{align}
H_0^{(\tau)}&=H_A^{(\tau)}+H_B^{(\tau)}, \notag\\
H_\nu^{(\tau)}
&=\sum_{i=1}^{b-1}Z_{\nu,i}Z_{\nu,i+1}
+0.809\sum_{i=1}^{b}Z_{\nu,i}
+\bigl(0.9045+s(\tau)\bigr)\sum_{i=1}^{b}X_{\nu,i},
\qquad \nu\in\{A,B\},
\label{eq:fig2-numerics-H}
\end{align}
where $X_{\nu,i}$ and $Z_{\nu,i}$ are Pauli operators, and identity
operators on the other chain are implicit in the sum defining
$H_0^{(\tau)}$. 
The parameter choices obey the reference~\cite{PhysRevE.90.052105}. 
The boundary interaction is
\begin{align}
V=Z_{A,1}\otimes Z_{B,1}.
\label{eq:fig2-numerics-V}
\end{align}
The total evolution time is $t=1$.

For $M>0$, the spatially uniform transverse-field modulation is
piecewise constant on $M$ equal intervals:
\begin{align}
s(\tau)=s_k\in\{-1,+1\},
\qquad \frac{k-1}{M}\le\tau<\frac{k}{M},
\qquad k=1,\ldots,M.
\label{eq:fig2-numerics-drive}
\end{align}
The two signs have equal probabilities. The sign list is generated
using NumPy~2.3.5 and the PCG64 generator through
\begin{verbatim}
np.random.default_rng(11).choice([-1,1], size=M)
\end{verbatim}
The generator is reinitialized with seed $11$ for every $M$, so the
lists for different $M$ share prefixes. Both chains receive exactly
the same list. For each $M>0$, the plotted data use one drive
realization, without disorder averaging. Here $M$ counts prescribed
intervals, including adjacent intervals with equal signs, and is not
the number of sign changes. The static reference $M=0$ means
$s(\tau)=0$ throughout; it is not the sample-mean field of a driven
realization. Changing $M$ changes the physical drive history through
both the sign list and the interval width $1/M$, rather than merely
refining the numerical integration.

\subsection{First-order state and Schmidt-rank criterion}
\label{sec:fig2-numerics-state}

Let
\begin{align}
U_{\nu,0\to\tau}
&=\mathcal T\exp\!\left[-i\int_0^\tau H_\nu^{(x)}\,dx\right],
\qquad \nu\in\{A,B\}, \notag\\
U_{A|B,0\to\tau}
&=U_{A,0\to\tau}\otimes U_{B,0\to\tau},
\label{eq:fig2-numerics-U}
\end{align}
and take the initial product state
\begin{align}
|\psi_0\rangle=|a\rangle_A\otimes|a\rangle_B,
\qquad |a\rangle=|+_x\rangle^{\otimes b}.
\end{align}
We evaluate the unnormalized state
\begin{align}
|\chi\rangle
=\int_0^1
U_{A|B,0\to\tau}^{\dagger}V
U_{A|B,0\to\tau}|\psi_0\rangle\,d\tau.
\label{eq:fig2-numerics-chi}
\end{align}
No normalization or additional link-strength or distance prefactor is
applied. Up to the overall factor $-i$, this is the first-order
interaction-picture correction associated with $V$. Its
Schr\"odinger-picture counterpart is
$-iU_{A|B,0\to1}|\chi\rangle$ and therefore has the same Schmidt
coefficients, since the final evolution is a product unitary.
Only this first-order contribution, not the full evolved state, is
analyzed here.

Identifying the single-chain computational bases, define
\begin{align}
v(\tau)=U_{A,0\to\tau}^{\dagger}Z_{A,1}
U_{A,0\to\tau}|a\rangle_A.
\end{align}
The identical drives and product form of $V$ then give
\begin{align}
|\chi\rangle=\int_0^1v(\tau)\otimes v(\tau)\,d\tau,
\qquad
C=\int_0^1v(\tau)v(\tau)^T\,d\tau,
\label{eq:fig2-numerics-C}
\end{align}
where $C$ is the coefficient matrix across $A|B$. The transpose in
this expression is not a Hermitian conjugate. For $b=9$, $C$ is a
$512\times512$ matrix.

Writing its descending Schmidt coefficients as
$\lambda_1\ge\lambda_2\ge\cdots\ge0$, we define
\begin{align}
\epsilon_D
&=\left(\sum_{s>D}\lambda_s^2\right)^{1/2},
\qquad
D_\epsilon
=\min\bigl\{D\in\{0,\ldots,2^b\}:\epsilon_D\le\epsilon\bigr\},
\label{eq:fig2-numerics-rank}\\
L_\epsilon&=\ln(1/\epsilon),
\qquad L_D=\ln(1/\epsilon_D).
\label{eq:fig2-numerics-L}
\end{align}
All logarithms are natural. The tolerance is an absolute state-vector
$2$-norm error, with no rescaling by $\|\chi\|_2$. Thus $D_\epsilon$
in this section refers to compression of $|\chi\rangle$, not to the
operator-norm approximation rank of the full propagator. The maximum
Schmidt rank is $2^b=512$.

\subsection{Plotted data and fixed guide curves}
\label{sec:fig2-numerics-guides}

\paragraph{Required rank and pulse dependence.}
Main-text Fig.~2b plots $\ln D_\epsilon$ against $L_\epsilon$ for
$M=0,32,64,128,256,512$. The solid step curves are obtained directly
from the saved Schmidt spectra. A higher curve means that more
Schmidt terms are required at the same absolute error tolerance.
Main-text Fig.~2c plots $D_\epsilon$ against
$M=0,32,64,128,256,512,1024$ at $L_\epsilon=80,120,150$.
The corresponding ranks are listed in
Table~\ref{tab:fig2-numerics-data}. Connecting lines guide the eye;
they are neither additional computed data nor fits. The static
reference is the $M=0$ point on each curve. The weak dependence at
larger $M$ refers only to the sampled range, and monotonicity in $M$
is not imposed.

\begin{table}[tb]
\centering
\caption{Schmidt-rank targets in main-text Fig.~2c and numerical
quadrature settings for $b=9$. Here $q$ is the number of
Gauss--Legendre nodes per constant-field interval. All listed
spectra use a working precision of $640$ bits; the static case
$M=0$ uses one integration interval.}
\label{tab:fig2-numerics-data}
\begin{tabular}{r r r r r}
\hline\hline
$M$ & $q$ & $D_{\epsilon=e^{-80}}$
& $D_{\epsilon=e^{-120}}$ & $D_{\epsilon=e^{-150}}$ \\
\hline
0    & 128 & 21 & 30  & 36  \\
32   & 40  & 60 & 130 & 191 \\
64   & 32  & 66 & 158 & 256 \\
128  & 24  & 62 & 170 & 295 \\
256  & 24  & 62 & 178 & 339 \\
512  & 24  & 62 & 179 & 356 \\
1024 & 20  & 60 & 176 & 351 \\
\hline\hline
\end{tabular}
\end{table}

\paragraph{Calibration of the guide curves.}
The static dotted guide uses the comparison model
\begin{align}
L_D=a_0+c_0D,
\qquad
a_0=-5.070387184721221,
\quad c_0=4.202899208860632,
\label{eq:fig2-numerics-fit-static}
\end{align}
fitted to the separate $b=7$, $M=0$ spectrum over integer ranks
$D=8,\ldots,24$. The driven dashed guide uses
\begin{align}
L_D=a+c(\ln D)^2,
\qquad
a=4.072793217419958,
\quad c=4.338195687364548,
\label{eq:fig2-numerics-fit-driven}
\end{align}
fitted to the separate $b=6$, $M=64$ spectrum with seed $11$ over
$D=8,\ldots,32$. Both fits minimize the unweighted sum of squared
residuals in $L_D$, not in $\ln D$. Their root-mean-square residuals
in $L_D$ are $0.461876166709412$ and $0.831318286932846$,
respectively.

The plotted inverse functions are
\begin{align}
\ln D=\ln\!\left(\frac{L_\epsilon-a_0}{c_0}\right)
\quad\text{(static)},
\qquad
\ln D=\sqrt{\frac{L_\epsilon-a}{c}}
\quad\text{(driven)}.
\label{eq:fig2-numerics-fit-inverse}
\end{align}
Each guide starts at its predicted $D=8$ value and extends to
$L_\epsilon=150$. The coefficients are held fixed for all displayed
$b=9$ data, without refitting. The selected smaller-system spectra
and finite-rank windows specify this comparison; the choices are not
prescribed by theory and are not claimed to be unique or optimal.
These fitting models correspond to $D_\epsilon\sim L_\epsilon$ and
$D_\epsilon\sim\exp[\mathrm{const.}\sqrt{L_\epsilon}]$,
respectively. They are finite-window guides, not tests establishing
asymptotic scaling.

\paragraph{Cross-size consistency check.}
Over the respective calibration windows, the maximum absolute
difference in $L_D$ between the calibration spectrum and the
corresponding $b=9$ spectrum is approximately $0.282$ for the static
case and $0.0178$ for the driven $M=64$ case. This compares the
spectra themselves, rather than the fitted curves. It is a
consistency check within the stated windows, not a criterion used
to select their endpoints or a claim that finite-size effects are
absent outside those windows.

\subsection{Numerical implementation}
\label{sec:fig2-numerics-implementation}

Each constant-field propagator is evaluated in its energy eigenbasis,
without a Trotter decomposition. The time integral in
Eq.~\eqref{eq:fig2-numerics-C} is evaluated by composite
Gauss--Legendre quadrature with the per-interval orders in
Table~\ref{tab:fig2-numerics-data}. The main spectra use
$640$-bit working precision, approximately $193$ decimal digits,
with python-flint~0.9.0 and mpmath~1.3.0. The recorded
calculation and analysis environment also uses Python~3.12.14,
NumPy~2.3.5, SciPy~1.17.0, and Matplotlib~3.10.8.
The calibration spectra use $(p,q)=(640,128)$ for the static fit
and $(512,24)$ for the driven fit, where $p$ is the working
precision in bits.

\paragraph{Initialization of the Hamiltonian coefficients.}
The coefficients are initialized directly from decimal strings after
setting the working precision. Specifically, the strings
\texttt{'0.9045'}, \texttt{'0.809'}, and \texttt{'1'} are
converted to Arb numbers after setting \texttt{ctx.prec}; no
intermediate double-precision conversion is used for the Hamiltonian
coefficients. The same initialization is used in the full-space and
reflection-sector implementations, and the field parameters are
also stored as strings in the result JSON files. This is
arbitrary-precision numerical evaluation, not exact rational
arithmetic. The stated $640$-bit precision is a working precision
and does not guarantee $193$-digit accuracy in the final spectrum.

\paragraph{Reflection sectors and spectral extraction.}
For $M=32,512,1024$, a reflection-symmetry decomposition is used.
Both the $272$-dimensional even sector and the $240$-dimensional
odd sector are retained, including all cross-sector blocks of the
final $512\times512$ coefficient matrix $C$. No reflection sector
is discarded. The Schmidt coefficients are obtained as the square
roots of the eigenvalues of $CC^\dagger$.

At working precision $p$, computed eigenvalues at or below
$2^{-p+80}$, or with imaginary parts whose magnitudes exceed
$10^{-10}$ times their positive real parts, are treated as unresolved.
Logarithmic Schmidt coefficients are saved as $50$-digit decimal
strings, and their tails are evaluated using logarithmic summation.
The adopted unresolved-tail floor does not change any displayed
target rank. Target ranks reconstructed from the saved spectra
were also checked using $100$-digit Decimal arithmetic. This checks
the tail evaluation and rank extraction from the stored spectra,
rather than independently recomputing those spectra.

\subsection{Error control and validation}
\label{sec:fig2-numerics-validation}

\paragraph{Quadrature error.}
The exact-arithmetic composite-quadrature bound used to choose $q$
is
\begin{align}
\|\delta\chi\|_2
&\le
\left(\frac{4h}{n}\right)^{2q}
\frac{(q!)^4}{(2q+1)\bigl((2q)!\bigr)^3},
\label{eq:fig2-numerics-quadrature}\\
n&=\max(1,M),
\qquad
h=(b-1)+b\left[
\max_{0\le\tau<1}|0.9045+s(\tau)|+0.809
\right].
\notag
\end{align}
Here $\|\delta\chi\|_2$ is the state-vector $2$-norm error.
Evaluating this bound for the settings in
Table~\ref{tab:fig2-numerics-data} gives a quadrature error below
$e^{-206}$ for every $b=9$ case in the table, smaller than the
minimum target tolerance $e^{-150}$. For $M=32$, the natural logarithms
of the bound are $-272.180844$ at $q=40$ and $-343.676734$ at
$q=48$. These are exact-arithmetic quadrature bounds; they do not
include numerical diagonalization or roundoff errors.

\paragraph{Higher-precision check.}
The $M=32$ calculation was repeated at $768$-bit working precision
with $q=48$, compared with the main calculation at $640$ bits and
$q=40$. Writing $C_{p,q}$ for the corresponding coefficient matrix,
the recorded difference is
\begin{align}
\ln\|C_{640,40}-C_{768,48}\|_F=-387.810977.
\label{eq:fig2-numerics-precision-check}
\end{align}
All extracted ranks agree on the grid
$L_\epsilon=1,1.1,\ldots,150$. The $211$ saved values of
$\ln\lambda_s$ satisfying $\ln\lambda_s\ge-160$ agree at the
stored $50$-digit precision. This is a numerical consistency check,
not a rigorous interval certificate for the complete calculation.
In particular, this higher-precision comparison concerns $M=32$;
it should not be confused with the all-case quadrature bound above.

\paragraph{Reproducibility record.}
The complete sign lists, spectra, per-case numerical settings,
quadrature bounds, validation records, and plotting code are
collected in the numerical archive
\texttt{b9\_snapshot\_progress\_10.zip}.

\section{Use of generative artificial intelligence}

Generative-AI tools were used extensively in the preparation of this manuscript and, in several parts, in the technical development of the proofs. Supplementary Sections I–IV and VI–X were based on drafts and derivations originally written by the authors and were subsequently reorganized and extensively rewritten with GPT Pro 5.6. Supplementary Sections XI and XII were drafted with GPT Pro 5.6 from detailed proof roadmaps and algorithmic constructions prepared by the authors, and were subsequently reviewed and revised by the authors. Supplementary Section V, which concerns elementary properties of buffered Schmidt rank, was developed through iterative discussions with GPT Pro 5.6. The initial idea underlying Appendix B was suggested by Claude Fable 5.1 and was subsequently developed, checked, and incorporated by the authors. The extension to specific time-dependent dynamics (presented in Sec.~VII B, Sec. XI A, and Sec. XII A) originated from conceptual ideas of the authors and was developed through iterative interaction with GPT Astra 6, which assisted in constructing detailed derivations and quantitative estimates. 

The numerical study presented in Supplementary Sec.~XIII was based on
the model, physical questions, and quantities of interest specified by
the authors. GPT Astra 6 was used extensively in the subsequent
computational implementation, including writing and debugging the
numerical code, choosing numerical parameters, carrying out convergence
and high-precision consistency checks, and analyzing and presenting the
resulting data.

GPT Pro 5.6 and GPT Astra 6 were also used extensively for algebraic and quantitative cross-checks throughout the work. All mathematical statements, proofs, and scientific conclusions were reviewed by the authors. The authors take full responsibility for the accuracy, integrity, and conclusions of the manuscript, and no scientific conclusion was accepted solely on the basis of an AI-generated output.

\appendix

\section{Lower and upper bounds on the decay of Schmidt coefficients}

\begin{lemma}\label{lemma:Renyi_lower_upper}
Let $(\lambda_s)_{s\ge1}$ be a nonincreasing sequence of positive numbers
satisfying
\begin{align}
    \sum_{s=1}^{\infty}\lambda_s^2=1.
\end{align}
Suppose that, for every $0<\alpha<1/2$,
\begin{align}
    T_\alpha:=\sum_{s=1}^{\infty}\lambda_s^{2\alpha}
    =e^{c_\alpha/\alpha^\nu},
    \qquad c_\alpha\in[c_1,c_2],
\end{align}
where $\nu>0$ and $c_1,c_2$ are constants independent of $\alpha$, with
\begin{align}
    0<c_1\le c_2,
    \qquad c_1\le 2^{-\nu}\log 3.
\end{align}
The last condition can always be enforced by decreasing the lower-bound
constant $c_1$ to $\min\{c_1,2^{-\nu}\log 3\}$.
Then, for every integer $m\ge3$,
\begin{align}
    e^{-\widetilde c_2(\log m)^{1+1/\nu}}
    \le\lambda_m
    \le e^{-\widetilde c_1(\log m)^{1+1/\nu}},
\end{align}
where
\begin{align}
    \widetilde c_1
    :=\frac{1}{4(2c_2)^{1/\nu}},
    \qquad
    \widetilde c_2
    :=\frac{2^{2\nu+1/\nu+1}c_2}{c_1^{1+1/\nu}}.
\end{align}
All logarithms are natural.
\end{lemma}

\subsection{Proof of Lemma~\ref{lemma:Renyi_lower_upper}}

\textit{Upper bound.}
By monotonicity, for every $0<\alpha<1/2$,
\begin{align}
    m\lambda_m^{2\alpha}\le T_\alpha
    \le e^{c_2/\alpha^\nu},
\end{align}
and hence
\begin{align}
    \lambda_m
    \le \exp\left[
    -\frac{\log m}{2\alpha}
    +\frac{c_2}{2\alpha^{\nu+1}}
    \right].
\end{align}
If $\log m>2^{\nu+1}c_2$, choose
\begin{align}
    \alpha=\left(\frac{2c_2}{\log m}\right)^{1/\nu}
    \in(0,1/2).
\end{align}
Then
\begin{align}
    \lambda_m
    \le \exp\left[-\frac{\log m}{4\alpha}\right]
    =\exp\left[-\widetilde c_1(\log m)^{1+1/\nu}\right].
\end{align}
For the remaining case $\log m\le2^{\nu+1}c_2$, normalization and
monotonicity give
\begin{align}
    m\lambda_m^2\le\sum_{s=1}^m\lambda_s^2\le1.
\end{align}
Since $\widetilde c_1(\log m)^{1/\nu}\le1/2$ in this case,
\begin{align}
    \lambda_m\le m^{-1/2}
    \le\exp\left[-\widetilde c_1(\log m)^{1+1/\nu}\right].
\end{align}
Thus the upper bound holds for every integer $m\ge3$.

\textit{Lower bound.}
Fix an integer $m\ge3$ and choose
\begin{align}
    q&:=\frac{2}{c_1\widetilde c_1^\nu}
       =\frac{2^{2\nu+2}c_2}{c_1}\ge1,
    \\
    \alpha&:=\left(\frac{c_1}{2\log m}\right)^{1/\nu}
       =\frac{1}{\widetilde c_1(q\log m)^{1/\nu}}.
\end{align}
The imposed bound on $c_1$ ensures that
\begin{align}
    0<\alpha
    \le 2^{-1-1/\nu}<\frac12.
\end{align}
Moreover,
\begin{align}
    \log(m^q)
    =\left(\frac{1}{\widetilde c_1\alpha}\right)^\nu,
    \qquad
    e^{c_1/\alpha^\nu}=m^2.
    \label{cond_m^q_alpha}
\end{align}
For every integer $s>m^q$, the upper bound already proved gives
\begin{align}
    \lambda_s^{2\alpha}
    &\le
    \exp\left[-2\widetilde c_1\alpha(\log s)^{1+1/\nu}\right]
    \notag\\
    &\le
    s^{-2\widetilde c_1\alpha(q\log m)^{1/\nu}}
    =s^{-2}.
\end{align}
Consequently, without requiring $m^q$ to be an integer,
\begin{align}
    \sum_{s>m^q}\lambda_s^{2\alpha}
    \le\sum_{s=2}^{\infty}s^{-2}
    \le\int_1^\infty x^{-2}\,dx=1.
\end{align}
Since normalization implies $\lambda_s\le1$, splitting the sum at $m$
and $m^q$ yields
\begin{align}
    T_\alpha
    &=\sum_{s<m}\lambda_s^{2\alpha}
      +\sum_{m\le s\le m^q}\lambda_s^{2\alpha}
      +\sum_{s>m^q}\lambda_s^{2\alpha}
    \notag\\
    &\le m+1+m^q\lambda_m^{2\alpha}.
\end{align}
Using $T_\alpha\ge e^{c_1/\alpha^\nu}=m^2$, we obtain
\begin{align}
    \lambda_m^{2\alpha}
    &\ge \frac{m^2-m-1}{m^q}
    \ge \frac{m^2-2m}{m^q}
    \ge m^{1-q}
    \ge m^{-q},
    \label{lower_bound_lambda_m2alpha}
\end{align}
where $m\ge3$ was used. Finally,
\begin{align}
    \lambda_m
    &\ge \exp\left[-\frac{q}{2\alpha}\log m\right]
    \notag\\
    &=\exp\left[
    -\frac{2^{2\nu+1/\nu+1}c_2}{c_1^{1+1/\nu}}
    (\log m)^{1+1/\nu}
    \right]
    =\exp\left[-\widetilde c_2(\log m)^{1+1/\nu}\right].
\end{align}
\hfill\(\square\)

\section{Free fermions: absence of driven amplification of R\'enyi entanglement}
\label{subsec:renyi-free-fermions}

For a pure fermionic Gaussian state, the entanglement across a cut is completely
determined by a matrix of two-point correlations
\cite{Bravyi2005,Peschel2003,Peschel_2009}.  A useful further
reduction is to the singular values of the block connecting the two sides
of the cut.  We first explain this algebraic reduction and then use locality
to bound those singular values.  We work throughout with general bilinear
Hamiltonians, including pairing terms.

\begin{prop}[R\'enyi bound for general fermionic bilinear dynamics]
\label{prop:renyi-free-fermion-bound}
Let $\Lambda\subset\mathbb Z$ be a finite interval with a fixed number
$r_{\mathrm f}$ of complex fermion modes per site.  Consider
\begin{align}
 H(\tau)
 &=\sum_{x,y\in\Lambda}\sum_{a,b=1}^{r_{\mathrm f}}
 h_{xa,yb}(\tau)c_{x,a}^\dagger c_{y,b}
 \notag\\
 &\quad+\frac12\sum_{x,y\in\Lambda}\sum_{a,b=1}^{r_{\mathrm f}}
 \left[
 \Delta_{xa,yb}(\tau)c_{x,a}^\dagger c_{y,b}^\dagger
 +\mathrm{h.c.}
 \right],
 \qquad h^\dagger=h,\quad \Delta^T=-\Delta.
 \label{eq:free-fermion-general-bilinear}
\end{align}
Introduce the Majorana operators
\begin{align}
 \gamma_{x,2a-1}&:=c_{x,a}+c_{x,a}^\dagger,
 &
 \gamma_{x,2a}&:=i(c_{x,a}^\dagger-c_{x,a}),
\end{align}
and let $\gamma_x$ collect the $2r_{\mathrm f}$ operators at site $x$.
Up to an additive scalar, the Hamiltonian is
\begin{align}
 H(\tau)=\frac{i}{4}\sum_{x,y\in\Lambda}
 \gamma_x^T K_{xy}(\tau)\gamma_y,
 \qquad K(\tau)\ \text{real},\quad K(\tau)^T=-K(\tau).
 \label{eq:free-fermion-majorana-generator}
\end{align}
Assume that $K(\tau)$ is measurable in $\tau$ and, for some $\kappa>0$,
satisfies the uniform bound (short-range hopping/pairing)
\begin{align}
 g_\kappa:=
 \operatorname*{ess\,sup}_{\tau\ge0}\sup_{x\in\Lambda}
 \sum_{y\in\Lambda}e^{\kappa|x-y|}\|K_{xy}(\tau)\|<\infty.
 \label{eq:free-fermion-weighted-locality}
\end{align}
Here $\kappa$ and an upper bound on $g_\kappa$ are independent of the chain
length.  This condition includes finite-range and uniformly exponentially
decaying hopping and pairing, with $\kappa$ smaller than the decay rate
in the latter case.

Start from a product of on-site pure fermionic Gaussian states, including
any occupation-number product state, and consider the cut
$A=\Lambda\cap(-\infty,0]$, $B=\Lambda\cap[1,\infty)$.
There is a constant $C$, depending only on $r_{\mathrm f}$, $\kappa$, and
the local-strength bound, such that
\begin{align}
 E_\alpha(\psi_t)\le C\left(t+\frac1\alpha\right),
 \qquad t\ge0,\quad 0<\alpha<1.
 \label{eq:free-fermion-renyi-bound}
\end{align}
The bound is uniform in the chain length and in the time dependence of the
coefficients.  No logarithmic factor in $1/\alpha$ is needed.
\end{prop}

Pairing is important for the choice of correlation matrix.  The normal
matrix $C_{ij}:=\langle c_j^\dagger c_i\rangle$ is a projection for a Slater
determinant, but need not be one for a pure Gaussian state with pairing.
For example, the two-mode state
$\cos\theta\,|00\rangle+\sin\theta\,|11\rangle$, with $0<\theta<\pi/2$,
has $C=\sin^2\theta\,I$ and $C_{AB}=0$, although the two modes are
entangled.  The missing information is in the anomalous correlations.
The Majorana covariance matrix used below includes both normal and
anomalous correlations.

\textit{Proof.}

\textit{Step 1: a purity identity that includes pairing.}
Define the real antisymmetric covariance matrix
\begin{align}
 \Gamma_{ij}(t):=\frac{i}{2}
 \langle\psi_t|[\gamma_i,\gamma_j]|\psi_t\rangle,
 \label{eq:free-fermion-majorana-covariance}
\end{align}
where $i,j$ combine the site and Majorana indices.
The initial product structure means
$\Gamma(0)=\bigoplus_x\Gamma_x(0)$.
For an occupation-number product state, each complex mode contributes
\begin{align}
 \begin{pmatrix}
 0&2n_{x,a}-1\\
 1-2n_{x,a}&0
 \end{pmatrix},
 \qquad n_{x,a}\in\{0,1\},
\end{align}
whose square is $-I_2$.  An on-site pure Gaussian state has the same
property after an on-site canonical change of modes.  Thus
$\Gamma(0)^2=-I$.

The Heisenberg evolution is linear:
\begin{align}
 \gamma(t)=O(t)\gamma(0),\qquad
 \partial_tO(t)=K(t)O(t),\qquad O(0)=I.
\end{align}
Here $O(t)$ is absolutely continuous, and the differential equation holds
for almost every $t$.  Since $K^T=-K$, the propagator $O(t)$ is orthogonal.
Consequently,
\begin{align}
 \Gamma(t)=O(t)\Gamma(0)O(t)^T,\qquad
 \Gamma(t)^2=-I.
 \label{eq:free-fermion-gaussian-purity}
\end{align}
Quadratic evolution preserves Gaussianity, so this covariance continues
to determine the state~\cite{Bravyi2005}.

\textit{Step 2: why entanglement reduces to the singular values of
the cross block.}
At the chosen cut, write
\begin{align}
 \Gamma=
 \begin{pmatrix}
 \Gamma_A&\Gamma_{AB}\\
 -\Gamma_{AB}^T&\Gamma_B
 \end{pmatrix}.
\end{align}
Taking the $AA$ block of Eq.~\eqref{eq:free-fermion-gaussian-purity} gives
\begin{align}
 \Gamma_A^2-\Gamma_{AB}\Gamma_{AB}^T=-I_A,
 \qquad
 \boxed{I_A+\Gamma_A^2=\Gamma_{AB}\Gamma_{AB}^T.}
 \label{eq:free-fermion-cross-block-purity}
\end{align}
The left-hand side measures the failure of the reduced covariance
$\Gamma_A$ to describe a pure state.  The identity says that this failure
is exactly determined by correlations with $B$.
In the number-conserving Slater case, the familiar version of the same
argument is $C_A-C_A^2=C_{AB}C_{BA}$, obtained from $C^2=C$.

To turn this matrix identity into entanglement probabilities, choose a real
orthogonal matrix $V_A$ that puts $\Gamma_A$ into its canonical form:
\begin{align}
 V_A\Gamma_A V_A^T
 =\bigoplus_{j=1}^{N_A}
 \begin{pmatrix}0&\nu_j\\-\nu_j&0\end{pmatrix},
 \qquad
 N_A=r_{\mathrm f}|A|,\quad 0\le\nu_j\le1.
 \label{eq:free-fermion-reduced-canonical-form}
\end{align}
Signs can be absorbed into a relabelling of the two occupations of each
mode.  The reduced state is Gaussian, and its covariance has no
correlations between distinct canonical modes.  Wick's rule therefore
makes the reduced density matrix a product of one-mode density matrices.
Up to a choice of basis and the occupation relabelling, it is
\begin{align}
 \rho_A\simeq\bigotimes_{j=1}^{N_A}
 \begin{pmatrix}1-q_j&0\\0&q_j\end{pmatrix},
 \qquad
 q_j=\frac{1-\nu_j}{2}\in[0,1/2].
 \label{eq:free-fermion-mode-factorization}
\end{align}
In particular, $q_j$ is the smaller probability of the $j$th
entanglement mode~\cite{Peschel2003,Peschel_2009}.

In this canonical basis, the eigenvalues of $I_A+\Gamma_A^2$ are
$1-\nu_j^2$, each repeated twice.  By
Eq.~\eqref{eq:free-fermion-cross-block-purity}, these are the squared
singular values of $\Gamma_{AB}$.  Order the physical modes so that
$s_1\ge s_2\ge\cdots$, where
\begin{align}
 \sigma_{2j-1}(\Gamma_{AB})
 =\sigma_{2j}(\Gamma_{AB})
 =:s_j=\sqrt{1-\nu_j^2}.
 \label{eq:free-fermion-majorana-double-counting}
\end{align}
Singular values are padded with zeros beyond the smaller matrix dimension.
The two equal singular values belong to the two Majorana components of
one complex fermion mode; they are counted only once in the entropy.
It follows that
\begin{align}
 q_j
 =\frac{1-\sqrt{1-s_j^2}}{2}
 =\frac{s_j^2}{2(1+\sqrt{1-s_j^2})},
 \qquad
 \boxed{\frac{s_j^2}{4}\le q_j\le\frac{s_j^2}{2}.}
 \label{eq:free-fermion-singular-value-to-probability}
\end{align}
Thus a small cross-block singular value produces a nearly pure
entanglement mode.  At the endpoints, $s_j=0$ gives a pure mode and
$s_j=1$ gives a maximally mixed mode.
In the paired two-mode example above, both Majorana singular values are
$|\sin2\theta|$, and the formula recovers
$q=\min\{\sin^2\theta,\cos^2\theta\}$ even though $C_{AB}=0$.

For clarity, the complete reduction is
\begin{align*}
 h_\alpha(q)
 :=\frac{\log\!\left[q^\alpha+(1-q)^\alpha\right]}{1-\alpha},
\end{align*}
\begin{align}
 \boxed{
 E_\alpha(\psi_t)
 =\sum_{j=1}^{N_A}
 h_\alpha\!\left(\frac{1-\sqrt{1-s_j^2}}{2}\right).
 }
 \label{eq:free-fermion-mode-renyi}
\end{align}
This is the point at which the many-body entropy problem becomes a
singular-value problem for the off-diagonal covariance block.
Gaussianity is essential: for a general non-Gaussian state, two-point
correlations do not determine the reduced density matrix.

\textit{Step 3: locality concentrates the cross correlations near the cut.}
The locality assumption directly gives a one-particle Lieb--Robinson
estimate.  Indeed, the triangle inequality for path lengths and
Eq.~\eqref{eq:free-fermion-weighted-locality} imply, for almost every
choice of $\tau_1,\ldots,\tau_n$,
\begin{align}
 \left\|\left[K(\tau_n)\cdots K(\tau_1)\right]_{xy}\right\|
 \le e^{-\kappa|x-y|}g_\kappa^n.
\end{align}
Summing the ordered Dyson series, whose $n$th integration region has
volume $t^n/n!$, yields
\begin{align}
 \|O_{xy}(t)\|\le e^{g_\kappa t-\kappa|x-y|}.
 \label{eq:free-fermion-weighted-propagation}
\end{align}
Only the uniform local bound enters; the drive can otherwise vary
arbitrarily in time.

Because the initial covariance is site-diagonal,
\begin{align}
 \Gamma_{xy}(t)
 =\sum_{z\in\Lambda}
 O_{xz}(t)\Gamma_z(0)O_{yz}(t)^T.
\end{align}
A correlation between $x$ and $y$ therefore requires propagation from
an initial site $z$ to both endpoints.  Since $\|\Gamma_z(0)\|=1$,
\begin{align}
 \|\Gamma_{xy}(t)\|
 \le e^{2g_\kappa t}
 \sum_{z\in\Lambda}
 e^{-\kappa(|x-z|+|y-z|)}.
\end{align}
Set $\mu:=\kappa/2$ and
$c_\mu:=\sum_{z\in\mathbb Z}e^{-\mu|z|}<\infty$.
Using $|x-z|+|y-z|\ge|x-y|$ in one half of the exponent gives
\begin{align}
 \sum_{z\in\Lambda}e^{-\kappa(|x-z|+|y-z|)}
 \le e^{-\mu|x-y|}
      \sum_{z\in\mathbb Z}e^{-\mu|x-z|}
 =c_\mu e^{-\mu|x-y|}.
\end{align}
Hence
\begin{align}
 \|\Gamma_{xy}(t)\|
 \le c_\mu e^{2g_\kappa t-\mu|x-y|}.
 \label{eq:free-fermion-covariance-locality}
\end{align}
For $x=-a\in A$ and $y=b+1\in B$, with $a,b\ge0$, the separation is
$a+b+1$: the distances from the two sides of the cut add.  Thus the
cross block has appreciable entries only near a boundary region of
width $O(t+1)$, followed by exponential spatial tails.

\textit{Step 4: a boundary strip gives a low-rank approximation.}
Let $P_A^{(m)}$ and $P_B^{(m)}$ retain the at most $m$ sites closest to
the cut on their respective sides, and define
\begin{align}
 M_m:=P_A^{(m)}\Gamma_{AB}P_B^{(m)}.
\end{align}
There are $2r_{\mathrm f}$ Majorana components per site, so
$\operatorname{rank}M_m\le2r_{\mathrm f}m$.
Every omitted entry has either $a\ge m$ or $b\ge m$.
For a $2r_{\mathrm f}\times2r_{\mathrm f}$ block, the squared Frobenius
norm is at most $2r_{\mathrm f}$ times the squared operator norm.
Equation~\eqref{eq:free-fermion-covariance-locality} therefore gives
\begin{align}
 \|\Gamma_{AB}-M_m\|_{\mathrm F}^2
 &\le
 2r_{\mathrm f}c_\mu^2e^{4g_\kappa t}
 \sum_{\substack{a,b\ge0\\a\ge m\ \mathrm{or}\ b\ge m}}
 e^{-2\mu(a+b+1)}
 \notag\\
 &\le
 \frac{4r_{\mathrm f}c_\mu^2e^{-2\mu}}
 {(1-e^{-2\mu})^2}
 e^{4g_\kappa t-2\mu m}.
\end{align}
The two geometric sums converge independently of the lengths of $A$
and $B$.  With
\begin{align}
 D_\mu:=
 \max\left\{1,\frac{2\sqrt{r_{\mathrm f}}c_\mu e^{-\mu}}
 {1-e^{-2\mu}}\right\},
\end{align}
the variational characterization of singular values
\cite{Bhatia1997} yields
\begin{align}
 \sigma_{2r_{\mathrm f}m+1}(\Gamma_{AB})
 &\le\|\Gamma_{AB}-M_m\|
 \le\|\Gamma_{AB}-M_m\|_{\mathrm F}
 \notag\\
 &\le D_\mu e^{2g_\kappa t-\mu m}.
\end{align}
Using the pair counting in
Eq.~\eqref{eq:free-fermion-majorana-double-counting}, this is exactly
\begin{align}
 s_{r_{\mathrm f}m+1}
 \le D_\mu e^{2g_\kappa t-\mu m},
 \qquad m\in\mathbb Z_{\ge0}.
 \label{eq:free-fermion-crossing-singular-values}
\end{align}
In words, retaining a strip of $m$ sites leaves only $r_{\mathrm f}m$
physical entanglement modes uncontrolled.  The bound on the remaining
modes decays exponentially with $m$ beyond a strip of width $O(t+1)$.

\textit{Step 5: summing the entanglement modes.}
Choose
\begin{align}
 m_t:=
 \left\lceil\frac{2g_\kappa t+\log D_\mu}{\mu}\right\rceil.
 \label{eq:free-fermion-boundary-width}
\end{align}
Monotonicity of $s_j$ and
Eq.~\eqref{eq:free-fermion-crossing-singular-values} imply
\begin{align}
 s_{r_{\mathrm f}(m_t+j)+a}&\le e^{-\mu j},
 &
 q_{r_{\mathrm f}(m_t+j)+a}&\le\tfrac12e^{-2\mu j},
 \qquad j\ge0,\quad 1\le a\le r_{\mathrm f}.
\end{align}
Missing modes are padded with $s_j=q_j=0$ and $\nu_j=1$.
Consequently, the nonnegative entanglement energies
\begin{align}
 \varepsilon_j
 :=\log\frac{1-q_j}{q_j}
 =\log\frac{1+\nu_j}{1-\nu_j}
\end{align}
obey
\begin{align}
 \varepsilon_{r_{\mathrm f}(m_t+j)+a}\ge2\mu j.
 \label{eq:free-fermion-linear-entanglement-energies}
\end{align}
Here $\varepsilon_j=\infty$ when $q_j=0$; the inequality follows from
$e^{-\varepsilon_j}=s_j^2/(1+\nu_j)^2\le s_j^2$.

Each of the first $r_{\mathrm f}m_t$ modes contributes at most $\log2$.
This bound already applies to the binary R\'enyi entropy itself and
does not carry a factor $1/(1-\alpha)$.
For the remaining modes and $0<\alpha\le1/2$, use
\begin{align}
 h_\alpha(q)
 &\le\frac{\log(1+q^\alpha)}{1-\alpha}
 \le2q^\alpha,
 \qquad 0\le q\le1/2.
\end{align}
It follows that
\begin{align}
 E_\alpha(\psi_t)
 &\le r_{\mathrm f}m_t\log2
      +2r_{\mathrm f}\sum_{j=0}^{\infty}e^{-2\mu\alpha j}
 \notag\\
 &=r_{\mathrm f}m_t\log2
      +\frac{2r_{\mathrm f}}{1-e^{-2\mu\alpha}}.
\end{align}
For any $0<\alpha<1$, set $\beta=\min\{\alpha,1/2\}$.
R\'enyi monotonicity gives $E_\alpha\le E_\beta$, while $2\beta\ge\alpha$.
Using $(1-e^{-x})^{-1}\le1+x^{-1}$, we obtain
\begin{align}
 E_\alpha(\psi_t)
 \le r_{\mathrm f}m_t\log2
      +2r_{\mathrm f}\left(1+\frac{1}{\mu\alpha}\right).
\end{align}
Since $m_t=O(t+1)$ with constants fixed by the local data, and
$1/\alpha\ge1$, this proves Eq.~\eqref{eq:free-fermion-renyi-bound}.
\hfill$\square$

\begin{remark}[What excludes the $\alpha^{-2}$ enhancement]
The proof controls two quantities separately: only $O(t+1)$ physical modes
can remain strongly entangled, and the smaller probabilities of the
remaining modes decay at least exponentially with their index.  The sum
of their R\'enyi contributions is therefore bounded by a geometric series
of order $1/\alpha$.  Consequently, at fixed time and fixed local data,
an $\Omega(\alpha^{-2})$ enhancement as $\alpha\downarrow0$ is excluded
for the local fermionic Gaussian dynamics and pure Gaussian product
inputs considered here, uniformly in the chain length.  This conclusion
includes hopping, pairing, and arbitrary measurable time dependence
satisfying Eq.~\eqref{eq:free-fermion-weighted-locality}.
\end{remark}


\bibliography{Supplemental_references}